\documentclass[aps, prb, twocolumn, superscriptaddress, floatfix,english,10pt]{revtex4-2}
\usepackage{graphicx}
\usepackage{mathrsfs} 
\usepackage{svg}
\usepackage{amsmath,mathtools,amssymb,amsfonts}
\usepackage{tikz}
\usepackage{comment}
\usepackage{braket}
\usepackage{pgfplots}
\usepackage[version=4]{mhchem}
\pgfplotsset{compat=1.17}
\usepackage{subcaption}
\usetikzlibrary{shapes.geometric}
\usetikzlibrary{shadows,patterns,perspective}
\usetikzlibrary {decorations,decorations.text}
\usetikzlibrary{decorations.pathreplacing}
\usepackage{float}
\usepackage{hyperref}
\usepackage{xcolor}

\definecolor{linkcolor}{RGB}{0, 0, 255}      % Blue color for links
\definecolor{citecolor}{RGB}{0, 128, 0}     % Green color for citations
\definecolor{urlcolor}{RGB}{255, 0, 0}       % Red color for URLs

\hypersetup{
    colorlinks=true,
    linkcolor=linkcolor,
    citecolor=citecolor,
    urlcolor=urlcolor,
    linktoc=all,  % Set links in the table of contents to be colored as well
    pdfborder={0 0 0}  % Removes the box around links
}

\DeclareCaptionJustification{justified}{\leftskip=0pt \rightskip=0pt \parfillskip=0pt plus 1fil}

\begin{document}
\definecolor{dy}{rgb}{0.9,0.9,0.4}
\definecolor{dr}{rgb}{0.95,0.65,0.55}
\definecolor{db}{rgb}{0.5,0.8,0.9}
\definecolor{dg}{rgb}{0.2,0.9,0.6}
\definecolor{BrickRed}{rgb}{0.8,0.3,0.3}
\definecolor{Navy}{rgb}{0.2,0.2,0.6}
\definecolor{DarkGreen}{rgb}{0.1,0.4,0.1}
%\title{Integrable and non-integrable boundary defects in spin-$\frac12$ anisotropic Heisenberg chain}

\title{Edge modes and boundary impurities in the anisotropic Heisenberg spin chain}

\author{Pradip Kattel }
\email{pradip.kattel@rutgers.edu}
\affiliation{Department of Physics, Center for Material Theory, Rutgers University,
Piscataway, New Jersey 08854, United States of America}

\author{Parameshwar R. Pasnoori}
\affiliation{Department of Physics, University of Maryland, College Park, MD 20742, United
States of America}
\affiliation{Laboratory for Physical Sciences, 8050 Greenmead Dr, College Park, MD 20740,
United States of America}
\author{J. H. Pixley}
\affiliation{Department of Physics, Center for Material Theory, Rutgers University,
Piscataway, New Jersey 08854, United States of America}
\affiliation{Center for Computational Quantum Physics, Flatiron Institute, 162 5th Avenue, New York, NY 10010}
\author{Natan Andrei}
\affiliation{Department of Physics, Center for Material Theory, Rutgers University,
Piscataway, New Jersey 08854, United States of America}

\begin{abstract}
We present a comprehensive analysis of boundary phenomena in a spin-$\frac{1}{2}$ anisotropic Heisenberg chain (XXZ-$\frac{1}{2}$) in the gapped antiferromagnetic phase, with a particular focus on the interplay between fractionalized spin-$\frac{1}{4} $ edge modes and a coupled spin-$\frac{1}{2}$ impurity at the edge. Employing a combination of Bethe Ansatz, exact diagonalization, and density matrix renormalization group (DMRG) methods, we explore the intricate phase diagram that emerges when the impurity is coupled either integrably or non-integrably to the chain. For integrable antiferromagnetic impurity couplings, we identify two distinct phases: the Kondo phase, where the impurity is screened by a multiparticle Kondo effect, and the antiferromagnetic bound mode phase, where an exponentially localized bound state screens the impurity in the ground state. When coupled ferromagnetically while maintaining integrability, the impurity behaves as a free spin-$\frac{1}{2}$, leading to either a ferromagnetic bound mode phase, where the impurity remains free in the ground state but may be screened at higher energy excitations or an unscreened (or local moment) phase where impurity remains unscreened in every eigenstate whereas for non-integrable ferromagnetic coupling, the impurity is not free.
In the case of non-integrable antiferromagnetic coupling, a third phase emerges, characterized by mid-gap excitations with two degenerate states below the mass gap on top of the Kondo and antiferromagnetic bound mode phases, further enriching the phase diagram. Our findings highlight the nuanced behavior of boundary impurities in gapped antiferromagnetic systems, offering new insights into Kondo effects and impurity screening in the presence of fractionalized edge modes and bulk antiferromagnetic order.
\end{abstract}
\maketitle

\section{Introduction}
The Kondo effect~\cite{kondo2012physics, hewson1997kondo}, which was discovered experimentally in metals containing dilute magnetic impurities, manifests itself as an increase in the resistivity as the temperature is lowered~\cite{de1934electrical}.  Magnetic impurities interact with the electrons of the metal via spin-exchange, whose effective strength increases at low temperatures, accounting for the observed increase in resistivity~\cite{kondo1964resistance}.  It also accounts for other effects, in particular, the quenching of the impurity spin at low temperatures. At high temperatures, on the other hand, the impurity behaves essentially as a free spin.  In other words, the Kondo effect can be described theoretically in terms of an RG flow from an (asymptotically free) UV fixed point to a non-trivial (strong-coupling) IR fixed point~\cite{anderson2018poor,nozieres1974fermi,nozieres1980kondo, wilson1975renormalization,affleck1995conformal} characterized by the decrease in ``ground state degeneracy" described by the universal $g-$function \cite{PhysRevLett.67.161,friedan2004boundary}.  The crossover from the UV fixed point (around which one has a local magnetic moment) to the IR fixed point (around which there is complete screening of the impurity magnetic moment) has been studied extensively by various analytical~\cite{Metlitski-Kondo} and numerical tools~\cite{wilson1975renormalization,krishna1980renormalizationa,krishna1980renormalizationb} including an exact solution via the Bethe Ansatz technique~\cite{andrei1980diagonalization,wiegmann1981exact}.

The impact of a single impurity on the low-energy physics of a quantum system is not limited to the Fermi sea of non-interacting electrons. There has been an immense effort in understanding the role of novel impurity-mediated features in strongly interacting systems like Luttinger liquids~\cite{schiller1995exact,furusaki1994kondo,frojdh1995kondo,lee1992kondo,KLLegger1998scaling}, superconductors\cite{KScuevas2001kondo,KSmuller1971kondo,KSborkowski1992kondo,KSsteglich2016foundations,Shiba,Yu,Rusinov}, spin liquids~\cite{kim2008kondo,florens2006kondo}, quantum chromodynamics~\cite{QCDhattori2015qcd,QCDsuenaga2020qcd,QCDkimura2019conformal,QCDozaki2016magnetically} and spin chains~\cite{XXZkondoDMRG,laflorencie2008kondo,kattel2023kondo,wang1997exact,frahm1997open,furusaki1998kondo}. Amongst these systems, the spin chains provide a simple platform to understand the effect of impurities in the strongly interacting systems. Since the effect is mediated by spin-exchange interaction, it is natural to expect that Kondo physics would arise in a system consisting of magnetic impurities coupled to an antiferromagnetic spin chain, often called a spin chain Kondo system. The case of gapless Heisenberg spin chain coupled with magnetic impurities has been studied extensively~\cite{andrei1984heisenberg, laflorencie2008kondo,schlottmann, wang1997exact, frahm1997open,kattel2023kondo, furusaki1998kondo,kattel2024kondo,XXZkondoDMRG}. 

Here, we study the effect of magnetic impurities coupled to the edges of a gapped anisotropic Heisenberg spin chain (XXZ chain).  The spin chain Hamiltonian is of the form
\begin{align}
    H&=\sum_{i=1}^{N_b-1}\frac{J}{2}(\sigma_i^x \sigma_{i+1}^x+\sigma_i^y \sigma_{i+1}^y+\Delta \sigma_i^z \sigma_{i+1}^z)\nonumber\\
    &+\frac{J_L}{2}(\sigma_L^x \sigma_{1}^x+\sigma_L^y \sigma_{1}^y+\Delta_L \sigma_L^z \sigma_{1}^z)\nonumber\\
    &+\frac{J_R}{2}(\sigma_{N_b}^x \sigma_{R}^x+\sigma_{N_b}^y \sigma_{R}^y+\Delta_R \sigma_{N_b}^z \sigma_{R}^z),
    \label{ham}
\end{align}
where there are two impurities described by $\vec\sigma_L$ and $\vec\sigma_R$ at the left and right ends of the spin chain, respectively, and the bulk sites are labeled 1 through $N_b$.  It is often easier to parameterize the anisotropy parameter $\Delta$ as $\cosh(\eta)$ so that when the crossing parameter $\eta\in \mathbb{R}$, the model is gapped. We will focus on the case where $J > 0$, meaning that the interaction within the bulk is antiferromagnetic. However, we will allow the impurity coupling $J_{L/R}$ to range from $-\infty$ to $\infty$, accommodating both antiferromagnetic and ferromagnetic scenarios.

The XXZ chain with boundary impurities described by Eq.~\eqref{ham}  is not integrable for arbitrary boundary couplings, 
unlike the XX chain~\cite{kattel2024kondo} or the isotropic XXX chain~\cite{kattel2023kondo} with boundary impurities. Instead, a particular relation between the bulk anisotropy $\Delta$ and the impurity parameters $\Delta_{R, L}$ and $J_{R,L}$ is required~\cite{hou1999integrability,chen1998integrability,hu1998two,Shu_Chen_1998} to maintain integrability as follows
\begin{equation}
   J_{q}=J\frac{\sinh^2(\eta)\cosh(d_q)}{\sinh^2(\eta)-\sinh^2(d_{q})} \quad \mathrm{and} \quad \Delta_{q}=\frac{\cosh(\eta)}{\cosh(d_{q})},
   \label{bulk-boundary}
   \end{equation}
where $J_{q}$ refers to $J_R$ or $J_L$, and there are two independent free parameters $d_L$ and $d_R$ at the boundary that control the strength of the impurity. 
We will use a combination of Bethe Ansatz~\cite{1931_Bethe_ZP_71,baxter2016exactly,sutherland2004beautiful,gaudin2014bethe,slavnov2022algebraic,franchini2017introduction,eckle2019models} and density matrix renormalization group (DMRG)~\cite{schollwock2005density,schollwock2011density} to probe the integrable case and use DMRG and exact diagonalization for the nonintegrable regimes of the model. In the integrable case,  the Bethe Ansatz equations are

\begin{widetext}
\begin{equation}
        \begin{gathered}
\left(\frac{\sin \frac{1}{2}\left(\lambda _j-{i \eta }\right)}{\sin \frac12\left(\lambda _j+{i \eta }\right)}\right)^{2 N_b}
\frac{\cos^2\frac12\left(\lambda _j+{i \eta }\right)}{\cos^2\frac12\left(\lambda _j-{i \eta }\right)}
=\prod_{\upsilon=\pm}\prod_{q=\{L,R\}}\frac{\sin\frac12\left(\lambda_j +\upsilon 2id_q+{i \eta }\right)}{\sin\frac12\left(\lambda_j +\upsilon 2id_q-{i \eta }\right)} \prod_{k=1(\neq j)}^M \frac{\sin\frac12 \left(\lambda_j+\upsilon\lambda_k-2i\eta\right)}{\sin\frac12 \left(\lambda_j+\upsilon\lambda_k+2i\eta\right) }.
\end{gathered}
     \label{BAE11}
\end{equation}
\end{widetext}
The equations describe the system in terms of the $M$ spin momenta $\lambda_j, \; j=1\cdots M$ with $M$ being the number of down spins.
Given a solution for the spin momenta, the energy and spin component of the eigenstate are given by
\begin{align}
E&=\left[(N_b-1) J+\sum_{q=\{R,L \}}\frac{J \sinh ^2 \eta}{\left(\sinh ^2 \eta-\sinh ^2 d_q\right)}\right] \cosh \eta\nonumber\\
&+2 J \sinh \eta \sum_{j=1}^M \frac{\sinh \eta}{\cos  \lambda_j-\cosh \eta}\\
S^z&=\frac{N_b+2}{2}-M
\end{align}
We show that in the integrable case, each impurity exhibits four distinct phases depending on the impurity parameter $d_q$ and the anisotropy. For antiferromagnetic impurity coupling, there are two distinct phases: the Kondo phase, where the impurity is screened by the Kondo effect, and the antiferromagnetic bound mode (ABM) phase, where the impurity is screened by a localized boundary bound mode, which can be removed to form an excitation where the impurity is unscreened at the cost of the dressed energy of the bound mode. Likewise, in the case of impurity ferromagnetic coupling, the model exhibits two boundary phases: the unscreened phase, where the impurity is essentially a free local moment, and the ferromagnetic bound mode (FBM) phase, where the impurity is unscreened in the ground state but a localized bound mode can screen the impurity in the excited states. When the integrability is broken, the antiferromagnetic impurity exhibits more diverse phases. Apart from the Kondo and ABM phases, there exists a unique phase in which midgap states are present in the model. The mid-gap states are present in the case of the boundary magnetic field, where the model remains integrable for arbitrary values of the boundary magnetic field ~\cite{kapustin1996surface,grijalva2019open,pasnoori2023spin}. 

Another important aspect of the gapped XXZ spin chain is the recently discovered existence of fractionalized spin $\pm 1/4$ edge states ~\cite{pasnoori2023spin}. The fractionalized edge spins can be identified with the strong zero modes when they are projected onto the low energy subspace spanned by the ground state and mid-gap states mode described in~\cite{fendley2016strong, fendley2012parafermionic,fendley2016strong,yates2020dynamics,yates2021strong,vasiloiu2019strong}. These fractional charges in XXZ chains were also studied in~\cite{zvyagin2021majorana,zvyagin2022charging,zvyagin2024strong}. More generally, the existence of fractionalized $S/2$ spin excitations was recently established for any generic antiferromagnetic gapped spin-S model with $U(1)$ symmetry and spontaneous or explicitly broken $\mathbb{Z}_2$ symmetry~\cite{kattel2024XXZ-S} where it was also shown that these fractionalized spins are robust to certain kind of disorder.  

Here we show that the fractional excitations survive in the presence of the boundary impurity in both integrable and non-integrable cases. We further show here, using DMRG results, that for the non-integrable case, mid-gap states appear in the spectrum, leading to a phase diagram that is richer than the integrable case. The appearance of midgap states in gapped systems in the presence of impurities is quite a general phenomenon that exists in several systems, including superconductors and super-fluid with impurities~\cite{Yu,Shiba,Rusinov,pasnoori2022rise,MGPhysRevA.83.061604}, two dimensional magnets~\cite{bauer2024scanning}, graphene~\cite{MGlang2014topologically,MGsachs2016midgap}, and other one dimensional systems~\cite{pasnoori2023spin,MGlang2014topologically}. The effect of impurities has been experimentally studied in a few quasi-one-dimensional systems~\cite{aczel2007impurity} where some aspects of Kondo physics, such as impurity susceptibility~\cite{chakhalian2004impurity}, have experimentally been observed.

Before we proceed further, we shall briefly discuss some aspects of the Kondo effect which helps to clarify some of the terminologies that follow in the remainder of the paper.

\section{The Kondo effect in spin chains}\label{Kondo-brief}

Let us briefly review the key features of the standard Kondo effect and examine how they manifest themselves in our model. The main aspect associated with the effect of a single spin-$\frac12$ magnetic impurity in a three-dimensional metal can be captured by perturbed $1+1$D chiral $SU(2)_1$ Wess–Zumino–Witten (WZW) model with a Hamiltonian density of the form
\begin{equation}
    \mathcal{H}=\frac{1}{6\pi} \mathcal J^a(t,x) \mathcal J^a(t,x) + \lambda S^\alpha \mathcal J^a(t,0),
    \label{hamdensK}
\end{equation}

where $\mathcal J^a(t,x)=\psi^{\dagger j \alpha}\frac{\vec\sigma_\alpha^\beta}{2}\psi_{j\beta}$ is the spin current. The charge current $\mathcal J=:\psi^{\dagger j \alpha}\psi_{j \alpha}:$ commutes with the spin current and decouples.  Thus, the Kondo Hamiltonian Eq.~\eqref{hamdensK} only involves the spin sector where the perturbative coupling $\lambda$ associated with the impurity (or defect)  coupling is marginally relevant triggering the boundary renormalization group (RG) flow from the UV theory where the spin-$\frac{1}{2}$ impurity is essentially free to the IR endpoint where the impurity spin is completely quenched. The ground state degeneracy ($g$), a quantity that measures the effective number of degrees of freedom associated with a boundary of a two-dimensional quantum field theory\cite{affleck1995conformal,affleck1991universal}, in the UV is $g=2$ 
as the impurity could be either up or down or equivalently, the boundary entropy is $S_{\mathrm{imp}}=\ln g=\ln 2$ whereas, in the IR, $g=1$ as its spin is screened such that $S_{\mathrm{imp}}=\ln g=0$. The crossover regime between these two fixed points is characterized by a dynamically generated non-perturbative scale $T_K$.

This boundary RG flow from the asymptotically free UV theory with $g=2$ to the strongly coupled IR theory with $g=1$ manifests itself through various physical quantities associated with the impurity, such as the impurity magnetization, which transitions from zero at vanishing magnetic field ($h$) to asymptotically reaching 0.5 as $ h \to \infty $. The impurity susceptibility remains constant at zero temperature but it follows Curie’s law at higher temperatures. Likewise, the resistivity shows an upturn as the temperature drops below $T_K$. These are a few of the effects resulting from this flow.

Before discussing the Kondo effect in our model Eq.~\eqref{ham}, let us briefly look at the isotropic limit $\Delta=1$ where  a single magnetic {($J_R=0$)} impurity interacts with an antiferromagnetic Heisenberg chain with $SU(2)$ symmetry is described by following Hamiltonian
\begin{equation}
    H=J\sum_{j=1}^{N-1}\vec\sigma_j\cdot\vec\sigma_{j+1}+J_\mathrm{imp}\vec\sigma_1\cdot\vec\sigma_L.
    \label{ham1impiso}
\end{equation} 

 Following standard steps: fermionizing the model ~\cite{jordan1993paulische}, taking the continuum limit~\cite{giamarchi2003quantum}, carrying out  non-abelian bosoniation~\cite{witten1984non,cabra2008field} and imposing the boundary conditions $\psi_L(x)=-\psi_R(x)$ for the bulk fermions, the Hamiltonian density  can be written in terms of spin currents as \cite{laflorencie2008kondo,giamarchi2003quantum}
\begin{equation}
   \mathcal H=\frac{1}{6\pi}\left[\vec{\mathcal J}(x)\right]^2  - \frac{\lambda_B}{2 \pi} \vec{\mathcal J}(x) \cdot \vec{\mathcal  J}(-x) + \lambda_K \vec{\mathcal J}(0) \cdot \vec{S}
    \label{bosxxx}
\end{equation}
Here $\lambda_K \propto J_{\mathrm{imp}}$ and $\lambda_B$ is a constant of order $J$. Notice that the bulk coupling constant $\lambda_B$ is marginally irrelevant, whereas the defect coupling constant $\lambda_K$ is marginally relevant. Thus, it is evident that at longer and longer distances and lower and lower energies where the bulk interaction proportional to $\lambda_B$ becomes smaller, the Hamiltonian density Eq.~\eqref{bosxxx} contains the same physics as that of the conventional Kondo Hamiltonian Eq.~\eqref{hamdensK}.

By exactly solving the model given by Eq.~\eqref{ham1impiso}  on the lattice, it was explicitly shown that the model exhibits Kondo physics for $0<J_{\mathrm{imp}}/J<\frac{4}{3}$ where the impurity is screened by the multi-particle Kondo effect \cite{kattel2023kondo,wang1997exact,frahm1997open}. Whereas when $J_{\mathrm{imp}}/J>\frac{4}{3}$, the continuum description given by Eq.~\eqref{bosxxx} breaks down as a high energy bound mode exponentially localized at the boundary appears in the spectrum. This single particle bound mode then screens the impurity in this regime, and it is, therefore, possible to unscreen the impurity by exciting this single bound mode.  Just like in the conventional Kondo physics, the effect arises due to the decrease in the ground state degeneracy from $g=2$ when the impurity is asymptotically free in the UV to $g=1$ when the impurity screened is quenched in the IR. Hence as in the conventional Kondo problem, the physical quantities such as magnetization and susceptibility associated with the impurity are different in the UV and the IR \cite{kattel2023kondo,kattel2024kondo,wang1997exact,frahm1997open,laflorencie2008kondo}.

Let us now briefly discuss the Kondo effect in our model Eq.~\eqref{ham} and stress some differences in the Kondo physics. We consider Eq.~\eqref{ham} in the gapped antiferromagnetic regime $\Delta>1$ such that our bulk is no longer a CFT, unlike the low energy of the isotropic Heisenberg chain and the conventional Kondo problem described by Eq.~\eqref{hamdensK}.  Because our system is gapped, the Kondo physics is different in our system compared to the gapless bulk in the conventional Kondo problem and the Kondo effect in the gapless Heisenberg chain. For instance, in the XXX chain, the gaplessness implies infinite correlation such that the effect of impurity propagates deep into the bulk depending on the ratio of the bulk and boundary couplings \cite{kattel2023kondo}. When the model is gapless ($\Delta\leq 1$), for vanishing boundary couplings ($J_{L/R}\to 0$), the Kondo temperature approaches zero~\cite{kattel2024kondo,kattel2023kondo} and the Kondo length diverges just like in the conventional Kondo problem~\cite{andrei1992integrable,wiegmann1983exact}. However, the situation is completely different in the current case primarily due to the existence of a gap in the bulk, which makes the correlation not only finite but also exponentially decaying~\cite{giamarchi2003quantum}. Thus, in the Kondo phase in the present model, the impurity is screened by gapped spinons, but its effect does not penetrate deep into the bulk as in the conventional Kondo problem or the gapless spin chain Kondo problems.

More prominent differences arise due to the emergence of the antiferromagnetic order in the bulk (see below) such that the local expectation values of the $z-$component of the spin $S^z_j$ at each site $j$ are nonzero, unlike in the isotropic case~\cite{kattel2023kondo,laflorencie2008kondo,frahm1997open} or the conventional Kondo problem~\cite{andrei1983solution,wiegmann1981exact}. 
Moreover, unlike in both the conventional Kondo problem and the isotropic Heisenberg chain, the ground state of the bulk in the case of the gapped XXZ chain is two-fold degenerate in the thermodynamics limit. When a single spin-$\frac{1}{2}$ impurity with an initial entropy of $\ln 2$ is coupled separately to each of the two vacua, the system evolves to a unique ground state characterized by a boundary entropy $S_{\mathrm{imp}} = 0$. For each of the vacua separately, this process reduces the ground-state degeneracy from $g=2$ to $g=1$, analogous to what occurs in the conventional Kondo problem and in the isotropic case where there is only a single vacuum before adding the impurity.

These fundamental distinctions give rise to unique behaviors that we will explore in detail in the subsequent sections. We will begin by examining the bulk physics in Section \ref{noimp} before introducing the impurity. Then, in Section \ref{1intimp-det}, we will discuss the four distinct phases exhibited by the impurity under integrable coupling. Finally, Section \ref{sec:nonintimp} will examine the effects of a single impurity attached with non-integrable antiferromagnetic and ferromagnetic coupling, revealing three impurity phases. A summary of key results for the integrable case is provided in Table \ref{tab:my-table1}. 
\begin{table}
\caption{In the gapped antiferromagnetic phase, the XXZ chain's bulk is two-fold degenerate. When a single free spin with $\ln 2$ entropy is integrably coupled to the XXZ chain via antiferromagnetic exchange coupling, the impurity exhibits either the Kondo or antiferromagnetic bound mode (ABM) phase. In the Kondo phase, the impurity is screened by a multiparticle cloud, whereas in the ABM phase, it is screened by a single particle bound mode exponentially localized at the edge. In both phases, the impurity's entropy reduces to $0$, indicating that the impurity is no longer free. On the other hand, when the impurity is coupled via ferromagnetic exchange coupling, it transitions to either the ferromagnetic bound mode (FBM) or unscreened phase, where the impurity's entropy remains $\ln 2$, showing that the impurity remains essentially free. The entropy reduction in the Kondo phase results from the many-body Kondo effect, while in the ABM phase, it is due to single-particle screening. In the FBM and unscreened (US) phases, the impurity remains unscreened, akin to a Kondo impurity coupled ferromagnetically to a metallic bath, where the coupling is irrelevant, leading to a trivial fixed point in the IR. Here $N$ denotes the total number of sites (including the bulk and impurity).
}
\label{tab:my-table1}
\begin{tabular}{|l|l|l|l|l|l|}
\hline
\centering
& Bulk & Kondo & ABM & FBM & US \\ \hline
\begin{tabular}[c]{@{}l@{}}Edge modes\\ (Even N)\end{tabular} & \begin{tabular}[c]{@{}l@{}}$+\frac{1}{4}$, $-\frac{1}{4}$\\ $-\frac{1}{4}$, $+\frac{1}{4}$\end{tabular} & \begin{tabular}[c]{@{}l@{}}$+\frac{1}{4}$, $-\frac{1}{4}$\\ $-\frac{1}{4}$, $+\frac{1}{4}$\end{tabular} & \begin{tabular}[c]{@{}l@{}}$+\frac{1}{4}$, $-\frac{1}{4}$\\ $-\frac{1}{4}$, $+\frac{1}{4}$\end{tabular} & \begin{tabular}[c]{@{}l@{}}$+\frac{1}{4}$, $-\frac{1}{4}$\\ $-\frac{1}{4}$, $+\frac{1}{4}$\\ $-\frac{3}{4}$, $-\frac{1}{4}$\\ $+\frac{3}{4}$, $+\frac{1}{4}$\end{tabular} & \begin{tabular}[c]{@{}l@{}}$+\frac{1}{4}$, $-\frac{1}{4}$\\ $-\frac{1}{4}$, $+\frac{1}{4}$\\ $-\frac{3}{4}$, $-\frac{1}{4}$\\ $+\frac{3}{4}$, $+\frac{1}{4}$\end{tabular} \\ \hline
\begin{tabular}[c]{@{}l@{}}Edge modes\\ (Odd N)\end{tabular} & \begin{tabular}[c]{@{}l@{}}$+\frac{1}{4}$, $+\frac{1}{4}$\\ $-\frac{1}{4}$, $-\frac{1}{4}$\end{tabular} & \begin{tabular}[c]{@{}l@{}}$+\frac{1}{4}$, $+\frac{1}{4}$\\ $-\frac{1}{4}$, $-\frac{1}{4}$\end{tabular} & \begin{tabular}[c]{@{}l@{}}$+\frac{1}{4}$, $+\frac{1}{4}$\\ $-\frac{1}{4}$, $-\frac{1}{4}$\end{tabular} & \begin{tabular}[c]{@{}l@{}}$+\frac{1}{4}$, $+\frac{1}{4}$\\ $-\frac{1}{4}$, $-\frac{1}{4}$\\ $-\frac{3}{4}$, $+\frac{1}{4}$\\ $+\frac{3}{4}$, $-\frac{1}{4}$\end{tabular} & \begin{tabular}[c]{@{}l@{}}$+\frac{1}{4}$, $+\frac{1}{4}$\\ $-\frac{1}{4}$, $-\frac{1}{4}$\\ $-\frac{3}{4}$, $+\frac{1}{4}$\\ $+\frac{3}{4}$, $-\frac{1}{4}$\end{tabular} \\ \hline
& Bulk + imp & Kondo & ABM & FBM & US \\ \hline
\begin{tabular}[c]{@{}l@{}}GS\\ Degeneracy\end{tabular}                              & 2*2                                                                     & 2                                                                       & 2                                                                     & 4                                                                                   & 4                                                                                   \\ \hline
\begin{tabular}[c]{@{}l@{}}Imp. entropy\\ in each vacua\\ (UV $\to$ IR)\end{tabular} & \begin{tabular}[c]{@{}l@{}}$\ln 2$\\ (before\\ coupling)\end{tabular} & \begin{tabular}[c]{@{}l@{}}$\ln 2 \to 0$\\ (Kondo\\ effect)\end{tabular} & \begin{tabular}[c]{@{}l@{}}$\ln 2 \to 0$\\ (Bound\\ mode)\end{tabular} & \begin{tabular}[c]{@{}l@{}}$\ln 2 \to \ln 2$\\ (free\\  impurity)\end{tabular} & \begin{tabular}[c]{@{}l@{}}$\ln 2 \to \ln 2$\\ (free\\  impurity)\end{tabular} \\ \hline
\end{tabular}
\end{table}

\begin{figure}[H]
    \centering
    \includegraphics[width=\linewidth]{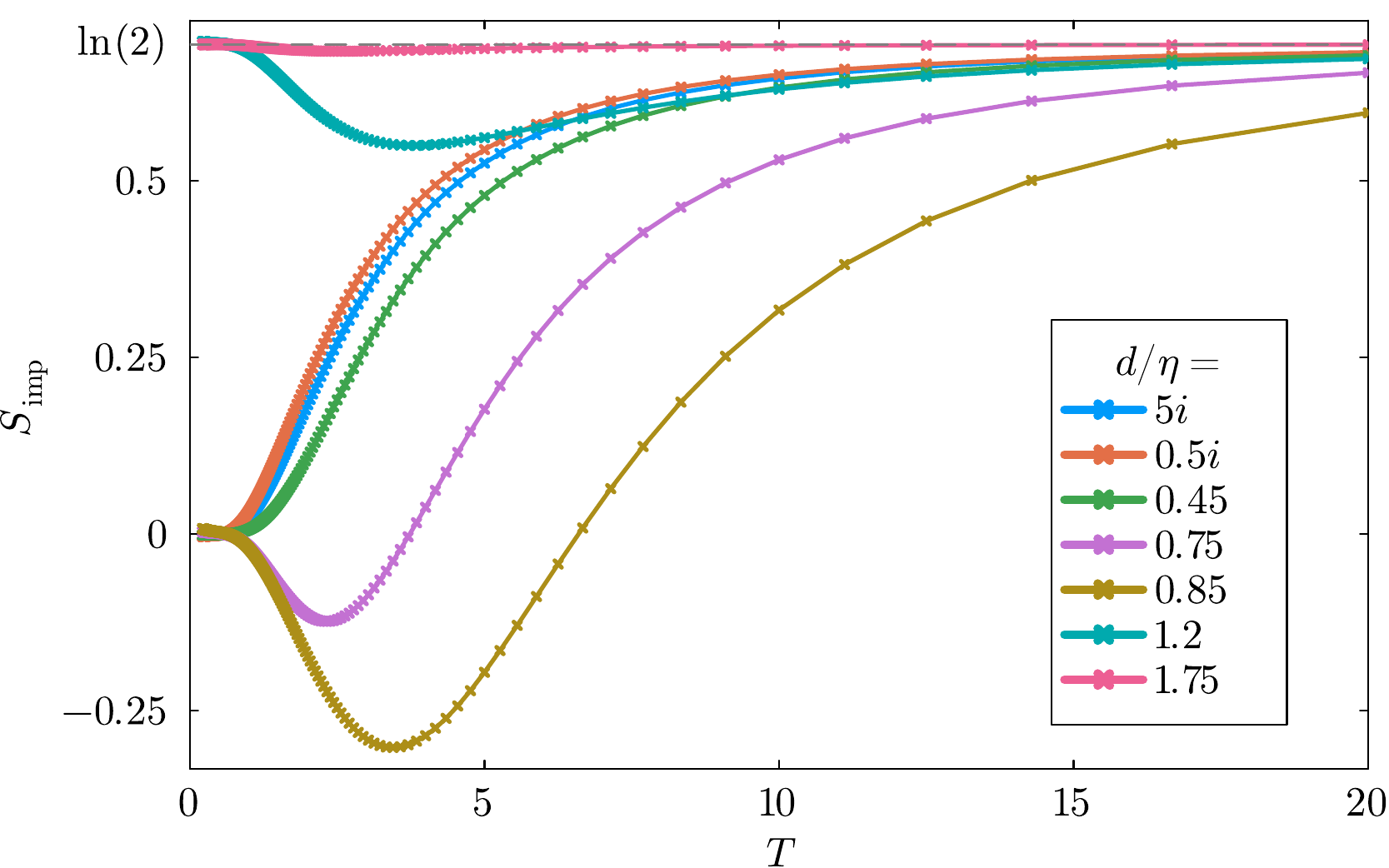}
\caption{Impurity entropy ($S_{\rm imp}$) vs. temperature ($T$) for various $d/\eta$, illustrating contrasting impurity screening. The plot shows $S_{\rm imp} = \ln(2)$ at high $T$, evolving with decreasing $T$: monotonic decrease to zero in the Kondo regime; non-monotonic behavior in bound mode phases (due to boundary scales); constant $S_{\rm imp} = \ln(2)$ in the unscreened (US) phase (impurity unscreened at all scales); and non-monotonic behavior in the FBM phase (due to a single high-energy screening mode). Results are for a bulk chain of $N_b = 100$ sites with $\eta = 2.5$ and a single impurity, obtained using the finite-temperature tensor network purification method by implementing in ITensors library).}
        \label{fig:impentint}
\end{figure}

{Impurity-related thermodynamic quantities reflect the boundary phase transitions, with the flow of impurity entropy ($S_{\rm imp}$) between the ultraviolet (UV) and infrared (IR) limits serving as a key indicator. As discussed in the last row of Table \ref{tab:my-table1}, $S_{\rm imp}$ decreases from $\ln(2)$ in the UV to 0 in the IR for both the Kondo and antiferromagnetic (AFM) phases (encompassing the entire range of antiferromagnetic impurity coupling). Conversely, $S_{\rm imp}$ remains at $\ln(2)$ in both the UV and IR limits for the ferromagnetic bound mode (FBM) and unscreened (US) phases (covering the entire ferromagnetic impurity coupling range). However, despite this shared UV/IR behavior, Fig. \ref{fig:impentint} reveals significant differences in the temperature dependence of $S_{\rm imp}$ between the two sub-phases within each coupling regime:

\begin{itemize}
    \item Kondo phase: $S_{\rm imp}$ decreases monotonically from $\ln(2)$ to 0 as there are no scales in the boundary. 
    \item Bound mode phase: $S_{\rm imp}$ decreases non-monotonically to 0 as there is a massive bound mode at the boundary. 
    \item Unscreened (US) phase: $S_{\rm imp}$ remains constant at $\ln(2)$ as impurity can not be screened at any scale. 
    \item Ferromagnetic bound mode (FBM) phase: $S_{\rm imp}$ is $\ln(2)$ at both zero and infinite temperatures but dips in the intermediate range as there exists a single particle bound mode that can screen impurity at high energy. 
\end{itemize}
These results were obtained using a finite-temperature tensor network method (purification algorithm ~\cite{feiguin2005finite}) implemented in ITensor library \cite{fishman2022itensor}) for 100 bulk sites. The impurity entropy is defined as the difference between the thermodynamic entropy of the chain with the impurity and the thermodynamic entropy of the impurity-free chain.

Before discussing the boundary phases in detail, we will briefly clarify the underlying bulk physics by considering the system without impurities, and then proceed to analyze the physics at the boundary.}

\section{ The bulk (No impurity attached)}\label{noimp}
Let us first consider the case when there are no impurities at the chain edges, $J_R=0=J_L$, to set the notation for the rest of the paper and briefly review some known results. In this case, the model is integrable via Bethe Ansatz for any values of anisotropy parameter \(\Delta\). Bethe Ansatz was first used to analytically solve the system with periodic boundary conditions in the isotropic case (\(\Delta = 1\))~\cite{1931_Bethe_ZP_71}, and the solution was later extended to include anisotropy along the $z-$direction~\cite{XXZPhysRev.112.309,XXZPhysRev.116.1089,XXZPhysRev.150.321,XXZPhysRev.150.327,XXZPhysRev.151.258,baxter1972partition}. In the regime where \(\Delta > 1\), the system exhibits both a continuous \(U(1)\) symmetry and a discrete \(\mathbb{Z}_2\) spin-flip symmetry. The \(\mathbb{Z}_2\) symmetry is spontaneously broken in the thermodynamic limit, and the ground state is a two-fold degenerate state with a total $S^z=0$ for an even number of sites and $S^z=\pm \frac{1}{2}$ for odd number of sites~\cite{XXZbabelon1983analysis,baxter1972partition,franchini2017introduction}. 

\begin{figure}
    \centering
    \includegraphics[width=\linewidth]{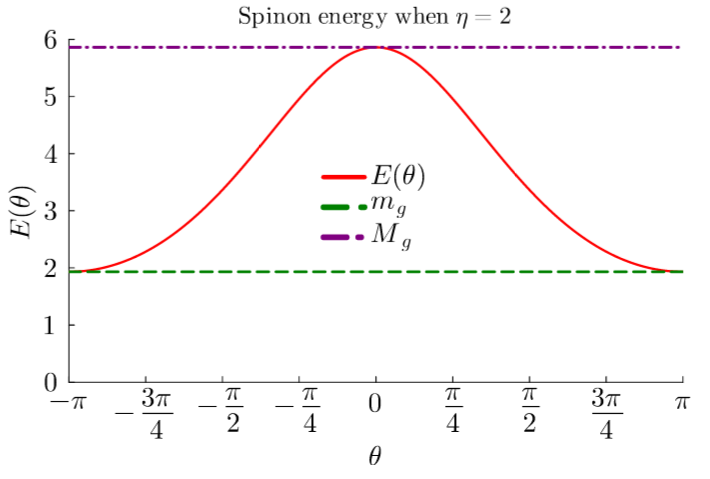}
    \caption{Energy of spinon for $\eta=2$. The green dashed line shows the minimum value of spinon energy $m_g=E(\pm \pi)$ where $E(\theta)$ is given in Eq.\eqref{spinonengeqn}, and the blue dotted-dashed line shows the maximum value of the spinon energy $M_g=E(0)$.}
    \label{fig:spinon-energy}
\end{figure}

Excitations above the ground states are constructed by adding an even number of spinons, quartets, strings  etc~\cite{destri1982analysis}. The fundamental excitations, the spinons,  are topological kink excitations interpolating between the two vacua~\cite{faddeev1981spin, Spinon-PhysRevLett.70.4003,Spinon-PhysRevLett.106.157205}. They carry $S^z=\frac{1}{2}$ and  are massive with energy,
\begin{equation}
    E(\theta)=J\sinh\eta \frac{\vartheta _1^{\prime }\left(0,e^{-\eta }\right) \vartheta _3\left(\frac{\theta }{2},e^{-\eta }\right)}{ \left(\vartheta _2\left(0,e^{-\eta }\right) \vartheta _4\left(\frac{\theta }{2},e^{-\eta }\right)\right)},
    \label{spinonengeqn}
\end{equation}
where $\vartheta_\alpha(u,q)$ for $\alpha=1,2,3,4$ are Jacobi theta functions~\cite{weisstein2000jacobi} and prime denotes the derivative with respect to the $u$ variable (see Appendix \ref{thetadef} for definitions of these functions). Notice that the maximum spinon energy, $M_g$, occurs at $\theta=0$, and its minimum value  $m_g$ occurs at $\theta=\pm \pi$, {which is non-zero as shown in Fig.~\ref{fig:spinon-energy}}. Thus the model is gapped with the mass gap given by $\mathcal{M}_g=2m_g$.

Notice that the model has spontaneously broken discrete spin flip $\mathbb{Z}_2$ symmetry, due to which the ground state is two-fold degenerate $S^z=0$ for even $N_b$ and $S^z=\pm \frac{1}{2}$ for odd $N_b$. Moreover, an antiferromagnetic order develops in the bulk where the staggered magnetization $(-1)^j \sigma$ where $\sigma$ is given by~\cite{baxter1973spontaneous,izergin1999spontaneous}
\begin{equation}
    \sigma= \frac{1}{2}\left(\prod_{m=1}^\infty\left( \frac{1-e^{-2m\eta}}{1+e^{-2m\eta}}\right) \right)^2,
    \label{bulkstgmag}
\end{equation}
and in the boundary, there are deviations due to the edge effects as discussed below.

\begin{figure}[htbp]
  \centering
  \begin{subfigure}{0.49\columnwidth}
    \centering
\includegraphics[width=\columnwidth]{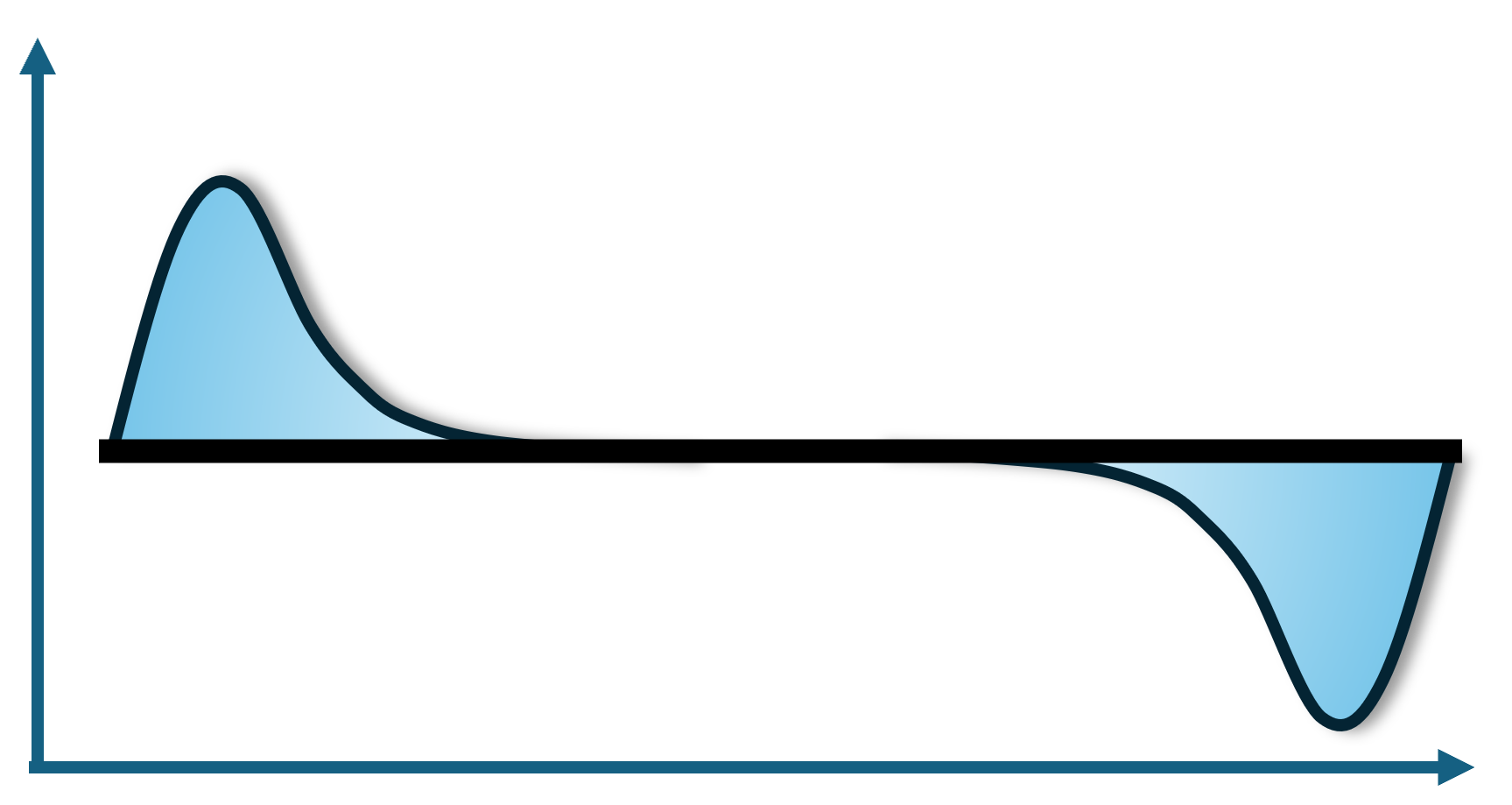}
  \end{subfigure}
  %\vspace{1em} % Add some vertical space between subfigures
  \begin{subfigure}{0.49\columnwidth}
    \centering
    \includegraphics[width=\columnwidth]{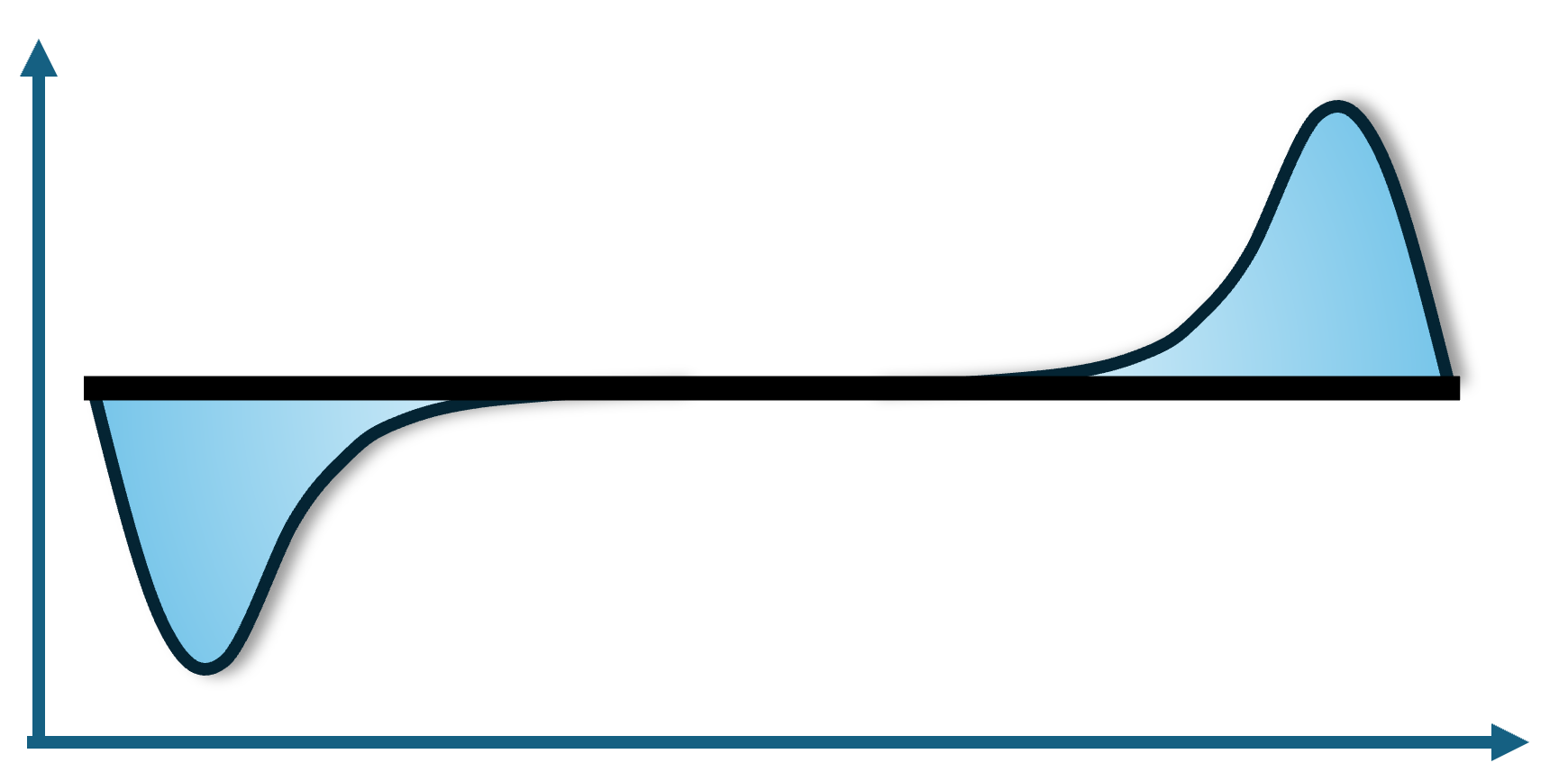}
  \end{subfigure}
  \begin{tikzpicture}[overlay, remember picture]
\node[black] at (-0.55,1) {-$\frac{1}{4}$};
\node[black] at (-3.65,1.65) {$\frac{1}{4}$};
\node[black] at (-2.0,0.2) {sites};
\node[black] at (-4.65,2.5) {$a)$};
\node[black] at (-4.125,2.7) {$\braket{S^z}$};
\node[black] at (-4,0.2) {$1$};
\node[black] at (-0.5,0.2) {$N$};
\node[black] at (-0.125,2.5) {$b)$};
\node[black] at (2,0.2) {sites};
\node[black] at (0.4,0.2) {$1$};
\node[black] at (3.85,0.22) {$N$};
\node[black] at (0.3,2.7) {$\braket{S^z}$};
\node[black] at (0.6,1.05) {-$\frac{1}{4}$};
\node[black] at (3.7,1.7) {$\frac{1}{4}$};
  \end{tikzpicture}
  \caption{Schematic representation of the spin accumulation in the two-fold degenerate ground state when the total number of bulk sites $N_b$ is even. a) The total spin is $S^z=0$ such that the left edge carries $S^z=\frac{1}{4}$ and the right edge carries $S^z=-\frac{1}{4}$. b) The total spin accumulation is $S^z=0$ such that left edge carries $S^z=-\frac{1}{4}$ and the right edge carries $S^z=\frac{1}{4}$.}
  \label{fig:updn-dnup}
\end{figure}

\textit{Edge modes}: A recent discovery indicates that integrable anisotropic Heisenberg $XXZ-\frac{1}{2}$ chain in the gapped antiferromagnetic regime hosts fractionalized edge spin~\cite{pasnoori2023spin}. These, when projected onto the low-energy subspace spanned by the ground state, can be
identified with the strong zero energy mode discussed in~\cite{fendley2016strong}.
 It was established that the ground state for an even number of sites with $S^z=0$ contains sharply localized $\pm\frac{1}{4}$ edge modes such that we label the two-fold degenerate ground state shown in Fig.~\ref{fig:updn-dnup} as
\begin{equation}
    \ket{GS}^E\equiv \left| \pm \frac{1}{4}, \mp \frac{1}{4}\right\rangle \quad \text{(for even $N_b$)},
    \label{evenEM}
\end{equation}

\begin{figure}[htbp]
  \centering
  \begin{subfigure}{0.49\columnwidth}
    \centering
\includegraphics[width=\columnwidth]{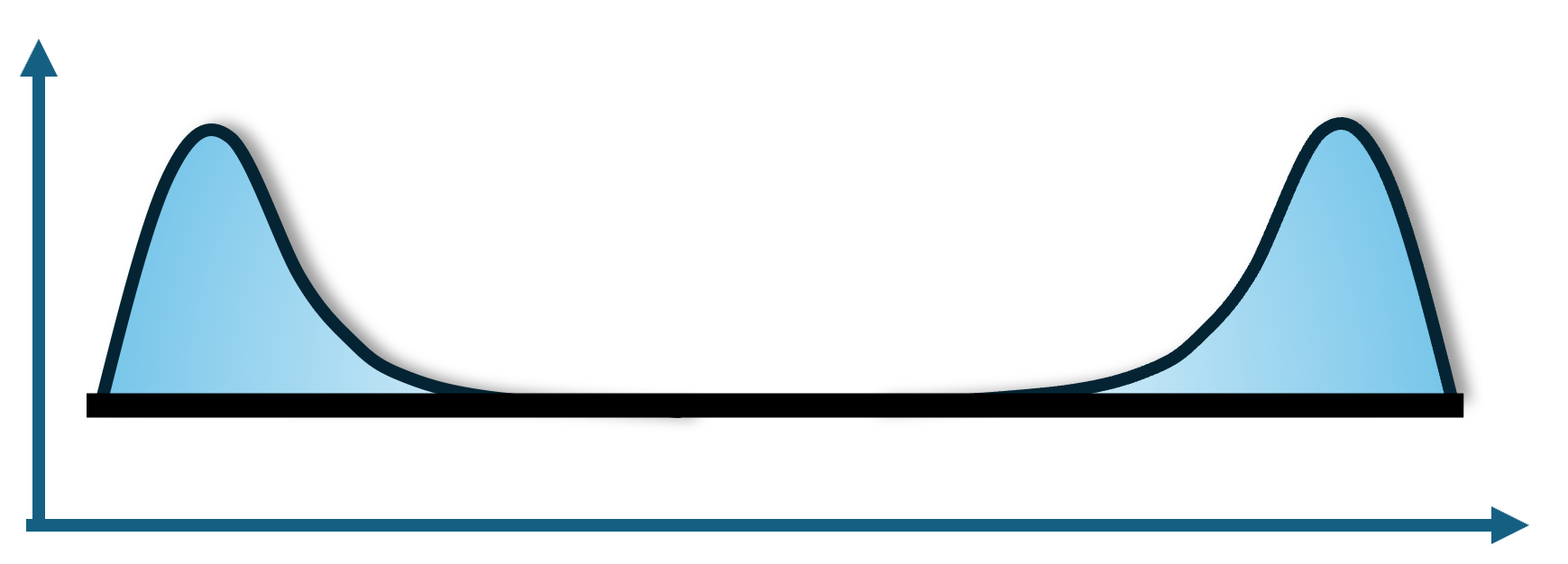}
  \end{subfigure}
  %\vspace{1em} % Add some vertical space between subfigures
  \begin{subfigure}{0.49\columnwidth}
    \centering
    \includegraphics[width=\columnwidth]{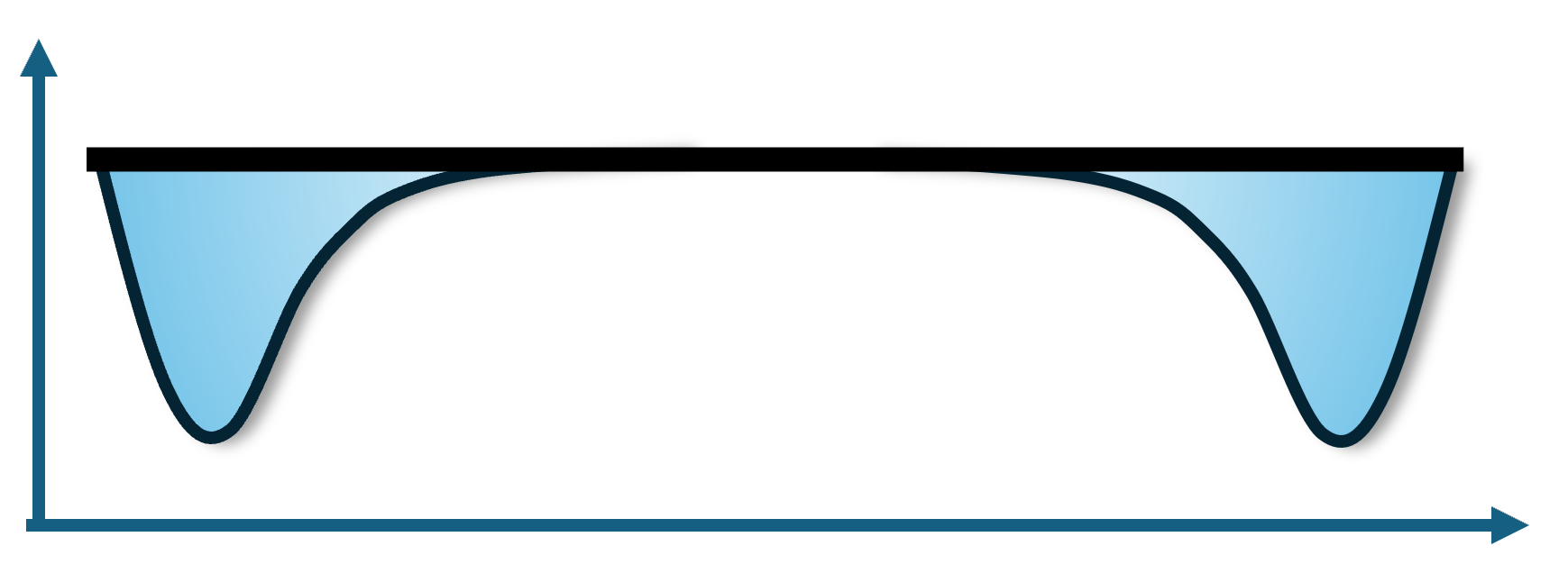}
  \end{subfigure}
  \begin{tikzpicture}[overlay, remember picture]
\node[black] at (-0.725,1.15) {$\frac{1}{4}$};
\node[black] at (-3.675,1.15) {$\frac{1}{4}$};
\node[black] at (-2.0,0.2) {sites};
\node[black] at (-4.5,1.5) {$a)$};
\node[black] at (-4.125,1.9) {$\braket{S^z}$};
\node[black] at (-3.98,0.2) {$1$};
\node[black] at (-0.35,0.2) {$N$};
\node[black] at (-0.125,1.5) {$b)$};
\node[black] at (2,0.2) {sites};
\node[black] at (0.35,0.2) {$1$};
\node[black] at (3.85,0.2) {$N$};
\node[black] at (0.3,1.9) {$\braket{S^z}$};
\node[black] at (0.6,1.15) {-$\frac{1}{4}$};
\node[black] at (3.6,1.15) {-$\frac{1}{4}$};
  \end{tikzpicture}
  \caption{Schematic representation of the spin accumulation in two-fold degenerate ground state when the total number of bulk sites $N_b$ is odd. The spin accumulations at the two edges point in the same direction such that total spin a) $S^z=\frac{1}{2}$ and b) $S^z=-\frac{1}{2}$ in the ground state.}
  \label{fig:upup-dndn}
\end{figure}

Likewise, the doubly degenerate ground state $S^z=\pm \frac{1}{2}$ for the odd number of case shown in Fig.~\ref{fig:upup-dndn} can be labelled by the two edge modes as
\begin{equation}
    \ket{GS}^O\equiv \left| \pm \frac{1}{4}, \pm \frac{1}{4}\right\rangle \quad \text{(for odd $N_b$)}.
    \label{oddEM}
\end{equation}

It is important to note that the edge localized fractional $S^z=\frac14$ excitations in the $XXZ$ chain emerge due to many-body interactions \cite{pasnoori2023spin,kattel2024edge} and these excitations are not simple averages but rather sharp quantum observables.

To understand this better, we look at the spin profile in the ground state using DMRG. The spin profile of the XXZ chain has the form
\begin{equation}
    S^z_j=(-1)^j\sigma+\Delta(S^z)(j),
        \label{sspinprofile}
\end{equation}
where $\sigma$ is the exact staggered magnetization of the XXZ chain in the thermodynamic limit defined in Eq.~\eqref{bulkstgmag} and $\Delta(S^z)(j)$ is the deviation from the bulk antiferromagnetic order due to the presence of the impurity and open boundary. Due to the gap in the bulk, all the correlations exponentially decay, and hence the deviation of the bulk is localized exponentially at the edges \textit{i.e.}
\begin{equation}
    \Delta(S^z)(j)= \Delta S^z_L(j)+ \Delta S^z_R(j),
\end{equation}
where $\Delta S^z_L(j)$ is localized near the left edge $j=1$ and $\Delta S^z_R(j)$ is localized near the right edge $j=N_b$. We shall verify the existence of fractionalized spin operators $\hat S^z_L$ and $\hat S^z_R$ at the left and right edges characterized by fractional expectation values $\pm\frac{1}{4}$ ~\cite{pasnoori2023spin}.

Following~\cite{jackiw1983fluctuations,kivelson1982fractional,pasnoori2023spin,kattel2024edge}, we define the fractional edge spin operators as  
\begin{align}
    \hat{S}^z_L &= \lim_{\alpha \to 0} \lim_{N \to \infty} \hat{S}^z_L(N, \alpha) = \lim_{\alpha \to 0} \lim_{N \to \infty} \sum_{j=1}^N e^{-\alpha j} S^z_j, 
    \label{leftspop}
\end{align}
with $\hat{S}^z_R$ defined similarly, replacing $\exp\{-\alpha j\}$ with $\exp\{-\alpha(N+1-j)\}$. These operators have quantized expectation values $\pm \frac{S}{2}$ in the ground state. The variance $\delta \hat{S}^2_{L/R}$ vanishes in the ground state, indicating that these fractional operators are sharp observables.

To compute the variance numerically, we adopt the Ansatz from \cite{pasnoori2023spin,kattel2024edge} in the thermodynamic limit:
\begin{equation}
    \delta S^2_{L/R}(N, \alpha) = \delta S^2_{L/R}(\infty, \alpha) - A \alpha e^{-B \alpha N},
    \label{ansatzvar}
\end{equation}
where $\delta S^2_{L/R} = \lim_{\alpha \to 0} \delta S^2_{L/R}(\infty, \alpha)$. {This Ansatz was verified for various $N$ using Hamiltonian Eq.~\eqref{ham} with $J_R=J_L=0$, in \cite{pasnoori2023spin}.}
 We shall analyze how these fractional excitations behave in the presence of an impurity, and we shall show the validity of this ansatz even in the presence of an impurity.  As we shall see, the effect of a single quantum impurity in an already strongly interacting many-body spin chain gives rise to several interesting phenomena discussed below.

\section{One integrable impurity}\label{1intimp-det}
\begin{figure}
    \centering
 \includegraphics[width=\linewidth]{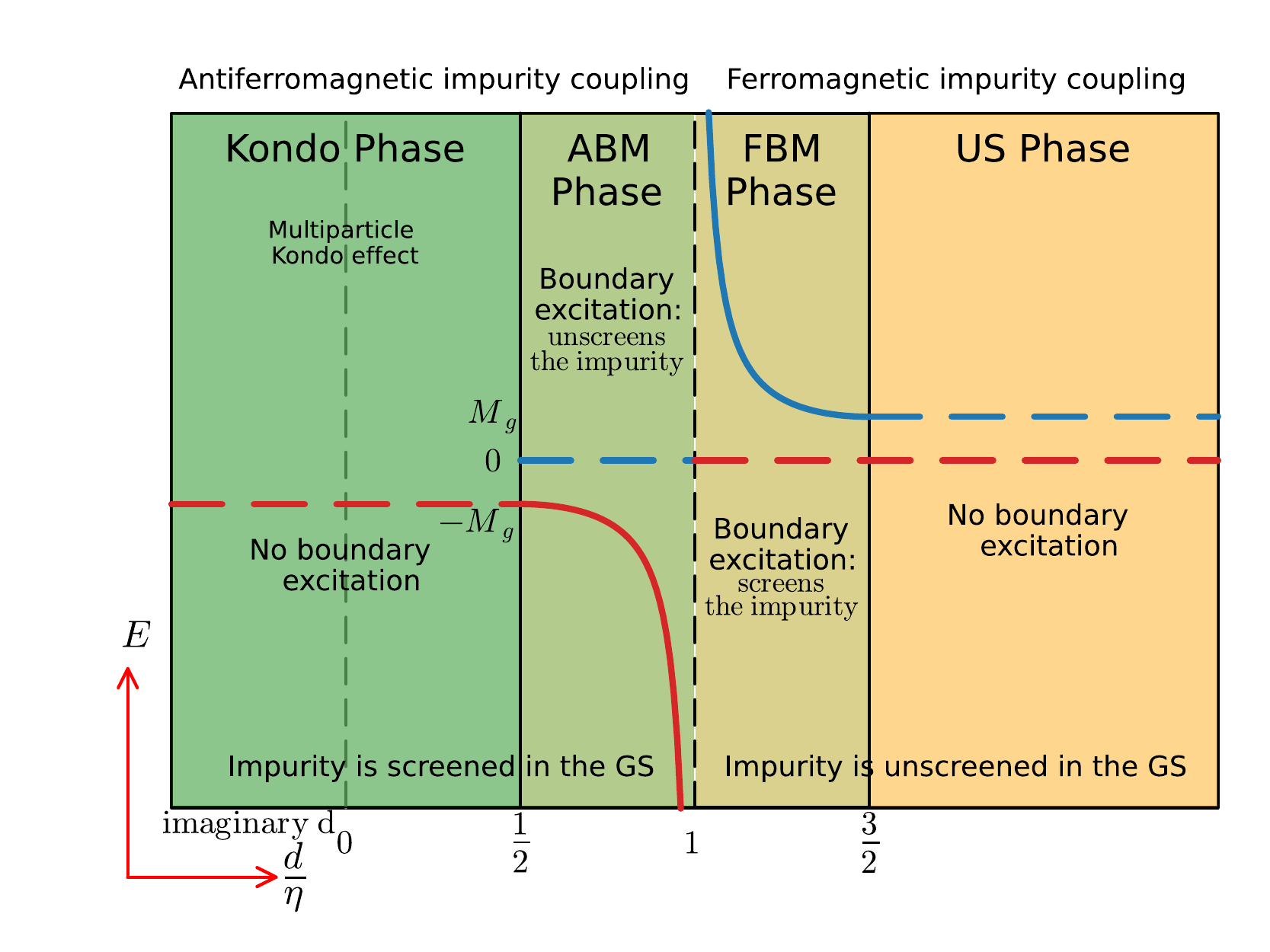}
\caption{Phase diagram for Hamiltonian \eqref{ham1} when the impurity parameter $d$ takes either purely imaginary or real values. 
In the Kondo phase, where $d$ is either purely imaginary or satisfies $0 < d < \frac{\eta}{2}$ (with the gray dashed line in the phase diagram distinguishing these regimes), the impurity is screened by a multiparticle cloud of spinons.
 In the antiferromagnetic bound mode phase ($\frac{\eta}{2}<d<\eta$), the red curve is the energy of the bound mode, which exists in the ground state and screens the impurity, and the blue line is the state in which the impurity is unscreened. The dashed line at $d=\eta$ represents the impurity coupling to the right of the line, i.e., for $d>\eta$ is ferromagnetic where, as for $d<\eta$ is antiferromagnetic.  In the ferromagnetic bound mode ($\eta<d<\frac{3\eta}{2}$) phase, the red line represents the ground state where the impurity is unscreened, and the blue curve is the high-energy state containing the bound mode where the impurity spin is screened. Finally, when $d>\frac{3\eta}{2}$, the impurity is unscreened at every energy scale. The phase diagram is similar to that of the isotropic model studied in \cite{kattel2023kondo}; however, the detail of impurity behavior is different in each phase.}
\label{fig:PD1}
\end{figure}

The Hamiltonian Eq.~\eqref{ham} is not integrable for arbitrary values of boundary couplings $J_R$ and $J_L$ and boundary anisotropy parameters $\Delta_R$ and $\Delta_L$. In the integrable limit, the boundary couplings and boundary anisotropy parameters are given by Eq.~\eqref{bulk-boundary}\cite{hou1999integrability,chen1998integrability,hu1998two,Shu_Chen_1998}.
Consider the case with one integrable impurity attached, which is defined through  $d_R=0$ and $d_L=d$; the Hamiltonian then takes the form
\begin{align}
    H&=\sum_{i=1}^{N_b-1}\frac{J}{2}(\sigma_i^x  \sigma_{i+1}^x+\sigma_i^y  \sigma_{i+1}^y+\Delta \sigma_i^z  \sigma_{i+1}^z)\nonumber\\
    &+\frac{J_{\mathrm{imp}}}{2}(\sigma_{\mathrm{imp}}^x  \sigma_{1}^x+\sigma_{\mathrm{imp}}^y  \sigma_{1}^y+\Delta_{\mathrm{imp}} \sigma_{\mathrm{imp}}^z  \sigma_{1}^z),
    \label{ham1}
\end{align}
In this case, the impurity exhibits four distinct phases, depending on the value of the impurity parameter $d$, as shown in Fig.~\ref{fig:PD1}. There are two distinct phases when the impurity coupling is antiferromagnetic: the Kondo phase and the antiferromagnetic bound mode phase. Likewise, when the impurity coupling is ferromagnetic, there are two distinct phases: the ferromagnetic bound mode phase and the unscreened phase. The phase diagram is exactly the same as in the isotropic limit studied in \cite{kattel2023kondo}. However, each phase has unique features that are different from the isotropic case, primarily because of a mass gap in the spectrum and the bulk magnetic order in the present case.

Before discussing the four distinct phases, let us first focus on the aspect of impurity physics that only depends on the sign of the impurity coupling. We shall begin by studying the case where the impurity is integrably coupled to the XXZ chain with antiferromagnetic coupling.

\subsection{Antiferromagnetic boundary coupling}
When the impurity parameter takes either purely imaginary values or real values in the range $0<d<\eta$, the impurity coupling is antiferromagnetic. In this range, the impurity exhibits two distinct phases: the Kondo phase and the antiferromagnetic bound mode (ABM) phase discussed below.

When $d$ is purely imaginary or when it takes real values in the range $0<d<\frac{\eta}{2}$, the impurity is screened by the many-body Kondo effect, and a few particle excitations can not unscreen the impurity. Thus, we call it the \textit{Kondo phase}.  When $\frac{\eta}{2}<d<\eta$, the impurity coupling is antiferromagnetic, and the impurity is screened by a single particle bound mode exponentially localized at the edge of the chain and hence the name \textit{antiferromagnetic bound mode} (ABM) phase. The single particle bound mode is described by a purely imaginary root of the Bethe Ansatz equations. In this phase, it is possible to unscreen the impurity by a single particle excitation by exciting this single particle bound mode. We shall discuss the difference between the two phases in much detail later.

In both the Kondo and 
%antiferromagnetic bound mode 
ABM
regime, 
the impurity coupling is antiferromagnetic, and the impurity is screened in the ground state.  All the excitations above the ground state in the Kondo phase are bulk excitations constructed by adding an even number of spinons, bulk string solutions, quartets, etc. \cite{destri1982analysis} such that in all of the low-lying excited states, the impurity is always anti-aligned with the quarter mode. In contrast, the excitations above the ground state in the ABM phase not only include bulk excitations but also unique boundary excitations. {Boundary excitations involve unoccupying the mode described by the imaginary solution of the Bethe Ansatz equations, which creates a massive excitation with energy greater than the maximum energy of the spinon $M_g$. In this state, the impurity becomes unscreened, and when the effectively free spin-$\frac12$ spin aligns with the effective quarter mode at the edge, it forms an effective three-quarter mode, as shown below.

Before discussing the detailed differences between the Kondo and ABM phases, we shall first focus on the aspects that are common to both phases \textit{i.e.} the aspect of the impurity physics that depends only on the sign of the impurity coupling but not on its strength.

 When a free spin-$\frac{1}{2}$ impurity, which initially has an entropy of $\ln 2$ 
 is coupled antiferromagnetically to the XXZ chain in its gapped antiferromagnetic phase, the entropy of the impurity reduces to zero in either of the two degenerate vacua as its spin gets screened. In the language of RG, the impurity is asymptotically free in the UV, hence it has $\ln 2$ entropy. However, when coupled to the XXZ chain with two-fold degenerate vacua, the defect coupling initiates a boundary RG flow. In the IR, this flow results in the screening of the impurity, reducing its entropy to zero as it becomes screened in each of the two degenerate vacua. The detailed behavior of this process depends on whether the total number of sites in the chain is even or odd. To understand this effect fully, let's examine both cases in detail.

\subsubsection{Interplay between the edge modes and the antiferromagnetically coupled impurity}\label{imp-effects-1}
Let us first consider a single impurity coupled to the left end of the XXZ chain containing an even {(E)} number of bulk sites with $\ket{GS}^E_{UV}\equiv \left| \pm \frac{1}{4}, \mp \frac{1}{4}\right\rangle$ with total spin $S^z=0$ as shown in Fig.~\ref{fig:updn-dnup}. The full set of Bethe Ansatz equations must be solved for the even number of bulk sites and the impurity. The results reveal that, due to the antiferromagnetic nature of the impurity-bulk coupling, the impurity effectively anti-aligns with the quarter mode to minimize its energy. 
As a result, the impurity effectively flips the direction of the quarter mode, leading to a two-fold degenerate ground state with edge modes that now point in the same direction, as shown in Fig.~\ref{fig:upup-dndn-res1} \textit{i.e.}  $\ket{GS}_{IR}^E\equiv \left| \mp \frac{1}{4}, \mp \frac{1}{4}\right\rangle$ resulting in states with total spin $S^z=\pm \frac{1}{2}$.
It is important to note that the fractional edge mode initially arises from complex many-body interactions, and hence, the effective flipping of the edge mode to point in the opposite direction due to the impurity is also a complex many-body process, given that the edge mode is an exponentially localized object.
\begin{figure}[htbp]
  \centering
  \begin{subfigure}{0.49\columnwidth}
    \centering
\includegraphics[width=\columnwidth]{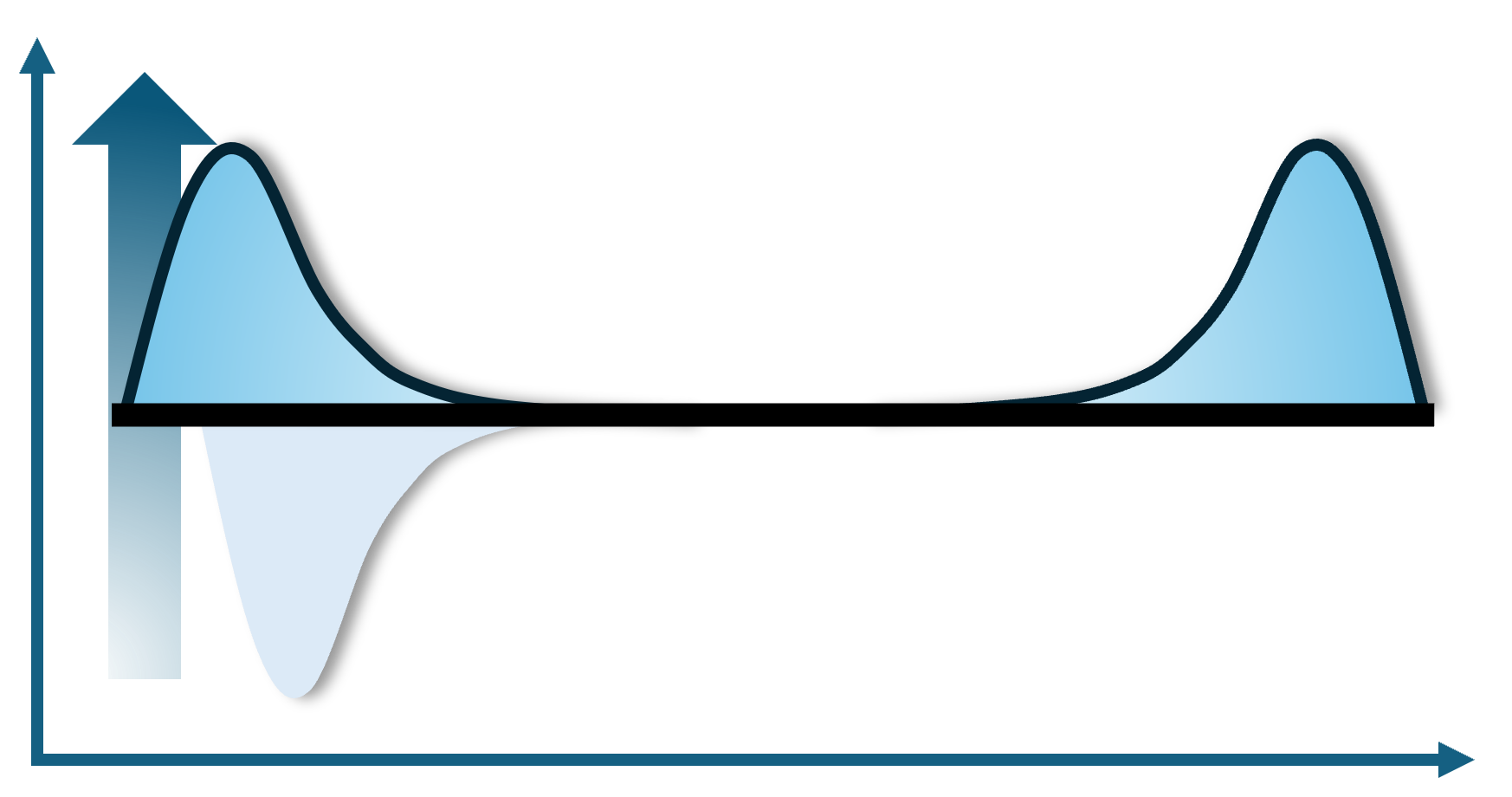}
  \end{subfigure}
  %\vspace{1em} % Add some vertical space between subfigures
  \begin{subfigure}{0.49\columnwidth}
    \centering
    \includegraphics[width=\columnwidth]{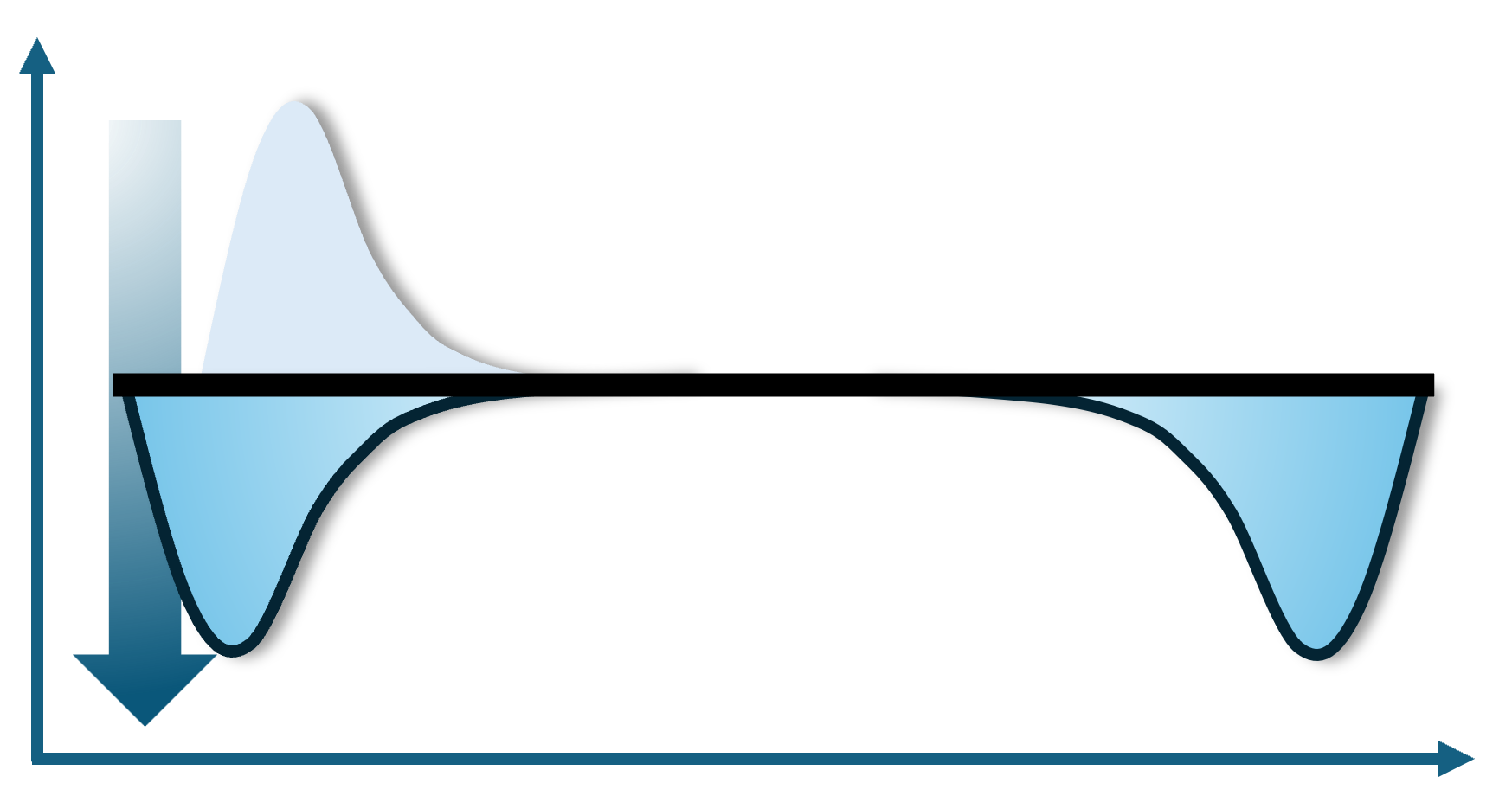}
  \end{subfigure}
  \begin{tikzpicture}[overlay, remember picture]
\node[black] at (-3.5,1.05) {-$\frac{1}{4}$};
\node[black] at (-3.6,1.7) {$\frac{1}{4}$};
\node[black] at (-0.65,1.7) {$\frac{1}{4}$};
\node[black] at (-2.0,0.2) {sites};
\node[black] at (-4.5,2.2) {$a)$};
\node[black] at (-4.125,2.6) {$\braket{S^z}$};
\node[black] at (-4,0.2) {$L$};
\node[black] at (-3.6,0.2) {$1$};
\node[black] at (-0.45,0.2) {$N$};
\node[black] at (-0.125,2.2) {$b)$};
\node[black] at (2,0.2) {sites};
\node[black] at (0.4,0.2) {$L$};
\node[black] at (0.8,0.2) {$1$};
\node[black] at (3.85,0.2) {$N$};
\node[black] at (0.3,2.6) {$\braket{S^z}$};
\node[black] at (0.725,1.1) {-$\frac{1}{4}$};
\node[black] at (3.75,1.1) {-$\frac{1}{4}$};
\node[black] at (0.9,1.8) {$\frac{1}{4}$};
  \end{tikzpicture}
  \caption{When a spin-$\frac{1}{2}$ impurity with entropy $\ln 2$ is antiferromagnetically coupled to an $XXZ$ chain with an even number of bulk sites, the impurity's spin and the quarter mode at the edge of the XXZ chain—arising from many-body interactions—rearrange to form an effective net quarter mode pointing in the opposite direction, as shown in the degenerate vacua (a) and (b). This many-body interaction, in turn, reduces the impurity's entropy to zero. The arrow represents an initially free spin-$\frac{1}{2}$ impurity that is coupled to the chain. The quarter mode that was present in the chain before coupling (depicted in faint color) disappears, and at the same time, due to the complex many-body interactions between the impurity and the quarter mode, a new quarter mode (shown in blue) forms, pointing in the opposite direction.
  }
  \label{fig:upup-dndn-res1}
\end{figure}

Let us now consider the case where a single spin-$\frac{1}{2}$ impurity is coupled antiferromagnetically to the XXZ chain in its gapped antiferromagnetic phase with an odd (O) number of bulk sites with $\ket{GS}^O_{UV}\equiv \left| \pm \frac{1}{4}, \pm \frac{1}{4}\right\rangle$ as shown in Fig.
\ref{fig:upup-dndn}. Again, carrying out a detailed calculation indicates that the impurity prefers to anti-align with the exponentially localized quarter mode to lower the energy and it forms two-fold degenerate ground state with edge modes pointing in the opposite direction as shown in Fig.~\ref{fig:updn-dnup-res1} \textit{i.e.}$\ket{GS}_{IR}^O\equiv \left| \mp \frac{1}{4}, \pm \frac{1}{4}\right\rangle$. As before, the effect of the impurity is to effectively flip the sign of the edge mode at the left end of the chain via a many-body effect. Thus, the asymptotically free impurity with entropy $\ln 2$ when coupled to the two-fold degenerate ground state $\ket{GS}^O_{UV}\equiv \left| \pm \frac{1}{4}, \pm \frac{1}{4}\right\rangle$, is essentially anti-aligned with the exponentially localized edge mode at left edge thereby resulting in vanishing impurity entropy. 
\begin{figure}[htbp]
  \centering
  \begin{subfigure}{0.49\columnwidth}
    \centering
\includegraphics[width=\columnwidth]{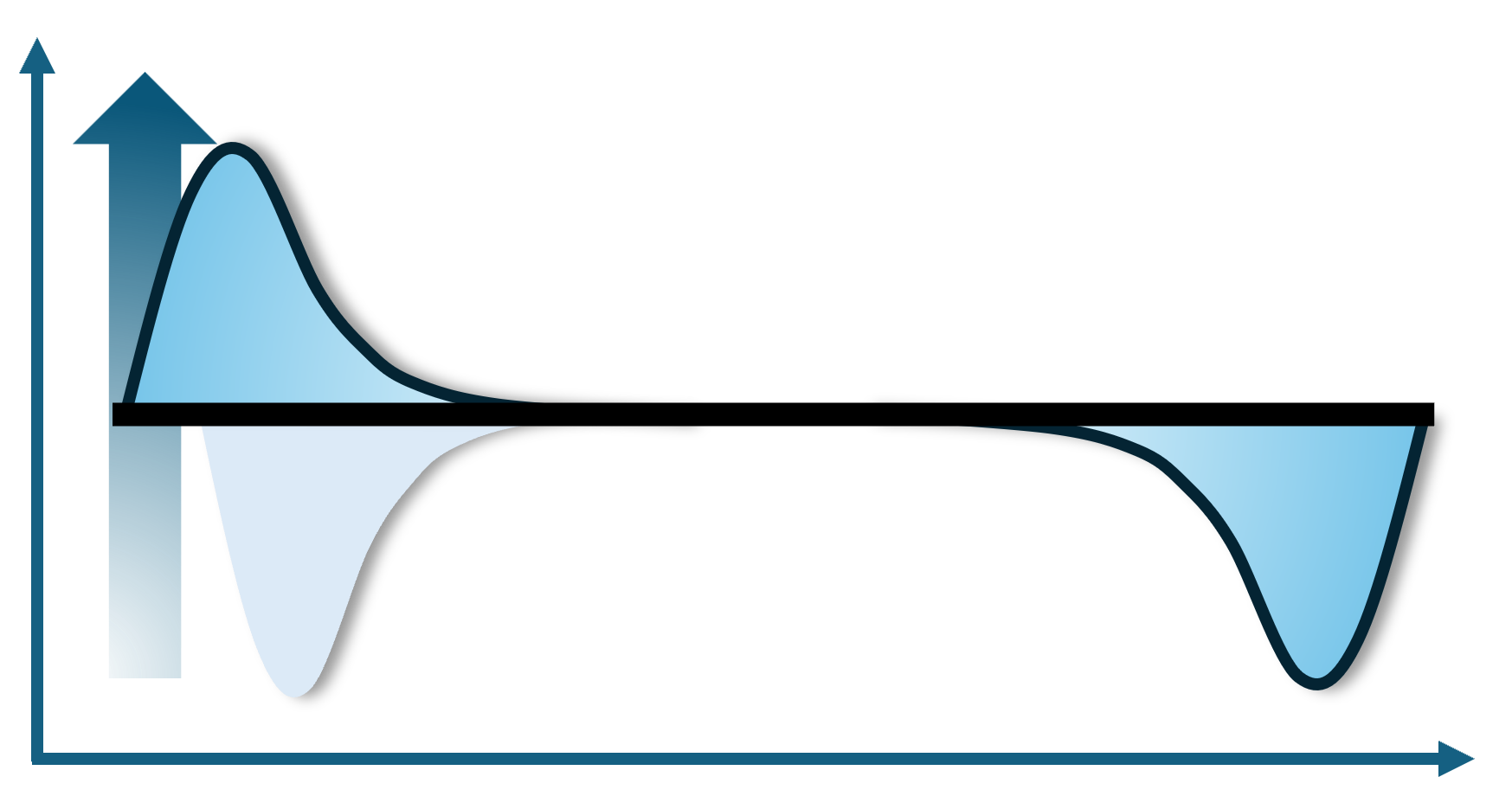}
  \end{subfigure}
  %\vspace{1em} % Add some vertical space between subfigures
  \begin{subfigure}{0.49\columnwidth}
    \centering
    \includegraphics[width=\columnwidth]{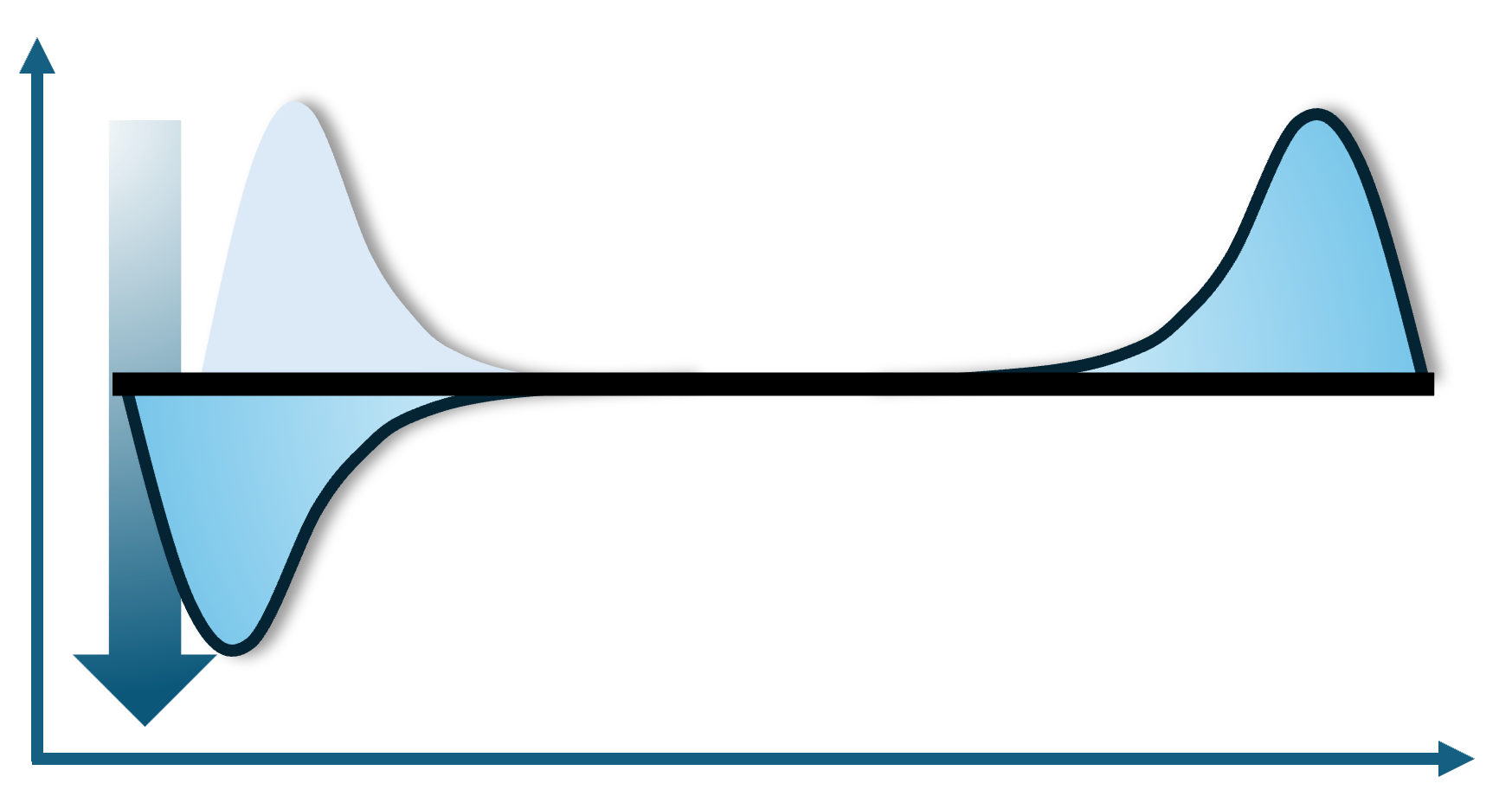}
  \end{subfigure}
  \begin{tikzpicture}[overlay, remember picture]
\node[black] at (-3.45,1.05) {-$\frac{1}{4}$};
\node[black] at (-3.6,1.75) {$\frac{1}{4}$};
\node[black] at (-0.6,1.05) {-$\frac{1}{4}$};
\node[black] at (-2.0,0.2) {sites};
\node[black] at (-4.5,2.3) {$a)$};
\node[black] at (-4.125,2.6) {$\braket{S^z}$};
\node[black] at (-4,0.2) {$L$};
\node[black] at (-3.6,0.2) {$1$};
\node[black] at (-0.45,0.2) {$N$};
\node[black] at (-0.125,2.3) {$b)$};
\node[black] at (2,0.2) {sites};
\node[black] at (0.4,0.2) {$L$};
\node[black] at (0.8,0.2) {$1$};
\node[black] at (3.85,0.2) {$N$};
\node[black] at (0.3,2.6) {$\braket{S^z}$};
\node[black] at (0.7,1.1) {-$\frac{1}{4}$};
\node[black] at (3.65,1.8) {$\frac{1}{4}$};
\node[black] at (0.95,1.8) {$\frac{1}{4}$};
  \end{tikzpicture}
  \caption{When a spin-$\frac{1}{2}$ impurity is antiferromagnetically coupled to an XXZ chain with an odd number of bulk sites, the impurity effectively anti-aligns with the exponentially localized quarter spin accumulation at the left edge of the chain, thereby resulting in a net quarter mode pointing in the opposite direction via a many-body process. The arrow represents the spin-$\frac{1}{2}$ impurity that is initially free (with $\ln 2$ entropy). The faint-colored spin quarter mode, which was present in the clean chain, disappears, and at the same time, the impurity, together with the quarter mode, now forms an effective quarter mode that points in the opposite direction. As a result, the ground state features quarter modes at the edges that point in opposite directions in both vacua, as shown in a) and b), and the impurity entropy is thus vanishing in each of the two degenerate vacua.}
  \label{fig:updn-dnup-res1}
\end{figure}

 \textit{General set up:} In order to study the nature of the edge mode appearing in the XXZ chain (in contrast to the ones that appear in topological spin chains) and its interplay with the impurity, we now briefly discuss the effect of impurity in a more general setup. Recently, some of us showed that in a generic antiferromagnetic spin-$S$ chain, there exists fractionalized $S/2$ edge modes as long as there is $U(1)$ symmetry, antiferromagnetic order in the bulk due to either spontaneous or explicit breaking of discrete $\mathbb{Z}_2$ symmetry and there is finite energy gap~\cite{kattel2024edge}. Now, if we antiferromagnetically couple a spin-$\frac{1}{2}$ impurity at the edge of the antiferromagnetic spin-$S$ chain hosting $\frac{S}{2}$ edge mode,  the interaction between the impurity and the spin chain results in a $\frac{S-1}{2}$ exponentially localized edge mode. For example, in a spin-$\frac{3}{2}$ XXZ chain with the anisotropy parameter $\Delta>1$ and an even number of sites, there exists fractionalized edge modes $\pm \frac{3}{4}, \mp \frac{3}{4}$ at the two edges in its two-fold degenerate ground states. When a spin-$\frac{1}{2}$ impurity is coupled to the left edge, the resultant ground state is still two-fold with edge modes $\pm \frac{1}{4},\mp \frac{3}{4}$. A particularly interesting case is when the spin-$\frac{1}{2}$ impurity is antiferromagnetically coupled to the left edge of a gapped spin-1 XXZ chain with spin-$\frac{1}{2}$ exponentially localized edge modes~\cite{kattel2024edge} in its antiferromagnetic phase \textit{i.e.} $\Delta \gtrapprox  1.185$~\cite{yu2021closing}. In this case, the half-edge mode at the left of the chain disappears, and since the half-edge mode at the right edge can point up or down, the resultant state with edge mode $0,\pm \frac{1}{2}$ is two-fold degenerate. 

At this point, it is important to understand this edge modes in spin-$1$ XXZ chain behave differently compared to the topological edge mode in the isotropic spin-1 XXX chain (Haldane chain). In the spin-1 Haldane chain, the spin-$\frac{1}{2}$ edge modes at the two ends can independently point up or down such that the resultant ground state is four-fold degenerate, and when a single spin-$\frac{1}{2}$ impurity is attached to its left edge with antiferromagnetic exchange coupling, the spin-$\frac{1}{2}$ impurity forms singlet with the spin-$\frac{1}{2}$ impurity thereby reducing the ground state degeneracy of the bulk from four to two. Likewise, if two spin-$\frac{1}{2}$ impurities are antiferromagnetically coupled to the two ends of a spin-1 Haldane chain, the impurities at each edge form a singlet state with the effective spin-$\frac{1}{2}$ edge modes. This effect removes the four-fold degeneracy of the Haldane chain, resulting in a unique ground state for the system. However, the two-fold degeneracy inherent in the spin-1 XXZ chain cannot be lifted by coupling one or two spin-$\frac{1}{2}$ impurities to the edges of the chain. In both scenarios, the resultant ground state remains two-fold degenerate. This shows that the $\frac{S}{2}$ edge modes discussed in ~\cite{kattel2024edge} in case of spin-$S$ antiferromagnetic chain and therefore the quarter-mode in spin-$\frac{1}{4}$ XXZ chain first found in ~\cite{pasnoori2023spin}, are not free effective $\frac{S}{2}$ particles as in the case of the topological chain but rather more complex many-body effective modes. This was already evident from the fact that unlike in Haldane chain where the effective spin-$\frac{1}{2}$ modes can independently point in either up or down direction for both even and an odd number of total sites, the two spin-$\frac{1}{2}$ edge modes at the edges of spin-1 $XXZ$ chain align in the same direction if the total number of sites is odd and they anti-align when the total number of sites is even thereby only giving size to two-fold degenerate vacua. We shall discuss this difference along with numerical results in more detail in Appendix \ref{diff-haldane-xxz}.

So far, we briefly discussed the impurity physics that is common to both the Kondo and ABM phases; we are now ready to discuss the impurity behavior that is unique in each of these two phases.  We shall defer the mathematical details of Bethe Ansatz equations to Appendix \ref{BAdets} and only describe the results here.
% \jp{JP: Why is this here? You then discuss the Kondo phase below after saying you have already done so... Before, it was common physics to both Kondo and ABM now it is about the things that are unique to each}

\subsubsection{The Kondo phase}

When the impurity parameter $d$ is purely imaginary or when it takes the real values between $0<d<\frac{\eta}{2}$, the impurity is in the Kondo phase as shown in the phase diagram Fig.~\ref{fig:PD1}. In this phase, the impurity is screened by a multiparticle cloud of spinons. Thus, the ground state is a two-fold degenerate $S^z=0$ state when $N_b$ is odd and $S^z=\pm \frac{1}{2}$ when $N_b$ is even. Unlike the conventional Kondo setup, where a localized magnetic impurity interacts with a non-interacting Fermi sea, or the spin chain Kondo problem in the isotropic and free fermion cases studied in~\cite{kattel2023kondo,kattel2024kondo}, the present case features bulk magnetic order. This makes it harder to identify impurity screening by analyzing the spectrum.
In the isotropic Heisenberg chain, the total spin in the bulk before attaching the impurity is $S^z_j=0$ \textit{i.e.} the z-component of spin identically vanishes in every site when the number of bulk sites is even, and after attaching the impurity, the total spin becomes $S^z=\pm \frac{1}{2}$ \cite{kattel2023kondo} however this total $S^z$ is not due to the impurity spin, but it is rather carried by a spinon induced which could be pointing up or down. The spin of the impurity is completely quenched either by the many-body Kondo effect (for small values of antiferromagnetic impurity couplings), 
{and hence its entropy decreases from $\ln 2$ before coupling to $0$ after it is coupled to the chain with antiferromagnetic coupling. }
In the present case, as discussed before, the bulk has magnetic order \textit{i.e.} $S^z_j\neq 0$, and the ground state is two-fold degenerate. When the magnetic impurity is coupled to this chain with antiferromagnetic coupling, the impurity spin is screened, but this does not mean that the magnetization vanishes at the impurity site, as we shall see later. Before coupling to the bulk, the impurity has $\ln 2$ entropy as it can freely point in either up or down direction, but as it is antiferromagnetically coupled to the XXZ chain with a two-fold degenerate ground state, the impurity spin is quenched such that its entropy decreases to $0$ in each of the degenerate vacua. 
The effect of many-body screening is visible in various ground-state physical quantities. For example, the ratio of the impurity spinon density of states (DOS) to the bulk spinon density $R(E)$ has a characteristic peak as shown in Fig.\eqref{fig:spinon-DOS}.

\begin{figure}
    \centering
\includegraphics[width=\linewidth]{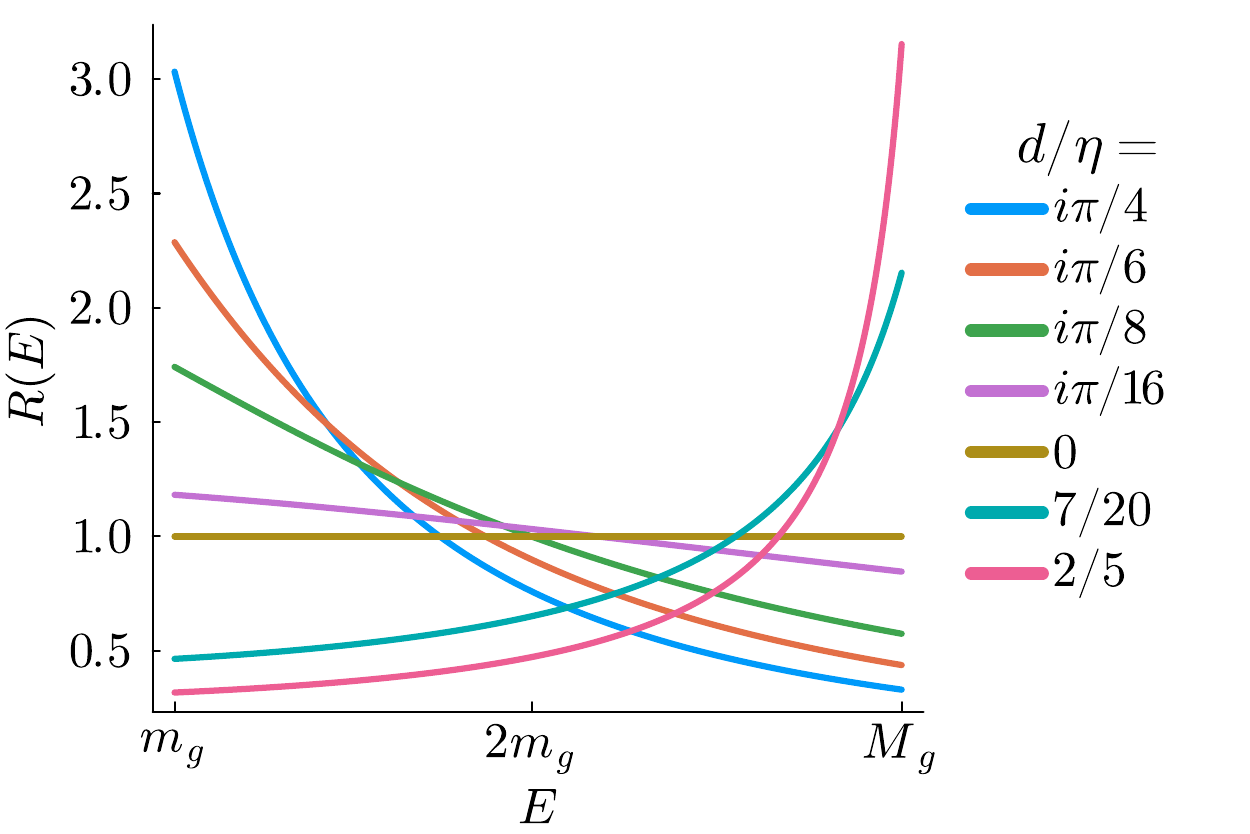}
    \caption{The plot offers a visualization of the spectral
weight of the spinons in the Kondo cloud screening impurity throughout the Kondo phase \textit{i.e.} when $d$ is purely imaginary or when $0<\frac{d}{\eta}<\frac{1}{2}$.
This spectral weight, denoted by the function
$R(E)$ is the $R$ defined in Eq.~\eqref{reeqn} expressed as a function of spinon's energy, is computed using the Bethe
ansatz method. The spinon DOS peaks at $E=m_g$ when $d$ is imaginary, showing that the spinons screening the impurity have low energy. When $d=0$, the impurity becomes part of the bulk, and hence the spinon DOS becomes a flat line showing that the spinons of every energy scale participate in forming the singlet and finally, when $0<\frac{d}{\eta}<\frac{1}{2}$, the spinon DOS peaks at $E=M_g$ showing that the high energy spinon participates in the screening of the impurity.}
    \label{fig:spinon-DOS}
\end{figure}

The ratio of the impurity contribution to the density of states to the bulk contribution to the density is given by
\begin{align}
    R(\lambda)&=\frac{N}{2}\frac{  \left| \frac{\rho_{{0}^{\mathrm{bulk}}}(\lambda)}{E'(\lambda)} \right|}{ \left| \frac{\rho_{{0}^{\mathrm{imp}}}(\lambda)}{E'(\lambda)} \right|}\nonumber\\
    &=\frac{1}{2} \frac{f(\lambda+2id)+f(\lambda-2id)}{f(\lambda)}
\end{align}
where $f(\lambda)=\frac{\vartheta _1^{\prime }\left(0,e^{-\eta }\right) \vartheta _3\left(\frac{\lambda }{2},e^{-\eta }\right)}{\vartheta _2\left(0,e^{-\eta }\right) \vartheta _4\left(\frac{\lambda }{2},e^{-\eta }\right)}$ and $\rho_0^\mathrm{bulk}(\lambda)$ and $\rho_0^\mathrm{imp}(\lambda)$ are the ground state root density in the Kondo phase given in Eq.\eqref{bulkpart} and Eq.\eqref{imppart} respectively. Using the expression for the energy of the spinon Eq.\eqref{spinonengeqn}, we rewrite the above equation in terms of the spinon energy as
\begin{equation}
    R(E(\lambda))=\frac{1}{2} \frac{E(\lambda+2id)+E(\lambda-2id)}{E(\lambda)}.
     \label{reeqn}
\end{equation}
Notice that the positive spectral weight $R(E)$ in Fig.~\ref{fig:spinon-DOS} is not contributed by a single mode but rather by spinons of various energies. When $d/\eta$ is purely imaginary, the maximum spectral weight is towards the low energy spinon with $E\to m_g$ whereas as $d$ becomes real and takes values close to $d/\eta\to \frac{1}{2}$, the spectral weight shifts to spinons with maximum energy $E\to M_g$. As we shall see later in the ABM case, the positive weight comes only from a single energy mode which shows that the screening in that case is by a single particle mode rather than many-body as in this phase.

Likewise, the local magnetization at the impurity site $\sigma^z_L$ in the presence of a global magnetic field given by adding a Zeeman term $-\sum_j h\sigma^z_j$ where $j$ (here runs over all the bulk and impurity sites) to the Hamiltonian Eq.~\eqref{ham1} is a dependent quantity $d$ that does not change when the magnetic field $h$ is varied from $h=0$ and $h=\mathcal{M}_g$ due to the presence of the finite mass gap in the model. However, $\sigma^z_L$ increases smoothly as $h$ increases from the mass gap $h=\mathcal{M}_g$ and reaches the maximum value $0.5$ at $h_c=2J(1+\mathrm{cosh}\eta)$ where the entire spin chain fully polarizes and hence has the total spin $S^z=\frac{N}{2}$~\cite{franchini2017introduction,korepin1997quantum}. For a representative case of $\eta=2$, the impurity magnetization is obtained from DMRG implemented in the ITensors library~\cite{fishman2022itensor} and shown in Fig.~\ref{fig:Kondo-mag}. Notice that this quantity described here is the magnetization at the impurity site where both the impurity and bulk degrees of freedom contribute, and the non-zero value at $h\to 0$ is due to the magnetic order in the bulk of the XXZ chain. The smooth growth of this quantity such that it reaches the maximum value of 0.5 exactly at $h=h_c$ shows that the screening of the impurity is due to many-body effects. When the impurity is screened by a single particle bound mode in the ABM phase, as shown later in Fig.~\ref{fig:bm_mag}, the local magnetization at the impurity site jumps suddenly at $h=h^\star>h_c$. Notice that the entire chain polarizes at $h=h_c$ except this single particle bound mode, and hence, this shows that the screening truly has a single particle effect.
\begin{figure}
    \centering
    \includegraphics[width=\linewidth]{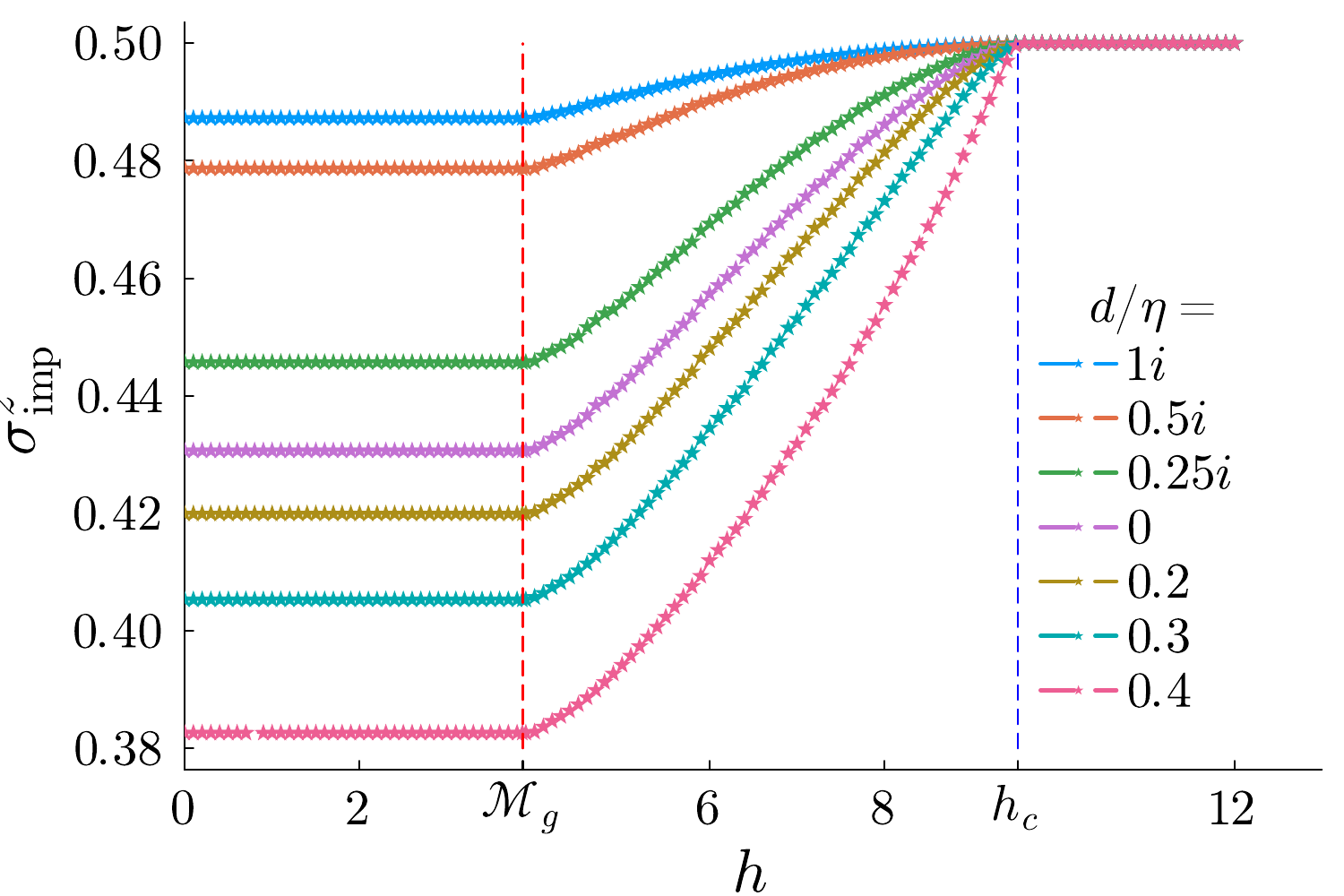}
    \caption{Local impurity magnetization for various values of the impurity coupling parameter $d$ in the Kondo phase when $\eta=2$ and a total number of sites (bulk and impurity) is $N=500$. The impurity magnetization is constant when the external magnetic field evolves from $h=0$ to $h=\mathcal{M}_g$. However, when $h$ changes from $h=\mathcal{M}_g$ (shown in dashed red vertical line) to $h=h_c=2J(1+\mathrm{cosh}(\eta)$ (shown in dashed blue vertical line), the impurity magnetization smoothly changes from some finite $d$ dependent value to $\frac{1}{2}$. The data is obtained using DMRG, where all the calculations are performed by setting the truncation cut-off of the singular values at $10^{-10}$ and performing 100 sweeps to ensure convergence for every data point. }
    \label{fig:Kondo-mag}
\end{figure}

We shall now show some results related to the effect of the impurity in the Kondo phase on the quarter mode and also show the validity of ansatz Eq.\eqref{ansatzvar} for various \(N\) values, $d_R=0$ and various values of $d_L$ in the Kondo phase for the Hamiltonian Eq.~\eqref{ham1}, as shown in the representation in Fig.~\ref{explocalization}, where it fits the data well.

\begin{figure}
    \centering
    \includegraphics[width=\linewidth]{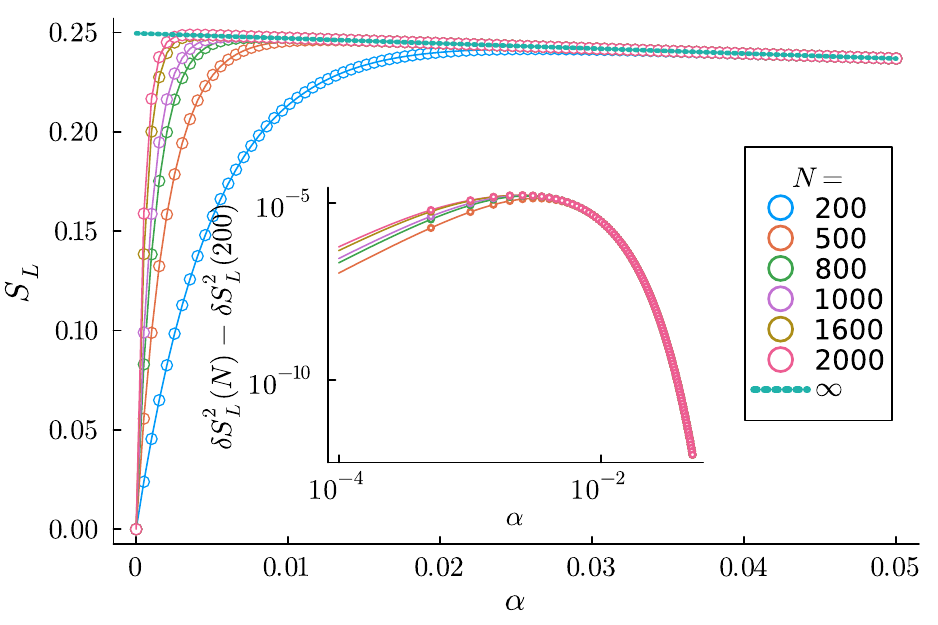}
    \caption{The exponentially localized $+\frac{1}{4}$ edge modes at the left end in one of the two degenerate ground states in the Kondo phase with impurity at the left end $d_L=0.3$ and $d_R=0$ when $\eta=1.75$. The inset shows that the variance fits well with our ansatz Eq.~\eqref{ansatzvar}; thus, it vanishes in the thermodynamic limit. The right edge (not shown in the figure) contains $\mathcal{S}_R=-\frac{1}{4}$ edge mode. Moreover, the other degenerate ground state contains $-\frac{1}{4}$ edge mode at the left and the right edge contains $\frac{1}{4}$.     }
    \label{explocalization}
\end{figure}

We find that irrespective of the value of $d_L$ in the Kondo phase, the two-fold ground state for odd $N_b$ contains the fractional edge modes of opposite polarization \textit{i.e.} $\ket{GS}^O= \left| \pm \frac{1}{4}, \mp \frac{1}{4}\right\rangle$ and for even $N_b$ the ground state have the two-fold degenerate ground state with the edge mode that points in the same direction \textit{i.e.}  $\ket{GS}^E= \left| \pm \frac{1}{4}, \pm \frac{1}{4}\right\rangle$.  

The excited states in the Kondo phase can be constructed
above the ground state by adding an even number of spinons,
bulk strings, quartets, etc.~\cite{destri1982analysis}, which forms a single tower of excited states where the impurity is screened by the Kondo cloud. The schematic of low-lying excited states and their total spin in the $z-$direction is shown in Fig.~\ref{fig:Kexcitedodd}.
\begin{figure}
    \centering
    \includegraphics[width=\linewidth]{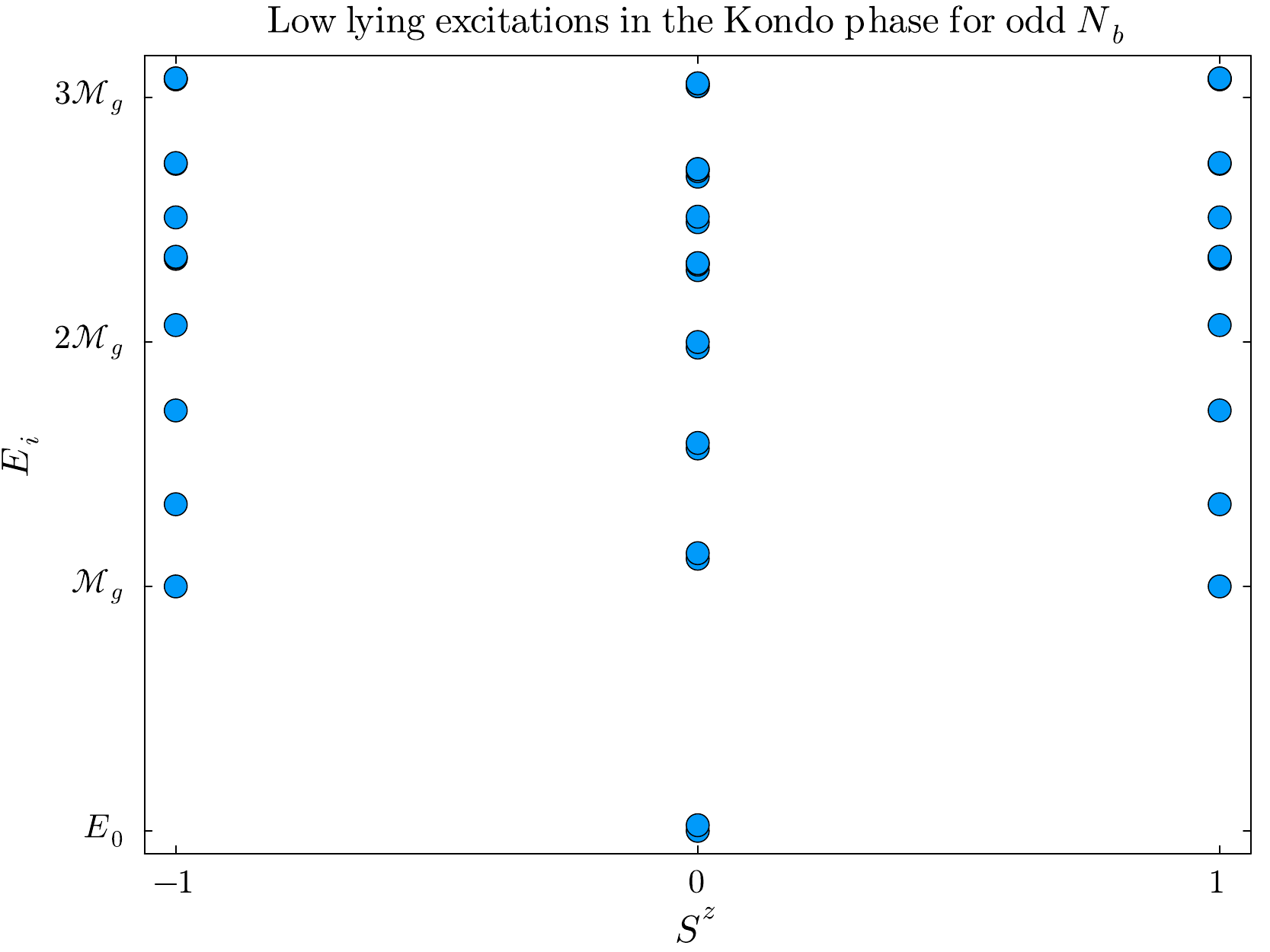}
    \caption{Schematic of low-lying excitation in the Kondo phase showing the two-fold degenerate $S^z=0$ ground state and a mass gap of $\mathcal{M}_g$ for an odd number of total bulk sites $N_b$. The data is obtained from the exact diagonalization of the Hamiltonian with $N_b=11$ such that the finite size effect shows the spitting of the energy of the two vacua, which are degenerate in the thermodynamic limit.   }
    \label{fig:Kexcitedodd}
\end{figure}

\subsubsection{The antiferromagnetic bound mode phase}

When the impurity parameter $d$ takes real values in the range $\frac{\eta}{2}<d<\eta$, the impurity is in the antiferromagnetic bound mode  (ABM) phase as shown in the phase diagram Fig.~\ref{fig:PD1}. In this phase, the impurity is screened by a single particle bound mode exponentially localized at the left edge of the spin chain. 
Thus, the ground state is a two-fold degenerate $S^z=0$ state.

The impurity bound mode is described by a purely imaginary root of the Bethe equation 
\begin{equation}
    \lambda_{d}=\pm i \left(\eta-2d\right),
\end{equation}
and has an energy given by

\begin{align}
    E_{d}&=J \sinh (\eta )\sum_{\omega=-\infty}^\infty e^{-2 \eta  | \omega | } \text{sech}(\eta   \omega  ) \cosh ( \omega   (2 d-\eta ))\nonumber\\
    &+\frac{J \sinh ^2(\eta )}{\sinh (d) \sinh (d-\eta )}.
    \label{Ed_eng}
\end{align}
Such purely imaginary solutions of Bethe Ansatz equations are common in various integrable models and in general, they describe boundary excitations that describe various novel boundary phenomena~\cite{kapustin1996surface,pasnoori2020kondo,pasnoori2021boundary,pasnoori2022rise,kattel2023exact,rylands2020exact,wang1997exact,frahm1997open,kattel2023exact,kattel2023kondo,kattel2024dissipation,kattel2024kondo,Kattel:2024lot}.
\begin{figure}
    \centering
    \includegraphics[width=\linewidth]{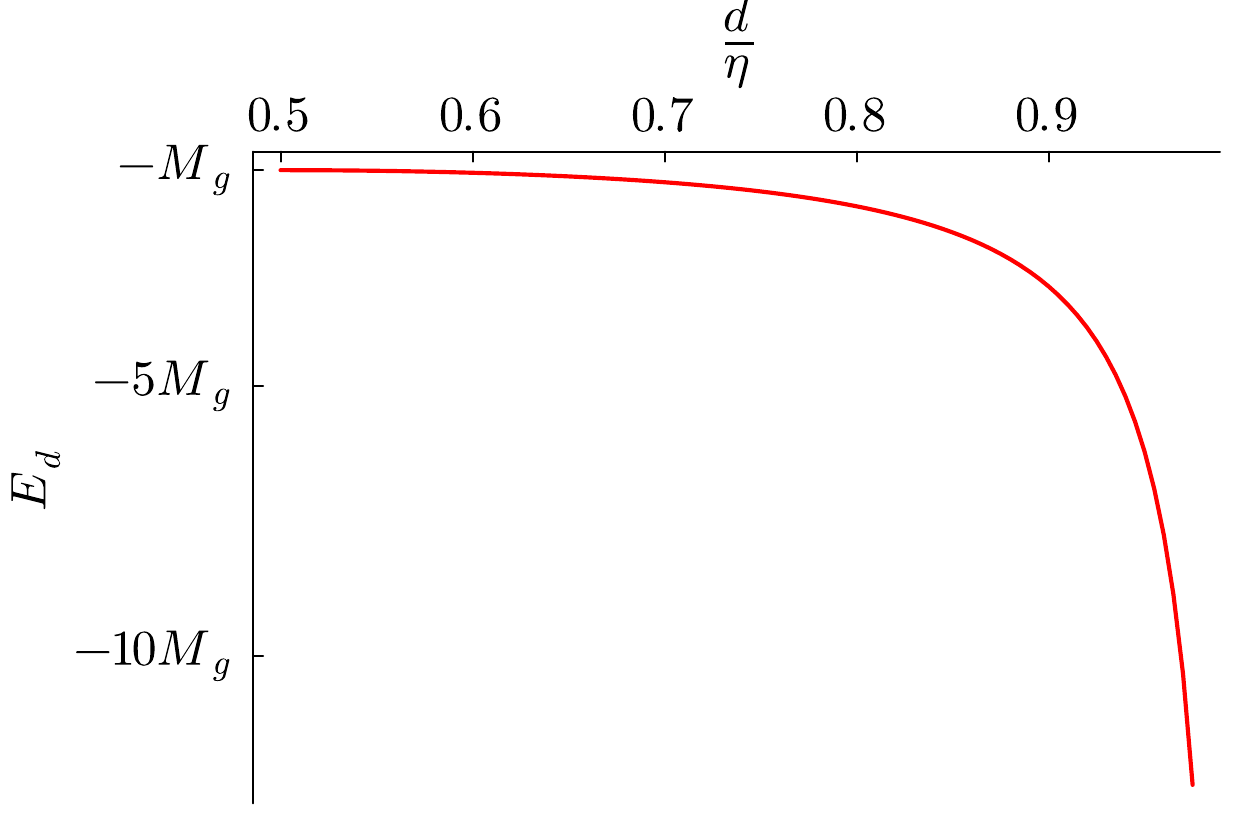}
    \caption{Energy of the impurity bound mode exponentially localized at the left end given by Eq.~\eqref{Ed_eng} is shown in the units of maximum energy of the spinon ${M}_g$  for impurity parameter $\eta=2$. Notice that the maximum value of $E_d$  is equal to the negative of the maximum energy of the spinon and occurs at $d=\frac{\eta}{2}$.}
    \label{fig:Ed_eng}
\end{figure}

As shown in Fig.~\ref{fig:Ed_eng}, the energy of the exponentially localized impurity bound mode is negative and ranges between $-\infty<E_d<-M_g$ in the antiferromagnetic bound mode regime. Thus, the ground state contains this root of the Bethe Equation on top of all other real roots and the trivial boundary string solution of the form 
\begin{equation}
    \lambda_{bs}=\pi \pm i\eta,
\end{equation}
which has zero energy \textit{i.e.} $E_{bs}=0$ (See Appendix \ref{kondo-bm-dets} for details). This solution plays an important role in describing the edge localized quarter modes as shown in~\cite{pasnoori2023spin,kattel2024XXZ-S}.

In this regime, the ratio of the impurity to the bulk density of states becomes
\begin{equation}
    R_d(E)=R(E)+\delta(E-E_d),
\end{equation}
where $R(E)$ given by Eq.~\eqref{reeqn} is negative in this phase. The only positive contribution to the spectral weight comes from the second term, which is due to the impurity boundary string. This demonstrates that the impurity is screened by a single particle bound mode with energy $E_d$ in this regime. 

Now, we shall study the model in the presence of a global magnetic field term $-h\sum_j\sigma^z_j $ where $j$ runs over all sites, including the bulk and impurity sites. Unlike in the Kondo phase, there is a characteristic jump in the local magnetization at the impurity site in the antiferromagnetic bound mode phase. Since the energy of the bound mode is larger than the energy of the spinons in the bulk, when the magnetic field reaches $h_c$ where all the bulk degrees of freedom are fully polarized, the spin at the impurity site is not yet fully polarized. 
When the magnetic field matches the energy of the bound mode in the presence of the field, which differs from the zero-field energy given in Eq.\eqref{Ed_eng}, it disrupts the singlet formed between the bound mode and the impurity. This causes the impurity magnetization to abruptly jump to 0.5. However, we do not have an analytic expression for the bound mode energy in the presence of the magnetic field. As shown in Fig.~\ref{fig:bm_mag}, the magnetization is some $d$ dependent value when $h=0$ which does not change when $h$ increases to the mass gap $h=\mathcal{M}_g$. However, between $h=\mathcal{M}_g$ and $h=h_c$, the impurity magnetization increases smoothly, but it does not reach $0.5$ at $h_c$ like in the Kondo phase. As the magnetic field is further increased, the local magnetization at the impurity site remains constant up until $h=h^\star(d)$, at which point the singlet is polarized, and the magnetization abruptly jumps to $0.5$.
\begin{figure}
    \centering
    \includegraphics[width=\linewidth]{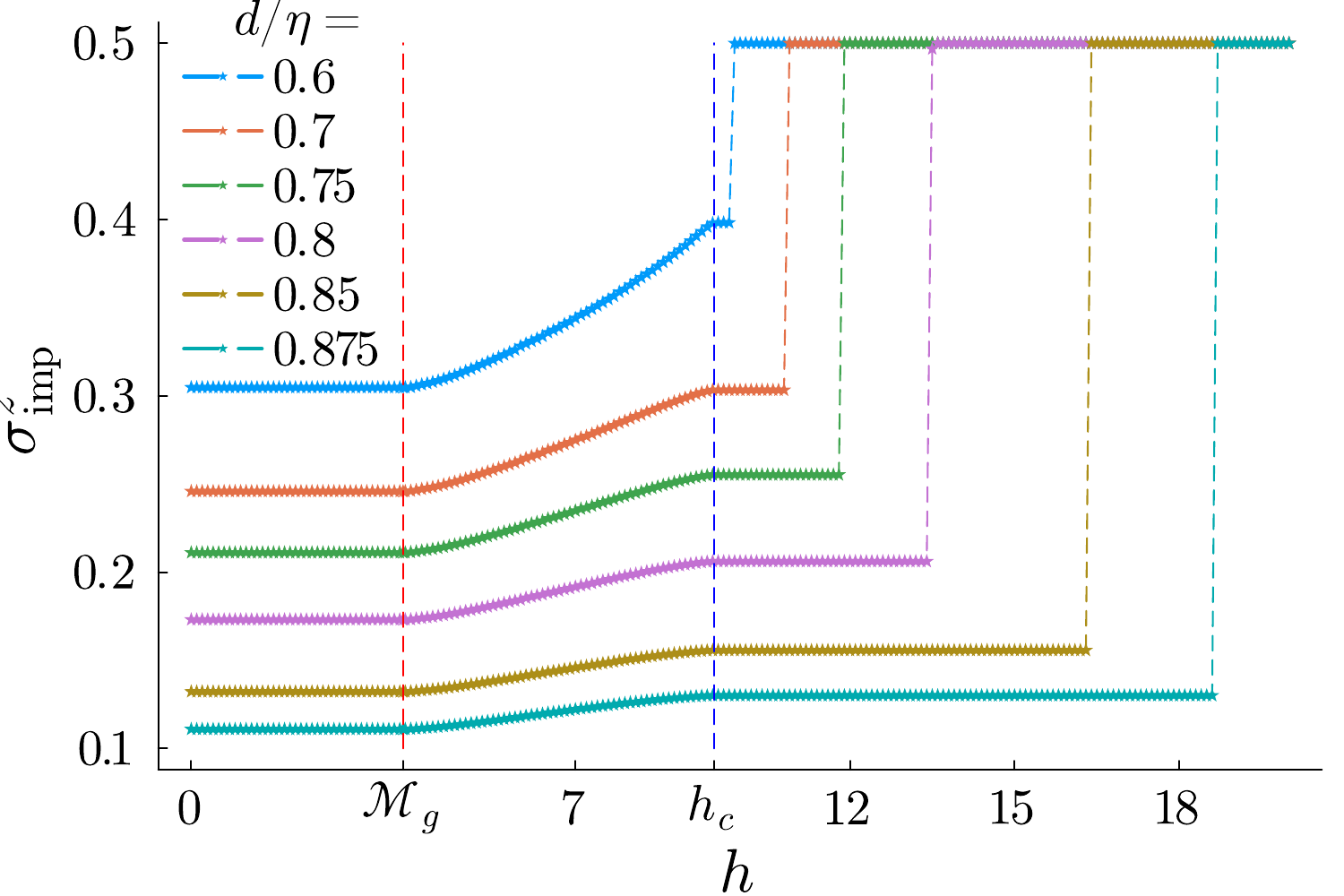}
    \caption{Local impurity magnetization for various values of impurity parameter in the antiferromagnetic bound mode phase when $\eta=2$ for a total number of sites (bulk and impurity) $N=500$. The impurity magnetization is constant when the external magnetic field evolves from $h=0$ to $h=\mathcal{M}_g$ due to the presence of the mass gap in the spectrum. However, when $h$ changes from $h=\mathcal{M}_g$ (shown by vertical red dashed line) to $h=h_c=2J(1+\mathrm{cosh}(\eta))$ (marked by a vertical dashed blue line), the impurity magnetization smoothly changes from some finite $d$ dependent value to some other $d$ value at $h=h_c$ where the local magnetization is smaller than 0.5. When $h$ is further increased, the impurity magnetization remains constant up until a critical value $h^\star(d)$, where it abruptly jumps to 0.5. The data is obtained by using DMRG, where all calculations are performed by setting truncation cut-off of the singular values at $10^{-10}$ and performing 100 sweeps to ensure convergence for every data point.
    }
    \label{fig:bm_mag}
\end{figure}

\begin{figure}
    \centering
\includegraphics[width=\linewidth]{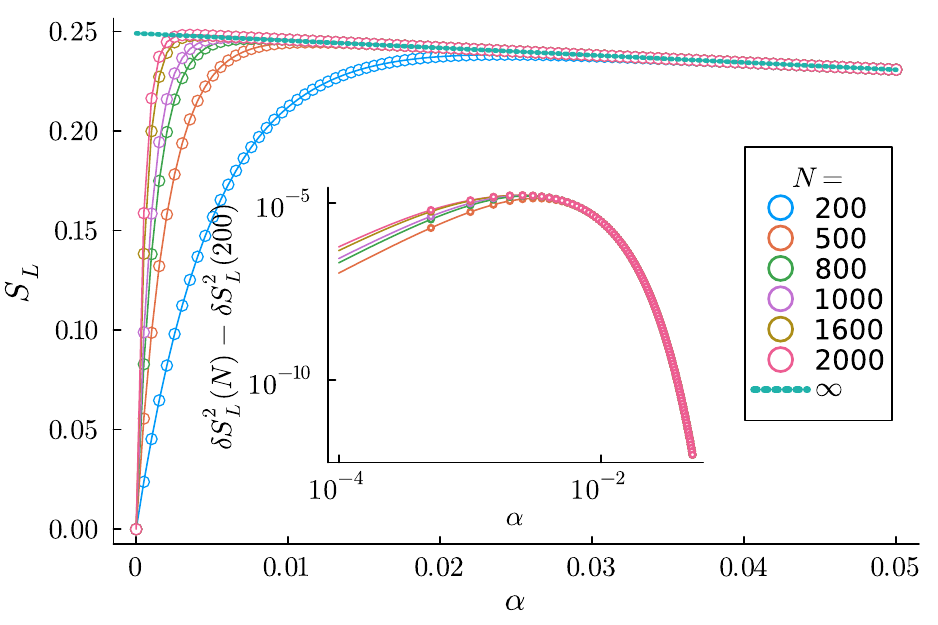}
    \caption{The exponentially localized $+\frac{1}{4}$ edge modes at the left end in one of the two degenerate ground states in the ABM phase with impurity at the left end $d_L=1.4875$ and $d_R=0$ when $\eta=1.75$. The inset shows that the variance vanishes in the thermodynamic limit as the data fits well with ansatz Eq.\eqref{ansatzvar}. The right edge (not shown in the figure) contains $\mathcal{S}_R=-\frac{1}{4}$ edge mode. Moreover, the other degenerate ground state contains $-\frac{1}{4}$ edge mode at the left and the right edge contains $\frac{1}{4}$. }
    \label{fig:Leftacc-ABM}
\end{figure}

Just like in the Kondo phase, we find that irrespective of the value of $d_L$ within the antiferromagnetic bound mode phase, the two-fold ground state for odd $N_b$ contains the fractional edge modes of opposite polarization \textit{i.e.} $\ket{GS}^O= \left| \pm \frac{1}{4}, \mp \frac{1}{4}\right\rangle$ and for even $N_b$ the ground state is two-fold degenerate with an edge mode that points in the same direction \textit{i.e.}  $\ket{GS}^E= \left| \pm \frac{1}{4}, \pm \frac{1}{4}\right\rangle$.
A representative case of the spin accumulation in the left end for $\eta=1.5$ and $d=1.4875$ is shown in Fig.~\ref{fig:Leftacc-ABM}. 

\begin{figure}
    \centering
    \includegraphics[width=\linewidth]{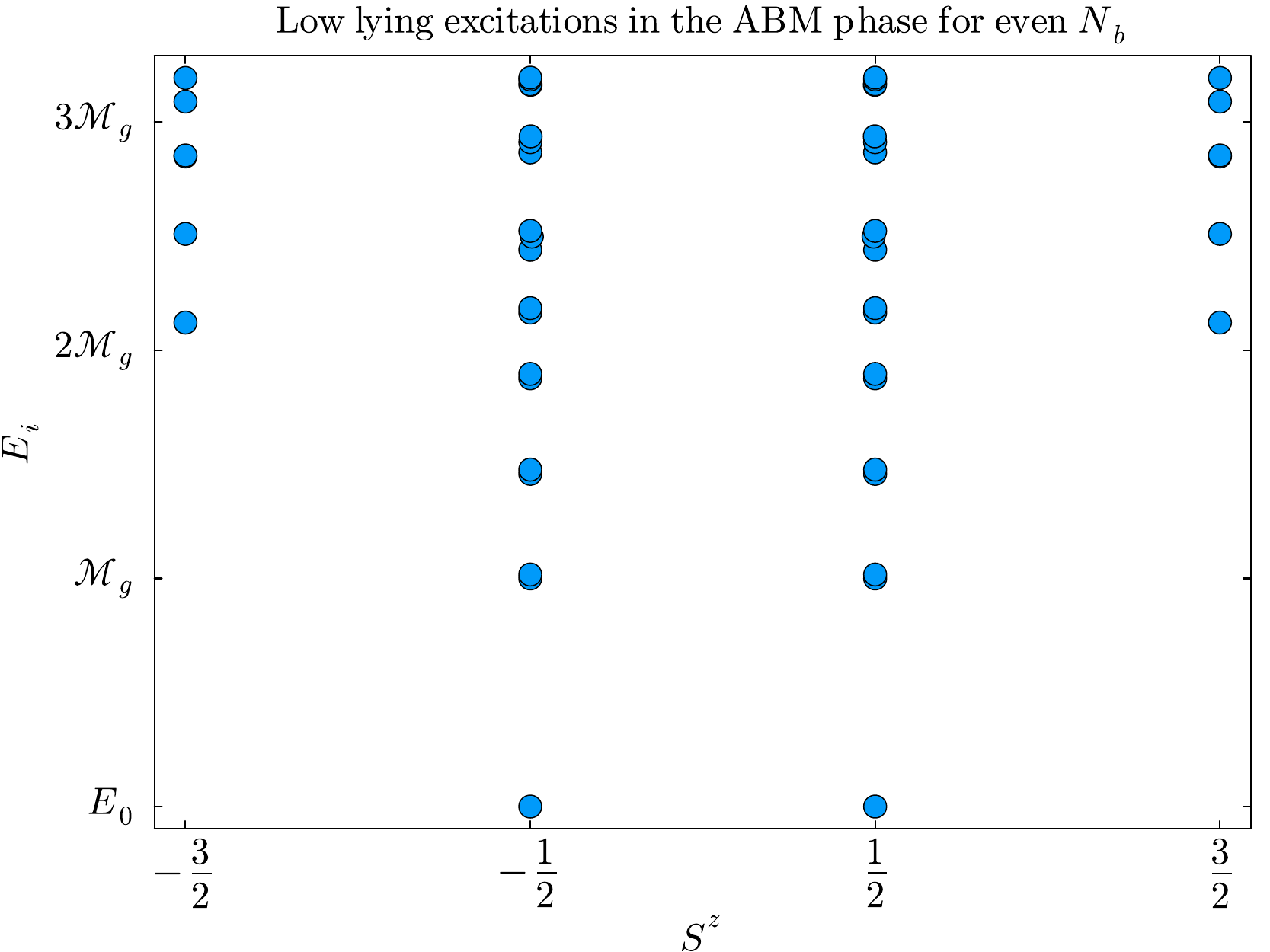}
    \caption{Schematic of low-lying excitation in the antiferromagnetic bound mode phase showing the two-fold degenerate $S^z=\pm \frac{1}{2}$ ground state and a mass gap of $\mathcal{M}_g$ for even number of total bulk sites $N_b$. The data is obtained from the exact diagonalization of the Hamiltonian with $N_b=10$.}
    \label{fig:abm-exstates}
\end{figure}

The bulk excitations are constructed as usual by adding an even number of spinons,
bulk strings, quartets, etc. However, there is a unique boundary excitation which is constructed by removing the boundary string solution $\lambda_d$. Removal of the impurity string costs an energy $E_d$ given by Eq.~\eqref{Ed_eng}, which is always larger than the maximum energy of a single spinon $M_g$.  The unscreened state {(US), denoted} $\ket{US}$ obtained by removing the impurity string is four-fold degenerate in the thermodynamic limit when the total number of bulk sites $N_b$ is even with spin $S^z=\pm \frac{1}{2}$. But when the total number of bulk sites $N_b$ is odd, the unscreened state $\ket{US}$ is four-fold degenerate with spin $S^z=\pm 1$ and two $S^z=0$ states. Fig.~\ref{fig:abm-exstates} shows a schematic of the two-fold degenerate ground state and low-lying excitations with mass gap $\mathcal{M}_g$. The state with unique boundary excitations where impurities are screened is possible above $E=2M_g$, which is not explicitly shown in the schematic.

\subsection{Ferromagnetic boundary coupling}
When the impurity parameter takes real values in the range $d>1$, the impurity coupling becomes ferromagnetic such that the impurity exhibits two distinct boundary phases: the ferromagnetic bound mode (FBM) phase and the unscreened phase, as shown in the phase diagram Fig.~\ref{fig:PD1}. When the impurity parameter $d$ takes value in the range $1<d<\frac{3}{2}$, the impurity is unscreened in the ground state, but there exists a high energy single particle bound mode in the spectrum which screens the impurity; therefore, we name it \textit{ferromagnetic bound mode phase}. Finally, when the impurity parameter takes the values in the range $d>\frac{3}{2}$, the impurity can no longer be screened such that we call it the \textit{unscreened phase}. 

In the ground state of both the ferromagnetic bound mode phase and unscreened phase, the unscreened impurity at the edge truly behaves as a free spin. Just like a free spin with $\ln 2$ entropy, the impurity integrably coupled to the antiferromagnetic gapped XXZ chain with ferromagnetic coupling also has $\ln 2$ entropy as it could align or anti-align with the quarter mode at the edge as both of them result in the state with equal energies in the thermodynamic limit. When it anti-aligns with the quarter edge mode, it effectively flips the sign of the exponentially localized quarter mode, whereas when it aligns with the edge mode, it forms an effective three-quarter edge mode pointing in the same direction. This seemingly surprising result is proven via Bethe Ansatz by explicitly constructing this four-fold degenerate ground state as discussed in Appendix~\ref{BAdets}, and here we shall provide an independent numerical proof of four-fold degeneracy from exact diagonalization as shown in Fig.~\ref{fig:fbm-exstates} and Fig.~\ref{fig:us-exstates} for odd and even number of bulk sites respectively.

 However, we can easily understand the reason for the four-fold degeneracy once we notice that this behavior is consistent with the conventional Kondo problem where the magnetic defect is coupled ferromagnetically, the perturbative ferromagnetic Kondo coupling flows to the trivial IR fixed point where the impurity is not screened, and hence it has $\ln 2$ entropy. The existence of the four-fold degenerate ground state also proves that the physics of the impurity attached to the XXZ chain is not a simple two-particle physics between the effective 1/4 mode and the spin-$\frac{1}{2}$ impurity as a two-particle Hamiltonian with ferromagnetic coupling would favor only spin alignment and the state where the two anti-align would be a high energy state. 

 As before, the distinction between the two phases will be discussed in much detail later. Here we shall focus on the aspects of the impurity physics that are common to both the FBM phase and the unscreened phase.

\subsubsection{Interplay between the edge modes and the ferromagnetically coupled impurity}

\begin{figure}
  \centering
  \begin{subfigure}{0.49\columnwidth}
    \centering
\includegraphics[width=\columnwidth]{updnres1.png}
  \end{subfigure}
  %\vspace{1em} % Add some vertical space between subfigures
  \begin{subfigure}{0.49\columnwidth}
    \centering
    \includegraphics[width=\columnwidth]{dnupres1.png}
  \end{subfigure}

{\color{white}{-------------------------}}\\
{\color{white}{-------------------------}}
  
    \begin{subfigure}{0.49\columnwidth}
    \centering
\includegraphics[width=\columnwidth]{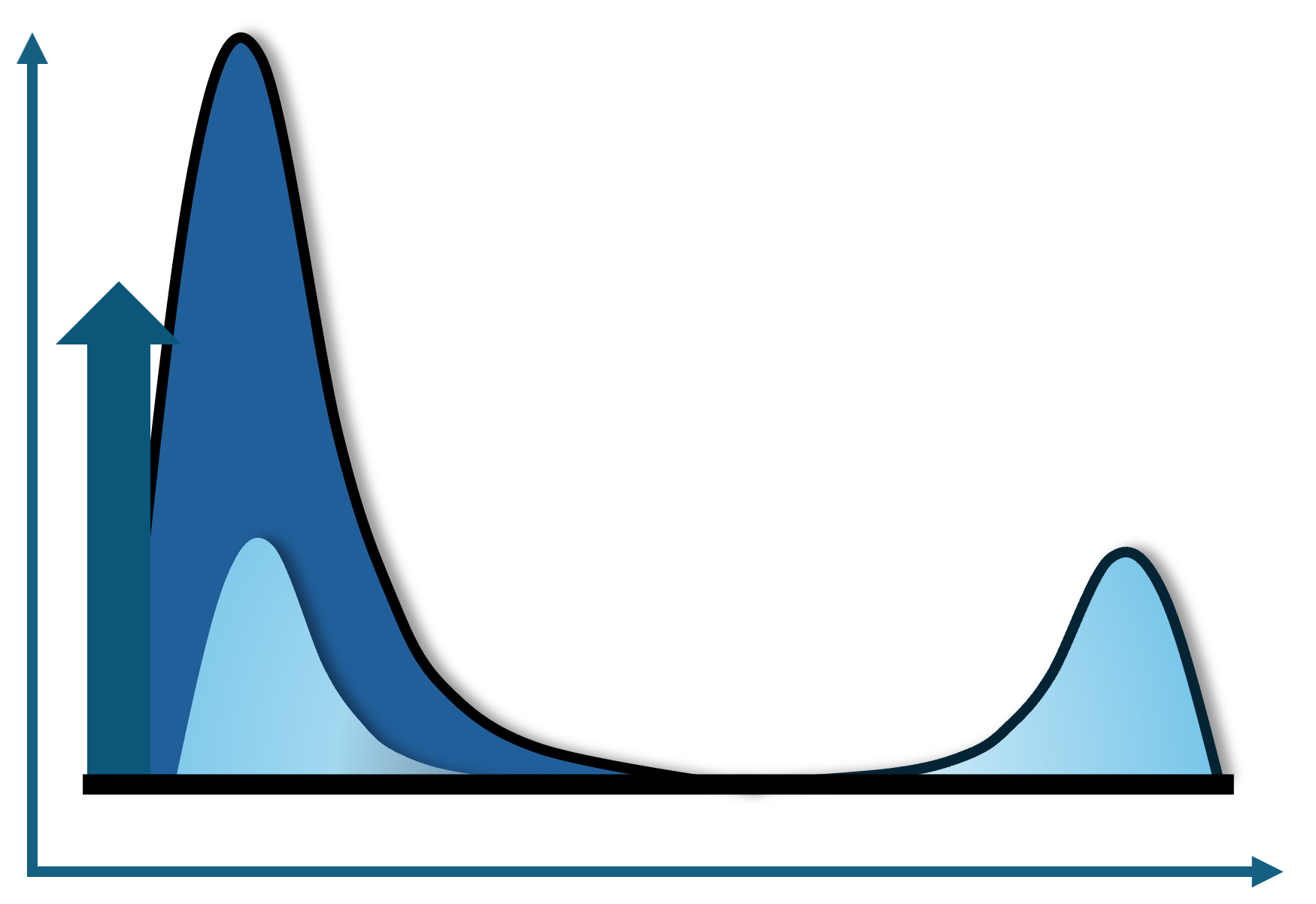}
  \end{subfigure}
  %\vspace{1em} % Add some vertical space between subfigures
  \begin{subfigure}{0.49\columnwidth}
    \centering
    \includegraphics[width=\columnwidth]{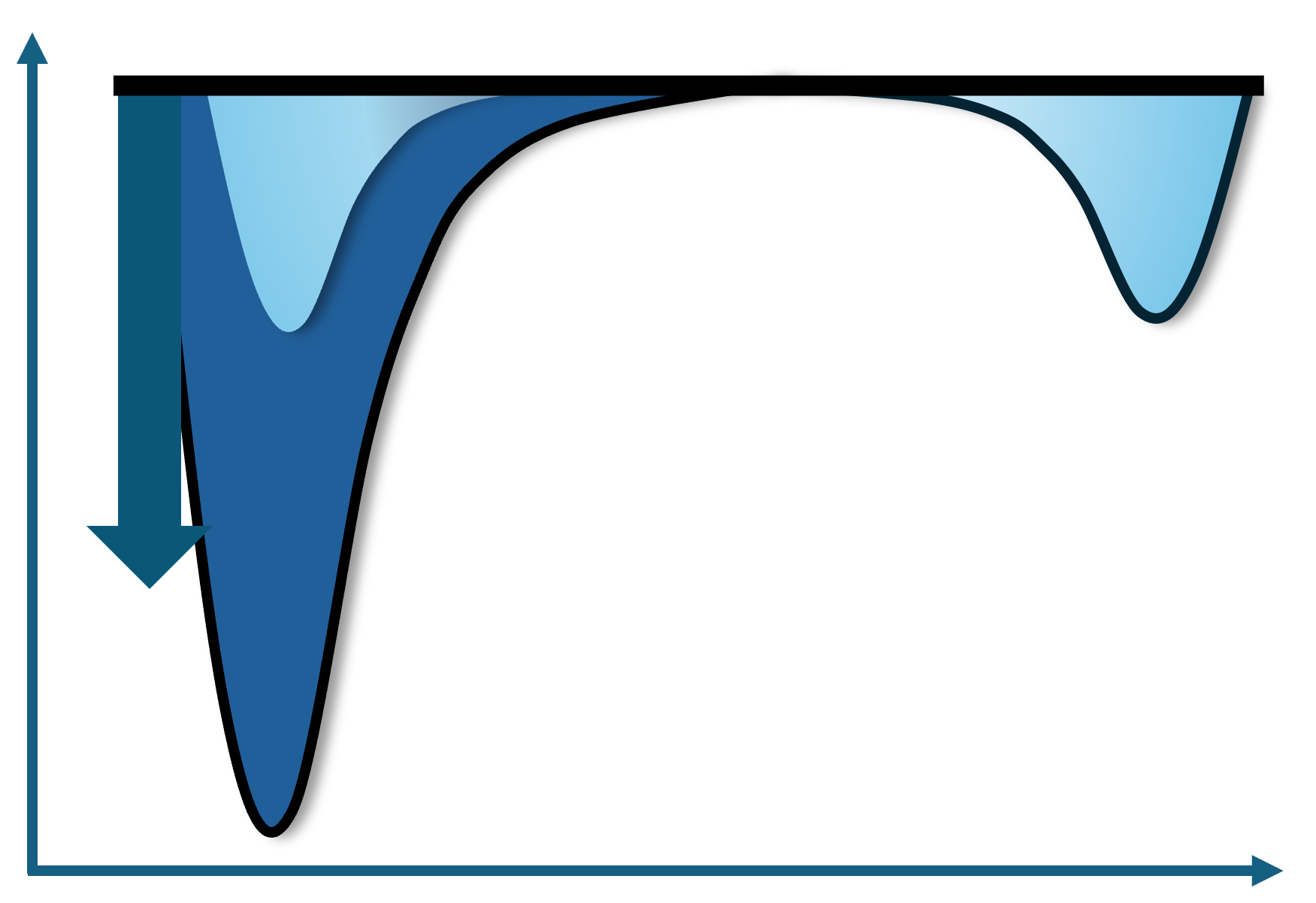}
  \end{subfigure}
  \begin{tikzpicture}[overlay, remember picture]
    \node at (-2.25,0.2) {sites};
    \node at (2.25,0.2) {sites};
    \node at (-3.8,0.2) {$L$};
    \node at (-3.4,0.2) {$1$};
    \node at (-0.45,0.2) {$N$};
    \node[black] at (0.4,0.2) {$L$};
    \node[black] at (0.8,0.2) {$1$};
    \node[black] at (3.85,0.2) {$N$};
    \node[black] at (-0.125,2.95) {$d)$};
\node[black] at (-0.125,6.0) {$b)$};
\node[black] at (-4.5,2.95) {$c)$};
\node[black] at (-4.5,6.0) {$a)$};
\node[black] at (-4.125,6.4) {$\braket{S^z}$};
\node[black] at (0.3,3.35) {$\braket{S^z}$};
\node[black] at (-4.125,3.35) {$\braket{S^z}$};
\node[black] at (0.3,6.4) {$\braket{S^z}$};
\node[black] at (-3.65,5.5) {$\frac{1}{4}$};
\node[black] at (-3.45,4.775) {-$\frac{1}{4}$};
\node[black] at (-0.6,4.775) {-$\frac{1}{4}$};
\node[black] at (-0.65,1.05) {$\frac{1}{4}$};
\node[black] at (3.775,5.5) {$\frac{1}{4}$};
\node[black] at (0.7,4.925) {-$\frac{1}{4}$};
\node[black] at (0.9,5.625) {$\frac{1}{4}$};
\node[black] at (3.775,2.65) {-$\frac{1}{4}$};
\node[black] at (-3.5,2.35) {{\color{white}{$\frac{3}{4}$}}};
\node[black] at (0.95,1.25) {{\color{white}{-$\frac{3}{4}$}}};
\node[black] at (0.925,2.75) {-$\frac{1}{4}$};
\node[black] at (-3.5,1.05) {$\frac{1}{4}$};
  \end{tikzpicture}
  \caption{The four-fold degenerate ground state a) and b) with total spin $S^z=0$, c) with $S^z=1$, and d) with total spin $S^z=-1$. In a) and b), the impurity anti-aligns with the impurity, thereby effectively flipping the sign of the edge mode, just like in the case of antiferromagnetic boundary coupling explained earlier. In these cases, the faint white color quarter mode that exists in the UV and the spin-$\frac{1}{2}$ impurity at the edge combine to form a new effective quarter mode pointing in the opposite direction shown in the sky blue color.
  However, in c) and d), the impurity and quarter edge modes align to form an effective three-quarter mode at the left end of the chain. In c) and d), the sky blue colored $\frac{1}{4}$ edge mode that exists in the $UV$ combines with the impurity with spin-$\frac{1}{2}$ shown as an arrow to form an effective three-quarter mode shown in the dark blue color. 
  }
    \label{fig:spin0and1only}
  \end{figure}
The interplay between the quarter edge mode and unbounded free spin-$\frac{1}{2}$ impurity depends on the parity of the total number of sites. Let us first consider a single impurity coupled to the left end of an XXZ chain with an odd number of bulk sites, where the ground state is denoted as $\ket{GS}^O\equiv \left| \pm \frac{1}{4}, \pm \frac{1}{4}\right\rangle$, as illustrated in Fig.~\ref{fig:updn-dnup}. The impurity spin can now align or anti-align with the quarter mode, thereby forming a four-fold degenerate ground state $\ket{GS}^E_{FM}\equiv \left| \pm \frac{1}{4}, \mp \frac{1}{4}\right\rangle$ and $\left| \pm \frac{3}{4}, \pm \frac{1}{4}\right\rangle$ as shown in Fig.~\ref{fig:spin0and1only}. 

As mentioned earlier, the impurity behaves as a completely free spin-$\frac{1}{2}$ impurity with $\ln 2$ entropy, and hence, it can point either in an up or down direction without costing any energy. Let us now consider a single impurity coupled to the left end of an XXZ chain with an even number of bulk sites, where the ground state is denoted as $\ket{GS}^E\equiv \left| \pm \frac{1}{4}, \mp \frac{1}{4}\right\rangle$, as illustrated in Fig.~\ref{fig:updn-dnup}. The impurity spin can now align or anti-align with the quarter mode thereby forming a four-fold degenerate ground state $\ket{GS}^O_{FM}\equiv \left| \mp \frac{1}{4}, \mp \frac{1}{4}\right\rangle$ and $\left| \pm \frac{3}{4}, \mp \frac{1}{4}\right\rangle$ as shown in Fig.~\ref{fig:spin12only}.

\begin{figure*}
\centering
    \begin{subfigure}[b]{0.47\textwidth}
        \includegraphics[width=\textwidth]{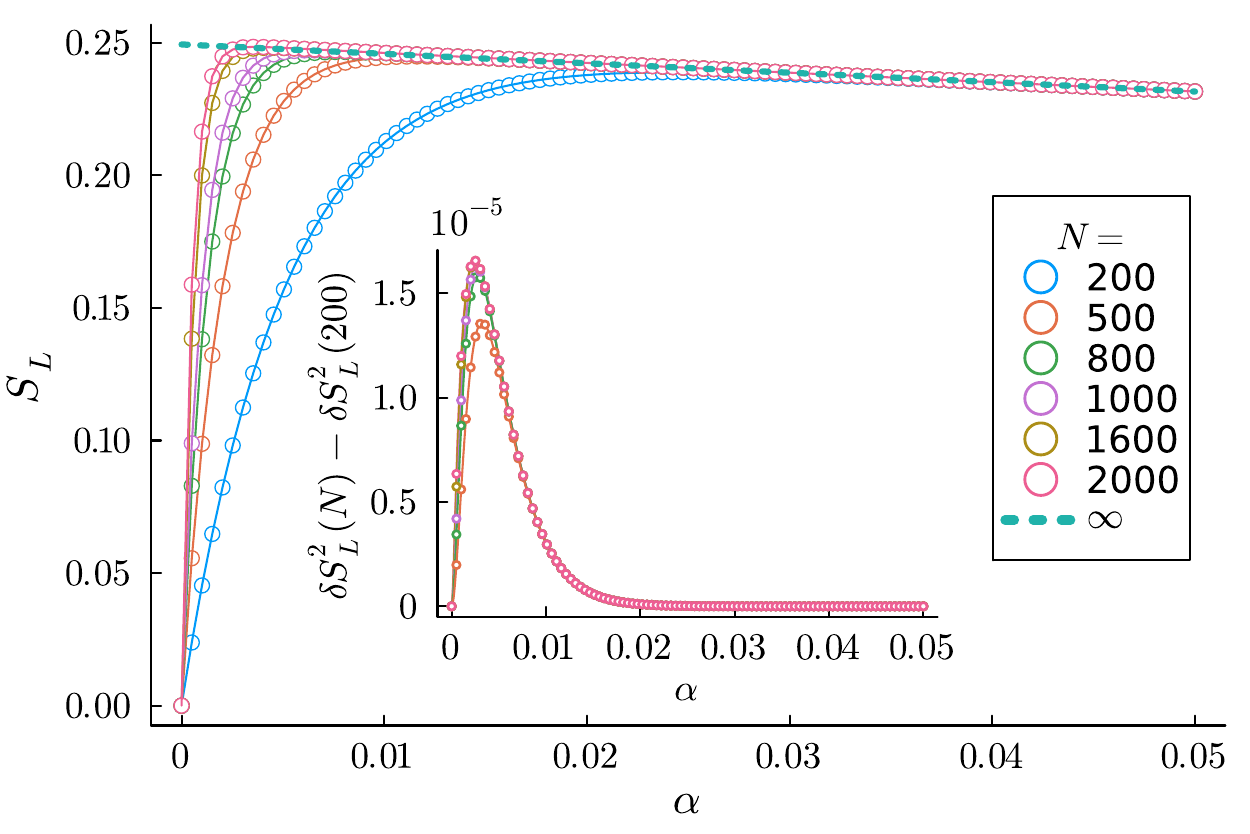}
        \label{fig:sub1}
    \end{subfigure}
    \begin{subfigure}[b]{0.47\textwidth}
        \includegraphics[width=\textwidth]{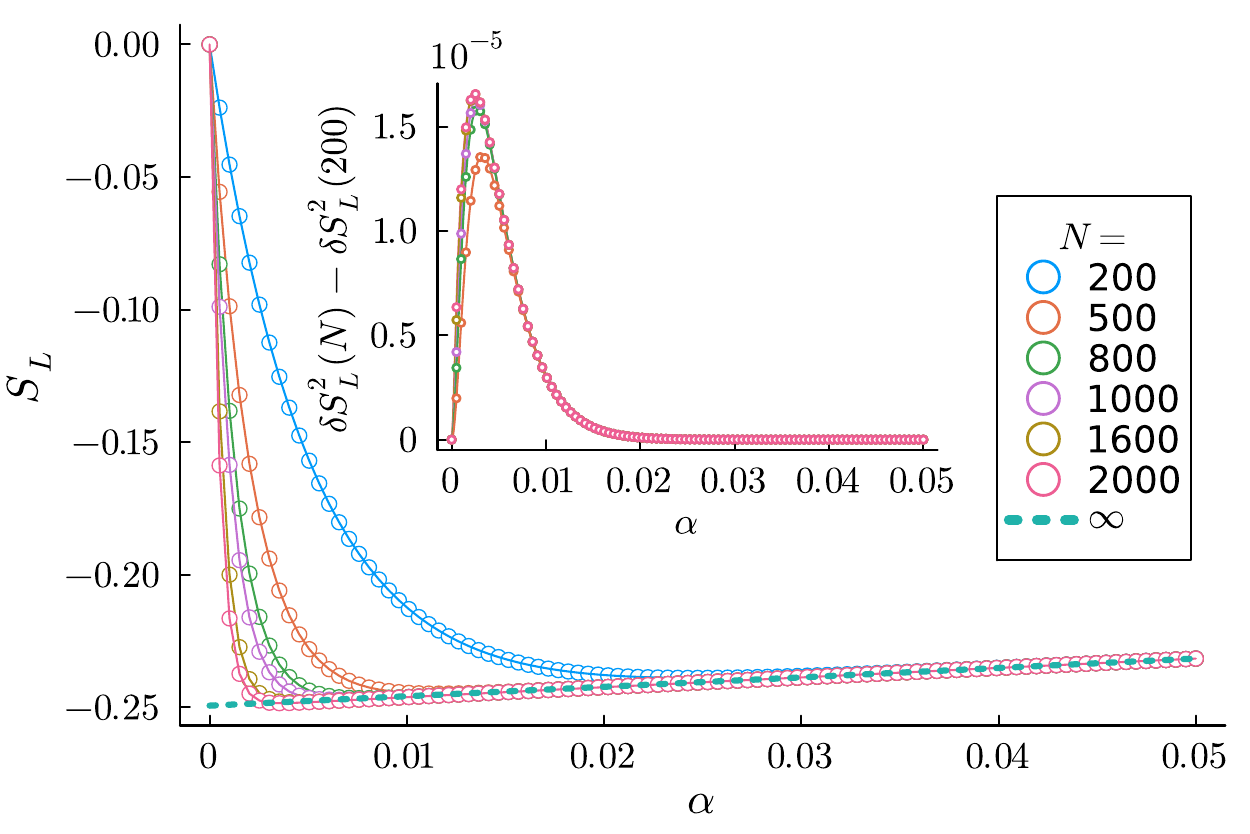}
        \label{fig:sub2}
    \end{subfigure}

    \begin{subfigure}[b]{0.47\textwidth}
        \includegraphics[width=\textwidth]{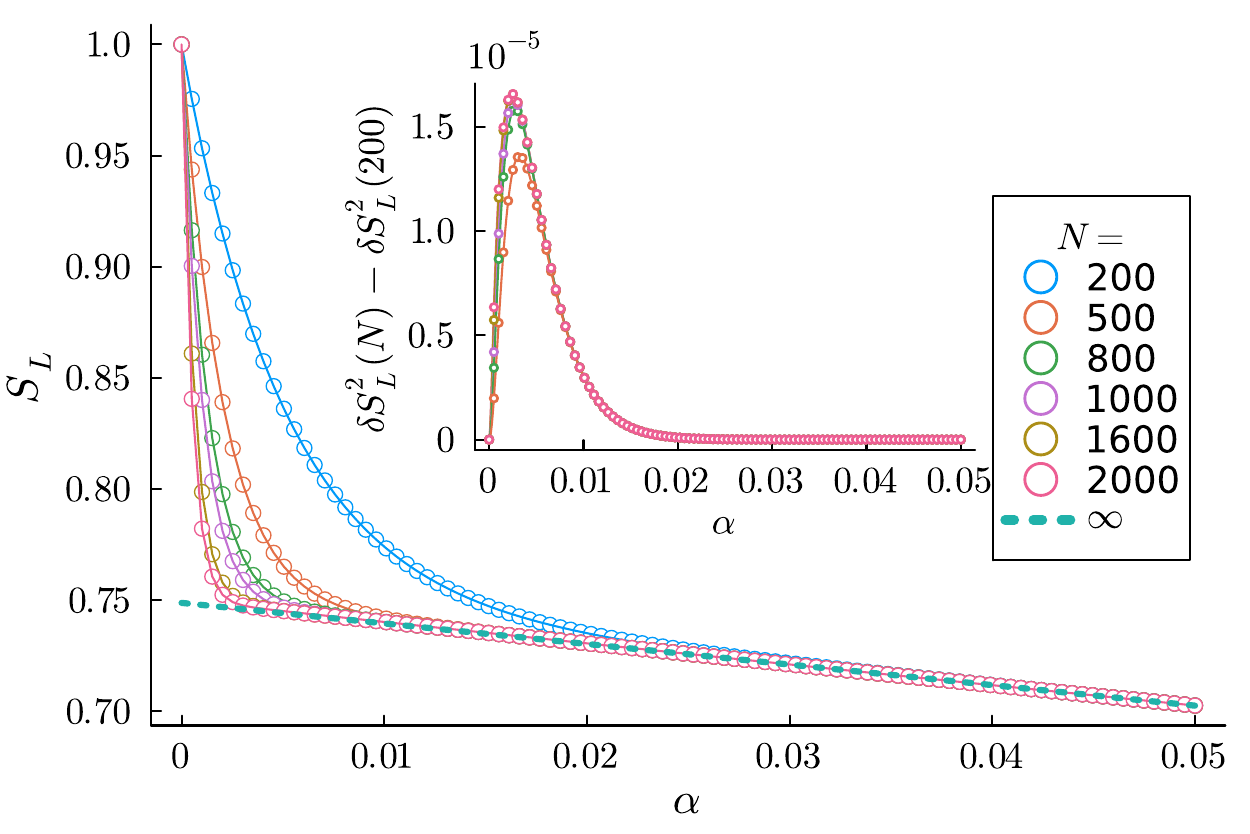}
        \label{fig:sub3}
    \end{subfigure}
    \begin{subfigure}[b]{0.47\textwidth}
        \includegraphics[width=\textwidth]{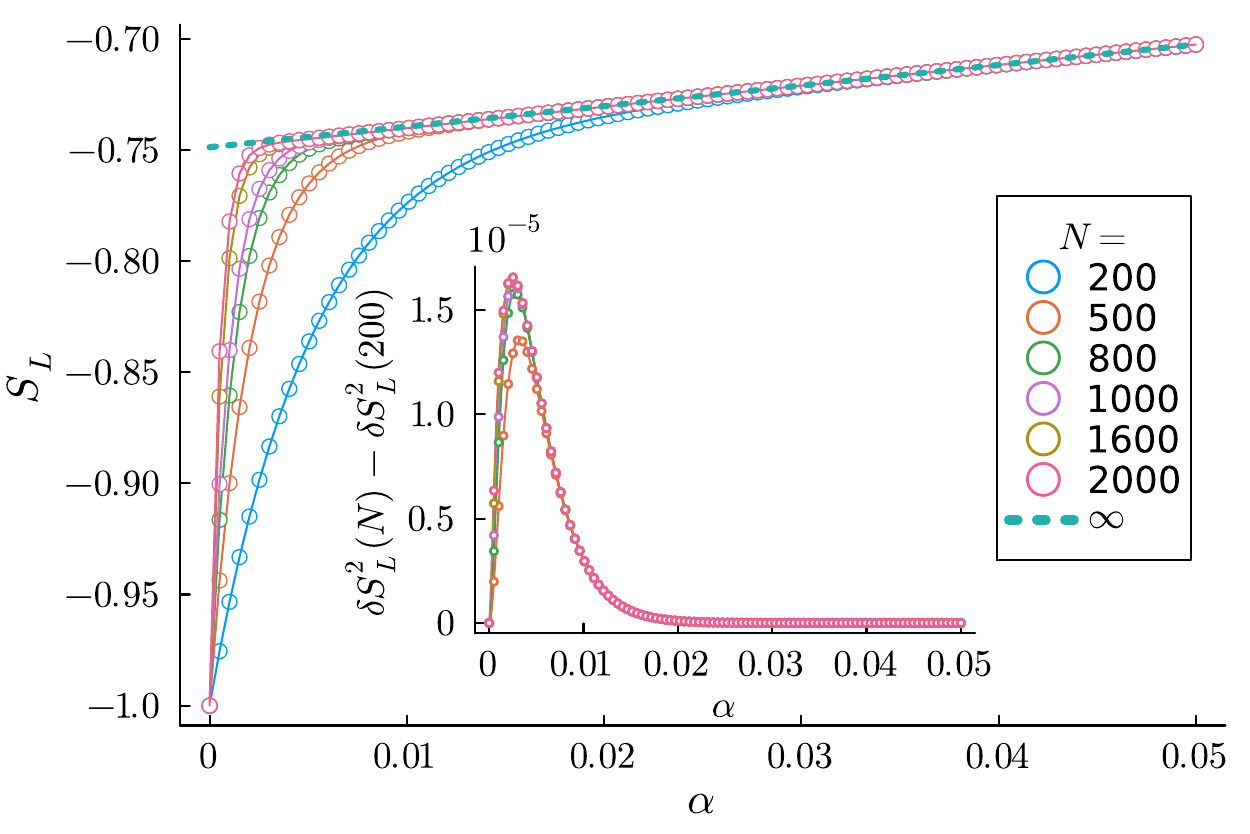}
        \label{fig:sub4}
    \end{subfigure}
    
    \caption{The left fractional spin accumulation on the four-fold degenerate ground state when $N_b$ is odd. When the quarter mode at the edge of the spin chain and the impurity spin anti-align, there is an effective quarter mode left at the edges, whereas when the quarter mode and the impurity spin align, it forms a three-quarter edge localized mode at the edges. The main inset shows the fit of the variance with the ansatz Eq.\eqref{ansatzvar}, and the side inset shows the spin deviation from the bulk antiferromagnetic order $\sigma$ at the left edge for $N=200$.
    }
    \label{fig:odd4acc}
\end{figure*}

\begin{figure}
  \centering
  \begin{subfigure}{0.49\columnwidth}
    \centering
\includegraphics[width=\columnwidth]{upupres1.png}
  \end{subfigure}
  %\vspace{1em} % Add some vertical space between subfigures
  \begin{subfigure}{0.49\columnwidth}
    \centering
    \includegraphics[width=\columnwidth]{dndnres1.png}
  \end{subfigure}

{\color{white}{-------------------------}}
  
    \begin{subfigure}{0.49\columnwidth}
    \centering
\includegraphics[width=\columnwidth]{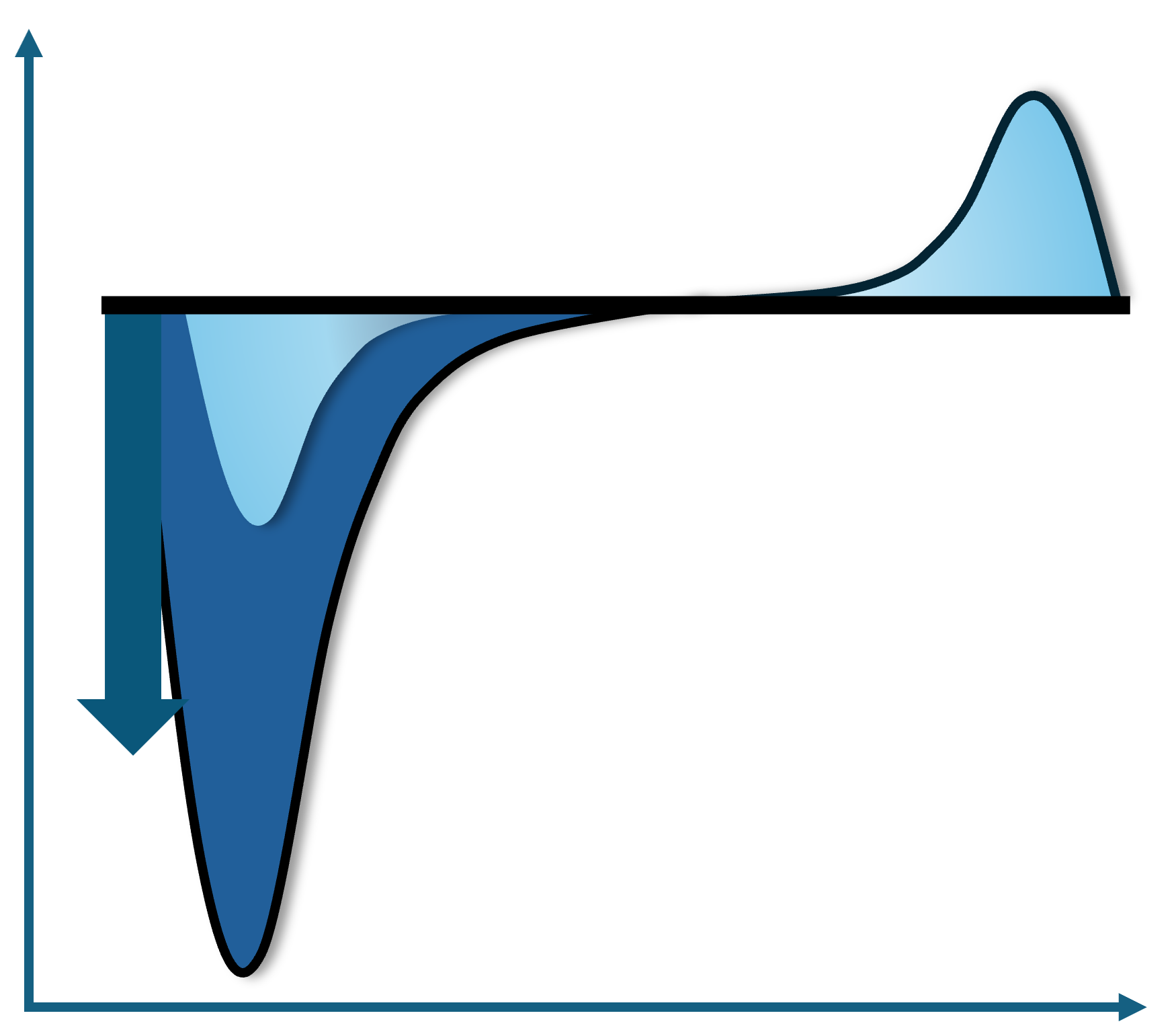}
  \end{subfigure}
  %\vspace{1em} % Add some vertical space between subfigures
  \begin{subfigure}{0.49\columnwidth}
    \centering
    \includegraphics[width=\columnwidth]{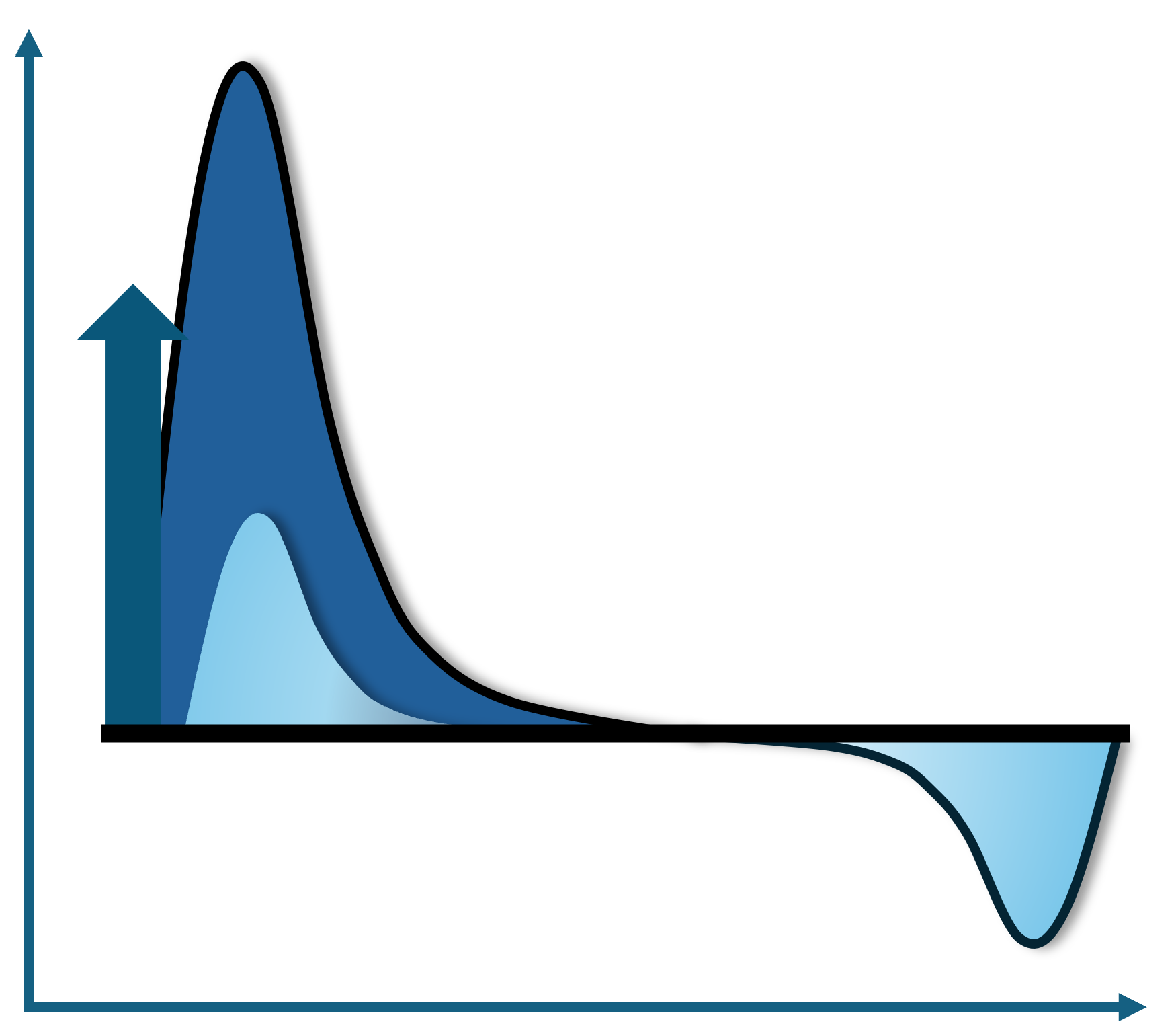}
  \end{subfigure}
  \begin{tikzpicture}[overlay, remember picture]
    \node at (-2.25,0.2) {sites};
    \node at (2.25,0.2) {sites};
    \node at (-3.8,0.2) {$L$};
    \node at (-3.4,0.2) {$1$};
    \node at (-0.45,0.2) {$N$};
    \node[black] at (0.4,0.2) {$L$};
    \node[black] at (0.8,0.2) {$1$};
    \node[black] at (3.85,0.2) {$N$};
    \node[black] at (-0.125,3.85) {$d)$};
\node[black] at (-0.125,6.4) {$b)$};
\node[black] at (-4.5,3.85) {$c)$};
\node[black] at (-4.5,6.4) {$a)$};
\node[black] at (-4.125,6.8) {$\braket{S^z}$};
\node[black] at (0.3,4.15) {$\braket{S^z}$};
\node[black] at (-4.125,4.15) {$\braket{S^z}$};
\node[black] at (0.3,6.8) {$\braket{S^z}$};
\node[black] at (-3.65,5.9) {$\frac{1}{4}$};
\node[black] at (-3.45,5.275) {-$\frac{1}{4}$};
\node[black] at (-0.6,5.9) {$\frac{1}{4}$};
\node[black] at (-0.6,3.25) {$\frac{1}{4}$};
\node[black] at (3.7,5.325) {-$\frac{1}{4}$};
\node[black] at (0.7,5.325) {-$\frac{1}{4}$};
\node[black] at (0.9,5.95) {$\frac{1}{4}$};
\node[black] at (3.775,1.05) {-$\frac{1}{4}$};
\node[black] at (-3.4,1.35) {{\color{white}{-$\frac{3}{4}$}}};
\node[black] at (0.9,3.25) {{\color{white}{$\frac{3}{4}$}}};
\node[black] at (0.925,1.75) {$\frac{1}{4}$};
\node[black] at (-3.425,2.75) {-$\frac{1}{4}$};
  \end{tikzpicture}
  \caption{The four-fold degenerate ground state a) and d) with total spin $S^z=\frac{1}{2}$ and b) and c) with total spin $S^z=-\frac{1}{2}$. In a) and b), the impurity anti-aligns with the impurity, thereby effectively flipping the sign of the exponentially localized quarter edge mode, whereas in c) and d), the impurity and quarter edge modes align to form effective three-quarter mode at the left end of the chain. In c) and d), the sky blue colored $\frac{1}{4}$ edge mode that exists in the $UV$ combines with the impurity with spin-$\frac{1}{2}$ shown as an arrow to form an effective three-quarter mode shown in the dark blue color.}
    \label{fig:spin12only}
  \end{figure}

\subsubsection{The ferromagnetic bound mode phase}

\begin{figure}
    \centering
    \includegraphics[width=\linewidth]{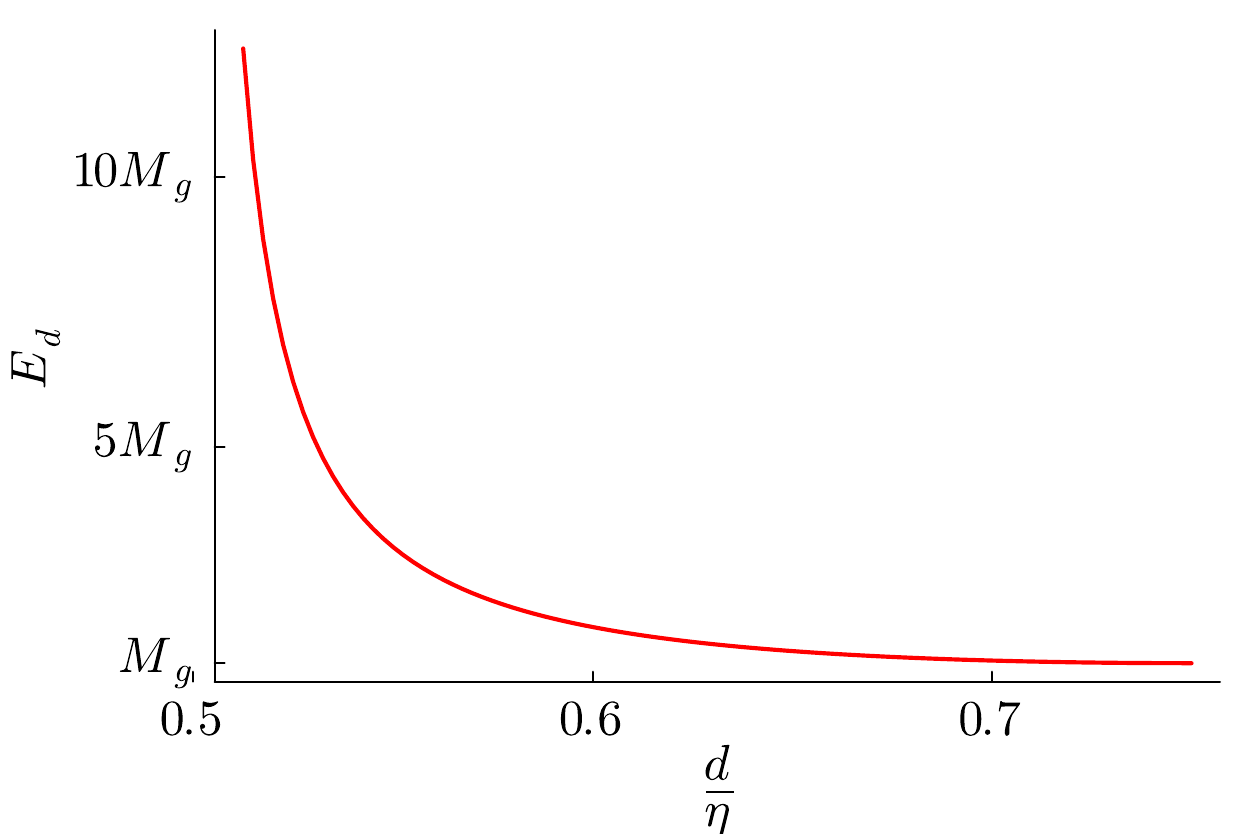}
    \caption{Energy of the impurity bound mode exponentially localized at the left end given by Eq.~\eqref{Ed_eng} is shown in the units maximum energy of the spinon $M_g$ for impurity parameter $\eta=2$. Notice that the minimum value of $E_d$  is equal to the maximum energy of the spinon and occurs at $d=\frac{3\eta}{2}$.}
    \label{fig:Edd_eng}
\end{figure}

When the impurity parameter $d$ is greater than the anisotropy parameter $\eta$, the impurity coupling is ferromagnetic. Moreover, when $d$ takes the value in the range $\eta<d<\frac{3}{2}\eta$, a bound mode exponentially localized at the edge is formed, and hence the impurity is in the ferromagnetic bound mode (FBM) phase. This bound mode, just like in the antiferromagnetic bound mode regime, is described by the purely imaginary solution $\lambda_d$ of the Bethe Ansatz equations and has energy $E_d$ given by Eq.~\eqref{Ed_eng}. $E_d$ takes positive values in this regime which is always greater than the maximum energy of the spinon $M_g$ as shown in Fig.~\ref{fig:Edd_eng}.

Since the energy of the impurity boundary string is positive, the bound mode does not exist in the ground state in this phase. Thus, the ground state contains an unscreened impurity. The ground state is four-fold degenerate with $Sz=\pm \frac{1}{2}$  when the total number of bulk sites $N_b$ is even where the quarter edge modes on the two ends are pointing in the same direction \textit{i.e.} $\ket{GS}^E=\ket{\pm \frac{1}{4}, \pm \frac{1}{4}}$. When the total number of bulk sites $N_b$ is odd, the ground state is four-fold degenerate: two $S^z=0$ states where the quarter modes at the two edges are anti-aligned \textit{i.e} $\ket{GS}^O_{0}=\ket{\pm \frac{1}{4}, \mp \frac{1}{4}}$ and the two states where $S^z=\pm 1$ where the edge mode combined with the free impurity to form $\pm \frac{3}{4}$ edge spin accumulation. Thus, the two degenerate ground states have spin accumulation $\ket{GS}^O_{\pm 1}=\ket{\pm \frac{3}{4}, \pm \frac{1}{4}}$. Some representative plots of the edge spin accumulation for the case of odd $N_b$ are shown in Fig.~\ref{fig:odd4acc}.

\begin{figure}
    \centering
    \includegraphics[width=\linewidth]{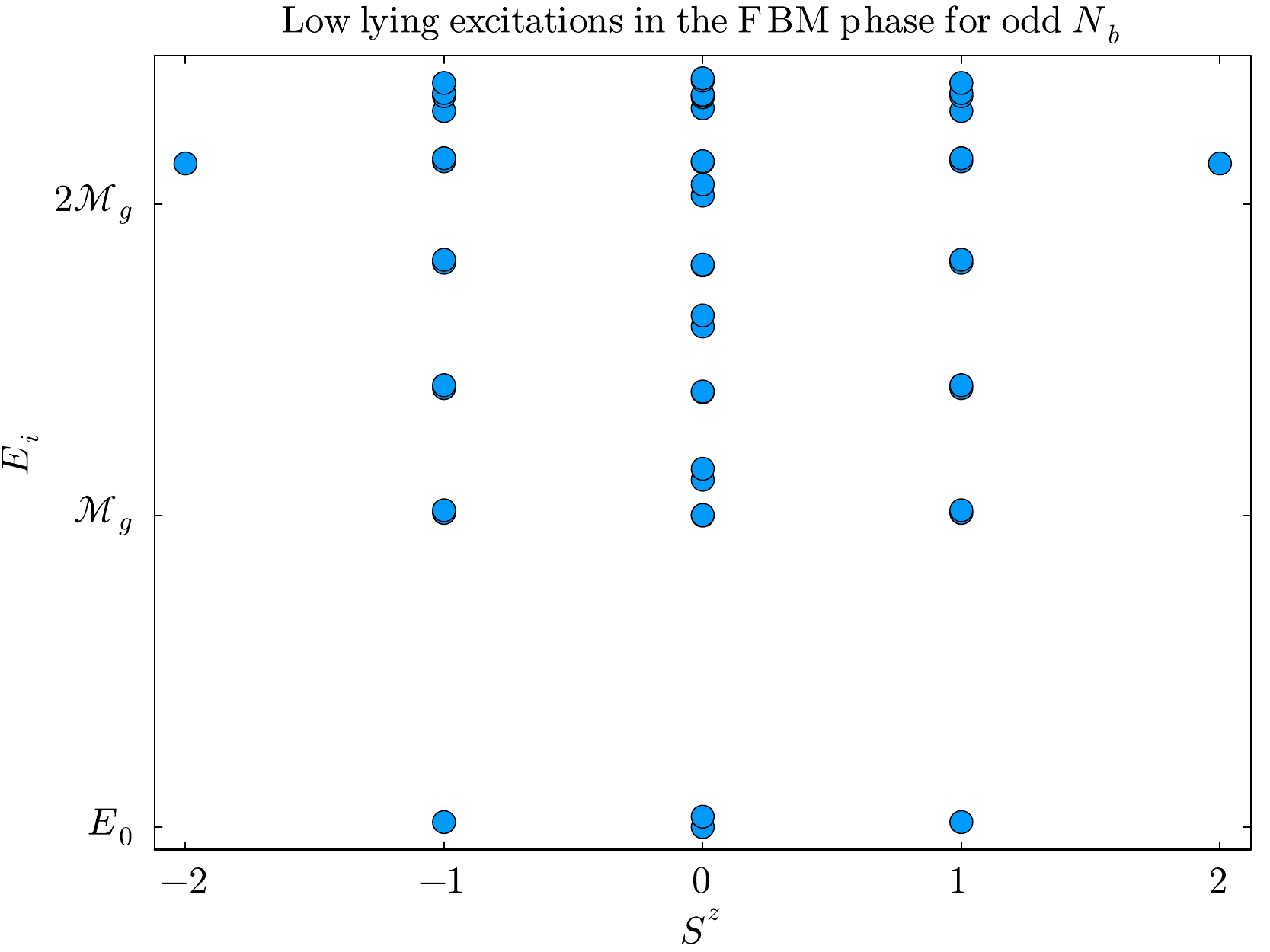}
    \caption{Schematic of low-lying excitation in the ferromagnetic bound mode phase showing the four-fold degenerate $S^z=\pm 1,0,0$ ground state and a mass gap of $\mathcal{M}_g$ for odd number of total bulk sites $N_b$. The data is obtained from the exact diagonalization of the Hamiltonian with $N_b=11$.}
    \label{fig:fbm-exstates}
\end{figure}

The highest weight state $S^z=1$ (and the descendants in the multiplet $S^z=0$ and $S^z=-1$) is constructed by considering all real roots of the Bethe Ansatz equations as described in Appendix~\ref{BAdets}. Whereas the fourth degenerate solution is constructed by adding the boundary string solution and higher order boundary string solution on top of the real solutions, the details construction follows the same way as in isotropic case detailed in \cite{kattel2023kondo} and briefly mentioned in Appendix~\ref{BAdets}.

Apart from the usual bulk excitations constructed by adding an even number of spinons, bulk strings, quartets, etc \cite{destri1982analysis}, a unique boundary excitation is possible in phase which is obtained by adding the impurity string solution $\lambda_d$. It costs $E_d$ energy to create such an excitation and impurity is screened in such an excited state. Fig.~\ref{fig:fbm-exstates} shows a schematic of the four-fold degenerate ground state and low-lying excitations with mass gap $\mathcal{M}_g$ when $N_b$ is odd.

\subsubsection{The unscreened phase}

When the impurity parameter takes the value in the range $d>\frac{3}{2}\eta$, the model is in the unscreened phase, as shown in the phase diagram Fig. \ref{fig:PD1}. In this phase, the impurity boundary bound string ceases to have finite mass and hence there exists no bound mode that screens the impurity. Thus, it is not possible to screen the impurity at any energy scale in this parametric regime. Since the impurity essentially becomes a free local moment, we call this the unscreened phase. The ground state is  four-fold degenerate with $S^z=\pm \frac{1}{2}$ and the boundary edge modes of the from $\ket{\pm \frac{1}{4}, \pm \frac{1}{4}}$ where the total number of bulk sites $N_b$ is even whereas the ground state is four-fold degenerate with $S^z=\pm$ and two $S^z=0$ states with the edge spin accumulation of the from $\ket{\pm \frac{1}{4},\pm \frac{1}{4}}$ and $\ket{\pm \frac{3}{4}, \pm \frac{3}{4}}$ as shown in Fig.~\ref{fig:odd4acc} when  the total number of bulk sites $N_b$ is odd.

\begin{figure}[H]
    \centering
    \includegraphics[width=\linewidth]{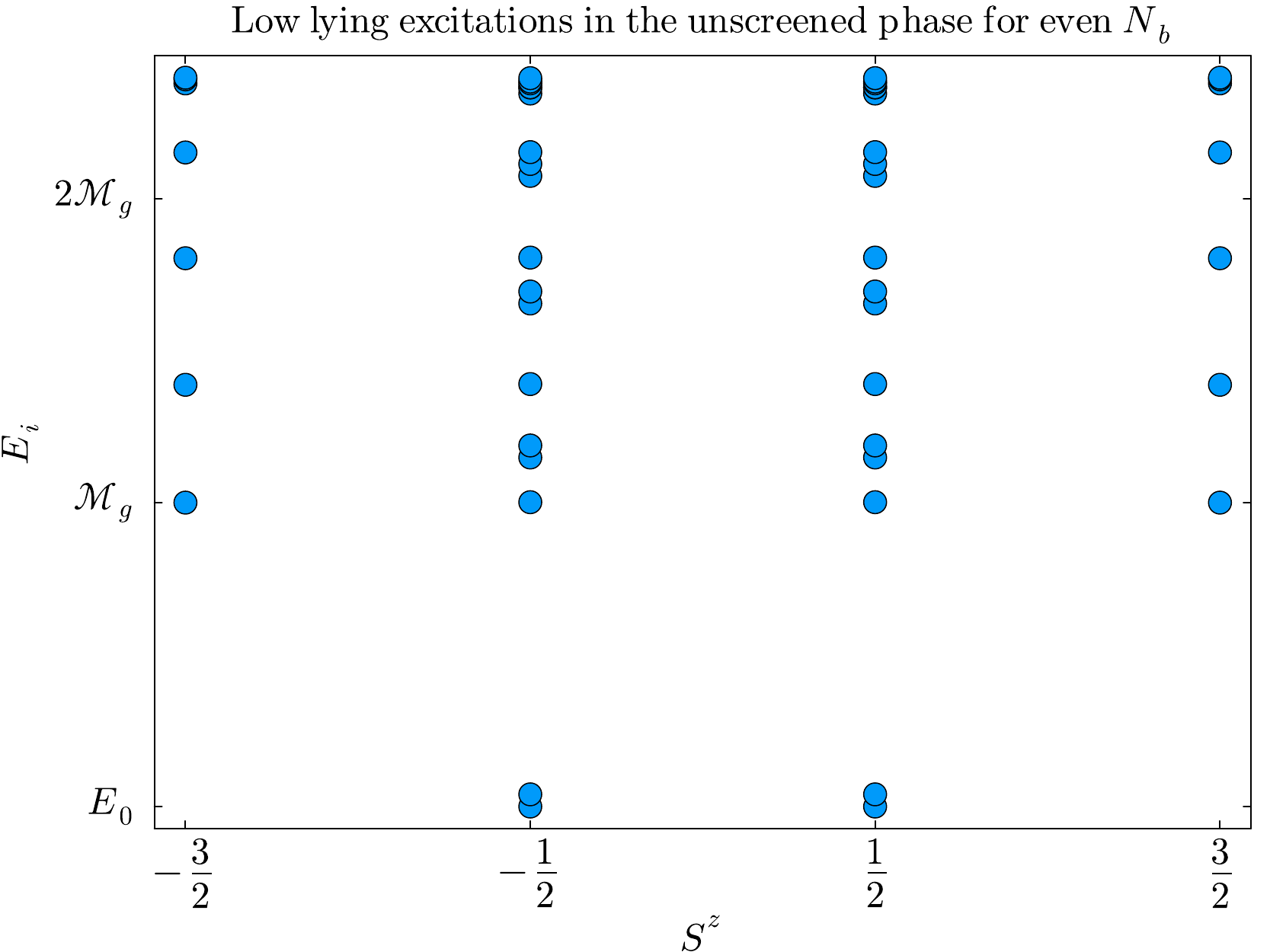}
    \caption{Schematic of low lying excitation in the unscreened phase showing the four-fold degenerate $S^z=\pm \frac{1}{2}$ ground state and a mass gap of $\mathcal{M}_g$ for even number of total bulk sites $N_b$. The data is obtained from the exact diagonalization of the Hamiltonian with $N_b=10$.}
    \label{fig:us-exstates}
\end{figure}

The four-fold degenerate ground state with $S^z=\pm \frac{1}{2}$, and excited states separated by a mass gap $\mathcal{M}_g$ is shown in the schematic Fig.~\ref{fig:us-exstates} when the total number of bulk sites $N_b$ is even.

We will briefly describe the effect of having two impurities, one at each edge, in Appendix \ref{sec:intlin2imp}. In this scenario, each impurity at the ends can independently reside in one of four distinct phases: the Kondo phase, the ABM phase, the FBM phase, or the US phase. Consequently, the system as a whole can exhibit a total of 16 possible impurity phase combinations as shown in the phase diagram Fig.~\ref{PDeven}.
 \begin{figure}
\includegraphics[width=\linewidth]{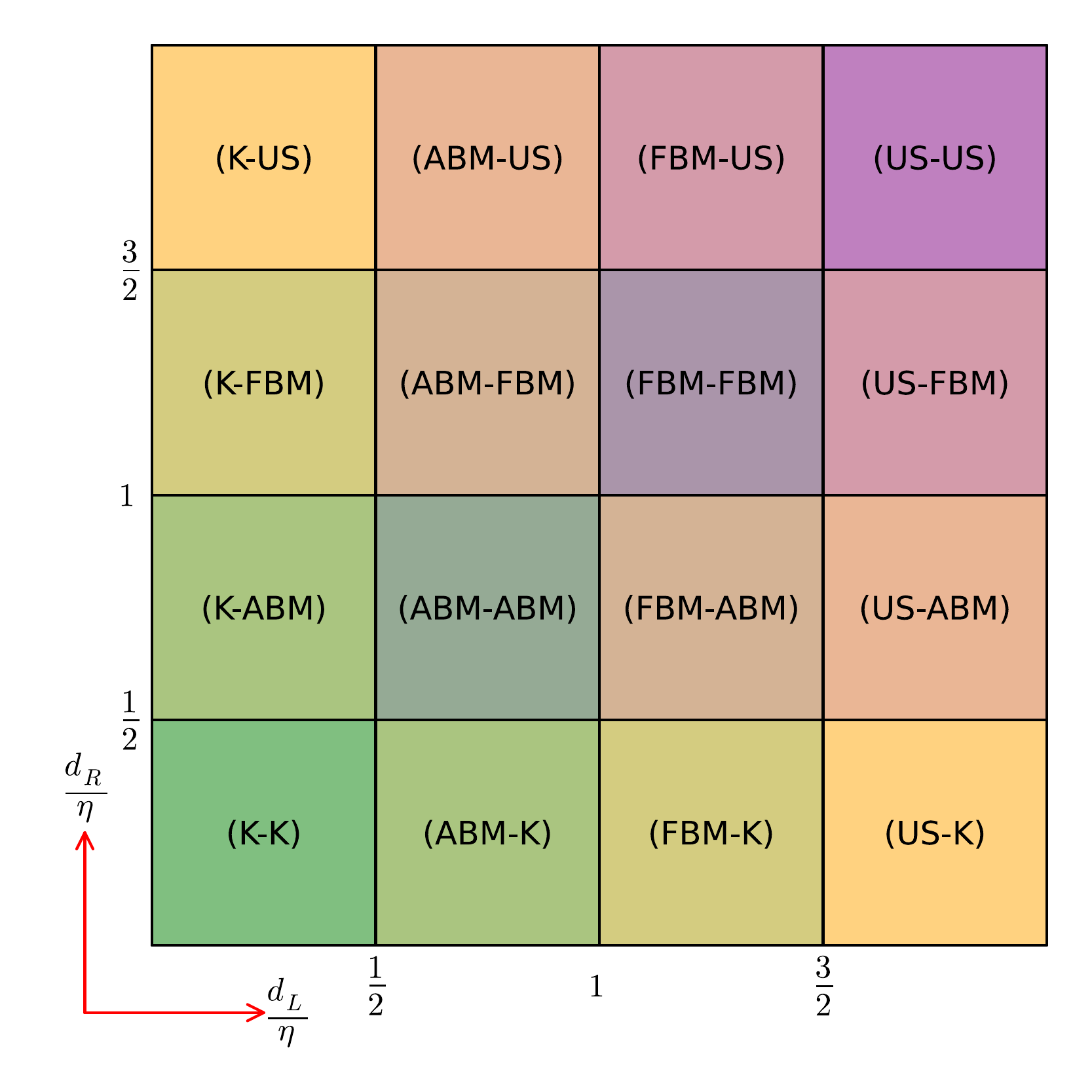}
\caption{Phase diagram for the Hamiltonian in Eq.~\eqref{ham} with two impurities, where a total of 16 phases arise due to the four independent phases possible at each edge of the chain, as shown in Fig.~\ref{fig:PD1}. Here, $K$, $ABM$, $FBM$, and $US$ correspond to the Kondo phase, the antiferromagnetic bound mode phase, the ferromagnetic bound mode phase, and the unscreened phase, respectively. A phase labeled $(X, Y)$ indicates that the left impurity is in phase $X$, and the right impurity is in phase $Y$, where $X$ and $Y$ independently belong to $\{K, ABM, FBM, US\}$. Phases with the same color are related by spatial reflection symmetry $L \leftrightarrow R$.}
\label{PDeven}
 \end{figure}

\section{Breaking boundary integrability: one nonintegrable impurity}\label{sec:nonintimp}
As described earlier, the integrable Hamiltonian Eq.~\eqref{ham} has a restricted form where a single variable, the impurity parameter $d$, controls both the impurity coupling $J_{\mathrm{imp}}$ and the boundary anisotropy parameter $\Delta_{\mathrm{imp}}$. Thus, in this section, we consider the Hamiltonian of the form
\begin{align}
    H&=\sum_{i=1}^{N_b-1}\frac{J}{2}(\sigma_i^x  \sigma_{i+1}^x+\sigma_i^y  \sigma_{i+1}^y+\Delta \sigma_i^z  \sigma_{i+1}^z)\nonumber\\
    &+\frac{J_{\mathrm{imp}}}{2}(\sigma_{\mathrm{imp}}^x  \sigma_{1}^x+\sigma_{\mathrm{imp}}^y  \sigma_{1}^y+\Delta \sigma_{\mathrm{imp}}^z  \sigma_{1}^z),
    \label{ham1-nonint}
\end{align}
where a single impurity  $\vec \sigma_{\mathrm{imp}}$ is situated at the left end of the spin chain interacting with the spin chain via anisotropic Heisenberg coupling. Note that both the bulk and the impurity have the same anisotropy parameter $\Delta$, and hence the ratio of the impurity coupling $J_{\mathrm{imp}}$ to the bulk coupling $J$ is the single free parameter in the model.  As stated earlier, the motivation for exploring this form of Hamiltonian is to explore the possibility of novel mid-gap states that were missing in the integrable limit. The model described by Hamiltonian Eq.~\ref{ham1-nonint} is no longer integrable, and hence, we do not have exact analytic means to explore the boundary physics anymore. However, in this section, using DMRG and exact diagonalization calculations, we illustrate that the boundary physics when the impurity coupling $J_{\mathrm{imp}}$ is antiferromagnetic is richer than the integrable case. We shall also discuss very briefly about the non-integrable ferromagnetic coupling.

\subsection{Antiferromagnetic boundary coupling}

As we learned from the integrable case, one of the signatures of the existence of the bound mode is the characteristic jump in the local magnetization at the impurity site. Since this quantity is the easiest to compute numerically via DMRG, we shall first explore the impurity phase diagram by computing the local magnetization at the impurity site. The local impurity magnetization $\braket{\sigma^z_{\mathrm{imp}}}$ for an odd number of total bulk site $N_b=499$ (equivalently, the total number of sites $N=500$) and the crossing parameter $\eta=2$, computed via DMRG is shown in Fig.~\ref{mag-nonint}.
\begin{figure}
    \centering\includegraphics[width=\linewidth]{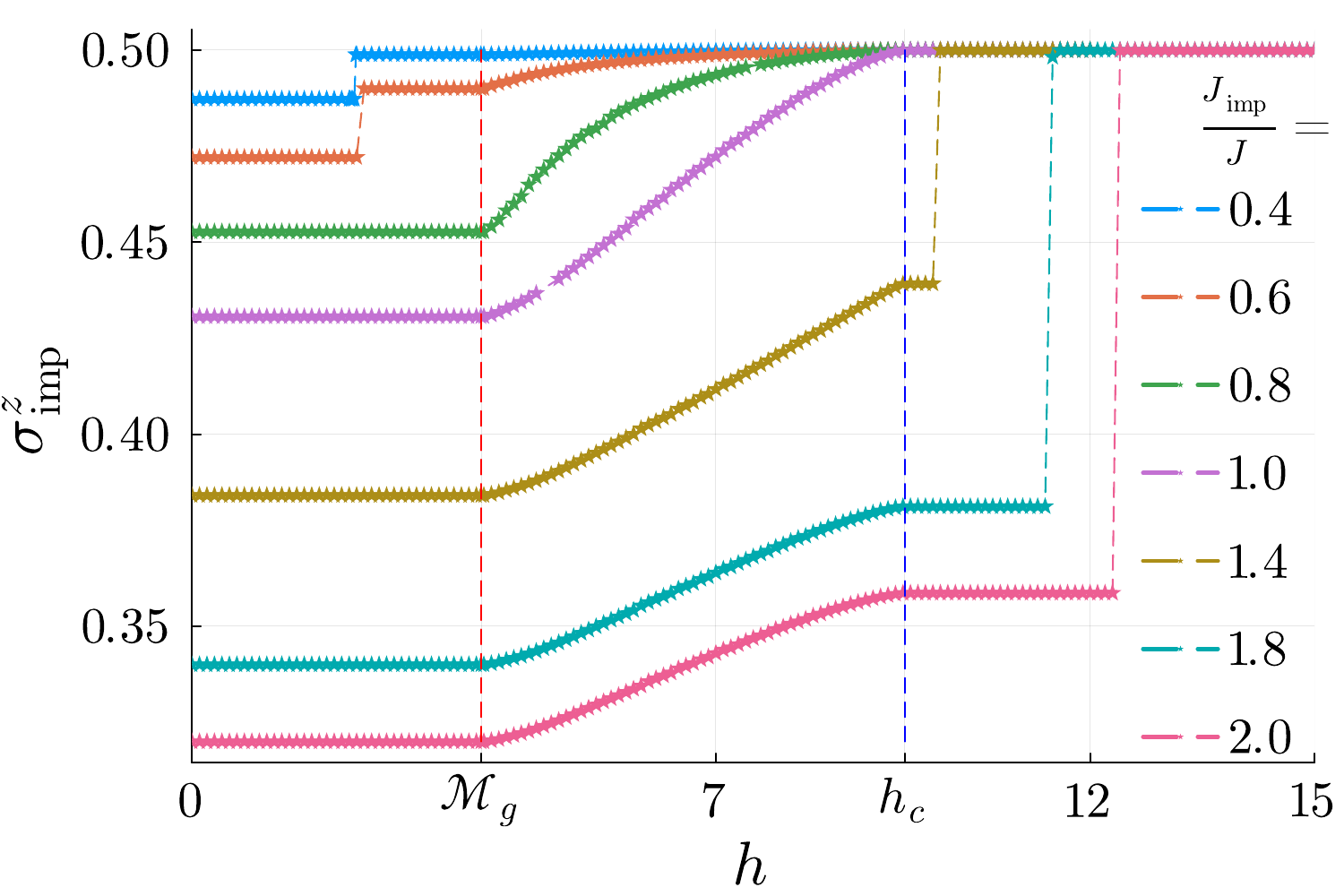}
    \caption{Local impurity magnetization for various values of the ratio of $\frac{J_{\mathrm{imp}}}{J}$ for Hamiltonian Eq.~\eqref{ham1-nonint} for antiferromagnetic boundary and bulk couplings computed for $\eta=2$ and total number of sites (bulk and impurity) $N=500$. For a small value of the ratio of bulk to boundary couplings $0<\frac{J_{\mathrm{imp}}}{J}\ll 1$, the impurity magnetization jumps when $0<h^\star_1(\eta)<\mathcal{M}_g$ which shows the existence of midgap state. Likewise, for intermediate values of the ratio of boundary to bulk couplings $\frac{J_{\mathrm{imp}}}{J}\approx 1$, magnetization is constant when the external magnetic field evolves from $h=0$ to $h=\mathcal{M}_g$ due to the absence of mid gap states in the spectrum. However, when $h$ changes from $h=\mathcal{M}_g$ to $h=h_c=2(1+\mathrm{cosh}(\eta))J$, the impurity magnetization smoothly changes from some finite $d$ dependent value to some other $d$ value at $h=h_c$ where the local magnetization of every site becomes 0.5. Finally, for a large value of the ratio of the boundary to bulk coupling $\frac{J_{\mathrm{imp}}}{J}\gg 1$, impurity magnetization is constant when the external magnetic field evolves from $h=0$ to $h=\mathcal{M}_g$ due to the presence of the mass gap in the spectrum. However, when $h$ changes from $h =\mathcal{M}_g$ to $h = h_c$, the impurity magnetization smoothly changes from some finite value to some other finite value at $h = h_c$ where the local magnetization is smaller than 0.5. When $h$ is further increased, the impurity magnetization remains constant up
until a critical value $h^\star_2(\eta)$, where it abruptly jumps to
0.5 because of the existence of the bound mode, which has energy higher than the maximum value of spinon energy $M_g$. The data is obtained by using DMRG, where all calculations are performed by setting the truncation cut-off of the singular values at
$10^{-10}$ and performing 100 sweeps to ensure convergence for every data points.}
    \label{mag-nonint}
\end{figure}

The local magnetization curve at the impurity site shows that for smaller values of the ratio of the boundary to the bulk coupling, the magnetization jumps when $0<h=h^*_1(\eta)<\mathcal{M}_g$, which shows that there exists midgap state in the spectrum. For intermediate values of the boundary coupling, the impurity is screened by a multiparticle cloud of bulk spinon excitations, and hence, the magnetization curve does not jump for $h<\mathcal{M}_g$ as there are midgap states. Moreover, between $h=\mathcal{M}_g$ and $h=h_c=2J(1+\cosh(\eta)$, the local magnetization grows smoothly, and exactly at $h=h_c$, it becomes 0.5, at which point every spin in the chain is fully polarized. Finally, for larger values of the ratio of the boundary to bulk couplings, the impurity magnetization does not jump when $h<\mathcal{M}_g$ due to the absence of midgap states. Moreover, between $h=\mathcal{M}_g$ and $h=h_c$, the impurity magnetization curve smoothly grows, however, it does not reach $0.5$ exactly at $h=h_c$ because there is a localized bound mode in the spectrum which has energy $E(h)>M_g$, the maximum value of the energy of a single spinon. Thus, when $h$ is further increased from $h=h_c$, the magnetization is again a constant up until $h=E(h)$, where it abruptly jumps to 0.5 exactly since the last remaining mode polarizes when its energy in the presence of the magnetic field is equal to the strength of the magnetic field. Notice that just as in the integrable case, $E(h)$ is the energy of the bound mode in the presence of magnetic field which is different from the energy of the bound mode when $h=0$.  

\begin{figure}
    \centering
    \includegraphics[width=\linewidth]{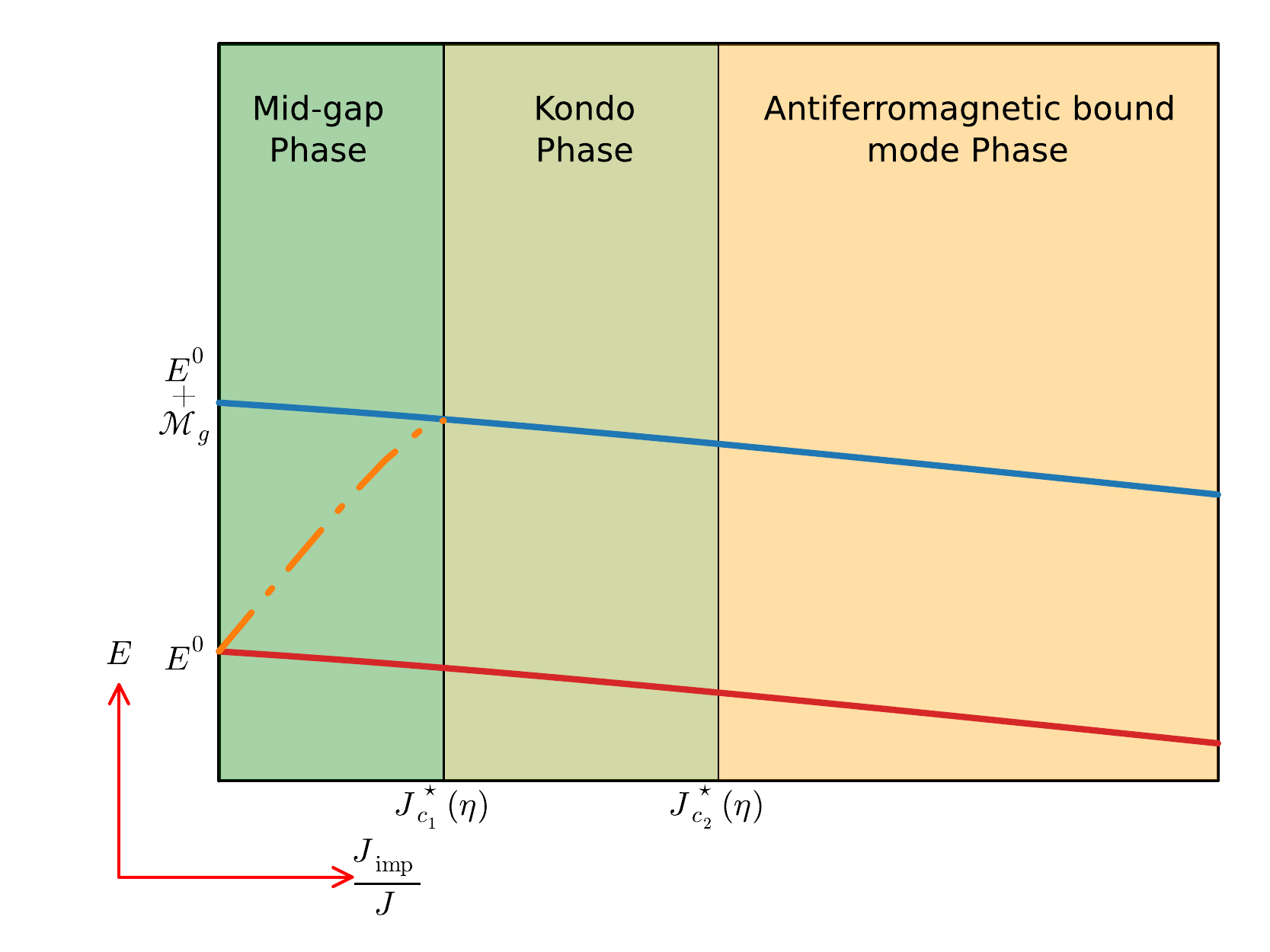}
    \caption{The three boundary phases exhibited by Hamiltonian Eq.~\eqref{ham1-nonint}. For small values of the ratio of the boundary to bulk coupling $0<J_{\mathrm{imp}}/J<J^\star_{c_1}(\eta)$, there is a mid-gap state whose energy is shown in the orange dash-dot line. For the ratio of the couplings between $J^\star_{c_1}(\eta)<J_{\mathrm{imp}}/J<J^\star_{c_2}(\eta)$, the impurity is screened by multiparticle Kondo effect and finally, for $J_{\mathrm{imp}}/J>J^\star_{c_2}(\eta)$, there exists a bound mode in the ground state whose energy is always higher than the maximum energy of the spinon. The two critical values of the ratio of the coupling depend on the crossing parameter $\eta$. The numerically obtained values of these two critical couplings as a function of $\eta$ are shown in Fig.~\ref{pbdiag}. The red line is the ground state of the model which depends on the ratio of the bulk and boundary couplings, the anisotropy parameter, and the system size. Likewise, the blue line is the energy of the two spinons excitation above the ground state, which in the thermodynamic limit is independent of the ratio of the couplings but depends on the crossing parameter $\eta$.}
    \label{fig:pd-non-int}
\end{figure}

Since the model is no longer integrable, it is extremely difficult to analytically compute the exact locations of the boundary phases. Thus, based on the magnetization curve, we sketch a schematic phase diagram of the model given by Eq.~\eqref{ham1-nonint} in Fig.~\ref{fig:pd-non-int}. The phase boundary for small values of crossing parameter $\eta$ is computed numerically and shown in Fig.~\ref{pbdiag}. Here $J^\star_{c_1}(\eta)$ is the value of the impurity coupling $J_{\mathrm{imp}}$ at which the impurity phase changes from the mid-gap phase where a midgap state appears in the spectrum to the Kondo phase where the impurity is screened by a multiparticle Kondo cloud. Likewise, $J^\star_{c_2}(\eta)$ is the value of the impurity coupling $J_{\mathrm{imp}}$ at which point the impurity changes from being screened by the Kondo cloud to being screened by the single particle bound mode as shown in the phase diagram Fig.~\ref{fig:pd-non-int}.
\begin{figure}
    \centering
    \includegraphics[width=\linewidth]{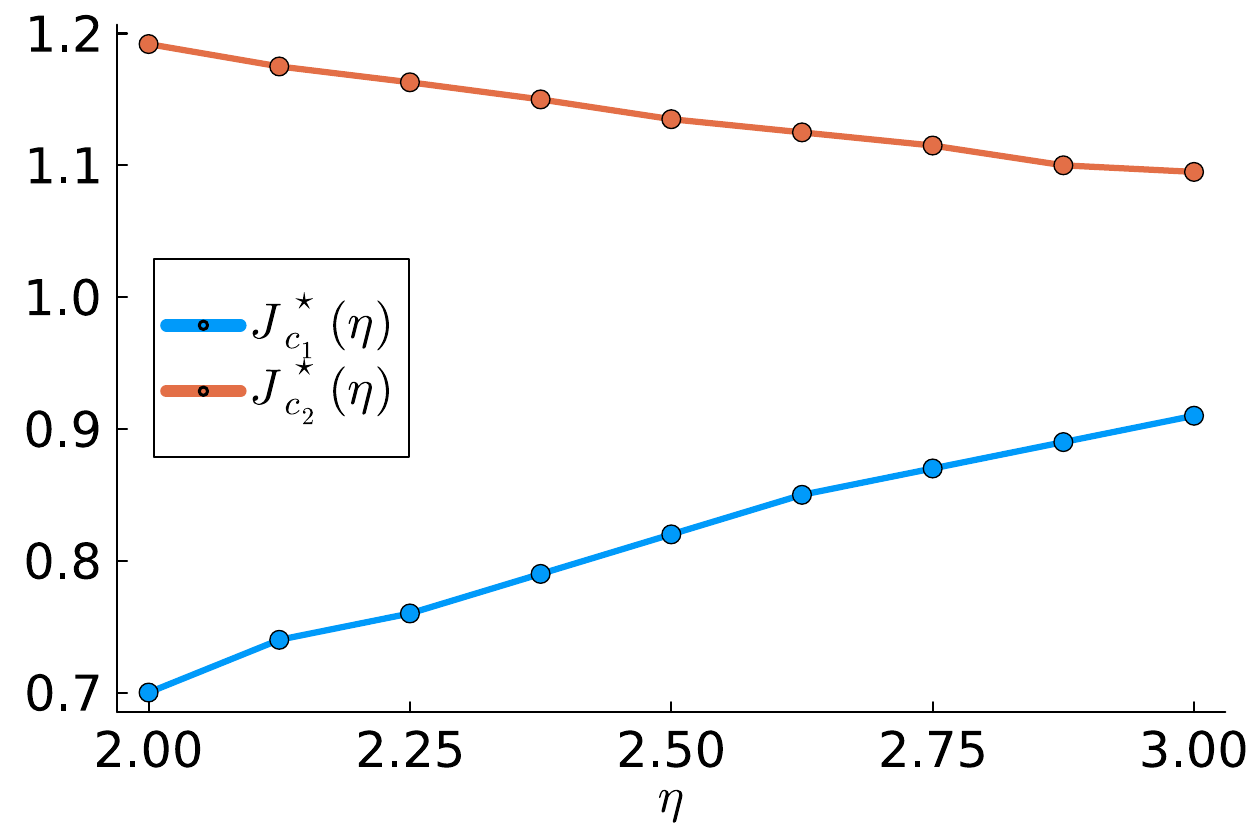}
    \begin{tikzpicture}[overlay, remember picture]
    \node at (2,2) {Mid-gap phase};
     \node at (2,4) {Kondo phase};
     \node at (2,5.75) {ABM phase};
    \end{tikzpicture}
    \caption{The phase boundary differentiating the three impurity phases shown in the phase diagram Fig.~\ref{fig:pd-non-int} is computed for various values of the crossing parameter $\eta$. There exists mid-gap state in the spectrum when $0<J_{\mathrm{imp}}/J<J^\star_{c_1}(\eta)$. When $J^\star_{c_1}(\eta)<J_{\mathrm{imp}}/J<J^\star_{c_2}(\eta)$, the impurity is in the Kondo phase where it is screened by many-body Kondo cloud. Finally, when $J_{\mathrm{imp}}/J>J^\star_{c_2}(\eta)$, the impurity is screened by a single particle bound mode formed at the edge of the chain. These phase boundaries are obtained numerically via DMRG.
    }
    \label{pbdiag}
\end{figure}

\begin{figure}
    \centering
    \includegraphics[width=\linewidth]{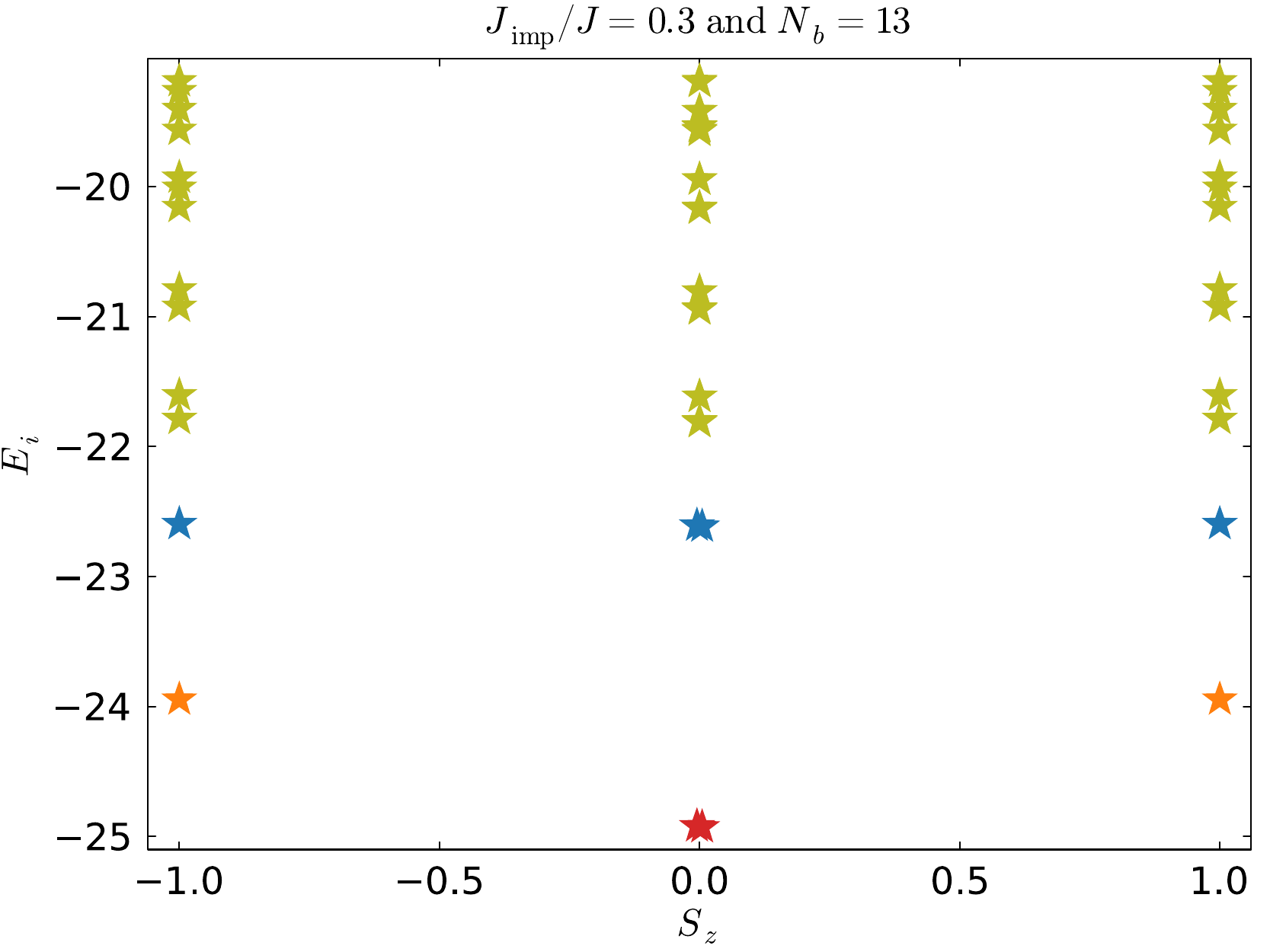}
    \caption{50 lowest lying excitations of Hamiltonian Eq.~\eqref{ham1-nonint} in the mid-gap phase with model parameters $N_b=13, \eta=2$ and $J_{\mathrm{imp}}/{J}=0.3$. The two-fold degenerate ground state with $S^z=0$ is shown in red color, the two-fold degenerate $S^z=\pm 1$ mid-gap states are shown in orange color, four-fold 2-spinon excitations, which form degenerate singlet and triplet states are shown in the blue colors, and all other higher excited states are shown in olive color.}
    \label{fig:ED-spec}
\end{figure}

We shall now show the existence of the mid-gap state by explicitly computing the spectrum of the model for small system sizes using the exact diagonalization method. Taking the model parameter $\frac{J_{\mathrm{imp}}}{J}=0.3, \eta=2$ and $N_b=11$, we compute the eigenvalues and the spin in each state and show the result for states with 50 lowest energies in Fig.~\ref{fig:ED-spec}. The low-lying excitations in the Kondo and antiferromagnetic bound mode phase have a similar structure as those of the integrable cases shown in Fig.~\ref{fig:Kexcitedodd} and Fig.~\ref{fig:abm-exstates} respectively.

\begin{figure}
    \centering
    \includegraphics[width=\linewidth]{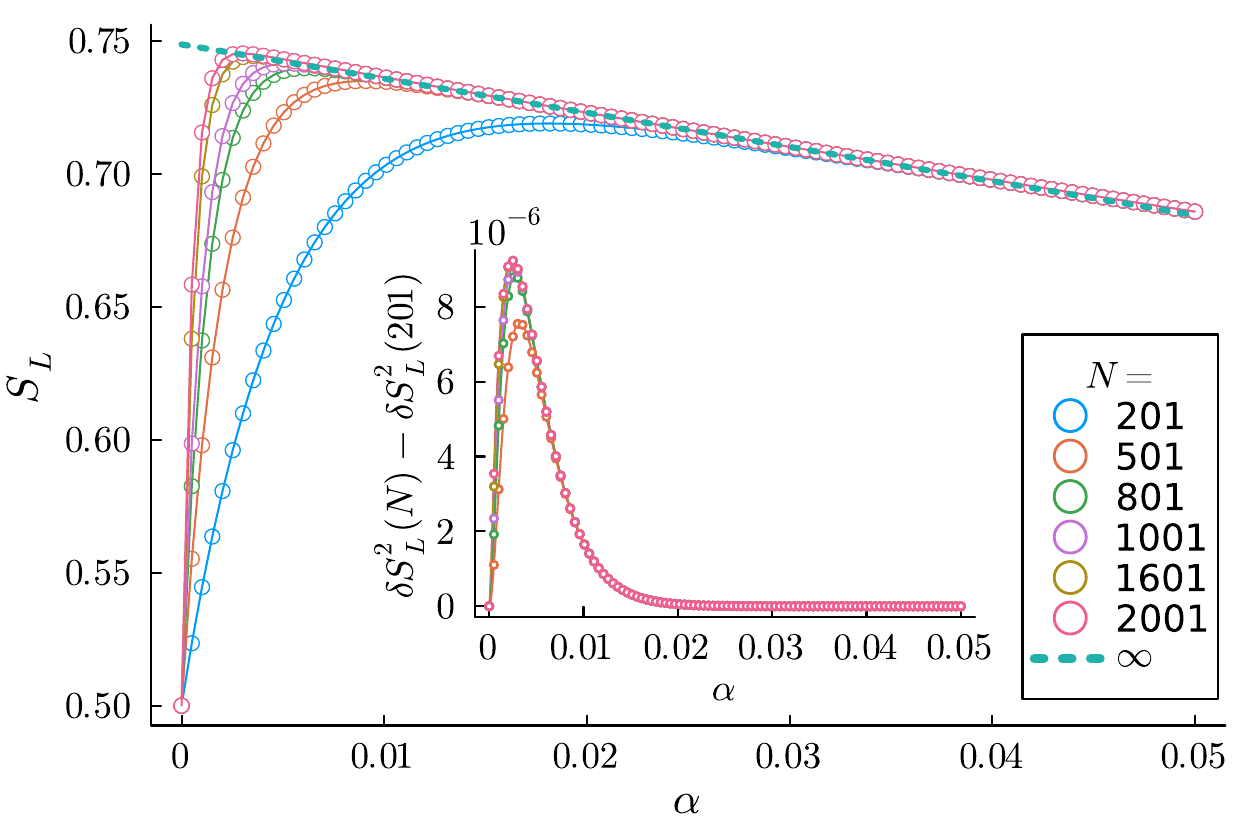}
    \caption{Left localized $\frac{3}{4}$ mode in one of the two-fold degenerate mid-gap states when $N_b$ is even and $J_{\mathrm{imp}}/J=0.5$. The right spin accumulation is $-\frac{1}{4}$ in this state such that the total spin is $S^z=\frac{1}{2}$.}
    \label{fig:non-int-midgap}
\end{figure}

Using DMRG, we compute the spin profile for the entire parametric range when the impurity coupling is antiferromagnetic and find that in all three phases, there are quarter modes at the edges of the spin chain. For odd $N_b$, the quarter modes in the ground state point in opposite direction \textit{i.e.} $\ket{GS}^O=\ket{\pm \frac{1}{4},\mp \frac{1}{4}}$ and when $N_b$ is even, the quarter modes in the ground state point in opposite direction \textit{i.e.} $\ket{GS}^E=\ket{\pm \frac{1}{4},\pm \frac{1}{4}}$ such that the total $S^z$ of the two-fold degenerate ground state is $0$ for the former case and $\pm \frac{1}{2}$ for the latter case. There is a two-fold degenerate mid-gap state state with $S^z=\pm 1$ with edge spin accumulation $\ket{\pm \frac{3}{4},\pm \frac{1}{4}}$ when $N_b$ is odd and $S^z=\pm \frac{1}{2}$ with edge spin anti-aligned \textit{i.e.} $\ket{\pm \frac{3}{4}, \mp \frac{1}{4}}$ when $N_b$ is even. The left edge spin accumulation in one of the two-fold degenerate $S^z=\pm \frac{1}{2}$ is shown in Fig.~\ref{fig:non-int-midgap}.

\subsection{Ferromagnetic boundary coupling}
As we saw in both integrable and non-integrable cases, when the impurity coupling is antiferromagnetic, bound modes with negative energies appear in the ground state. In contrast, for ferromagnetic coupling, bound modes at the edges have positive energies and exist in higher energy states. Studying these high-energy bound modes with DMRG is challenging, and hence, the detailed phases in the case of nonintegrable impurity ferromagnetically coupled to the XXZ chain will be addressed in future work. Here, we briefly highlight that the ferromagnetic non-integrable case is significantly different from the integrable case. In integrable cases, irrespective of the value of ferromagnetic boundary coupling strength, the ground state is always four-fold degenerate, which shows that the impurity is essentially free, and hence, it can align or anti-align with the edge mode that it is coupled to. However, this is not the case when the impurity is coupled non-integrably. With non-integrable coupling, the impurity can only align with the edge mode, thereby forming a three-quarter effective edge mode, as anti-aligning to form a quarter mode in the opposite direction costs energy.  

We show the representative spectrum of Hamiltonian Eq.~\eqref{ham1-nonint} for $J_{\mathrm{imp}}/J=-3.0$, and $\eta=2.0$ for the total number of bulk sites $N_b=13$ in Fig.~\ref{fig:fmc-nonint}. Notice that unlike in the case of the integrable coupling, the ground state is no longer four-fold degenerate rather, it is a two-fold degenerate $S^z=\pm 1$ state with edges modes $\pm \frac{3}{4}, \pm \frac{1}{4}$ \textit{i.e.} the state in which the quarter edge modes at the left edge of the chain align with the spin-$\frac{1}{2}$ impurity when the total number of sites is even. Likewise, the ground state is two-fold degenerate state with $S^z=\pm \frac{1}{2}$ with edge modes  $\pm \frac{3}{4}, \mp \frac{1}{4}$ when the total number of sites is odd. 
\begin{figure}[H]
    \centering
    \includegraphics[width=0.5\textwidth]{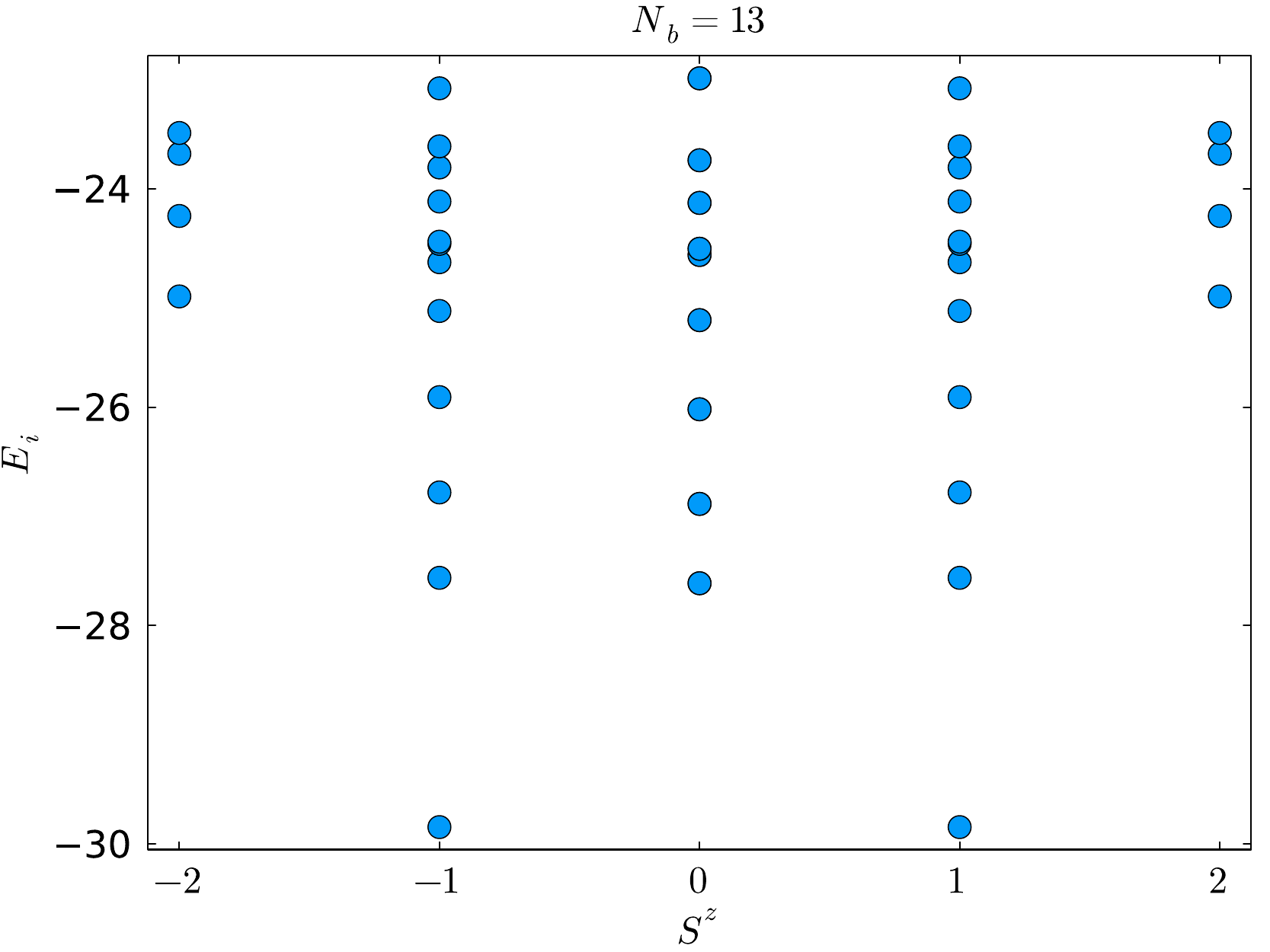}
    \caption{50 lowest lying excitations of Hamiltonian Eq.~\eqref{ham1-nonint} in the mid-gap phase with model parameters $N_b=11, \eta=2$ and $J_{\mathrm{imp}}/{J}=-3$. The ground state is two-fold degenerate which shows that the impurity is not completely free as in the integrable case.}
    \label{fig:fmc-nonint}
\end{figure}

{Complete impurity screening at low temperatures for non-integrable coupling is suggested by the two-fold degenerate ground state, observed regardless of whether the impurity coupling is ferromagnetic or antiferromagnetic. This low-temperature behavior is not unexpected for antiferromagnetic coupling, as the Kondo term becomes strongly coupled in the IR limit. The surprising finding is that, in contrast to the fine-tuned integrable ferromagnetic case, non-integrable ferromagnetic coupling also seems to drive the system towards impurity screening.  To prove this, we employ the purification method to calculate the impurity entropy by subtracting the entropy of the impurity-free chain from the entropy of the chain containing the impurity. The subsequent decrease of impurity entropy from $\ln(2)$ in the UV to 0 in the IR, irrespective of the impurity coupling as shown in Fig.~\ref{fig:non-int-ent}, demonstrates impurity screening at low temperatures.
\begin{figure}[H]
    \centering
    \includegraphics[width=\linewidth]{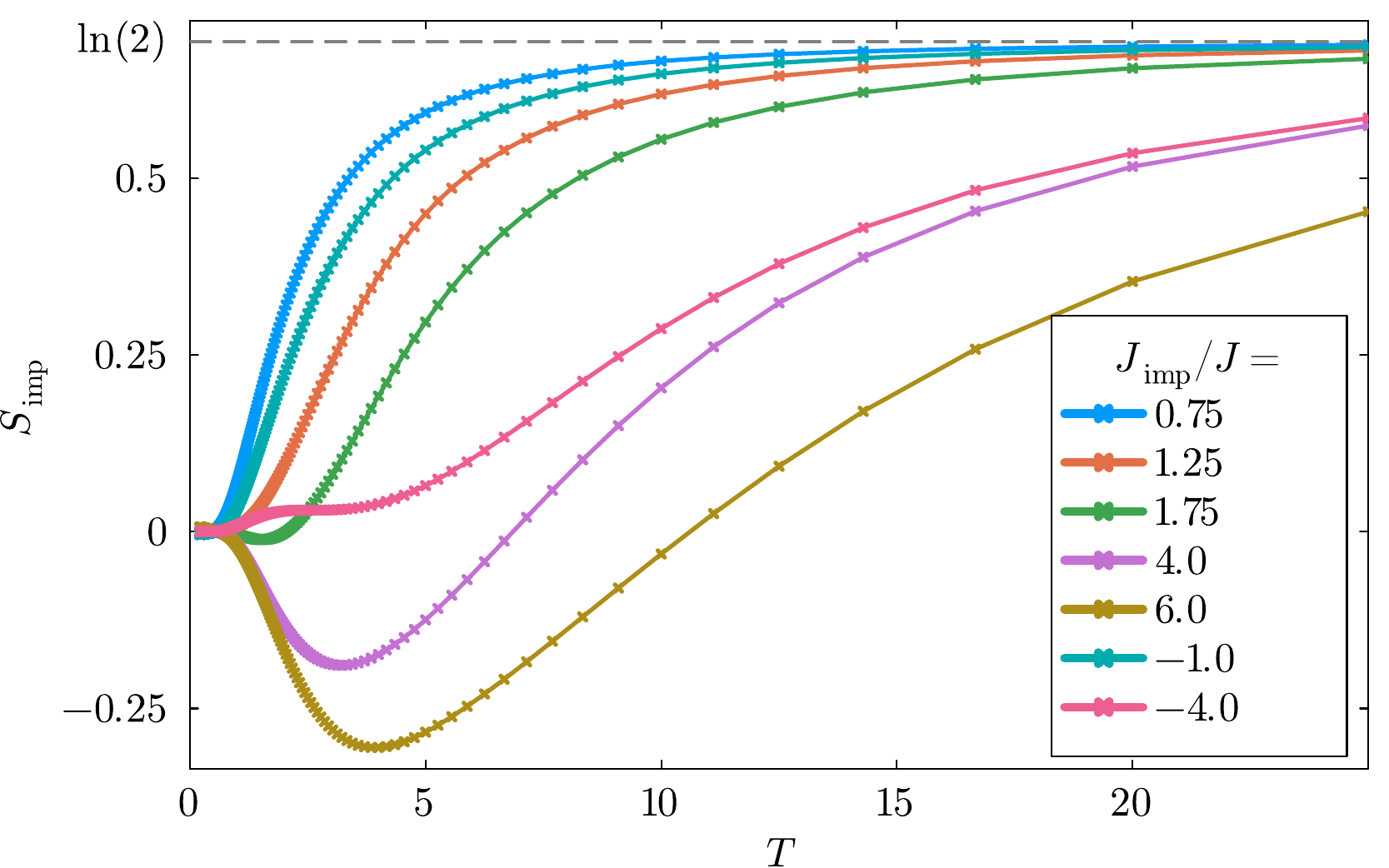}
    \caption{The impurity entropy is $\ln(2)$ in the UV and 0 in the IR irrespective of the sign of the impurity coupling.}
    \label{fig:non-int-ent}
\end{figure}

The temperature dependence of the impurity entropy, as depicted in Fig.~\ref{fig:non-int-ent}, is monotonic when the impurity resides in the Mid-gap or Kondo phase. In contrast, the antiferromagnetic bound mode phase shows a non-monotonic temperature dependence. Significantly, the ferromagnetic coupling regime is characterized by the impurity entropy tending to 0 at low temperatures. This is in stark contrast to the integrable ferromagnetic coupling case (Fig. \ref{fig:impentint}), where the low-temperature impurity entropy is $\ln(2)$.}

\section{Conclusion and outlook}
We considered the spin-$\frac{1}{2}$ anisotropic Heisenberg chain with integrable and nonintegrable boundary impurities and analyzed them using a combination of \textit{Bethe Ansatz} and DMRG.  We found that the system exhibits different phases depending on the impurity shift parameters $d_i$, which relates the bulk and boundary couplings as shown in Eq.\eqref{bulk-boundary},
in the case of integrable impurities and the ratio of boundary and bulk couplings in the case of nonintegrable impurities. 

In the case of one integrable impurity interacting antiferromagnetically, there exist two phases, namely the Kondo phase and the antiferromagnetic bound mode phase.
The Kondo phase occurs when the ratio of the impurity parameter $d$ is either purely imaginary or when it takes real values between $0<d<\frac{\eta}{2}$. The impurity is screened in the ground state due to the Kondo effect where the ratio of the impurity and the bulk density of states $R(E)$ has a characteristic peak near $m_g=E(\pm \pi)$, 
the minimum possible value of single spinon energy, for imaginary values of $d$ and a peak near $M_g=E(0)$, 
the maximum possible value of single spinon energy, for real values of $d<\frac{\eta}{2}$. The energy of the spinon $E(\theta)$ is given by Eq.\eqref{spinonengeqn}.  Eventually, as $d$ is increased, one enters the antiferromagnetic bound mode phase, which exists in the parametric range $\frac{\eta}{2}<d<\eta$ as shown in the phase diagram Fig.~\ref{fig:PD1}. The ground state in this phase contains an exponentially localized bound mode which screens the impurity. This mode can be removed, and thereby, the impurity can be unscreened. This process costs energy greater than $M_g$, the magimum energy of the spinon. In the Kondo phase, excitations can be built on top of the ground state and one finds that these excitations form a single tower. In the antiferromagnetic bound mode phase, excitations in the bulk can be built on top of the ground state in which the impurity is screened and also on top of the state in which the impurity is unscreened, and hence one obtains two distinct towers of excited states.

Similarly, when the integrable impurity interacts ferromagnetically, the system exhibits two phases: the ferromagnetic bound mode phase and the unscreened phase.  These phases correspond to the impurity parameter ranges $\eta<d<\frac{3\eta}{2}$ and $d>\frac{3\eta}{2}$ , respectively. In both the ferromagnetic bound mode and unscreened phases, the impurity remains unscreened in the ground state. However, unlike the unscreened phase, the ferromagnetic bound mode phase allows the impurity to be screened by introducing an exponentially localized bound mode at the cost of $E_d$ given by Eq.\eqref{Ed_eng}, the energy of the impurity boundary string solution. In the ferromagnetic bound mode phase, excitations can be constructed in the bulk on top of the ground state where the impurity is unscreened. Similarly, excitations can also be built on top of the state where the impurity is screened, leading to two towers of excited states. In contrast, the unscreened phase features a single tower corresponding to the ground state where the impurity is unscreened. The Hilbert space in the antiferromagnetic bound mode and ferromagnetic bound mode phases, characterized by the presence of multiple towers, undergoes what is known as ``Hilbert space fragmentation" \cite{pasnoori2023boundary,kattel2023kondo}.

When two integrable impurities are considered,  they are independent in the thermodynamic limit, especially due to the presence of the mass gap in the bulk, which makes all the correlations in the model fall off exponentially. Each impurity can be in any of the four phases corresponding to the one-impurity case. Thus, the model with two boundary impurities exhibits a total of sixteen possible phases. 

Finally, we consider a case where a single non-integrable impurity interacts with the chain antiferromagnetically. In this scenario, we observe, in addition to the Kondo and antiferromagnetic bound mode phases, a novel phase characterized by mid-gap states in the spectrum. {{Surprisingly, in contrast to the expectation that ferromagnetic coupling typically leads to unscreened impurities (as seen in the integrable case), we find that even with non-integrable ferromagnetic coupling, the impurity also becomes screened.}}

There are some interesting questions to ask about how these quarter edge modes, mid-gap states, and high energy bound states affect the dynamics and non-equilibrium aspects of the model. We shall leave these questions for future work.

\section*{Acknowledgement}
We thank Yicheng Tang for helpful suggestions and useful discussions at various stages of the project. J.H.P. is partially supported by the  NSF Career Grant No.~DMR- 1941569.

\bibliography{ref}

\begin{thebibliography}{113}%
\makeatletter
\providecommand \@ifxundefined [1]{%
 \@ifx{#1\undefined}
}%
\providecommand \@ifnum [1]{%
 \ifnum #1\expandafter \@firstoftwo
 \else \expandafter \@secondoftwo
 \fi
}%
\providecommand \@ifx [1]{%
 \ifx #1\expandafter \@firstoftwo
 \else \expandafter \@secondoftwo
 \fi
}%
\providecommand \natexlab [1]{#1}%
\providecommand \enquote  [1]{``#1''}%
\providecommand \bibnamefont  [1]{#1}%
\providecommand \bibfnamefont [1]{#1}%
\providecommand \citenamefont [1]{#1}%
\providecommand \href@noop [0]{\@secondoftwo}%
\providecommand \href [0]{\begingroup \@sanitize@url \@href}%
\providecommand \@href[1]{\@@startlink{#1}\@@href}%
\providecommand \@@href[1]{\endgroup#1\@@endlink}%
\providecommand \@sanitize@url [0]{\catcode `\\12\catcode `\$12\catcode
  `\&12\catcode `\#12\catcode `\^12\catcode `\_12\catcode `\%12\relax}%
\providecommand \@@startlink[1]{}%
\providecommand \@@endlink[0]{}%
\providecommand \url  [0]{\begingroup\@sanitize@url \@url }%
\providecommand \@url [1]{\endgroup\@href {#1}{\urlprefix }}%
\providecommand \urlprefix  [0]{URL }%
\providecommand \Eprint [0]{\href }%
\providecommand \doibase [0]{https://doi.org/}%
\providecommand \selectlanguage [0]{\@gobble}%
\providecommand \bibinfo  [0]{\@secondoftwo}%
\providecommand \bibfield  [0]{\@secondoftwo}%
\providecommand \translation [1]{[#1]}%
\providecommand \BibitemOpen [0]{}%
\providecommand \bibitemStop [0]{}%
\providecommand \bibitemNoStop [0]{.\EOS\space}%
\providecommand \EOS [0]{\spacefactor3000\relax}%
\providecommand \BibitemShut  [1]{\csname bibitem#1\endcsname}%
\let\auto@bib@innerbib\@empty
%</preamble>
\bibitem [{\citenamefont {Kondo}(2012)}]{kondo2012physics}%
  \BibitemOpen
  \bibfield  {author} {\bibinfo {author} {\bibfnamefont {J.}~\bibnamefont
  {Kondo}},\ }\href@noop {} {\emph {\bibinfo {title} {The physics of dilute
  magnetic alloys}}}\ (\bibinfo  {publisher} {Cambridge University Press},\
  \bibinfo {year} {2012})\BibitemShut {NoStop}%
\bibitem [{\citenamefont {Hewson}(1997)}]{hewson1997kondo}%
  \BibitemOpen
  \bibfield  {author} {\bibinfo {author} {\bibfnamefont {A.~C.}\ \bibnamefont
  {Hewson}},\ }\href@noop {} {\emph {\bibinfo {title} {The Kondo problem to
  heavy fermions}}},\ \bibinfo {number} {2}\ (\bibinfo  {publisher} {Cambridge
  university press},\ \bibinfo {year} {1997})\BibitemShut {NoStop}%
\bibitem [{\citenamefont {De~Haas}\ \emph {et~al.}(1934)\citenamefont
  {De~Haas}, \citenamefont {De~Boer},\ and\ \citenamefont {Van~den
  Berg}}]{de1934electrical}%
  \BibitemOpen
  \bibfield  {author} {\bibinfo {author} {\bibfnamefont {W.}~\bibnamefont
  {De~Haas}}, \bibinfo {author} {\bibfnamefont {J.}~\bibnamefont {De~Boer}},\
  and\ \bibinfo {author} {\bibfnamefont {G.}~\bibnamefont {Van~den Berg}},\
  }\bibfield  {title} {\bibinfo {title} {The electrical resistance of gold,
  copper and lead at low temperatures},\ }\href@noop {} {\bibfield  {journal}
  {\bibinfo  {journal} {Physica}\ }\textbf {\bibinfo {volume} {1}},\ \bibinfo
  {pages} {1115} (\bibinfo {year} {1934})}\BibitemShut {NoStop}%
\bibitem [{\citenamefont {Kondo}(1964)}]{kondo1964resistance}%
  \BibitemOpen
  \bibfield  {author} {\bibinfo {author} {\bibfnamefont {J.}~\bibnamefont
  {Kondo}},\ }\bibfield  {title} {\bibinfo {title} {Resistance minimum in
  dilute magnetic alloys},\ }\href@noop {} {\bibfield  {journal} {\bibinfo
  {journal} {Progress of theoretical physics}\ }\textbf {\bibinfo {volume}
  {32}},\ \bibinfo {pages} {37} (\bibinfo {year} {1964})}\BibitemShut {NoStop}%
\bibitem [{\citenamefont {Anderson}(2018)}]{anderson2018poor}%
  \BibitemOpen
  \bibfield  {author} {\bibinfo {author} {\bibfnamefont {P.}~\bibnamefont
  {Anderson}},\ }\bibfield  {title} {\bibinfo {title} {A poor man’s
  derivation of scaling laws for the kondo problem},\ }in\ \href@noop {} {\emph
  {\bibinfo {booktitle} {Basic Notions Of Condensed Matter Physics}}}\
  (\bibinfo  {publisher} {CRC Press},\ \bibinfo {year} {2018})\ pp.\ \bibinfo
  {pages} {483--488}\BibitemShut {NoStop}%
\bibitem [{\citenamefont {Nozieres}(1974)}]{nozieres1974fermi}%
  \BibitemOpen
  \bibfield  {author} {\bibinfo {author} {\bibfnamefont {P.}~\bibnamefont
  {Nozieres}},\ }\bibfield  {title} {\bibinfo {title} {A “fermi-liquid”
  description of the kondo problem at low temperatures},\ }\href@noop {}
  {\bibfield  {journal} {\bibinfo  {journal} {Journal of low temp{\'e}rature
  physics}\ }\textbf {\bibinfo {volume} {17}},\ \bibinfo {pages} {31} (\bibinfo
  {year} {1974})}\BibitemShut {NoStop}%
\bibitem [{\citenamefont {Nozieres}\ and\ \citenamefont
  {Blandin}(1980)}]{nozieres1980kondo}%
  \BibitemOpen
  \bibfield  {author} {\bibinfo {author} {\bibfnamefont {P.}~\bibnamefont
  {Nozieres}}\ and\ \bibinfo {author} {\bibfnamefont {A.}~\bibnamefont
  {Blandin}},\ }\bibfield  {title} {\bibinfo {title} {Kondo effect in real
  metals},\ }\href@noop {} {\bibfield  {journal} {\bibinfo  {journal} {Journal
  de Physique}\ }\textbf {\bibinfo {volume} {41}},\ \bibinfo {pages} {193}
  (\bibinfo {year} {1980})}\BibitemShut {NoStop}%
\bibitem [{\citenamefont {Wilson}(1975)}]{wilson1975renormalization}%
  \BibitemOpen
  \bibfield  {author} {\bibinfo {author} {\bibfnamefont {K.~G.}\ \bibnamefont
  {Wilson}},\ }\bibfield  {title} {\bibinfo {title} {The renormalization group:
  Critical phenomena and the kondo problem},\ }\href@noop {} {\bibfield
  {journal} {\bibinfo  {journal} {Reviews of modern physics}\ }\textbf
  {\bibinfo {volume} {47}},\ \bibinfo {pages} {773} (\bibinfo {year}
  {1975})}\BibitemShut {NoStop}%
\bibitem [{\citenamefont {Affleck}(1995)}]{affleck1995conformal}%
  \BibitemOpen
  \bibfield  {author} {\bibinfo {author} {\bibfnamefont {I.}~\bibnamefont
  {Affleck}},\ }\bibfield  {title} {\bibinfo {title} {Conformal field theory
  approach to the kondo effect},\ }\href@noop {} {\bibfield  {journal}
  {\bibinfo  {journal} {arXiv preprint cond-mat/9512099}\ } (\bibinfo {year}
  {1995})}\BibitemShut {NoStop}%
\bibitem [{\citenamefont {Affleck}\ and\ \citenamefont
  {Ludwig}(1991{\natexlab{a}})}]{PhysRevLett.67.161}%
  \BibitemOpen
  \bibfield  {author} {\bibinfo {author} {\bibfnamefont {I.}~\bibnamefont
  {Affleck}}\ and\ \bibinfo {author} {\bibfnamefont {A.~W.~W.}\ \bibnamefont
  {Ludwig}},\ }\bibfield  {title} {\bibinfo {title} {Universal noninteger
  ``ground-state degeneracy'' in critical quantum systems},\ }\href
  {https://doi.org/10.1103/PhysRevLett.67.161} {\bibfield  {journal} {\bibinfo
  {journal} {Phys. Rev. Lett.}\ }\textbf {\bibinfo {volume} {67}},\ \bibinfo
  {pages} {161} (\bibinfo {year} {1991}{\natexlab{a}})}\BibitemShut {NoStop}%
\bibitem [{\citenamefont {Friedan}\ and\ \citenamefont
  {Konechny}(2004)}]{friedan2004boundary}%
  \BibitemOpen
  \bibfield  {author} {\bibinfo {author} {\bibfnamefont {D.}~\bibnamefont
  {Friedan}}\ and\ \bibinfo {author} {\bibfnamefont {A.}~\bibnamefont
  {Konechny}},\ }\bibfield  {title} {\bibinfo {title} {Boundary entropy of
  one-dimensional quantum systems at low temperature},\ }\href@noop {}
  {\bibfield  {journal} {\bibinfo  {journal} {Physical review letters}\
  }\textbf {\bibinfo {volume} {93}},\ \bibinfo {pages} {030402} (\bibinfo
  {year} {2004})}\BibitemShut {NoStop}%
\bibitem [{\citenamefont {Krishnan}\ and\ \citenamefont
  {Metlitski}(2024)}]{Metlitski-Kondo}%
  \BibitemOpen
  \bibfield  {author} {\bibinfo {author} {\bibfnamefont {A.}~\bibnamefont
  {Krishnan}}\ and\ \bibinfo {author} {\bibfnamefont {M.~A.}\ \bibnamefont
  {Metlitski}},\ }\bibfield  {title} {\bibinfo {title} {The kondo impurity in
  the large spin limit},\ }\href@noop {} {\bibfield  {journal} {\bibinfo
  {journal} {arXiv preprint 2408.12650}\ } (\bibinfo {year}
  {2024})}\BibitemShut {NoStop}%
\bibitem [{\citenamefont {Krishna-Murthy}\ \emph
  {et~al.}(1980{\natexlab{a}})\citenamefont {Krishna-Murthy}, \citenamefont
  {Wilkins},\ and\ \citenamefont {Wilson}}]{krishna1980renormalizationa}%
  \BibitemOpen
  \bibfield  {author} {\bibinfo {author} {\bibfnamefont {H.}~\bibnamefont
  {Krishna-Murthy}}, \bibinfo {author} {\bibfnamefont {J.}~\bibnamefont
  {Wilkins}},\ and\ \bibinfo {author} {\bibfnamefont {K.}~\bibnamefont
  {Wilson}},\ }\bibfield  {title} {\bibinfo {title} {Renormalization-group
  approach to the anderson model of dilute magnetic alloys. i. static
  properties for the symmetric case},\ }\href@noop {} {\bibfield  {journal}
  {\bibinfo  {journal} {Physical Review B}\ }\textbf {\bibinfo {volume} {21}},\
  \bibinfo {pages} {1003} (\bibinfo {year} {1980}{\natexlab{a}})}\BibitemShut
  {NoStop}%
\bibitem [{\citenamefont {Krishna-Murthy}\ \emph
  {et~al.}(1980{\natexlab{b}})\citenamefont {Krishna-Murthy}, \citenamefont
  {Wilkins},\ and\ \citenamefont {Wilson}}]{krishna1980renormalizationb}%
  \BibitemOpen
  \bibfield  {author} {\bibinfo {author} {\bibfnamefont {H.}~\bibnamefont
  {Krishna-Murthy}}, \bibinfo {author} {\bibfnamefont {J.}~\bibnamefont
  {Wilkins}},\ and\ \bibinfo {author} {\bibfnamefont {K.}~\bibnamefont
  {Wilson}},\ }\bibfield  {title} {\bibinfo {title} {Renormalization-group
  approach to the anderson model of dilute magnetic alloys. ii. static
  properties for the asymmetric case},\ }\href@noop {} {\bibfield  {journal}
  {\bibinfo  {journal} {Physical Review B}\ }\textbf {\bibinfo {volume} {21}},\
  \bibinfo {pages} {1044} (\bibinfo {year} {1980}{\natexlab{b}})}\BibitemShut
  {NoStop}%
\bibitem [{\citenamefont {Andrei}(1980)}]{andrei1980diagonalization}%
  \BibitemOpen
  \bibfield  {author} {\bibinfo {author} {\bibfnamefont {N.}~\bibnamefont
  {Andrei}},\ }\bibfield  {title} {\bibinfo {title} {Diagonalization of the
  kondo hamiltonian},\ }\href@noop {} {\bibfield  {journal} {\bibinfo
  {journal} {Physical Review Letters}\ }\textbf {\bibinfo {volume} {45}},\
  \bibinfo {pages} {379} (\bibinfo {year} {1980})}\BibitemShut {NoStop}%
\bibitem [{\citenamefont {Wiegmann}(1981)}]{wiegmann1981exact}%
  \BibitemOpen
  \bibfield  {author} {\bibinfo {author} {\bibfnamefont {P.}~\bibnamefont
  {Wiegmann}},\ }\bibfield  {title} {\bibinfo {title} {Exact solution of the sd
  exchange model (kondo problem)},\ }\href@noop {} {\bibfield  {journal}
  {\bibinfo  {journal} {Journal of Physics C: Solid State Physics}\ }\textbf
  {\bibinfo {volume} {14}},\ \bibinfo {pages} {1463} (\bibinfo {year}
  {1981})}\BibitemShut {NoStop}%
\bibitem [{\citenamefont {Schiller}\ and\ \citenamefont
  {Ingersent}(1995)}]{schiller1995exact}%
  \BibitemOpen
  \bibfield  {author} {\bibinfo {author} {\bibfnamefont {A.}~\bibnamefont
  {Schiller}}\ and\ \bibinfo {author} {\bibfnamefont {K.}~\bibnamefont
  {Ingersent}},\ }\bibfield  {title} {\bibinfo {title} {Exact results for the
  kondo effect in a luttinger liquid},\ }\href@noop {} {\bibfield  {journal}
  {\bibinfo  {journal} {Physical Review B}\ }\textbf {\bibinfo {volume} {51}},\
  \bibinfo {pages} {4676} (\bibinfo {year} {1995})}\BibitemShut {NoStop}%
\bibitem [{\citenamefont {Furusaki}\ and\ \citenamefont
  {Nagaosa}(1994)}]{furusaki1994kondo}%
  \BibitemOpen
  \bibfield  {author} {\bibinfo {author} {\bibfnamefont {A.}~\bibnamefont
  {Furusaki}}\ and\ \bibinfo {author} {\bibfnamefont {N.}~\bibnamefont
  {Nagaosa}},\ }\bibfield  {title} {\bibinfo {title} {Kondo effect in a
  tomonaga-luttinger liquid},\ }\href@noop {} {\bibfield  {journal} {\bibinfo
  {journal} {Physical review letters}\ }\textbf {\bibinfo {volume} {72}},\
  \bibinfo {pages} {892} (\bibinfo {year} {1994})}\BibitemShut {NoStop}%
\bibitem [{\citenamefont {Fr{\"o}jdh}\ and\ \citenamefont
  {Johannesson}(1995)}]{frojdh1995kondo}%
  \BibitemOpen
  \bibfield  {author} {\bibinfo {author} {\bibfnamefont {P.}~\bibnamefont
  {Fr{\"o}jdh}}\ and\ \bibinfo {author} {\bibfnamefont {H.}~\bibnamefont
  {Johannesson}},\ }\bibfield  {title} {\bibinfo {title} {Kondo effect in a
  luttinger liquid: Exact results from conformal field theory},\ }\href@noop {}
  {\bibfield  {journal} {\bibinfo  {journal} {Physical review letters}\
  }\textbf {\bibinfo {volume} {75}},\ \bibinfo {pages} {300} (\bibinfo {year}
  {1995})}\BibitemShut {NoStop}%
\bibitem [{\citenamefont {Lee}\ and\ \citenamefont
  {Toner}(1992)}]{lee1992kondo}%
  \BibitemOpen
  \bibfield  {author} {\bibinfo {author} {\bibfnamefont {D.-H.}\ \bibnamefont
  {Lee}}\ and\ \bibinfo {author} {\bibfnamefont {J.}~\bibnamefont {Toner}},\
  }\bibfield  {title} {\bibinfo {title} {Kondo effect in a luttinger liquid},\
  }\href@noop {} {\bibfield  {journal} {\bibinfo  {journal} {Physical review
  letters}\ }\textbf {\bibinfo {volume} {69}},\ \bibinfo {pages} {3378}
  (\bibinfo {year} {1992})}\BibitemShut {NoStop}%
\bibitem [{\citenamefont {Egger}\ and\ \citenamefont
  {Komnik}(1998)}]{KLLegger1998scaling}%
  \BibitemOpen
  \bibfield  {author} {\bibinfo {author} {\bibfnamefont {R.}~\bibnamefont
  {Egger}}\ and\ \bibinfo {author} {\bibfnamefont {A.}~\bibnamefont {Komnik}},\
  }\bibfield  {title} {\bibinfo {title} {Scaling and criticality of the kondo
  effect in a luttinger liquid},\ }\href@noop {} {\bibfield  {journal}
  {\bibinfo  {journal} {Physical Review B}\ }\textbf {\bibinfo {volume} {57}},\
  \bibinfo {pages} {10620} (\bibinfo {year} {1998})}\BibitemShut {NoStop}%
\bibitem [{\citenamefont {Cuevas}\ \emph {et~al.}(2001)\citenamefont {Cuevas},
  \citenamefont {Yeyati},\ and\ \citenamefont
  {Mart{\'\i}n-Rodero}}]{KScuevas2001kondo}%
  \BibitemOpen
  \bibfield  {author} {\bibinfo {author} {\bibfnamefont {J.}~\bibnamefont
  {Cuevas}}, \bibinfo {author} {\bibfnamefont {A.~L.}\ \bibnamefont {Yeyati}},\
  and\ \bibinfo {author} {\bibfnamefont {A.}~\bibnamefont
  {Mart{\'\i}n-Rodero}},\ }\bibfield  {title} {\bibinfo {title} {Kondo effect
  in normal-superconductor quantum dots},\ }\href@noop {} {\bibfield  {journal}
  {\bibinfo  {journal} {Physical Review B}\ }\textbf {\bibinfo {volume} {63}},\
  \bibinfo {pages} {094515} (\bibinfo {year} {2001})}\BibitemShut {NoStop}%
\bibitem [{\citenamefont {M{\"u}ller-Hartmann}\ and\ \citenamefont
  {Zittartz}(1971)}]{KSmuller1971kondo}%
  \BibitemOpen
  \bibfield  {author} {\bibinfo {author} {\bibfnamefont {E.}~\bibnamefont
  {M{\"u}ller-Hartmann}}\ and\ \bibinfo {author} {\bibfnamefont
  {J.}~\bibnamefont {Zittartz}},\ }\bibfield  {title} {\bibinfo {title} {Kondo
  effect in superconductors},\ }\href@noop {} {\bibfield  {journal} {\bibinfo
  {journal} {Physical Review Letters}\ }\textbf {\bibinfo {volume} {26}},\
  \bibinfo {pages} {428} (\bibinfo {year} {1971})}\BibitemShut {NoStop}%
\bibitem [{\citenamefont {Borkowski}\ and\ \citenamefont
  {Hirschfeld}(1992)}]{KSborkowski1992kondo}%
  \BibitemOpen
  \bibfield  {author} {\bibinfo {author} {\bibfnamefont {L.~S.}\ \bibnamefont
  {Borkowski}}\ and\ \bibinfo {author} {\bibfnamefont {P.}~\bibnamefont
  {Hirschfeld}},\ }\bibfield  {title} {\bibinfo {title} {Kondo effect in
  gapless superconductors},\ }\href@noop {} {\bibfield  {journal} {\bibinfo
  {journal} {Physical Review B}\ }\textbf {\bibinfo {volume} {46}},\ \bibinfo
  {pages} {9274} (\bibinfo {year} {1992})}\BibitemShut {NoStop}%
\bibitem [{\citenamefont {Steglich}\ and\ \citenamefont
  {Wirth}(2016)}]{KSsteglich2016foundations}%
  \BibitemOpen
  \bibfield  {author} {\bibinfo {author} {\bibfnamefont {F.}~\bibnamefont
  {Steglich}}\ and\ \bibinfo {author} {\bibfnamefont {S.}~\bibnamefont
  {Wirth}},\ }\bibfield  {title} {\bibinfo {title} {Foundations of
  heavy-fermion superconductivity: lattice kondo effect and mott physics},\
  }\href@noop {} {\bibfield  {journal} {\bibinfo  {journal} {Reports on
  Progress in Physics}\ }\textbf {\bibinfo {volume} {79}},\ \bibinfo {pages}
  {084502} (\bibinfo {year} {2016})}\BibitemShut {NoStop}%
\bibitem [{\citenamefont {Shiba}(1968)}]{Shiba}%
  \BibitemOpen
  \bibfield  {author} {\bibinfo {author} {\bibfnamefont {H.}~\bibnamefont
  {Shiba}},\ }\bibfield  {title} {\bibinfo {title} {{Classical Spins in
  Superconductors}},\ }\href {https://doi.org/10.1143/PTP.40.435} {\bibfield
  {journal} {\bibinfo  {journal} {Progress of Theoretical Physics}\ }\textbf
  {\bibinfo {volume} {40}},\ \bibinfo {pages} {435} (\bibinfo {year}
  {1968})}\BibitemShut {NoStop}%
\bibitem [{\citenamefont {LUH}(1965)}]{Yu}%
  \BibitemOpen
  \bibfield  {author} {\bibinfo {author} {\bibfnamefont {Y.}~\bibnamefont
  {LUH}},\ }\bibfield  {title} {\bibinfo {title} {Bound state in
  superconductors with paramagnetic impurities},\ }\href
  {https://doi.org/10.7498/aps.21.75} {\bibfield  {journal} {\bibinfo
  {journal} {Acta Physica Sinica}\ }\textbf {\bibinfo {volume} {21}},\ \bibinfo
  {eid} {75} (\bibinfo {year} {1965})}\BibitemShut {NoStop}%
\bibitem [{\citenamefont {{Rusinov}}(1969)}]{Rusinov}%
  \BibitemOpen
  \bibfield  {author} {\bibinfo {author} {\bibfnamefont {A.~I.}\ \bibnamefont
  {{Rusinov}}},\ }\bibfield  {title} {\bibinfo {title} {{Superconductivity near
  a Paramagnetic Impurity}},\ }\href@noop {} {\bibfield  {journal} {\bibinfo
  {journal} {Soviet Journal of Experimental and Theoretical Physics Letters}\
  }\textbf {\bibinfo {volume} {9}},\ \bibinfo {pages} {85} (\bibinfo {year}
  {1969})}\BibitemShut {NoStop}%
\bibitem [{\citenamefont {Kim}\ and\ \citenamefont {Kim}(2008)}]{kim2008kondo}%
  \BibitemOpen
  \bibfield  {author} {\bibinfo {author} {\bibfnamefont {K.-S.}\ \bibnamefont
  {Kim}}\ and\ \bibinfo {author} {\bibfnamefont {M.~D.}\ \bibnamefont {Kim}},\
  }\bibfield  {title} {\bibinfo {title} {Kondo physics in the algebraic spin
  liquid},\ }\href@noop {} {\bibfield  {journal} {\bibinfo  {journal} {Journal
  of Physics: Condensed Matter}\ }\textbf {\bibinfo {volume} {20}},\ \bibinfo
  {pages} {125206} (\bibinfo {year} {2008})}\BibitemShut {NoStop}%
\bibitem [{\citenamefont {Florens}\ \emph {et~al.}(2006)\citenamefont
  {Florens}, \citenamefont {Fritz},\ and\ \citenamefont
  {Vojta}}]{florens2006kondo}%
  \BibitemOpen
  \bibfield  {author} {\bibinfo {author} {\bibfnamefont {S.}~\bibnamefont
  {Florens}}, \bibinfo {author} {\bibfnamefont {L.}~\bibnamefont {Fritz}},\
  and\ \bibinfo {author} {\bibfnamefont {M.}~\bibnamefont {Vojta}},\ }\bibfield
   {title} {\bibinfo {title} {Kondo effect in bosonic spin liquids},\
  }\href@noop {} {\bibfield  {journal} {\bibinfo  {journal} {Physical review
  letters}\ }\textbf {\bibinfo {volume} {96}},\ \bibinfo {pages} {036601}
  (\bibinfo {year} {2006})}\BibitemShut {NoStop}%
\bibitem [{\citenamefont {Hattori}\ \emph {et~al.}(2015)\citenamefont
  {Hattori}, \citenamefont {Itakura}, \citenamefont {Ozaki},\ and\
  \citenamefont {Yasui}}]{QCDhattori2015qcd}%
  \BibitemOpen
  \bibfield  {author} {\bibinfo {author} {\bibfnamefont {K.}~\bibnamefont
  {Hattori}}, \bibinfo {author} {\bibfnamefont {K.}~\bibnamefont {Itakura}},
  \bibinfo {author} {\bibfnamefont {S.}~\bibnamefont {Ozaki}},\ and\ \bibinfo
  {author} {\bibfnamefont {S.}~\bibnamefont {Yasui}},\ }\bibfield  {title}
  {\bibinfo {title} {Qcd kondo effect: quark matter with heavy-flavor
  impurities},\ }\href@noop {} {\bibfield  {journal} {\bibinfo  {journal}
  {Physical Review D}\ }\textbf {\bibinfo {volume} {92}},\ \bibinfo {pages}
  {065003} (\bibinfo {year} {2015})}\BibitemShut {NoStop}%
\bibitem [{\citenamefont {Suenaga}\ \emph {et~al.}(2020)\citenamefont
  {Suenaga}, \citenamefont {Suzuki},\ and\ \citenamefont
  {Yasui}}]{QCDsuenaga2020qcd}%
  \BibitemOpen
  \bibfield  {author} {\bibinfo {author} {\bibfnamefont {D.}~\bibnamefont
  {Suenaga}}, \bibinfo {author} {\bibfnamefont {K.}~\bibnamefont {Suzuki}},\
  and\ \bibinfo {author} {\bibfnamefont {S.}~\bibnamefont {Yasui}},\ }\bibfield
   {title} {\bibinfo {title} {Qcd kondo excitons},\ }\href@noop {} {\bibfield
  {journal} {\bibinfo  {journal} {Physical Review Research}\ }\textbf {\bibinfo
  {volume} {2}},\ \bibinfo {pages} {023066} (\bibinfo {year}
  {2020})}\BibitemShut {NoStop}%
\bibitem [{\citenamefont {Kimura}\ and\ \citenamefont
  {Ozaki}(2019)}]{QCDkimura2019conformal}%
  \BibitemOpen
  \bibfield  {author} {\bibinfo {author} {\bibfnamefont {T.}~\bibnamefont
  {Kimura}}\ and\ \bibinfo {author} {\bibfnamefont {S.}~\bibnamefont {Ozaki}},\
  }\bibfield  {title} {\bibinfo {title} {Conformal field theory analysis of the
  qcd kondo effect},\ }\href@noop {} {\bibfield  {journal} {\bibinfo  {journal}
  {Physical Review D}\ }\textbf {\bibinfo {volume} {99}},\ \bibinfo {pages}
  {014040} (\bibinfo {year} {2019})}\BibitemShut {NoStop}%
\bibitem [{\citenamefont {Ozaki}\ \emph {et~al.}(2016)\citenamefont {Ozaki},
  \citenamefont {Itakura},\ and\ \citenamefont
  {Kuramoto}}]{QCDozaki2016magnetically}%
  \BibitemOpen
  \bibfield  {author} {\bibinfo {author} {\bibfnamefont {S.}~\bibnamefont
  {Ozaki}}, \bibinfo {author} {\bibfnamefont {K.}~\bibnamefont {Itakura}},\
  and\ \bibinfo {author} {\bibfnamefont {Y.}~\bibnamefont {Kuramoto}},\
  }\bibfield  {title} {\bibinfo {title} {Magnetically induced qcd kondo
  effect},\ }\href@noop {} {\bibfield  {journal} {\bibinfo  {journal} {Physical
  Review D}\ }\textbf {\bibinfo {volume} {94}},\ \bibinfo {pages} {074013}
  (\bibinfo {year} {2016})}\BibitemShut {NoStop}%
\bibitem [{\citenamefont {Giuliano}\ \emph {et~al.}(2018)\citenamefont
  {Giuliano}, \citenamefont {Rossini},\ and\ \citenamefont
  {Trombettoni}}]{XXZkondoDMRG}%
  \BibitemOpen
  \bibfield  {author} {\bibinfo {author} {\bibfnamefont {D.}~\bibnamefont
  {Giuliano}}, \bibinfo {author} {\bibfnamefont {D.}~\bibnamefont {Rossini}},\
  and\ \bibinfo {author} {\bibfnamefont {A.}~\bibnamefont {Trombettoni}},\
  }\bibfield  {title} {\bibinfo {title} {From kondo effect to weak-link regime
  in quantum spin-$\frac{1}{2}$ spin chains},\ }\href
  {https://doi.org/10.1103/PhysRevB.98.235164} {\bibfield  {journal} {\bibinfo
  {journal} {Phys. Rev. B}\ }\textbf {\bibinfo {volume} {98}},\ \bibinfo
  {pages} {235164} (\bibinfo {year} {2018})}\BibitemShut {NoStop}%
\bibitem [{\citenamefont {Laflorencie}\ \emph {et~al.}(2008)\citenamefont
  {Laflorencie}, \citenamefont {S{\o}rensen},\ and\ \citenamefont
  {Affleck}}]{laflorencie2008kondo}%
  \BibitemOpen
  \bibfield  {author} {\bibinfo {author} {\bibfnamefont {N.}~\bibnamefont
  {Laflorencie}}, \bibinfo {author} {\bibfnamefont {E.~S.}\ \bibnamefont
  {S{\o}rensen}},\ and\ \bibinfo {author} {\bibfnamefont {I.}~\bibnamefont
  {Affleck}},\ }\bibfield  {title} {\bibinfo {title} {The kondo effect in spin
  chains},\ }\href@noop {} {\bibfield  {journal} {\bibinfo  {journal} {Journal
  of Statistical Mechanics: Theory and Experiment}\ }\textbf {\bibinfo {volume}
  {2008}},\ \bibinfo {pages} {P02007} (\bibinfo {year} {2008})}\BibitemShut
  {NoStop}%
\bibitem [{\citenamefont {Kattel}\ \emph
  {et~al.}(2024{\natexlab{a}})\citenamefont {Kattel}, \citenamefont {Pasnoori},
  \citenamefont {Pixley}, \citenamefont {Azaria},\ and\ \citenamefont
  {Andrei}}]{kattel2023kondo}%
  \BibitemOpen
  \bibfield  {author} {\bibinfo {author} {\bibfnamefont {P.}~\bibnamefont
  {Kattel}}, \bibinfo {author} {\bibfnamefont {P.~R.}\ \bibnamefont
  {Pasnoori}}, \bibinfo {author} {\bibfnamefont {J.~H.}\ \bibnamefont
  {Pixley}}, \bibinfo {author} {\bibfnamefont {P.}~\bibnamefont {Azaria}},\
  and\ \bibinfo {author} {\bibfnamefont {N.}~\bibnamefont {Andrei}},\
  }\bibfield  {title} {\bibinfo {title} {Kondo effect in the isotropic
  heisenberg spin chain},\ }\href {https://doi.org/10.1103/PhysRevB.109.174416}
  {\bibfield  {journal} {\bibinfo  {journal} {Phys. Rev. B}\ }\textbf {\bibinfo
  {volume} {109}},\ \bibinfo {pages} {174416} (\bibinfo {year}
  {2024}{\natexlab{a}})}\BibitemShut {NoStop}%
\bibitem [{\citenamefont {Wang}(1997)}]{wang1997exact}%
  \BibitemOpen
  \bibfield  {author} {\bibinfo {author} {\bibfnamefont {Y.}~\bibnamefont
  {Wang}},\ }\bibfield  {title} {\bibinfo {title} {Exact solution of the open
  heisenberg chain with two impurities},\ }\href@noop {} {\bibfield  {journal}
  {\bibinfo  {journal} {Physical Review B}\ }\textbf {\bibinfo {volume} {56}},\
  \bibinfo {pages} {14045} (\bibinfo {year} {1997})}\BibitemShut {NoStop}%
\bibitem [{\citenamefont {Frahm}\ and\ \citenamefont
  {Zvyagin}(1997)}]{frahm1997open}%
  \BibitemOpen
  \bibfield  {author} {\bibinfo {author} {\bibfnamefont {H.}~\bibnamefont
  {Frahm}}\ and\ \bibinfo {author} {\bibfnamefont {A.~A.}\ \bibnamefont
  {Zvyagin}},\ }\bibfield  {title} {\bibinfo {title} {The open spin chain with
  impurity: an exact solution},\ }\href@noop {} {\bibfield  {journal} {\bibinfo
   {journal} {Journal of Physics: Condensed Matter}\ }\textbf {\bibinfo
  {volume} {9}},\ \bibinfo {pages} {9939} (\bibinfo {year} {1997})}\BibitemShut
  {NoStop}%
\bibitem [{\citenamefont {Furusaki}\ and\ \citenamefont
  {Hikihara}(1998)}]{furusaki1998kondo}%
  \BibitemOpen
  \bibfield  {author} {\bibinfo {author} {\bibfnamefont {A.}~\bibnamefont
  {Furusaki}}\ and\ \bibinfo {author} {\bibfnamefont {T.}~\bibnamefont
  {Hikihara}},\ }\bibfield  {title} {\bibinfo {title} {Kondo effect in xxz spin
  chains},\ }\href@noop {} {\bibfield  {journal} {\bibinfo  {journal} {Physical
  Review B}\ }\textbf {\bibinfo {volume} {58}},\ \bibinfo {pages} {5529}
  (\bibinfo {year} {1998})}\BibitemShut {NoStop}%
\bibitem [{\citenamefont {Andrei}\ and\ \citenamefont
  {Johannesson}(1984)}]{andrei1984heisenberg}%
  \BibitemOpen
  \bibfield  {author} {\bibinfo {author} {\bibfnamefont {N.}~\bibnamefont
  {Andrei}}\ and\ \bibinfo {author} {\bibfnamefont {H.}~\bibnamefont
  {Johannesson}},\ }\bibfield  {title} {\bibinfo {title} {Heisenberg chain with
  impurities (an integrable model)},\ }\href@noop {} {\bibfield  {journal}
  {\bibinfo  {journal} {Physics Letters A}\ }\textbf {\bibinfo {volume}
  {100}},\ \bibinfo {pages} {108} (\bibinfo {year} {1984})}\BibitemShut
  {NoStop}%
\bibitem [{\citenamefont {Schlottmann}(1991)}]{schlottmann}%
  \BibitemOpen
  \bibfield  {author} {\bibinfo {author} {\bibfnamefont {P.}~\bibnamefont
  {Schlottmann}},\ }\bibfield  {title} {\bibinfo {title} {Impurity-induced
  critical behaviour in antiferromagnetic heisenberg chains},\ }\href
  {https://doi.org/10.1088/0953-8984/3/34/008} {\bibfield  {journal} {\bibinfo
  {journal} {Journal of Physics: Condensed Matter}\ }\textbf {\bibinfo {volume}
  {3}},\ \bibinfo {pages} {6617} (\bibinfo {year} {1991})}\BibitemShut
  {NoStop}%
\bibitem [{\citenamefont {Kattel}\ \emph
  {et~al.}(2024{\natexlab{b}})\citenamefont {Kattel}, \citenamefont {Tang},
  \citenamefont {Pixley},\ and\ \citenamefont {Andrei}}]{kattel2024kondo}%
  \BibitemOpen
  \bibfield  {author} {\bibinfo {author} {\bibfnamefont {P.}~\bibnamefont
  {Kattel}}, \bibinfo {author} {\bibfnamefont {Y.}~\bibnamefont {Tang}},
  \bibinfo {author} {\bibfnamefont {J.~H.}\ \bibnamefont {Pixley}},\ and\
  \bibinfo {author} {\bibfnamefont {N.}~\bibnamefont {Andrei}},\ }\bibfield
  {title} {\bibinfo {title} {The kondo effect in the quantum xx spin chain},\
  }\href@noop {} {\bibfield  {journal} {\bibinfo  {journal} {Journal of Physics
  A: Mathematical and Theoretical}\ } (\bibinfo {year}
  {2024}{\natexlab{b}})}\BibitemShut {NoStop}%
\bibitem [{\citenamefont {Hou}\ \emph {et~al.}(1999)\citenamefont {Hou},
  \citenamefont {Shi}, \citenamefont {Yue},\ and\ \citenamefont
  {Zhao}}]{hou1999integrability}%
  \BibitemOpen
  \bibfield  {author} {\bibinfo {author} {\bibfnamefont {B.}~\bibnamefont
  {Hou}}, \bibinfo {author} {\bibfnamefont {K.}~\bibnamefont {Shi}}, \bibinfo
  {author} {\bibfnamefont {R.}~\bibnamefont {Yue}},\ and\ \bibinfo {author}
  {\bibfnamefont {S.}~\bibnamefont {Zhao}},\ }\bibfield  {title} {\bibinfo
  {title} {Integrability of the heisenberg chains with boundary impurities and
  their bethe ansatz},\ }\href@noop {} {\bibfield  {journal} {\bibinfo
  {journal} {Journal of Physics A: Mathematical and General}\ }\textbf
  {\bibinfo {volume} {32}},\ \bibinfo {pages} {7623} (\bibinfo {year}
  {1999})}\BibitemShut {NoStop}%
\bibitem [{\citenamefont {Chen}\ \emph
  {et~al.}(1998{\natexlab{a}})\citenamefont {Chen}, \citenamefont {Wang},\ and\
  \citenamefont {Pu}}]{chen1998integrability}%
  \BibitemOpen
  \bibfield  {author} {\bibinfo {author} {\bibfnamefont {S.}~\bibnamefont
  {Chen}}, \bibinfo {author} {\bibfnamefont {Y.}~\bibnamefont {Wang}},\ and\
  \bibinfo {author} {\bibfnamefont {F.-C.}\ \bibnamefont {Pu}},\ }\bibfield
  {title} {\bibinfo {title} {Integrability of spin chain with boundary
  impurities},\ }\href@noop {} {\bibfield  {journal} {\bibinfo  {journal}
  {Physics Letters A}\ }\textbf {\bibinfo {volume} {247}},\ \bibinfo {pages}
  {176} (\bibinfo {year} {1998}{\natexlab{a}})}\BibitemShut {NoStop}%
\bibitem [{\citenamefont {Hu}\ and\ \citenamefont {Pu}(1998)}]{hu1998two}%
  \BibitemOpen
  \bibfield  {author} {\bibinfo {author} {\bibfnamefont {Z.-N.}\ \bibnamefont
  {Hu}}\ and\ \bibinfo {author} {\bibfnamefont {F.-C.}\ \bibnamefont {Pu}},\
  }\bibfield  {title} {\bibinfo {title} {Two magnetic impurities in a spin
  chain},\ }\href@noop {} {\bibfield  {journal} {\bibinfo  {journal} {Physical
  Review B}\ }\textbf {\bibinfo {volume} {58}},\ \bibinfo {pages} {R2925}
  (\bibinfo {year} {1998})}\BibitemShut {NoStop}%
\bibitem [{\citenamefont {Chen}\ \emph
  {et~al.}(1998{\natexlab{b}})\citenamefont {Chen}, \citenamefont {Wang},\ and\
  \citenamefont {Pu}}]{Shu_Chen_1998}%
  \BibitemOpen
  \bibfield  {author} {\bibinfo {author} {\bibfnamefont {S.}~\bibnamefont
  {Chen}}, \bibinfo {author} {\bibfnamefont {Y.}~\bibnamefont {Wang}},\ and\
  \bibinfo {author} {\bibfnamefont {F.-C.}\ \bibnamefont {Pu}},\ }\bibfield
  {title} {\bibinfo {title} {The open xxz chain with boundary impurities},\
  }\href {https://doi.org/10.1088/0305-4470/31/20/005} {\bibfield  {journal}
  {\bibinfo  {journal} {Journal of Physics A: Mathematical and General}\
  }\textbf {\bibinfo {volume} {31}},\ \bibinfo {pages} {4619} (\bibinfo {year}
  {1998}{\natexlab{b}})}\BibitemShut {NoStop}%
\bibitem [{\citenamefont {Bethe}(1931)}]{1931_Bethe_ZP_71}%
  \BibitemOpen
  \bibfield  {author} {\bibinfo {author} {\bibfnamefont {H.~A.}\ \bibnamefont
  {Bethe}},\ }\bibfield  {title} {\bibinfo {title} {Zur {T}heorie der
  {M}etalle. i. {E}igenwerte und {E}igenfunktionen der linearen {A}tomkette},\
  }\href {https://doi.org/10.1007/BF01341708} {\bibfield  {journal} {\bibinfo
  {journal} {Zeit. f\"ur Physik}\ }\textbf {\bibinfo {volume} {71}},\ \bibinfo
  {pages} {205} (\bibinfo {year} {1931})}\BibitemShut {NoStop}%
\bibitem [{\citenamefont {Baxter}(2016)}]{baxter2016exactly}%
  \BibitemOpen
  \bibfield  {author} {\bibinfo {author} {\bibfnamefont {R.~J.}\ \bibnamefont
  {Baxter}},\ }\href@noop {} {\emph {\bibinfo {title} {Exactly solved models in
  statistical mechanics}}}\ (\bibinfo  {publisher} {Elsevier},\ \bibinfo {year}
  {2016})\BibitemShut {NoStop}%
\bibitem [{\citenamefont {Sutherland}(2004)}]{sutherland2004beautiful}%
  \BibitemOpen
  \bibfield  {author} {\bibinfo {author} {\bibfnamefont {B.}~\bibnamefont
  {Sutherland}},\ }\href@noop {} {\emph {\bibinfo {title} {Beautiful models: 70
  years of exactly solved quantum many-body problems}}}\ (\bibinfo  {publisher}
  {World Scientific},\ \bibinfo {year} {2004})\BibitemShut {NoStop}%
\bibitem [{\citenamefont {Gaudin}(2014)}]{gaudin2014bethe}%
  \BibitemOpen
  \bibfield  {author} {\bibinfo {author} {\bibfnamefont {M.}~\bibnamefont
  {Gaudin}},\ }\href@noop {} {\emph {\bibinfo {title} {The bethe
  wavefunction}}}\ (\bibinfo  {publisher} {Cambridge University Press},\
  \bibinfo {year} {2014})\BibitemShut {NoStop}%
\bibitem [{\citenamefont {Slavnov}(2022)}]{slavnov2022algebraic}%
  \BibitemOpen
  \bibfield  {author} {\bibinfo {author} {\bibfnamefont {N.}~\bibnamefont
  {Slavnov}},\ }\href@noop {} {\emph {\bibinfo {title} {Algebraic Bethe ansatz
  and correlation functions: an advanced course}}}\ (\bibinfo  {publisher}
  {World Scientific},\ \bibinfo {year} {2022})\BibitemShut {NoStop}%
\bibitem [{\citenamefont {Franchini}\ \emph {et~al.}(2017)\citenamefont
  {Franchini} \emph {et~al.}}]{franchini2017introduction}%
  \BibitemOpen
  \bibfield  {author} {\bibinfo {author} {\bibfnamefont {F.}~\bibnamefont
  {Franchini}} \emph {et~al.},\ }\href@noop {} {\emph {\bibinfo {title} {An
  introduction to integrable techniques for one-dimensional quantum
  systems}}},\ Vol.\ \bibinfo {volume} {940}\ (\bibinfo  {publisher}
  {Springer},\ \bibinfo {year} {2017})\BibitemShut {NoStop}%
\bibitem [{\citenamefont {Eckle}(2019)}]{eckle2019models}%
  \BibitemOpen
  \bibfield  {author} {\bibinfo {author} {\bibfnamefont {H.-P.}\ \bibnamefont
  {Eckle}},\ }\href@noop {} {\emph {\bibinfo {title} {Models of Quantum Matter:
  A First Course on Integrability and the Bethe Ansatz}}}\ (\bibinfo
  {publisher} {Oxford University Press},\ \bibinfo {year} {2019})\BibitemShut
  {NoStop}%
\bibitem [{\citenamefont {Schollw{\"o}ck}(2005)}]{schollwock2005density}%
  \BibitemOpen
  \bibfield  {author} {\bibinfo {author} {\bibfnamefont {U.}~\bibnamefont
  {Schollw{\"o}ck}},\ }\bibfield  {title} {\bibinfo {title} {The density-matrix
  renormalization group},\ }\href@noop {} {\bibfield  {journal} {\bibinfo
  {journal} {Reviews of modern physics}\ }\textbf {\bibinfo {volume} {77}},\
  \bibinfo {pages} {259} (\bibinfo {year} {2005})}\BibitemShut {NoStop}%
\bibitem [{\citenamefont {Schollw{\"o}ck}(2011)}]{schollwock2011density}%
  \BibitemOpen
  \bibfield  {author} {\bibinfo {author} {\bibfnamefont {U.}~\bibnamefont
  {Schollw{\"o}ck}},\ }\bibfield  {title} {\bibinfo {title} {The density-matrix
  renormalization group in the age of matrix product states},\ }\href@noop {}
  {\bibfield  {journal} {\bibinfo  {journal} {Annals of physics}\ }\textbf
  {\bibinfo {volume} {326}},\ \bibinfo {pages} {96} (\bibinfo {year}
  {2011})}\BibitemShut {NoStop}%
\bibitem [{\citenamefont {Kapustin}\ and\ \citenamefont
  {Skorik}(1996)}]{kapustin1996surface}%
  \BibitemOpen
  \bibfield  {author} {\bibinfo {author} {\bibfnamefont {A.}~\bibnamefont
  {Kapustin}}\ and\ \bibinfo {author} {\bibfnamefont {S.}~\bibnamefont
  {Skorik}},\ }\bibfield  {title} {\bibinfo {title} {Surface excitations and
  surface energy of the antiferromagnetic xxz chain by the bethe ansatz
  approach},\ }\href@noop {} {\bibfield  {journal} {\bibinfo  {journal}
  {Journal of Physics A: Mathematical and General}\ }\textbf {\bibinfo {volume}
  {29}},\ \bibinfo {pages} {1629} (\bibinfo {year} {1996})}\BibitemShut
  {NoStop}%
\bibitem [{\citenamefont {Grijalva}\ \emph {et~al.}(2019)\citenamefont
  {Grijalva}, \citenamefont {De~Nardis},\ and\ \citenamefont
  {Terras}}]{grijalva2019open}%
  \BibitemOpen
  \bibfield  {author} {\bibinfo {author} {\bibfnamefont {S.}~\bibnamefont
  {Grijalva}}, \bibinfo {author} {\bibfnamefont {J.}~\bibnamefont
  {De~Nardis}},\ and\ \bibinfo {author} {\bibfnamefont {V.}~\bibnamefont
  {Terras}},\ }\bibfield  {title} {\bibinfo {title} {Open xxz chain and
  boundary modes at zero temperature},\ }\href@noop {} {\bibfield  {journal}
  {\bibinfo  {journal} {SciPost Physics}\ }\textbf {\bibinfo {volume} {7}},\
  \bibinfo {pages} {023} (\bibinfo {year} {2019})}\BibitemShut {NoStop}%
\bibitem [{\citenamefont {Pasnoori}\ \emph
  {et~al.}(2023{\natexlab{a}})\citenamefont {Pasnoori}, \citenamefont {Tang},
  \citenamefont {Lee}, \citenamefont {Pixley}, \citenamefont {Andrei},\ and\
  \citenamefont {Azaria}}]{pasnoori2023spin}%
  \BibitemOpen
  \bibfield  {author} {\bibinfo {author} {\bibfnamefont {P.~R.}\ \bibnamefont
  {Pasnoori}}, \bibinfo {author} {\bibfnamefont {Y.}~\bibnamefont {Tang}},
  \bibinfo {author} {\bibfnamefont {J.}~\bibnamefont {Lee}}, \bibinfo {author}
  {\bibfnamefont {J.}~\bibnamefont {Pixley}}, \bibinfo {author} {\bibfnamefont
  {N.}~\bibnamefont {Andrei}},\ and\ \bibinfo {author} {\bibfnamefont
  {P.}~\bibnamefont {Azaria}},\ }\bibfield  {title} {\bibinfo {title} {Spin
  fractionalization and zero modes in the spin-$\frac{1}{2}$ xxz chain with
  boundary fields},\ }\href@noop {} {\bibfield  {journal} {\bibinfo  {journal}
  {arXiv preprint arXiv:2312.05970}\ } (\bibinfo {year}
  {2023}{\natexlab{a}})}\BibitemShut {NoStop}%
\bibitem [{\citenamefont {Fendley}(2016)}]{fendley2016strong}%
  \BibitemOpen
  \bibfield  {author} {\bibinfo {author} {\bibfnamefont {P.}~\bibnamefont
  {Fendley}},\ }\bibfield  {title} {\bibinfo {title} {Strong zero modes and
  eigenstate phase transitions in the xyz/interacting majorana chain},\
  }\href@noop {} {\bibfield  {journal} {\bibinfo  {journal} {Journal of Physics
  A: Mathematical and Theoretical}\ }\textbf {\bibinfo {volume} {49}},\
  \bibinfo {pages} {30LT01} (\bibinfo {year} {2016})}\BibitemShut {NoStop}%
\bibitem [{\citenamefont {Fendley}(2012)}]{fendley2012parafermionic}%
  \BibitemOpen
  \bibfield  {author} {\bibinfo {author} {\bibfnamefont {P.}~\bibnamefont
  {Fendley}},\ }\bibfield  {title} {\bibinfo {title} {Parafermionic edge zero
  modes in zn-invariant spin chains},\ }\href@noop {} {\bibfield  {journal}
  {\bibinfo  {journal} {Journal of Statistical Mechanics: Theory and
  Experiment}\ }\textbf {\bibinfo {volume} {2012}},\ \bibinfo {pages} {P11020}
  (\bibinfo {year} {2012})}\BibitemShut {NoStop}%
\bibitem [{\citenamefont {Yates}\ \emph {et~al.}(2020)\citenamefont {Yates},
  \citenamefont {Abanov},\ and\ \citenamefont {Mitra}}]{yates2020dynamics}%
  \BibitemOpen
  \bibfield  {author} {\bibinfo {author} {\bibfnamefont {D.~J.}\ \bibnamefont
  {Yates}}, \bibinfo {author} {\bibfnamefont {A.~G.}\ \bibnamefont {Abanov}},\
  and\ \bibinfo {author} {\bibfnamefont {A.}~\bibnamefont {Mitra}},\ }\bibfield
   {title} {\bibinfo {title} {Dynamics of almost strong edge modes in spin
  chains away from integrability},\ }\href@noop {} {\bibfield  {journal}
  {\bibinfo  {journal} {Physical Review B}\ }\textbf {\bibinfo {volume}
  {102}},\ \bibinfo {pages} {195419} (\bibinfo {year} {2020})}\BibitemShut
  {NoStop}%
\bibitem [{\citenamefont {Yates}\ and\ \citenamefont
  {Mitra}(2021)}]{yates2021strong}%
  \BibitemOpen
  \bibfield  {author} {\bibinfo {author} {\bibfnamefont {D.~J.}\ \bibnamefont
  {Yates}}\ and\ \bibinfo {author} {\bibfnamefont {A.}~\bibnamefont {Mitra}},\
  }\bibfield  {title} {\bibinfo {title} {Strong and almost strong modes of
  floquet spin chains in krylov subspaces},\ }\href@noop {} {\bibfield
  {journal} {\bibinfo  {journal} {Physical Review B}\ }\textbf {\bibinfo
  {volume} {104}},\ \bibinfo {pages} {195121} (\bibinfo {year}
  {2021})}\BibitemShut {NoStop}%
\bibitem [{\citenamefont {Vasiloiu}\ \emph {et~al.}(2019)\citenamefont
  {Vasiloiu}, \citenamefont {Carollo}, \citenamefont {Marcuzzi},\ and\
  \citenamefont {Garrahan}}]{vasiloiu2019strong}%
  \BibitemOpen
  \bibfield  {author} {\bibinfo {author} {\bibfnamefont {L.~M.}\ \bibnamefont
  {Vasiloiu}}, \bibinfo {author} {\bibfnamefont {F.}~\bibnamefont {Carollo}},
  \bibinfo {author} {\bibfnamefont {M.}~\bibnamefont {Marcuzzi}},\ and\
  \bibinfo {author} {\bibfnamefont {J.~P.}\ \bibnamefont {Garrahan}},\
  }\bibfield  {title} {\bibinfo {title} {Strong zero modes in a class of
  generalized ising spin ladders with plaquette interactions},\ }\href@noop {}
  {\bibfield  {journal} {\bibinfo  {journal} {Physical Review B}\ }\textbf
  {\bibinfo {volume} {100}},\ \bibinfo {pages} {024309} (\bibinfo {year}
  {2019})}\BibitemShut {NoStop}%
\bibitem [{\citenamefont {Zvyagin}(2021)}]{zvyagin2021majorana}%
  \BibitemOpen
  \bibfield  {author} {\bibinfo {author} {\bibfnamefont {A.}~\bibnamefont
  {Zvyagin}},\ }\bibfield  {title} {\bibinfo {title} {Majorana zero modes in
  the interacting fermion chain without pairing},\ }\href@noop {} {\bibfield
  {journal} {\bibinfo  {journal} {Low Temperature Physics}\ }\textbf {\bibinfo
  {volume} {47}},\ \bibinfo {pages} {401} (\bibinfo {year} {2021})}\BibitemShut
  {NoStop}%
\bibitem [{\citenamefont {Zvyagin}(2022)}]{zvyagin2022charging}%
  \BibitemOpen
  \bibfield  {author} {\bibinfo {author} {\bibfnamefont {A.}~\bibnamefont
  {Zvyagin}},\ }\bibfield  {title} {\bibinfo {title} {Charging of majorana edge
  modes caused by interaction: Exact results},\ }\href@noop {} {\bibfield
  {journal} {\bibinfo  {journal} {Physical Review B}\ }\textbf {\bibinfo
  {volume} {105}},\ \bibinfo {pages} {115406} (\bibinfo {year}
  {2022})}\BibitemShut {NoStop}%
\bibitem [{\citenamefont {Zvyagin}(2024)}]{zvyagin2024strong}%
  \BibitemOpen
  \bibfield  {author} {\bibinfo {author} {\bibfnamefont {A.}~\bibnamefont
  {Zvyagin}},\ }\bibfield  {title} {\bibinfo {title} {Strong zero modes and
  edge states in the interacting fermion chain without pairing},\ }\href@noop
  {} {\bibfield  {journal} {\bibinfo  {journal} {Low Temperature Physics}\
  }\textbf {\bibinfo {volume} {50}},\ \bibinfo {pages} {316} (\bibinfo {year}
  {2024})}\BibitemShut {NoStop}%
\bibitem [{\citenamefont {Kattel}\ \emph
  {et~al.}(2024{\natexlab{c}})\citenamefont {Kattel}, \citenamefont {Tang},
  \citenamefont {Pixley},\ and\ \citenamefont {Andrei}}]{kattel2024XXZ-S}%
  \BibitemOpen
  \bibfield  {author} {\bibinfo {author} {\bibfnamefont {P.}~\bibnamefont
  {Kattel}}, \bibinfo {author} {\bibfnamefont {Y.}~\bibnamefont {Tang}},
  \bibinfo {author} {\bibfnamefont {J.}~\bibnamefont {Pixley}},\ and\ \bibinfo
  {author} {\bibfnamefont {N.}~\bibnamefont {Andrei}},\ }\bibfield  {title}
  {\bibinfo {title} {Edge spin fractionalization in one-dimensional spin-s
  quantum antiferromagnets},\ }\href@noop {} {\bibfield  {journal} {\bibinfo
  {journal} {arXiv preprint 2406.11955}\ } (\bibinfo {year}
  {2024}{\natexlab{c}})}\BibitemShut {NoStop}%
\bibitem [{\citenamefont {Pasnoori}\ \emph {et~al.}(2022)\citenamefont
  {Pasnoori}, \citenamefont {Andrei}, \citenamefont {Rylands},\ and\
  \citenamefont {Azaria}}]{pasnoori2022rise}%
  \BibitemOpen
  \bibfield  {author} {\bibinfo {author} {\bibfnamefont {P.~R.}\ \bibnamefont
  {Pasnoori}}, \bibinfo {author} {\bibfnamefont {N.}~\bibnamefont {Andrei}},
  \bibinfo {author} {\bibfnamefont {C.}~\bibnamefont {Rylands}},\ and\ \bibinfo
  {author} {\bibfnamefont {P.}~\bibnamefont {Azaria}},\ }\bibfield  {title}
  {\bibinfo {title} {Rise and fall of yu-shiba-rusinov bound states in
  charge-conserving s-wave one-dimensional superconductors},\ }\href@noop {}
  {\bibfield  {journal} {\bibinfo  {journal} {Physical Review B}\ }\textbf
  {\bibinfo {volume} {105}},\ \bibinfo {pages} {174517} (\bibinfo {year}
  {2022})}\BibitemShut {NoStop}%
\bibitem [{\citenamefont {Jiang}\ \emph {et~al.}(2011)\citenamefont {Jiang},
  \citenamefont {Baksmaty}, \citenamefont {Hu}, \citenamefont {Chen},\ and\
  \citenamefont {Pu}}]{MGPhysRevA.83.061604}%
  \BibitemOpen
  \bibfield  {author} {\bibinfo {author} {\bibfnamefont {L.}~\bibnamefont
  {Jiang}}, \bibinfo {author} {\bibfnamefont {L.~O.}\ \bibnamefont {Baksmaty}},
  \bibinfo {author} {\bibfnamefont {H.}~\bibnamefont {Hu}}, \bibinfo {author}
  {\bibfnamefont {Y.}~\bibnamefont {Chen}},\ and\ \bibinfo {author}
  {\bibfnamefont {H.}~\bibnamefont {Pu}},\ }\bibfield  {title} {\bibinfo
  {title} {Single impurity in ultracold fermi superfluids},\ }\href
  {https://doi.org/10.1103/PhysRevA.83.061604} {\bibfield  {journal} {\bibinfo
  {journal} {Phys. Rev. A}\ }\textbf {\bibinfo {volume} {83}},\ \bibinfo
  {pages} {061604} (\bibinfo {year} {2011})}\BibitemShut {NoStop}%
\bibitem [{\citenamefont {Bauer}\ \emph {et~al.}(2024)\citenamefont {Bauer},
  \citenamefont {Freitas}, \citenamefont {Andrade}, \citenamefont {Egger},\
  and\ \citenamefont {Pereira}}]{bauer2024scanning}%
  \BibitemOpen
  \bibfield  {author} {\bibinfo {author} {\bibfnamefont {T.}~\bibnamefont
  {Bauer}}, \bibinfo {author} {\bibfnamefont {L.~R.}\ \bibnamefont {Freitas}},
  \bibinfo {author} {\bibfnamefont {E.}~\bibnamefont {Andrade}}, \bibinfo
  {author} {\bibfnamefont {R.}~\bibnamefont {Egger}},\ and\ \bibinfo {author}
  {\bibfnamefont {R.~G.}\ \bibnamefont {Pereira}},\ }\bibfield  {title}
  {\bibinfo {title} {Scanning tunneling spectroscopy of magnetic quantum
  impurities in two-dimensional magnets},\ }\href@noop {} {\bibfield  {journal}
  {\bibinfo  {journal} {arXiv preprint arXiv:2405.19962}\ } (\bibinfo {year}
  {2024})}\BibitemShut {NoStop}%
\bibitem [{\citenamefont {Lang}\ and\ \citenamefont
  {Chen}(2014)}]{MGlang2014topologically}%
  \BibitemOpen
  \bibfield  {author} {\bibinfo {author} {\bibfnamefont {L.-J.}\ \bibnamefont
  {Lang}}\ and\ \bibinfo {author} {\bibfnamefont {S.}~\bibnamefont {Chen}},\
  }\bibfield  {title} {\bibinfo {title} {Topologically protected mid-gap states
  induced by impurity in one-dimensional superlattices},\ }\href@noop {}
  {\bibfield  {journal} {\bibinfo  {journal} {Journal of Physics B: Atomic,
  Molecular and Optical Physics}\ }\textbf {\bibinfo {volume} {47}},\ \bibinfo
  {pages} {065302} (\bibinfo {year} {2014})}\BibitemShut {NoStop}%
\bibitem [{\citenamefont {Sachs}\ \emph {et~al.}(2016)\citenamefont {Sachs},
  \citenamefont {Wehling}, \citenamefont {Katsnelson},\ and\ \citenamefont
  {Lichtenstein}}]{MGsachs2016midgap}%
  \BibitemOpen
  \bibfield  {author} {\bibinfo {author} {\bibfnamefont {B.}~\bibnamefont
  {Sachs}}, \bibinfo {author} {\bibfnamefont {T.}~\bibnamefont {Wehling}},
  \bibinfo {author} {\bibfnamefont {M.}~\bibnamefont {Katsnelson}},\ and\
  \bibinfo {author} {\bibfnamefont {A.}~\bibnamefont {Lichtenstein}},\
  }\bibfield  {title} {\bibinfo {title} {Midgap states and band gap
  modification in defective graphene/h-bn heterostructures},\ }\href@noop {}
  {\bibfield  {journal} {\bibinfo  {journal} {Physical Review B}\ }\textbf
  {\bibinfo {volume} {94}},\ \bibinfo {pages} {224105} (\bibinfo {year}
  {2016})}\BibitemShut {NoStop}%
\bibitem [{\citenamefont {Aczel}\ \emph {et~al.}(2007)\citenamefont {Aczel},
  \citenamefont {MacDougall}, \citenamefont {Rodriguez}, \citenamefont {Luke},
  \citenamefont {Russo}, \citenamefont {Savici}, \citenamefont {Uemura},
  \citenamefont {Dabkowska}, \citenamefont {Wiebe}, \citenamefont {Janik} \emph
  {et~al.}}]{aczel2007impurity}%
  \BibitemOpen
  \bibfield  {author} {\bibinfo {author} {\bibfnamefont {A.}~\bibnamefont
  {Aczel}}, \bibinfo {author} {\bibfnamefont {G.}~\bibnamefont {MacDougall}},
  \bibinfo {author} {\bibfnamefont {J.}~\bibnamefont {Rodriguez}}, \bibinfo
  {author} {\bibfnamefont {G.}~\bibnamefont {Luke}}, \bibinfo {author}
  {\bibfnamefont {P.}~\bibnamefont {Russo}}, \bibinfo {author} {\bibfnamefont
  {A.}~\bibnamefont {Savici}}, \bibinfo {author} {\bibfnamefont
  {Y.}~\bibnamefont {Uemura}}, \bibinfo {author} {\bibfnamefont
  {H.}~\bibnamefont {Dabkowska}}, \bibinfo {author} {\bibfnamefont
  {C.}~\bibnamefont {Wiebe}}, \bibinfo {author} {\bibfnamefont
  {J.}~\bibnamefont {Janik}}, \emph {et~al.},\ }\bibfield  {title} {\bibinfo
  {title} {Impurity-induced singlet breaking in sr cu 2 (b o 3) 2},\
  }\href@noop {} {\bibfield  {journal} {\bibinfo  {journal} {Physical Review
  B}\ }\textbf {\bibinfo {volume} {76}},\ \bibinfo {pages} {214427} (\bibinfo
  {year} {2007})}\BibitemShut {NoStop}%
\bibitem [{\citenamefont {Chakhalian}\ \emph {et~al.}(2004)\citenamefont
  {Chakhalian}, \citenamefont {Kiefl}, \citenamefont {Brewer}, \citenamefont
  {Dunsiger}, \citenamefont {Morris}, \citenamefont {Eggert}, \citenamefont
  {Affleck},\ and\ \citenamefont {Yamada}}]{chakhalian2004impurity}%
  \BibitemOpen
  \bibfield  {author} {\bibinfo {author} {\bibfnamefont {J.}~\bibnamefont
  {Chakhalian}}, \bibinfo {author} {\bibfnamefont {R.}~\bibnamefont {Kiefl}},
  \bibinfo {author} {\bibfnamefont {J.}~\bibnamefont {Brewer}}, \bibinfo
  {author} {\bibfnamefont {S.}~\bibnamefont {Dunsiger}}, \bibinfo {author}
  {\bibfnamefont {G.}~\bibnamefont {Morris}}, \bibinfo {author} {\bibfnamefont
  {S.}~\bibnamefont {Eggert}}, \bibinfo {author} {\bibfnamefont
  {I.}~\bibnamefont {Affleck}},\ and\ \bibinfo {author} {\bibfnamefont
  {I.}~\bibnamefont {Yamada}},\ }\bibfield  {title} {\bibinfo {title} {Impurity
  effects in quasi-one-dimensional s= 12 antiferromagnetic chain kcuf3 studied
  by muon spin rotation},\ }\href@noop {} {\bibfield  {journal} {\bibinfo
  {journal} {Journal of magnetism and magnetic materials}\ }\textbf {\bibinfo
  {volume} {272}},\ \bibinfo {pages} {979} (\bibinfo {year}
  {2004})}\BibitemShut {NoStop}%
\bibitem [{\citenamefont {Affleck}\ and\ \citenamefont
  {Ludwig}(1991{\natexlab{b}})}]{affleck1991universal}%
  \BibitemOpen
  \bibfield  {author} {\bibinfo {author} {\bibfnamefont {I.}~\bibnamefont
  {Affleck}}\ and\ \bibinfo {author} {\bibfnamefont {A.~W.}\ \bibnamefont
  {Ludwig}},\ }\bibfield  {title} {\bibinfo {title} {Universal noninteger
  ‘‘ground-state degeneracy’’in critical quantum systems},\ }\href@noop
  {} {\bibfield  {journal} {\bibinfo  {journal} {Physical Review Letters}\
  }\textbf {\bibinfo {volume} {67}},\ \bibinfo {pages} {161} (\bibinfo {year}
  {1991}{\natexlab{b}})}\BibitemShut {NoStop}%
\bibitem [{\citenamefont {Jordan}\ and\ \citenamefont
  {Wigner}(1993)}]{jordan1993paulische}%
  \BibitemOpen
  \bibfield  {author} {\bibinfo {author} {\bibfnamefont {P.}~\bibnamefont
  {Jordan}}\ and\ \bibinfo {author} {\bibfnamefont {E.~P.}\ \bibnamefont
  {Wigner}},\ }\href@noop {} {\emph {\bibinfo {title} {{\"U}ber das paulische
  {\"a}quivalenzverbot}}}\ (\bibinfo  {publisher} {Springer},\ \bibinfo {year}
  {1993})\BibitemShut {NoStop}%
\bibitem [{\citenamefont {Giamarchi}(2003)}]{giamarchi2003quantum}%
  \BibitemOpen
  \bibfield  {author} {\bibinfo {author} {\bibfnamefont {T.}~\bibnamefont
  {Giamarchi}},\ }\href@noop {} {\emph {\bibinfo {title} {Quantum physics in
  one dimension}}},\ Vol.\ \bibinfo {volume} {121}\ (\bibinfo  {publisher}
  {Clarendon press},\ \bibinfo {year} {2003})\BibitemShut {NoStop}%
\bibitem [{\citenamefont {Witten}(1984)}]{witten1984non}%
  \BibitemOpen
  \bibfield  {author} {\bibinfo {author} {\bibfnamefont {E.}~\bibnamefont
  {Witten}},\ }\bibfield  {title} {\bibinfo {title} {Non-abelian bosonization
  in two dimensions},\ }\href@noop {} {\bibfield  {journal} {\bibinfo
  {journal} {Communications in Mathematical Physics}\ }\textbf {\bibinfo
  {volume} {92}},\ \bibinfo {pages} {455} (\bibinfo {year} {1984})}\BibitemShut
  {NoStop}%
\bibitem [{\citenamefont {Cabra}\ and\ \citenamefont
  {Pujol}(2008)}]{cabra2008field}%
  \BibitemOpen
  \bibfield  {author} {\bibinfo {author} {\bibfnamefont {D.~C.}\ \bibnamefont
  {Cabra}}\ and\ \bibinfo {author} {\bibfnamefont {P.}~\bibnamefont {Pujol}},\
  }\bibfield  {title} {\bibinfo {title} {Field-theoretical methods in quantum
  magnetism},\ }\href@noop {} {\bibfield  {journal} {\bibinfo  {journal}
  {Quantum Magnetism}\ ,\ \bibinfo {pages} {253}} (\bibinfo {year}
  {2008})}\BibitemShut {NoStop}%
\bibitem [{\citenamefont {Andrei}(1992)}]{andrei1992integrable}%
  \BibitemOpen
  \bibfield  {author} {\bibinfo {author} {\bibfnamefont {N.}~\bibnamefont
  {Andrei}},\ }\bibfield  {title} {\bibinfo {title} {Integrable models in
  condensed matter physics},\ }\href@noop {} {\bibfield  {journal} {\bibinfo
  {journal} {Low-Dimensional Quantum Field Theories for Condensed Matter
  Physicists (World Scientific Publishing Co, 2013) pp}\ ,\ \bibinfo {pages}
  {457}} (\bibinfo {year} {1992})}\BibitemShut {NoStop}%
\bibitem [{\citenamefont {Wiegmann}\ and\ \citenamefont
  {Tsvelick}(1983)}]{wiegmann1983exact}%
  \BibitemOpen
  \bibfield  {author} {\bibinfo {author} {\bibfnamefont {P.}~\bibnamefont
  {Wiegmann}}\ and\ \bibinfo {author} {\bibfnamefont {A.}~\bibnamefont
  {Tsvelick}},\ }\bibfield  {title} {\bibinfo {title} {Exact solution of the
  anderson model: I},\ }\href@noop {} {\bibfield  {journal} {\bibinfo
  {journal} {Journal of Physics C: Solid State Physics}\ }\textbf {\bibinfo
  {volume} {16}},\ \bibinfo {pages} {2281} (\bibinfo {year}
  {1983})}\BibitemShut {NoStop}%
\bibitem [{\citenamefont {Andrei}\ \emph {et~al.}(1983)\citenamefont {Andrei},
  \citenamefont {Furuya},\ and\ \citenamefont
  {Lowenstein}}]{andrei1983solution}%
  \BibitemOpen
  \bibfield  {author} {\bibinfo {author} {\bibfnamefont {N.}~\bibnamefont
  {Andrei}}, \bibinfo {author} {\bibfnamefont {K.}~\bibnamefont {Furuya}},\
  and\ \bibinfo {author} {\bibfnamefont {J.}~\bibnamefont {Lowenstein}},\
  }\bibfield  {title} {\bibinfo {title} {Solution of the kondo problem},\
  }\href@noop {} {\bibfield  {journal} {\bibinfo  {journal} {Reviews of modern
  physics}\ }\textbf {\bibinfo {volume} {55}},\ \bibinfo {pages} {331}
  (\bibinfo {year} {1983})}\BibitemShut {NoStop}%
\bibitem [{\citenamefont {Feiguin}\ and\ \citenamefont
  {White}(2005)}]{feiguin2005finite}%
  \BibitemOpen
  \bibfield  {author} {\bibinfo {author} {\bibfnamefont {A.~E.}\ \bibnamefont
  {Feiguin}}\ and\ \bibinfo {author} {\bibfnamefont {S.~R.}\ \bibnamefont
  {White}},\ }\bibfield  {title} {\bibinfo {title} {Finite-temperature density
  matrix renormalization using an enlarged hilbert space},\ }\href@noop {}
  {\bibfield  {journal} {\bibinfo  {journal} {Physical Review B—Condensed
  Matter and Materials Physics}\ }\textbf {\bibinfo {volume} {72}},\ \bibinfo
  {pages} {220401} (\bibinfo {year} {2005})}\BibitemShut {NoStop}%
\bibitem [{\citenamefont {Fishman}\ \emph {et~al.}(2022)\citenamefont
  {Fishman}, \citenamefont {White},\ and\ \citenamefont
  {Stoudenmire}}]{fishman2022itensor}%
  \BibitemOpen
  \bibfield  {author} {\bibinfo {author} {\bibfnamefont {M.}~\bibnamefont
  {Fishman}}, \bibinfo {author} {\bibfnamefont {S.}~\bibnamefont {White}},\
  and\ \bibinfo {author} {\bibfnamefont {E.}~\bibnamefont {Stoudenmire}},\
  }\bibfield  {title} {\bibinfo {title} {The itensor software library for
  tensor network calculations},\ }\href@noop {} {\bibfield  {journal} {\bibinfo
   {journal} {SciPost Physics Codebases}\ ,\ \bibinfo {pages} {004}} (\bibinfo
  {year} {2022})}\BibitemShut {NoStop}%
\bibitem [{\citenamefont {Orbach}(1958)}]{XXZPhysRev.112.309}%
  \BibitemOpen
  \bibfield  {author} {\bibinfo {author} {\bibfnamefont {R.}~\bibnamefont
  {Orbach}},\ }\bibfield  {title} {\bibinfo {title} {Linear antiferromagnetic
  chain with anisotropic coupling},\ }\href
  {https://doi.org/10.1103/PhysRev.112.309} {\bibfield  {journal} {\bibinfo
  {journal} {Phys. Rev.}\ }\textbf {\bibinfo {volume} {112}},\ \bibinfo {pages}
  {309} (\bibinfo {year} {1958})}\BibitemShut {NoStop}%
\bibitem [{\citenamefont {Walker}(1959)}]{XXZPhysRev.116.1089}%
  \BibitemOpen
  \bibfield  {author} {\bibinfo {author} {\bibfnamefont {L.~R.}\ \bibnamefont
  {Walker}},\ }\bibfield  {title} {\bibinfo {title} {Antiferromagnetic linear
  chain},\ }\href {https://doi.org/10.1103/PhysRev.116.1089} {\bibfield
  {journal} {\bibinfo  {journal} {Phys. Rev.}\ }\textbf {\bibinfo {volume}
  {116}},\ \bibinfo {pages} {1089} (\bibinfo {year} {1959})}\BibitemShut
  {NoStop}%
\bibitem [{\citenamefont {Yang}\ and\ \citenamefont
  {Yang}(1966{\natexlab{a}})}]{XXZPhysRev.150.321}%
  \BibitemOpen
  \bibfield  {author} {\bibinfo {author} {\bibfnamefont {C.~N.}\ \bibnamefont
  {Yang}}\ and\ \bibinfo {author} {\bibfnamefont {C.~P.}\ \bibnamefont
  {Yang}},\ }\bibfield  {title} {\bibinfo {title} {One-dimensional chain of
  anisotropic spin-spin interactions. i. proof of bethe's hypothesis for ground
  state in a finite system},\ }\href {https://doi.org/10.1103/PhysRev.150.321}
  {\bibfield  {journal} {\bibinfo  {journal} {Phys. Rev.}\ }\textbf {\bibinfo
  {volume} {150}},\ \bibinfo {pages} {321} (\bibinfo {year}
  {1966}{\natexlab{a}})}\BibitemShut {NoStop}%
\bibitem [{\citenamefont {Yang}\ and\ \citenamefont
  {Yang}(1966{\natexlab{b}})}]{XXZPhysRev.150.327}%
  \BibitemOpen
  \bibfield  {author} {\bibinfo {author} {\bibfnamefont {C.~N.}\ \bibnamefont
  {Yang}}\ and\ \bibinfo {author} {\bibfnamefont {C.~P.}\ \bibnamefont
  {Yang}},\ }\bibfield  {title} {\bibinfo {title} {One-dimensional chain of
  anisotropic spin-spin interactions. ii. properties of the ground-state energy
  per lattice site for an infinite system},\ }\href
  {https://doi.org/10.1103/PhysRev.150.327} {\bibfield  {journal} {\bibinfo
  {journal} {Phys. Rev.}\ }\textbf {\bibinfo {volume} {150}},\ \bibinfo {pages}
  {327} (\bibinfo {year} {1966}{\natexlab{b}})}\BibitemShut {NoStop}%
\bibitem [{\citenamefont {Yang}\ and\ \citenamefont
  {Yang}(1966{\natexlab{c}})}]{XXZPhysRev.151.258}%
  \BibitemOpen
  \bibfield  {author} {\bibinfo {author} {\bibfnamefont {C.~N.}\ \bibnamefont
  {Yang}}\ and\ \bibinfo {author} {\bibfnamefont {C.~P.}\ \bibnamefont
  {Yang}},\ }\bibfield  {title} {\bibinfo {title} {One-dimensional chain of
  anisotropic spin-spin interactions. iii. applications},\ }\href
  {https://doi.org/10.1103/PhysRev.151.258} {\bibfield  {journal} {\bibinfo
  {journal} {Phys. Rev.}\ }\textbf {\bibinfo {volume} {151}},\ \bibinfo {pages}
  {258} (\bibinfo {year} {1966}{\natexlab{c}})}\BibitemShut {NoStop}%
\bibitem [{\citenamefont {Baxter}(1972)}]{baxter1972partition}%
  \BibitemOpen
  \bibfield  {author} {\bibinfo {author} {\bibfnamefont {R.~J.}\ \bibnamefont
  {Baxter}},\ }\bibfield  {title} {\bibinfo {title} {Partition function of the
  eight-vertex lattice model},\ }\href@noop {} {\bibfield  {journal} {\bibinfo
  {journal} {Annals of Physics}\ }\textbf {\bibinfo {volume} {70}},\ \bibinfo
  {pages} {193} (\bibinfo {year} {1972})}\BibitemShut {NoStop}%
\bibitem [{\citenamefont {Babelon}\ \emph {et~al.}(1983)\citenamefont
  {Babelon}, \citenamefont {De~Vega},\ and\ \citenamefont
  {Viallet}}]{XXZbabelon1983analysis}%
  \BibitemOpen
  \bibfield  {author} {\bibinfo {author} {\bibfnamefont {O.}~\bibnamefont
  {Babelon}}, \bibinfo {author} {\bibfnamefont {H.}~\bibnamefont {De~Vega}},\
  and\ \bibinfo {author} {\bibfnamefont {C.}~\bibnamefont {Viallet}},\
  }\bibfield  {title} {\bibinfo {title} {Analysis of the bethe ansatz equations
  of the xxz model},\ }\href@noop {} {\bibfield  {journal} {\bibinfo  {journal}
  {Nuclear Physics B}\ }\textbf {\bibinfo {volume} {220}},\ \bibinfo {pages}
  {13} (\bibinfo {year} {1983})}\BibitemShut {NoStop}%
\bibitem [{\citenamefont {Destri}\ and\ \citenamefont
  {Lowenstein}(1982)}]{destri1982analysis}%
  \BibitemOpen
  \bibfield  {author} {\bibinfo {author} {\bibfnamefont {C.}~\bibnamefont
  {Destri}}\ and\ \bibinfo {author} {\bibfnamefont {J.}~\bibnamefont
  {Lowenstein}},\ }\bibfield  {title} {\bibinfo {title} {Analysis of the
  bethe-ansatz equations of the chiral-invariant gross-neveu model},\
  }\href@noop {} {\bibfield  {journal} {\bibinfo  {journal} {Nuclear Physics
  B}\ }\textbf {\bibinfo {volume} {205}},\ \bibinfo {pages} {369} (\bibinfo
  {year} {1982})}\BibitemShut {NoStop}%
\bibitem [{\citenamefont {Faddeev}\ and\ \citenamefont
  {Takhtajan}(1981)}]{faddeev1981spin}%
  \BibitemOpen
  \bibfield  {author} {\bibinfo {author} {\bibfnamefont {L.}~\bibnamefont
  {Faddeev}}\ and\ \bibinfo {author} {\bibfnamefont {L.}~\bibnamefont
  {Takhtajan}},\ }\bibfield  {title} {\bibinfo {title} {What is the spin of a
  spin wave?},\ }\href@noop {} {\bibfield  {journal} {\bibinfo  {journal}
  {Physics Letters A}\ }\textbf {\bibinfo {volume} {85}},\ \bibinfo {pages}
  {375} (\bibinfo {year} {1981})}\BibitemShut {NoStop}%
\bibitem [{\citenamefont {Tennant}\ \emph {et~al.}(1993)\citenamefont
  {Tennant}, \citenamefont {Perring}, \citenamefont {Cowley},\ and\
  \citenamefont {Nagler}}]{Spinon-PhysRevLett.70.4003}%
  \BibitemOpen
  \bibfield  {author} {\bibinfo {author} {\bibfnamefont {D.~A.}\ \bibnamefont
  {Tennant}}, \bibinfo {author} {\bibfnamefont {T.~G.}\ \bibnamefont
  {Perring}}, \bibinfo {author} {\bibfnamefont {R.~A.}\ \bibnamefont
  {Cowley}},\ and\ \bibinfo {author} {\bibfnamefont {S.~E.}\ \bibnamefont
  {Nagler}},\ }\bibfield  {title} {\bibinfo {title} {Unbound spinons in the
  s=1/2 antiferromagnetic chain ${\mathrm{kcuf}}_{3}$},\ }\href
  {https://doi.org/10.1103/PhysRevLett.70.4003} {\bibfield  {journal} {\bibinfo
   {journal} {Phys. Rev. Lett.}\ }\textbf {\bibinfo {volume} {70}},\ \bibinfo
  {pages} {4003} (\bibinfo {year} {1993})}\BibitemShut {NoStop}%
\bibitem [{\citenamefont {Klauser}\ \emph {et~al.}(2011)\citenamefont
  {Klauser}, \citenamefont {Mossel}, \citenamefont {Caux},\ and\ \citenamefont
  {van~den Brink}}]{Spinon-PhysRevLett.106.157205}%
  \BibitemOpen
  \bibfield  {author} {\bibinfo {author} {\bibfnamefont {A.}~\bibnamefont
  {Klauser}}, \bibinfo {author} {\bibfnamefont {J.}~\bibnamefont {Mossel}},
  \bibinfo {author} {\bibfnamefont {J.-S.}\ \bibnamefont {Caux}},\ and\
  \bibinfo {author} {\bibfnamefont {J.}~\bibnamefont {van~den Brink}},\
  }\bibfield  {title} {\bibinfo {title} {Spin-exchange dynamical structure
  factor of the $s=1/2$ heisenberg chain},\ }\href
  {https://doi.org/10.1103/PhysRevLett.106.157205} {\bibfield  {journal}
  {\bibinfo  {journal} {Phys. Rev. Lett.}\ }\textbf {\bibinfo {volume} {106}},\
  \bibinfo {pages} {157205} (\bibinfo {year} {2011})}\BibitemShut {NoStop}%
\bibitem [{\citenamefont {Weisstein}(2000)}]{weisstein2000jacobi}%
  \BibitemOpen
  \bibfield  {author} {\bibinfo {author} {\bibfnamefont {E.~W.}\ \bibnamefont
  {Weisstein}},\ }\bibfield  {title} {\bibinfo {title} {Jacobi theta
  functions},\ }\href@noop {} {\bibfield  {journal} {\bibinfo  {journal}
  {https://mathworld. wolfram. com/}\ } (\bibinfo {year} {2000})}\BibitemShut
  {NoStop}%
\bibitem [{\citenamefont {Baxter}(1973)}]{baxter1973spontaneous}%
  \BibitemOpen
  \bibfield  {author} {\bibinfo {author} {\bibfnamefont {R.~J.}\ \bibnamefont
  {Baxter}},\ }\bibfield  {title} {\bibinfo {title} {Spontaneous staggered
  polarization of the f-model},\ }\href@noop {} {\bibfield  {journal} {\bibinfo
   {journal} {Journal of Statistical Physics}\ }\textbf {\bibinfo {volume}
  {9}},\ \bibinfo {pages} {145} (\bibinfo {year} {1973})}\BibitemShut {NoStop}%
\bibitem [{\citenamefont {Izergin}\ \emph {et~al.}(1999)\citenamefont
  {Izergin}, \citenamefont {Kitanine}, \citenamefont {Maillet},\ and\
  \citenamefont {Terras}}]{izergin1999spontaneous}%
  \BibitemOpen
  \bibfield  {author} {\bibinfo {author} {\bibfnamefont {A.}~\bibnamefont
  {Izergin}}, \bibinfo {author} {\bibfnamefont {N.}~\bibnamefont {Kitanine}},
  \bibinfo {author} {\bibfnamefont {J.~M.}\ \bibnamefont {Maillet}},\ and\
  \bibinfo {author} {\bibfnamefont {V.}~\bibnamefont {Terras}},\ }\bibfield
  {title} {\bibinfo {title} {Spontaneous magnetization of the xxz heisenberg
  spin-12 chain},\ }\href@noop {} {\bibfield  {journal} {\bibinfo  {journal}
  {Nuclear Physics B}\ }\textbf {\bibinfo {volume} {554}},\ \bibinfo {pages}
  {679} (\bibinfo {year} {1999})}\BibitemShut {NoStop}%
\bibitem [{\citenamefont {Kattel}\ \emph
  {et~al.}(2024{\natexlab{d}})\citenamefont {Kattel}, \citenamefont {Tang},
  \citenamefont {Pixley},\ and\ \citenamefont {Andrei}}]{kattel2024edge}%
  \BibitemOpen
  \bibfield  {author} {\bibinfo {author} {\bibfnamefont {P.}~\bibnamefont
  {Kattel}}, \bibinfo {author} {\bibfnamefont {Y.}~\bibnamefont {Tang}},
  \bibinfo {author} {\bibfnamefont {J.}~\bibnamefont {Pixley}},\ and\ \bibinfo
  {author} {\bibfnamefont {N.}~\bibnamefont {Andrei}},\ }\bibfield  {title}
  {\bibinfo {title} {Edge spin fractionalization in one-dimensional spin-$ s $
  quantum antiferromagnets},\ }\href@noop {} {\bibfield  {journal} {\bibinfo
  {journal} {arXiv preprint arXiv:2406.11955}\ } (\bibinfo {year}
  {2024}{\natexlab{d}})}\BibitemShut {NoStop}%
\bibitem [{\citenamefont {Jackiw}\ \emph {et~al.}(1983)\citenamefont {Jackiw},
  \citenamefont {Kerman}, \citenamefont {Klebanov},\ and\ \citenamefont
  {Semenoff}}]{jackiw1983fluctuations}%
  \BibitemOpen
  \bibfield  {author} {\bibinfo {author} {\bibfnamefont {R.}~\bibnamefont
  {Jackiw}}, \bibinfo {author} {\bibfnamefont {A.}~\bibnamefont {Kerman}},
  \bibinfo {author} {\bibfnamefont {I.}~\bibnamefont {Klebanov}},\ and\
  \bibinfo {author} {\bibfnamefont {G.}~\bibnamefont {Semenoff}},\ }\bibfield
  {title} {\bibinfo {title} {Fluctuations of fractional charge in soliton
  anti-soliton systems},\ }\href@noop {} {\bibfield  {journal} {\bibinfo
  {journal} {Nuclear Physics B}\ }\textbf {\bibinfo {volume} {225}},\ \bibinfo
  {pages} {233} (\bibinfo {year} {1983})}\BibitemShut {NoStop}%
\bibitem [{\citenamefont {Kivelson}\ and\ \citenamefont
  {Schrieffer}(1982)}]{kivelson1982fractional}%
  \BibitemOpen
  \bibfield  {author} {\bibinfo {author} {\bibfnamefont {S.}~\bibnamefont
  {Kivelson}}\ and\ \bibinfo {author} {\bibfnamefont {J.}~\bibnamefont
  {Schrieffer}},\ }\bibfield  {title} {\bibinfo {title} {Fractional charge, a
  sharp quantum observable},\ }\href@noop {} {\bibfield  {journal} {\bibinfo
  {journal} {Physical Review B}\ }\textbf {\bibinfo {volume} {25}},\ \bibinfo
  {pages} {6447} (\bibinfo {year} {1982})}\BibitemShut {NoStop}%
\bibitem [{\citenamefont {Yu}\ and\ \citenamefont {Lee}(2021)}]{yu2021closing}%
  \BibitemOpen
  \bibfield  {author} {\bibinfo {author} {\bibfnamefont {C.}~\bibnamefont
  {Yu}}\ and\ \bibinfo {author} {\bibfnamefont {J.-W.}\ \bibnamefont {Lee}},\
  }\bibfield  {title} {\bibinfo {title} {Closing of the haldane gap in a spin-1
  xxz chain},\ }\href@noop {} {\bibfield  {journal} {\bibinfo  {journal}
  {Journal of the Korean Physical Society}\ }\textbf {\bibinfo {volume} {79}},\
  \bibinfo {pages} {841} (\bibinfo {year} {2021})}\BibitemShut {NoStop}%
\bibitem [{\citenamefont {Korepin}\ \emph {et~al.}(1997)\citenamefont
  {Korepin}, \citenamefont {Korepin}, \citenamefont {Bogoliubov},\ and\
  \citenamefont {Izergin}}]{korepin1997quantum}%
  \BibitemOpen
  \bibfield  {author} {\bibinfo {author} {\bibfnamefont {V.~E.}\ \bibnamefont
  {Korepin}}, \bibinfo {author} {\bibfnamefont {V.~E.}\ \bibnamefont
  {Korepin}}, \bibinfo {author} {\bibfnamefont {N.}~\bibnamefont
  {Bogoliubov}},\ and\ \bibinfo {author} {\bibfnamefont {A.}~\bibnamefont
  {Izergin}},\ }\href@noop {} {\emph {\bibinfo {title} {Quantum inverse
  scattering method and correlation functions}}},\ Vol.~\bibinfo {volume} {3}\
  (\bibinfo  {publisher} {Cambridge university press},\ \bibinfo {year}
  {1997})\BibitemShut {NoStop}%
\bibitem [{\citenamefont {Pasnoori}\ \emph {et~al.}(2020)\citenamefont
  {Pasnoori}, \citenamefont {Rylands},\ and\ \citenamefont
  {Andrei}}]{pasnoori2020kondo}%
  \BibitemOpen
  \bibfield  {author} {\bibinfo {author} {\bibfnamefont {P.~R.}\ \bibnamefont
  {Pasnoori}}, \bibinfo {author} {\bibfnamefont {C.}~\bibnamefont {Rylands}},\
  and\ \bibinfo {author} {\bibfnamefont {N.}~\bibnamefont {Andrei}},\
  }\bibfield  {title} {\bibinfo {title} {Kondo impurity at the edge of a
  superconducting wire},\ }\href@noop {} {\bibfield  {journal} {\bibinfo
  {journal} {Physical Review Research}\ }\textbf {\bibinfo {volume} {2}},\
  \bibinfo {pages} {013006} (\bibinfo {year} {2020})}\BibitemShut {NoStop}%
\bibitem [{\citenamefont {Pasnoori}\ \emph {et~al.}(2021)\citenamefont
  {Pasnoori}, \citenamefont {Andrei},\ and\ \citenamefont
  {Azaria}}]{pasnoori2021boundary}%
  \BibitemOpen
  \bibfield  {author} {\bibinfo {author} {\bibfnamefont {P.~R.}\ \bibnamefont
  {Pasnoori}}, \bibinfo {author} {\bibfnamefont {N.}~\bibnamefont {Andrei}},\
  and\ \bibinfo {author} {\bibfnamefont {P.}~\bibnamefont {Azaria}},\
  }\bibfield  {title} {\bibinfo {title} {Boundary-induced topological and
  mid-gap states in charge conserving one-dimensional superconductors:
  Fractionalization transition},\ }\href@noop {} {\bibfield  {journal}
  {\bibinfo  {journal} {Physical Review B}\ }\textbf {\bibinfo {volume}
  {104}},\ \bibinfo {pages} {134519} (\bibinfo {year} {2021})}\BibitemShut
  {NoStop}%
\bibitem [{\citenamefont {Kattel}\ \emph {et~al.}(2023)\citenamefont {Kattel},
  \citenamefont {Pasnoori},\ and\ \citenamefont {Andrei}}]{kattel2023exact}%
  \BibitemOpen
  \bibfield  {author} {\bibinfo {author} {\bibfnamefont {P.}~\bibnamefont
  {Kattel}}, \bibinfo {author} {\bibfnamefont {P.~R.}\ \bibnamefont
  {Pasnoori}},\ and\ \bibinfo {author} {\bibfnamefont {N.}~\bibnamefont
  {Andrei}},\ }\bibfield  {title} {\bibinfo {title} {Exact solution of a
  non-hermitian-symmetric spin chain},\ }\href@noop {} {\bibfield  {journal}
  {\bibinfo  {journal} {Journal of Physics A: Mathematical and Theoretical}\
  }\textbf {\bibinfo {volume} {56}},\ \bibinfo {pages} {325001} (\bibinfo
  {year} {2023})}\BibitemShut {NoStop}%
\bibitem [{\citenamefont {Rylands}(2020)}]{rylands2020exact}%
  \BibitemOpen
  \bibfield  {author} {\bibinfo {author} {\bibfnamefont {C.}~\bibnamefont
  {Rylands}},\ }\bibfield  {title} {\bibinfo {title} {Exact boundary modes in
  an interacting quantum wire},\ }\href@noop {} {\bibfield  {journal} {\bibinfo
   {journal} {Physical Review B}\ }\textbf {\bibinfo {volume} {101}},\ \bibinfo
  {pages} {085133} (\bibinfo {year} {2020})}\BibitemShut {NoStop}%
\bibitem [{\citenamefont {Kattel}\ \emph
  {et~al.}(2024{\natexlab{e}})\citenamefont {Kattel}, \citenamefont {Zhakenov},
  \citenamefont {Pasnoori}, \citenamefont {Azaria},\ and\ \citenamefont
  {Andrei}}]{kattel2024dissipation}%
  \BibitemOpen
  \bibfield  {author} {\bibinfo {author} {\bibfnamefont {P.}~\bibnamefont
  {Kattel}}, \bibinfo {author} {\bibfnamefont {A.}~\bibnamefont {Zhakenov}},
  \bibinfo {author} {\bibfnamefont {P.~R.}\ \bibnamefont {Pasnoori}}, \bibinfo
  {author} {\bibfnamefont {P.}~\bibnamefont {Azaria}},\ and\ \bibinfo {author}
  {\bibfnamefont {N.}~\bibnamefont {Andrei}},\ }\bibfield  {title} {\bibinfo
  {title} {Dissipation driven phase transition in the non-hermitian kondo
  model},\ }\href@noop {} {\bibfield  {journal} {\bibinfo  {journal} {arXiv
  preprint arXiv:2402.09510}\ } (\bibinfo {year}
  {2024}{\natexlab{e}})}\BibitemShut {NoStop}%
\bibitem [{\citenamefont {Kattel}\ \emph
  {et~al.}(2024{\natexlab{f}})\citenamefont {Kattel}, \citenamefont {Pasnoori},
  \citenamefont {Pixley},\ and\ \citenamefont {Andrei}}]{Kattel:2024lot}%
  \BibitemOpen
  \bibfield  {author} {\bibinfo {author} {\bibfnamefont {P.}~\bibnamefont
  {Kattel}}, \bibinfo {author} {\bibfnamefont {P.~R.}\ \bibnamefont
  {Pasnoori}}, \bibinfo {author} {\bibfnamefont {J.~H.}\ \bibnamefont
  {Pixley}},\ and\ \bibinfo {author} {\bibfnamefont {N.}~\bibnamefont
  {Andrei}},\ }\bibfield  {title} {\bibinfo {title} {{A spin chain with
  non-Hermitian $\mathscr{PT}-$symmetric boundary couplings: exact solution,
  dissipative Kondo effect, and phase transitions on the edge}},\ }\href@noop
  {} {\bibfield  {journal} {\bibinfo  {journal} {arXiv preprint
  arXiv:2406.10334}\ } (\bibinfo {year} {2024}{\natexlab{f}})}\BibitemShut
  {NoStop}%
\bibitem [{\citenamefont {Pasnoori}\ \emph
  {et~al.}(2023{\natexlab{b}})\citenamefont {Pasnoori}, \citenamefont {Lee},
  \citenamefont {Pixley}, \citenamefont {Andrei},\ and\ \citenamefont
  {Azaria}}]{pasnoori2023boundary}%
  \BibitemOpen
  \bibfield  {author} {\bibinfo {author} {\bibfnamefont {P.~R.}\ \bibnamefont
  {Pasnoori}}, \bibinfo {author} {\bibfnamefont {J.}~\bibnamefont {Lee}},
  \bibinfo {author} {\bibfnamefont {J.}~\bibnamefont {Pixley}}, \bibinfo
  {author} {\bibfnamefont {N.}~\bibnamefont {Andrei}},\ and\ \bibinfo {author}
  {\bibfnamefont {P.}~\bibnamefont {Azaria}},\ }\bibfield  {title} {\bibinfo
  {title} {Boundary quantum phase transitions in the spin-1 2 heisenberg chain
  with boundary magnetic fields},\ }\href@noop {} {\bibfield  {journal}
  {\bibinfo  {journal} {Physical Review B}\ }\textbf {\bibinfo {volume}
  {107}},\ \bibinfo {pages} {224412} (\bibinfo {year}
  {2023}{\natexlab{b}})}\BibitemShut {NoStop}%
\bibitem [{\citenamefont {Wang}\ \emph {et~al.}(2015)\citenamefont {Wang},
  \citenamefont {Yang}, \citenamefont {Cao},\ and\ \citenamefont
  {Shi}}]{wang2015off}%
  \BibitemOpen
  \bibfield  {author} {\bibinfo {author} {\bibfnamefont {Y.}~\bibnamefont
  {Wang}}, \bibinfo {author} {\bibfnamefont {W.-L.}\ \bibnamefont {Yang}},
  \bibinfo {author} {\bibfnamefont {J.}~\bibnamefont {Cao}},\ and\ \bibinfo
  {author} {\bibfnamefont {K.}~\bibnamefont {Shi}},\ }\href@noop {} {\emph
  {\bibinfo {title} {Off-diagonal Bethe ansatz for exactly solvable models}}}\
  (\bibinfo  {publisher} {Springer},\ \bibinfo {year} {2015})\BibitemShut
  {NoStop}%
\bibitem [{\citenamefont {Nepomechie}(2003)}]{nepomechie2003functional}%
  \BibitemOpen
  \bibfield  {author} {\bibinfo {author} {\bibfnamefont {R.~I.}\ \bibnamefont
  {Nepomechie}},\ }\bibfield  {title} {\bibinfo {title} {Functional relations
  and bethe ansatz for the xxz chain},\ }\href@noop {} {\bibfield  {journal}
  {\bibinfo  {journal} {Journal of statistical physics}\ }\textbf {\bibinfo
  {volume} {111}},\ \bibinfo {pages} {1363} (\bibinfo {year}
  {2003})}\BibitemShut {NoStop}%
\end{thebibliography}%

\newpage

\begin{widetext}
The appendices serve to complement the main text by providing detailed discussions and derivations of concepts and results that could not be fully explored within the main body.

In Appendix \ref{thetadef}, we introduce and define the Jacobi elliptic theta functions, which play a key role in the expression for the energy of the spinon presented in Eq.\eqref{spinonengeqn}. 

Appendix \ref{int-proof} is dedicated to establishing the integrability of the Hamiltonian given in Eq.\eqref{ham}. Here, we rigorously derive the Bethe Ansatz equations, which are central to solving the model and exploring its rich physics.

In Appendix \ref{sec:intlin2imp}, we delve into a comprehensive description of the sixteen distinct boundary phases exhibited by the Hamiltonian in Eq.\eqref{ham}. This appendix provides a detailed phase-by-phase characterization, offering deeper insight into the interplay between boundary impurities and bulk interactions.

Appendix \ref{BAdets} focuses on the solutions of the Bethe Ansatz equations across different phases. This includes a meticulous examination of the solutions of the Bethe Ansatz equations in various parametric regimes, shedding light on the intricate phase structure of the model.

Finally, in Appendix \ref{diff-haldane-xxz}, we clarify the distinction between the edge modes associated with topological phases, such as those in the Haldane phase of the bilinear-biquadratic spin-1 model, and the edge modes found in the XXZ chain. To provide further context, we briefly study the effects of impurities at the Haldane point of the bilinear-biquadratic model, highlighting the unique physics that emerges in this regime.

Together, these appendices enrich the understanding of the model, offering a comprehensive supplement to the main text.
    \begin{appendix}
    \section{Jacobi Elliptic theta functions}
    \label{thetadef}
The Jacobi elliptic theta functions are defined as
\begin{align}
& \vartheta_1(u, q)=2 q^{1 / 4} \sum_{n=0}^{\infty}(-1)^n q^{n(n+1)} \sin ((2 n+1) u) . \\
& \vartheta_2(u, q)=2 q^{1 / 4} \sum_{n=0}^{\infty} q^{n(n+1)} \cos ((2 n+1) u) . \\
& \vartheta_3(u, q)=1+2 \sum_{n=1}^{\infty} q^{n^2} \cos (2 n u) . \\
& \vartheta_4(u, q)=1+2 \sum_{n=1}^{\infty}(-1)^n q^{n^2} \cos (2 n u) .
\end{align}
and $\vartheta'$ denotes the derivative with respect to the variable $u$.

\section{Hamiltonian and Bethe Ansatz Equations}\label{int-proof}

Consider the trigonometric six vertex R-matrix
\begin{equation}
   R(u)=\left(
\begin{array}{cccc}
 \frac{\sinh (\eta +u)}{\sinh(\eta )} & 0 & 0 & 0 \\
 0 & \frac{\sinh (u)}{\sinh(\eta )} & 1 & 0 \\
 0 & 1 & \frac{\sinh (u)}{\sinh(\eta )} & 0 \\
 0 & 0 & 0 & \frac{\sinh (\eta +u)}{\sinh(\eta )} \\
\end{array}
\right) 
\end{equation}

    which is  a solution of the Yang-Baxter equation
\begin{equation}
    R_{12}(u-v)R_{13}(u)R_{23}(v)=R_{23}(v)R_{13}(u)R_{12}(u-v).
\end{equation}
Notice that, the Yang-Baxter equation remains satisfied if we shift $u_i\to u_i-\theta_i$ where $\theta_i$ are arbitrary inhomogeneous parameters.

The R-matrix satisfies the following properties
\begin{align}
\text{Initial condition} &: R_{1,2}(0)=P_{1,2},\label{int}\\
\text{Unitary relation} &: R_{1,2}(u) R_{2,1}(-u)=-\frac{\sinh (u+\eta) \sinh (u-\eta)}{\sinh ^2 \eta} \times \mathrm{id}=(1-\mathrm{csch} ^2(\eta ) \sinh ^2(u))\times\mathrm{id}\label{unt}\\
\text{Crossing relation} &: R_{1,2}(u)=-\sigma_1^y R_{1,2}^{t_1}(-u-\eta) \sigma_1^y,\label{cs}\\
\text { Fusion condition} &:  R_{1,2}(-\eta)=-2 P_{1,2}^{(-)}\label{fsn} \\
\text{PT-symmetry} &: R_{1,2}(u)=R_{2,1}(u)=R_{1,2}^{t_1 t_2}(u),\\
Z_2-\text{symmetry} &: \sigma_1^\alpha \sigma_2^\alpha R_{1,2}(u)=R_{1,2}(u) \sigma_1^\alpha \sigma_2^\alpha, \text{ for } \alpha=x, y, z.
\end{align}

   Let us consider the following two single-row transfer matrices
 \begin{align}
T_0(u) &= R_{0,L}(u-b-\theta_L)R_{0,N_b}(u-\theta_{N_b})R_{0,N_b-1}(u-\theta_{N_b-1})\cdots R_{0,2}(u-\theta_2)R_{0,1}(u-\theta_1)R_{0,R}(u-d-\theta_R)\nonumber\\
\hat{T}_0(u) &= R_{0,R}(u+d+\theta_R)R_{0,1}(u+\theta_1)R_{0,2}(u+\theta_2)\cdots R_{0,N_b-1}(u+\theta_{N_b-1})R_{0,N_b}(u+\theta_{N_b})R_{0,L}(u+b+\theta_L)\nonumber
\end{align}

Now, we define the monodromy matrix
\begin{equation}
    \Xi(u)=T_0(u)\hat T_0(u)
\end{equation}
The trace of the monodromy matrix over the auxiliary space is defined as the double-row transfer matrix
\begin{equation}
    t(u)=\operatorname{tr}_0\Xi(u)
    \label{tmat}
\end{equation}
    
Now, we construct a Hamiltonian as
\begin{equation}
    H=J\sinh(\eta)\frac{\mathrm{d}}{\mathrm{d}\lambda}\ln t(\lambda)\Big\vert_{\lambda\to 0,\{\theta_i\}\to 0}-JN_b\cosh(\eta) -\frac{J \sinh ^2(\eta ) \cosh (\eta )}{\sinh ^2(\eta )-\sinh ^2(d)}-\frac{J \sinh ^2(\eta ) \cosh (\eta )}{\sinh ^2(\eta )-\sinh ^2(b)},
    \label{hameqn}
\end{equation}
 which gives
  as
\begin{align}
    H&=\sum_{i=1}^{N_b-1}J(\sigma_i^x \sigma_{i+1}^x+\sigma_i^y \sigma_{i+1}^y+\Delta \sigma_i^z \sigma_{i+1}^z)\nonumber\\
    &+J_R(\sigma_i^x \sigma_{i+1}^x+\sigma_i^y \sigma_{i+1}^y+\Delta_R \sigma_i^z \sigma_{i+1}^z)+J_L(\sigma_i^x \sigma_{i+1}^x+\sigma_i^y \sigma_{i+1}^y+\Delta_L \sigma_i^z \sigma_{i+1}^z)
\end{align}
where we introduced 
\begin{align}
   \Delta&=\cosh(\eta)\\
   J_R&=J\frac{\sinh^2(\eta)\cosh(d)}{\sinh^2(\eta)-\sinh^2(d)}\\
   J_L&=J\frac{\sinh^2(\eta)\cosh(b)}{\sinh^2(\eta)-\sinh^2(b)}\\
   \Delta_R&=\frac{\cosh(\eta)}{\cosh(d)}\\
    \Delta_L&=\frac{\cosh(\eta)}{\cosh(b)}
\end{align}

The eigenvalues $\Lambda(u)$ of the transfer matrix matrix $t(u)$ satisfy Baxter’s $T-Q$ relation~\cite{baxter1972partition,wang2015off,nepomechie2003functional} 
\begin{equation}
    \Lambda(u)=a(u) \frac{\bar Q(u-\eta)}{\bar Q(u)}+d(u)\frac{\bar Q(u+\eta)}{\bar Q(u)},
\end{equation}
where
\begin{equation}
    a(u)=\frac{\sin(u+\eta)}{\sin(2u+\eta)}\cos^2(u)\frac{\sin(u-b+\eta)\sin(u+b+\eta)}{\sin^2(\eta)}\frac{\sin(u-d+\eta)\sin(u+d+\eta)}{\sin^2(\eta)}
\end{equation}
and
\begin{equation}
    d(u)=a(-u-\eta)
\end{equation}
and the trigonometric $Q-$function is given by
\begin{equation}
    \bar Q(u)=\prod_{\ell=1}^M \frac{\sin(u-u_{\ell})\sin(u-u_{\ell}+\eta)}{\sin^2(\eta)}.
\end{equation}

Imposing the regularity condition on the $T-Q$ relation, we obtain the Bethe equation
\begin{align}
    &\frac{\cos^2(u)}{\cos^2(u+\eta)}\left(\frac{\sin(u_j+\eta)}{\sin(u_j)} \right)^{2N_b+1}\frac{\sin(u_j+b+\eta)}{\sin(u_j+b)}\frac{\sin(u_j-b+\eta)}{\sin(u_j-b)}\frac{\sin(u_j+d+\eta)}{\sin(u_j+d)}\frac{\sin(u_j-d+\eta)}{\sin(u_j-d)}\nonumber\\
    &=-\prod_{\ell=1}^M \frac{\sin(u_j-u_\ell+\eta)}{\sin(u_j-u_\ell-\eta)}\frac{\sin(u_j+u_\ell+2\eta)}{\sin(u_j+u_\ell)}
\end{align}
To write the equation more symmetrically, we introduce a new variable via relation $u_j=i\frac{\lambda_j}{2}-\frac{\eta}{2}$, such that the Bethe
ansatz equation becomes
\begin{equation}
     \begin{gathered}
\left(\frac{\sin \frac{1}{2}\left(\lambda _j-{i \eta }\right)}{\sin \frac12\left(\lambda _j+{i \eta }\right)}\right)^{2 N_b}
\frac{\cos^2\frac12\left(\lambda _j+{i \eta }\right)}{\cos^2\frac12\left(\lambda _j-{i \eta }\right)}\frac{\sin\frac12\left(\lambda_j -2i b-{i \eta }\right)}{\sin\frac12\left(\lambda_j -2i b+{i \eta }\right)}\frac{\sin\frac12\left(\lambda_j +2i b-{i \eta }\right)}{\sin\frac12\left(\lambda_j +2i b+{i \eta }\right)} \\
\frac{\sin\frac12\left(\lambda_j -2i d-{i \eta }\right)}{\sin\frac12\left(\lambda_j -2i d+{i \eta }\right)}\frac{\sin\frac12\left(\lambda_j +2i d-{i \eta }\right)}{\sin\frac12\left(\lambda_j +2i d+{i \eta }\right)}=\prod_{k=1(\neq j)}^M \frac{\sin\frac12 \left(\lambda_j-\lambda_k-2i\eta\right) \sin \frac12\left(\lambda_j+\lambda_k-2i\eta\right)}{\sin\frac12 \left(\lambda_j-\lambda_k+2i\eta\right) \sin \frac12\left(\lambda_j+\lambda_k+2i\eta\right)}
\end{gathered}
     \label{BAE}
\end{equation}
  and the energy is given by

   \begin{equation}
       E=2 J \sinh \eta \sum_{j=1}^M \frac{\sinh \eta}{\cos  \lambda_j-\cosh \eta}+\left[(N_b-1) J+\frac{J \sinh ^2 \eta}{\left(\sinh ^2 \eta-\sinh ^2 b\right)}+\frac{J \sinh ^2 \eta}{\left(\sinh ^2 \eta-\sinh ^2 d\right)}\right] \cosh \eta .
       \label{engeqn}
   \end{equation}

\section{Two impurities case}\label{sec:intlin2imp}
 Let us now briefly return to the Hamiltonian Eq.~\eqref{ham} and discuss the phase diagram when there are two impurities. As shown in the phase diagram Fig.~\ref{PDeven}, there are now sixteen total boundary phases, as there are four independent boundary phases on the two edges of the spin chain. The construction follows similar to the case of the isotropic chase studied in~\cite{kattel2023kondo} with the only major difference being that the model under consideration in this work is gapped, and hence the two impurities do not affect one another in the thermodynamic limit because all correlation in the model falls exponentially. 

\definecolor{dy}{rgb}{0.9,0.9,0.5}
\definecolor{dr}{rgb}{0.95,0.55,0.5}
\definecolor{db}{rgb}{0.5,0.8,0.9}
\definecolor{dg}{rgb}{0.3,0.9,0.75}
\definecolor{dm}{rgb}{0.3,0.6,0.85}
\definecolor{dn}{rgb}{1,0.7,0.6}

We provide a detailed construction of these sixteen phases in Appendix \ref{BAdets}, while this section offers only a brief overview, with progressively fewer details. As illustrated in the phase diagram in Fig.~\ref{fig:PD1}, a single impurity coupled integrably to the edge of the XXZ-$\frac{1}{2}$ chain can reside in one of four phases: the Kondo phase (K), the antiferromagnetic bound mode phase (ABM), the ferromagnetic bound mode phase (FBM), or the unscreened phase (US).

When two impurities are coupled integrably at opposite edges of the chain, each impurity can independently occupy any of these four phases. Consequently, the system can exist in one of sixteen possible configurations, as shown in the phase diagram in Fig.~\ref{PDeven}. In this diagram, the labels K, ABM, FBM, and US refer to the four phases, and a phase labeled $(X,Y)$ indicates that the left impurity is in phase $X$, while the right impurity is in phase $Y$, where $X$ and $Y$ can independently be any of the four phases.

\subsubsection{Kondo-Kondo phase}
When both left ($d_L$) and right ($d_R$) impurity parameters, that either purely imaginary values or real values between $0<d_i<\frac{\eta}{2}$, both impurities are screened by the Kondo effect. The ground state for even $N_b$ is two-fold degenerate where the fractional spin accumulations on the two ends are polarized in opposite direction \textit{i.e.} $\ket{GS}=\ket{\pm \frac{1}{4}, \mp \frac{1}{4}}$ whereas when $N_b$ is odd, the two-fold degenerate ground state has total spin $S^z=\pm\frac{1}{2}$ where the fractional edge spin accumulation polarize in the same direction \textit{i.e.} $\ket{GS}=\ket{\pm \frac{1}{4}, \pm \frac{1}{4}}$.

All the excited states can be constructed by adding even numbers of spinons, bulk string solutions, quartets etc~\cite{destri1982analysis}. There is no boundary excitation in the low energy in this phase.

\subsubsection{Kondo-ABM and ABM-Kondo phases}
The Kondo-ABM phase occurs when the left impurity parameter $d_L$ takes either imaginary values or real values between $0<d_R<\frac{\eta}{2}$ and the right impurity parameter $d_R$ takes real values between $\frac{\eta}{2}<d_R<\eta$. In the ground of the Kondo-ABM phase, the impurity coupled to the left edge is screened by the Kondo effect, whereas the impurity coupled to the right edge is screened by an exponentially localized bound state. The ground state for even $N_b$ is two-fold degenerate where the fractional spin accumulations on the two ends are polarized in opposite direction \textit{i.e.} $\ket{GS}=\ket{\pm \frac{1}{4}, \mp \frac{1}{4}}$ whereas when $N_b$ is odd, the two-fold degenerate ground state has total spin $S^z=\pm\frac{1}{2}$ where the fractional edge spin accumulation polarize in the same direction \textit{i.e.} $\ket{GS}=\ket{\pm \frac{1}{4}, \pm \frac{1}{4}}$.

The relationship between the ABM-Kondo phase and the Kondo-ABM phase is established via a space parity transformation. Exchanging $L$ and $R$ upholds the described results \textit{i.e.} in the ABM-Kondo phase, the impurity at the left edge is screened by an exponentially localized bound mode formed at the edge, whereas the right impurity is screened by multiparticle Kondo effect.

Apart from the usual bulk excitation constructed by adding an even number of spinons, bulk strings, quartets, etc., unique boundary excitations where impurities are unscreened are possible in this phase. The impurity boundary string solution describing the bound mode state screening the left or right impurity in the phases ABM-Kondo or Kondo-ABM phases, respectively, can be removed at the expense of energy 
\begin{align}
    E_{d_i}&=J \sinh (\eta )\sum_{\omega=-\infty}^\infty e^{-2 \eta  | \omega | } \text{sech}(\eta   \omega  ) \cosh ( \omega   (2 d_i-\eta ))\nonumber\\
    &+\frac{J \sinh ^2(\eta )}{\sinh (d_i) \sinh (d_i-\eta )},
\end{align}
and thereby unscreen the respective impurities creating a four-fold degenerate excited state with boundary excitation. Here, $i = \{L, R \}$ for ABM-Kondo and Kondo-ABM
phases respectively.

\subsubsection{Kondo-FBM and FBM-Kondo phases}

The Kondo-FBM phase occurs when the left impurity parameter $d_L$ takes either imaginary values or real values between $0<d_R<\frac{\eta}{2}$ and the right impurity parameter $d_R$ resides in the range $\eta<d_R<\frac{3\eta}{2}$. In the ground state of the Kondo-ABM phase, the impurity coupled to the left edge is screened by the Kondo effect, while the impurity coupled to the right edge remains unscreened. The ground state is four-fold degenerate with $S^z=\pm \frac{1}{2}$ where the edge fractional modes point in the same direction \textit{i.e.} $\ket{\pm \frac{1}{4},\pm \frac{1}{4}}$ where $N_b$ is odd. Likewise, the ground state is four-fold degenerate wit $S^z=0,\pm 1 $ when $N_b$ is even where for the two $S^z=0$ states, the edge modes polarize in the opposite direction $\ket{\pm \frac{1}{4}, \mp \frac{1}{4}}$ and for $S^z=\pm$ state, the edge modes polarize in the same direction $\ket{\pm \frac{3}{4}, \pm \frac{1}{4}}$.

The impurity in the right end can be screened by the bound mode at the expense of $E_{d_R}$ energy. Such a state would be two-fold degenerate with $S^z=0$ if $N_b$ is even and $S^z=\pm \frac{1}{2}$ if $N_b$ is odd.

The FBM-Kondo phase is related to the above-described Kondo-FBM phase by space parity transformation ($L\leftrightarrow R$). In this case, the ground state contains an unscreened left impurity while the right impurity is screened by an exponentially localized edge mode formed at the right edge of the spin chain. Similarly, all other results about the excited state follow by applying transformation $L\leftrightarrow R$.

\subsubsection{Kondo-US and US-Kondo phases}
The Kondo-US phase occurs when the left impurity parameter $d_L$ takes either purely imaginary values or real values in the range $0<d_L<\frac{\eta}{2}$ and the right impurity parameter $d_R$ resides in the range $d_R>\frac{3}{2}\eta$. The ground state in this phase is a four-fold degenerate state when the impurity in the left edge is screened by the Kondo effect and the right impurity is unscreened. The four-fold degenerate ground state has total spin $S^z=\pm \frac{1}{2}$ when $N_b$ is even where the fractional edge modes align in the same direction \textit{i.e.} $\ket{\pm \frac{1}{4}}, \pm \frac{1}{4}$ whereas the two out of the four-fold degenerate ground state has total spin $S^z=0$ where the fractional edge spin anti-align \textit{i.e.} $\ket{\pm \frac{1}{4}}, \mp \frac{1}{4}$ and in the remaining two state $S^z=\pm$ where the unscreened impurity combines with the quarter spin edge modes to form $\pm \frac{3}{4}$ edge spin accumulation such that the spin accumulation are of the from $\ket{\pm \frac{1}{4}}, \mp \frac{3}{4}$.

No boundary excitations are possible in this phase because bound mode no longer exists in this phase. Thus, all the excited states on top of the four-fold degenerate ground states are constructed by adding an even number of spinons, bulk string solutions, quartets, etc. Moreover, the impurity coupled to the right end of the spin chain can not be screened in this phase. The US-Kondo phase is related to the Kondo-US phase by the space parity transformation.

\subsubsection{ABM-ABM phase}

When both left and right impurity parameters take real values in the range $\frac{\eta}{2}<d_i<\eta$ where $i=\{L,R\}$, the model is in the ABM-ABM phase. The ground state in this phase is characterized by the screening of both impurities by exponentially localized bound mode formed at the two edges of the spin chain. The ground state is two-fold degenerate with $S^z=0$ when $N_b$ is even where edge quarter modes point in the opposite direction and it is $S^z=\pm \frac{1}{2}$ state when $N_b$ is odd where the edge quarter modes align in the same direction.

Apart from the usual bulk excitation constructed by adding an even number of spinons, bulk string solution, quartets, etc., unique boundary excitations are possible in this phase. For example, the bound mode solution screening left (right) impurities can be removed at the expense of energy $E_{d_L}$ ($E_{d_R}$) to unscreen the left (right) impurities. One can also remove both the bound mode solution, thereby unscreening both the impurities at the cost of $E_{d_L}+E_{d_R}$ energy.

\subsubsection{FBM-FBM phase}

When both left and right impurity parameters take real values in the range $\eta<d_i<\frac{3\eta}{2}$ where $i=\{L,R\}$, the model is in the FBM-FBM phase. The ground state in this phase has both of the impurities unscreened. The ground state is thus eight-fold degenerate for both even and odd number of bulk sites $N_b$.

Apart from the usual bulk excitation constructed by adding an even number of spinons, bulk strings, quartets, etc., unique boundary excitations where impurities are screened are possible in this phase. The bound mode solution screening left (right) impurities can be added at the expense of energy $E_{d_L}$ ($E_{d_R}$) to unscreen the left (right) impurities. One can also add both the bound mode solution, thereby unscreening both the impurities at the cost of $E_{d_L}+E_{d_R}$ energy.

\subsubsection{ABM-FBM and FBM-ABM phases}
The ABM-FBM phase occurs when the left impurity parameter $d_L$ takes real values in the range $\frac{\eta}{2}<d_L<\eta$ and the right impurity parameter $d_R$ resides in the range $\eta<d_R<\frac{3}{2}\eta$. The ground state in this phase is a four-fold degenerate state where the impurity in the left edge is screened by an exponentially localized bound mode formed at the left edge, and the right impurity is unscreened. 

The impurity at the left edge can be unscreened by removing the bound mode solution at the expense of energy $E_{d_L}$. Similarly, the impurity at the right edge can be screened by adding the bound mode solution
which costs energy $E_{d_R}$.
Moreover, One can simultaneously unscreen the left impurity and screen the right impurity at the expense of energy $E_{d_L}+E_{d_R}$. 

The FBM-ABM phase is related to the ABM-FBM
phase by space parity transformation. The above-described results follow through by applying the transformation $L \leftrightarrow R$.

\subsubsection{ABM-US and US-ABM phases}

The ABM-US phase occurs when the left impurity parameter $d_L$ takes real values in the range $\frac{\eta}{2}<d_L<\eta$ and the right impurity parameter $d_R$ resides in the range $d_R>\frac{3}{2}\eta$. The ground state in this phase is a four-fold degenerate state where the impurity in the left edge is screened by an exponentially localized bound mode formed at the left edge, and the right impurity is unscreened. 

The impurity at the right edge can not be screened in this phase. However, the impurity at the left edge can be unscreened by removing the bound mode solution $\lambda_{d_L}$, which costs energy $E_{d_L}$. 

The US-ABM phase is related to the ABM-US
phase by space parity transformation.

\subsubsection{FBM-US and US-FBM phases}
The FBM-US phase occurs when the left impurity parameter $d_L$ takes values in the range $\eta<d_L<\frac{3\eta}{2}$ and the right impurity parameter $d_R$ resides in the range $d_R>\frac{3}{2}\eta$. The ground state in this phase is an eight-fold degenerate state for both even and odd $N_b$ where both impurities are unscreened.

The impurity at the right edge can not be screened in this phase. However, the impurity at the left edge can be screened by adding the bound mode solution $\lambda_{d_L}$, which costs energy $E_{d_L}$. 

The US-FBM phase is related to the FBM-US
phase by space parity transformation.

\subsubsection{US-US phase}

The US-US phase occurs when both left impurity parameter $d_L$ and right impurity parameter $d_R$ take values in the range $d_i>\frac{3}{2}\eta$ where $i=\{L,R\}$. The ground state in this phase is an eight-fold degenerate state for both even and odd $N_b$ where both impurities are unscreened.

The impurities can not be screened in this phase. Thus, all the excited states are the bulk excitations constructed by adding an even number of spinons, bulk string solutions, quartets, etc.

    \section{Detailed solution of Bethe Ansatz equation}\label{BAdets}

    We shall describe the full solution of the Bethe Ansatz equations Eq.~\eqref{BAE} in this section for the case where $N_b$ is even. As shown in the phase diagram Fig.~\ref{PDeven}, the model described by Hamiltonian Eq.\eqref{ham} with two impurity exhibits total of sixteen impurity phases. We shall gradually decrease the amount of detail for each phase. There are sixteen phases because each impurity at the two ends of the chain can independently reside in one of four possible phases: the Kondo phase, the antiferromagnetic bound mode phase, the ferromagnetic bound mode phase, or the unscreened phase. This independence leads to $4\times 4=16$ possible combinations of impurity phases.

\subsection[Kondo-Kondo]{The Kondo-Kondo regime}
The Kondo-Kondo regime exists when the impurity parameters $b$ and $d$ take either purely imaginary values or real values between $0$ and $\frac{\eta}{2}$.

In this regime, Bethe equation takes the form
\begin{align}
   & \left(\frac{\sin \frac{1}{2}\left(\lambda _j-{i \eta }\right)}{\sin \frac12\left(\lambda _j+{i \eta }\right)}\right)^{2 N_b}
\frac{\cos^2\frac12\left(\lambda _j+{i \eta }\right)}{\cos^2\frac12\left(\lambda _j-{i \eta }\right)}\frac{\sin\frac12\left(\lambda_j -i(2 b+ \eta )\right)}{\sin\frac12\left(\lambda_j +i(2 b+ \eta )\right)}\frac{\sin\frac12\left(\lambda_j -i( \eta -2b)\right)}{\sin\frac12\left(\lambda_j +i( \eta +2b)\right)} \nonumber\\
&\frac{\sin\frac12\left(\lambda_j -i(2 d+ \eta )\right)}{\sin\frac12\left(\lambda_j +i(2 d+ \eta )\right)}\frac{\sin\frac12\left(\lambda_j -i( \eta -2d)\right)}{\sin\frac12\left(\lambda_j +i( \eta -2d)\right)}=\prod_{k=1(\neq j)}^M \frac{\sin\frac12 \left(\lambda_j-\lambda_k-2i\eta\right) \sin \frac12\left(\lambda_j+\lambda_k-2i\eta\right)}{\sin\frac12 \left(\lambda_j-\lambda_k+2i\eta\right) \sin \frac12\left(\lambda_j+\lambda_k+2i\eta\right)}
\end{align}

Taking $\ln$ on both sides, we write
\begin{align}
   (2N_b+1) \phi\left(\lambda_j,\eta\right)&-2\psi\left(\lambda_j,\eta\right)+\phi\left(\lambda_j ,{\eta+2b}\right)+\phi\left(\lambda_j ,\eta-2b\right)+\phi\left(\lambda_j ,\eta+2d\right)+\phi\left(\lambda_j ,\eta-2d\right)\nonumber\\
   &+\psi(\lambda_j,\eta)=\pi i I_j +\sum_{k}\left[\phi(\lambda_j+\lambda_k,2\eta)+\phi(\lambda_j-\lambda_k,2\eta) \right]
\end{align}

Differentiating and removing the solutions $\lambda=0$ and $\lambda=\pi$, we obtain
\begin{align}
    (2N_b+1) &a(\lambda ,\eta )-2 a\left(\lambda -\pi,\eta \right)+ a(\lambda ,\eta+2b )+ a(\lambda ,\eta-2b )+ a(\lambda ,\eta+2d )+ a(\lambda,\eta-2d )+a(\lambda-\pi,\eta)\nonumber\\
    &=2\pi \rho(\lambda)+\int \rho(\lambda) \left[a(\lambda-\lambda',2\eta)+a(\lambda+\lambda',2\eta) \right]\mathrm{d}\lambda'+2\pi\delta(\lambda-\pi)+2\pi\delta(\lambda)
\end{align}

The solution is immediate in the Fourier space
\begin{equation}
    \tilde\rho_{\ket{\frac12}}(\omega)=\frac{(2N_b+1)e^{-\eta|\omega|}+(-1)^\omega e^{-\eta|\omega|}-2(-1)^\omega e^{-\eta|\omega|}+e^{-\eta|\omega|}\left(e^{-2b |\omega|}+e^{2b |\omega|}+e^{-2d |\omega| }+e^{2d|\omega|}\right)-(1+(-1)^\omega)}{4\pi(1+e^{-2\eta|\omega|})}
\end{equation}
The total number of Bethe roots is given by
\begin{equation}
    M_{\ket{\frac12}}=\int_{-\pi}^\pi \rho_{\ket{\frac{1}{2}}}(\lambda)\mathrm{d}\lambda=2\pi \tilde \rho_{\ket{\frac12}}(0)=\frac{2N_b+2}{4}=\frac{N_b+1}{2}
\end{equation}

Since the total number of roots is not an integer when $N_b$ is even, thus, we need to add the complex solution $\lambda_{bs}$. \\

Adding the complex solution $\lambda_{bs}=\pi \pm i\eta$, we obtain
\begin{align}
   & (2N_b+1) a(\lambda ,\eta )-2 a\left(\lambda -\pi,\eta \right)+ a(\lambda ,\eta+2b )+ a(\lambda ,\eta-2b )+ a(\lambda ,\eta+2d )+ a(\lambda,\eta-2d )+a(\lambda-\pi,\eta)\nonumber\\
    &=2\pi \rho(\lambda)+\int \rho(\lambda) \left[a(\lambda-\lambda',2\eta)+a(\lambda+\lambda',2\eta) \right]\mathrm{d}\lambda'+a(\lambda-\pi,\eta)+a(\lambda-\pi,3\eta)+2\pi\delta(\lambda-\pi)+2\pi\delta(\lambda)
\end{align}
The solution in Fourier space becomes
\begin{align}
    \tilde\rho_{\ket{0}_{bs}}(\omega)&=\frac{(2N_b+1)e^{-\eta|\omega|}+(-1)^\omega e^{-\eta|\omega|}-2(-1)^\omega e^{-\eta|\omega|}+e^{-\eta|\omega|}\left(e^{-2b |\omega|}+e^{2b |\omega|}+e^{-2d |\omega| }+e^{2d|\omega|}\right)-(1+(-1)^\omega)}{4\pi(1+e^{-2\eta|\omega|})}\nonumber\\
    &-\frac{(-1)^\omega}{4\pi} \frac{e^{-\eta|\omega|}+e^{-3\eta|\omega|}}{(1+e^{-2\eta|\omega|})}
\end{align}

The total number of Bethe roots is given by
\begin{equation}
    M_{\ket{0}_{bs}}=1+\int_{-\pi}^\pi \rho_{\ket{0}}(\lambda)\mathrm{d}\lambda=1+2\pi \tilde \rho_{\ket{0}}(0)=\frac{N_b+2}{2}
\end{equation}

Hence the spin of this state is
\begin{equation}
    S_{\ket{0}_{bs}}=\frac{N_b+2}{2}-M_{\ket{0}_{bs}}=0
\end{equation}
The change induced in the density due to the string solution is
\begin{equation}
    \Delta \tilde \rho(\omega)=-\frac{(-1)^\omega}{4\pi} \frac{e^{-\eta|\omega|}+e^{-3\eta|\omega|}}{(1+e^{-2\eta|\omega|})}
\end{equation}

As shown above, the energy of this string solution vanishes. 

Hence the energy of this state is
\begin{equation}
    E_{\ket{0}_{bs}}=E_{ar}+\left[(N_b-1) J+\frac{J \sinh ^2 \eta}{\left(\sinh ^2 \eta-\sinh ^2 b\right)}+\frac{J \sinh ^2 \eta}{\left(\sinh ^2 \eta-\sinh ^2 d\right)}\right] \cosh \eta
\end{equation}

Notice that this is the ground state in this phase, and it is two-fold degenerate as one could add either the left or right boundary string $\pi\pm i\eta$ to construct it. 

Thus, the impurities in this phase are screened by the multiparticle Kondo effect. All the excitations in this phase are bulk excitations which are constructed by adding an even number of spinons, bulk strings, and quartets. There is a single tower of excitations in this phase.

Here, the bulk contribution to the root density is
\begin{equation}
    \rho_0^{bulk}(\omega)=\frac{N_b e^{-\eta|\omega|}}{2\pi(1+e^{-2\eta|\omega|})}
    \label{bulkpart}
\end{equation}
and the  single impurity contribution is
\begin{equation}
    \rho_{0}^{imp}(\omega)=\frac{e^{-\eta|\omega|(e^{-2 d | \omega | }+e^{2 d | \omega | })}}{4\pi(1+e^{-2\eta|\omega|})}
    \label{imppart}
\end{equation}

\subsection{Kondo-ABM and ABM-Kondo phases}\label{kondo-bm-dets}
Let us consider the case, when $0<b<\frac{\eta}{2}$, $\frac{\eta}{2}<d<\eta$ and $N_b$ is even. The case where $0<d<\frac{\eta}{2}$ and $\frac{\eta}{2}<b<\eta$ can then be obtained  by applying the transformation $L\leftrightarrow R$.
In this regime, Bethe equation takes the form
\begin{align}
   & \left(\frac{\sin \frac{1}{2}\left(\lambda _j-{i \eta }\right)}{\sin \frac12\left(\lambda _j+{i \eta }\right)}\right)^{2 N_b}
\frac{\cos^2\frac12\left(\lambda _j+{i \eta }\right)}{\cos^2\frac12\left(\lambda _j-{i \eta }\right)}\frac{\sin\frac12\left(\lambda_j -i(\eta-2b )\right)}{\sin\frac12\left(\lambda_j +i(\eta-2b )\right)}\frac{\sin\frac12\left(\lambda_j -i(2b+ \eta )\right)}{\sin\frac12\left(\lambda_j +i(2b+ \eta )\right)} \nonumber\\
&\frac{\sin\frac12\left(\lambda_j -i(2 d+ \eta )\right)}{\sin\frac12\left(\lambda_j +i(2 d+ \eta )\right)}\frac{\sin\frac12\left(\lambda_j +i(2d- \eta)\right)}{\sin\frac12\left(\lambda_j -i( 2d- \eta)\right)}=\prod_{k=1(\neq j)}^M \frac{\sin\frac12 \left(\lambda_j-\lambda_k-2i\eta\right) \sin \frac12\left(\lambda_j+\lambda_k-2i\eta\right)}{\sin\frac12 \left(\lambda_j-\lambda_k+2i\eta\right) \sin \frac12\left(\lambda_j+\lambda_k+2i\eta\right)}
\end{align}

Taking $\ln$ on both sides, we write
\begin{align}
   (2N_b+1) \phi\left(\lambda_j,\eta\right)&-2\psi\left(\lambda_j,\eta\right)+\phi\left(\lambda_j ,{\eta+2b}\right)+\phi\left(\lambda_j ,\eta-2b\right)+\phi\left(\lambda_j ,\eta+2d\right)-\phi\left(\lambda_j ,2d-\eta\right)\nonumber\\
   &+\psi(\lambda_j,\eta)=\pi i I_j +\sum_{k}\left[\phi(\lambda_j+\lambda_k,2\eta)+\phi(\lambda_j-\lambda_k,2\eta) \right]
\end{align}

Differentiating and removing the solutions $\lambda=0$ and $\lambda=\pi$, we obtain
\begin{align}
    (2N_b+1) &a(\lambda ,\eta )-2 a\left(\lambda -\pi,\eta \right)+ a(\lambda,\eta+2b )+ a(\lambda ,\eta-2b )+ a(\lambda ,\eta+2d )- a(\lambda,2d-\eta )+a(\lambda-\pi,\eta)\nonumber\\
    &=2\pi \rho(\lambda)+\int \rho(\lambda) \left[a(\lambda-\lambda',2\eta)+a(\lambda+\lambda',2\eta) \right]\mathrm{d}\lambda'+2\pi\delta(\lambda)+2\pi\delta(\lambda-\pi)
\end{align}
Solving the above integral equation, we write the solution density in Fourier space
\begin{align}
    \tilde\rho_{\ket{1}}(\omega)&=\frac{(2N_b+1)e^{-\eta|\omega|}+(-1)^\omega e^{-\eta|\omega|}-2(-1)^\omega e^{-\eta|\omega|}-(1+(-1)^\omega)}{4\pi(1+e^{-2\eta|\omega|})}\nonumber\\
    &+ \frac{e^{-(\eta-2b)|\omega|}+e^{-(\eta+2b)|\omega|}+e^{-(\eta+2d)|\omega|}-e^{-(2d-\eta)|\omega|}}{4\pi(1+e^{-2\eta|\omega|})}
\end{align}
The total number of Bethe roots is given by
\begin{equation}
    M_{\ket{1}}=\int_{-\pi}^\pi \rho_{\ket{1}}(\lambda)\mathrm{d}\lambda=2\pi \tilde \rho_{\ket{1}}(0)=\frac{N_b}{2}
\end{equation}

Thus, the total spin of the state is
\begin{equation}
    S^z_{\ket{1}}=\frac{N_b+2}{2}-M_{\ket 1}=1
\end{equation}
The energy of this state is
\begin{equation}
    E_{\ket1}=E_{ar_{\ket{1}}}+\left[(N_b-1) J+\frac{J \sinh ^2 \eta}{\left(\sinh ^2 \eta-\sinh ^2 b\right)}+\frac{J \sinh ^2 \eta}{\left(\sinh ^2 \eta-\sinh ^2 d\right)}\right] \cosh \eta
\end{equation}

where the energy of all real roots is given by
\begin{equation}
    E_{ar_{\ket{1}}}=-4\pi J \sinh \eta \sum_{\omega=-\infty}^\infty \tilde \rho_{\ket{1}}(\omega)\tilde a(\omega,\eta) 
\end{equation}

Because of the $\mathbb{Z}_2$ symmetry there is a state $\ket{-1}$ which is degenerate to state $\ket 1$.\\

However, this state described by all real roots is not the ground state. Adding the complex root $\lambda_{bs}$ and the purely imaginary root $\lambda_d$, we obtain Bethe equation of the form

\begin{align}
   &(2N_b+1) \phi\left(\lambda_j,\eta\right)-2\psi\left(\lambda_j,\eta\right)+\phi\left(\lambda_j ,\eta+2b\right)+\phi\left(\lambda_j ,\eta-2b\right)+\phi\left(\lambda_j ,\eta+2d\right)-\phi\left(\lambda_j ,2d-\eta\right)\nonumber\\
   &+\psi(\lambda_j,\eta)=\pi i I_j +\sum_{k}\left[\phi(\lambda_j+\lambda_k,2\eta)+\phi(\lambda_j-\lambda_k,2\eta) \right]+\phi(\lambda_j,2d+\eta)+\phi(\lambda_j,3\eta-2d)
\end{align}

Differentiating and removing the solutions $\lambda=0$ and $\lambda=\pi$, we obtain
\begin{align}
    &(2N_b+1) a(\lambda ,\eta )-2 a\left(\lambda -\pi,\eta \right)+ a(\lambda ,\eta+2b )+ a(\lambda ,\eta-2b )+ a(\lambda ,\eta+2d )- a(\lambda,2d-\eta )+a(\lambda-\pi,\eta)\nonumber\\
    &=2\pi \rho(\lambda)+\int \rho(\lambda) \left[a(\lambda-\lambda',2\eta)+a(\lambda+\lambda',2\eta) \right]\mathrm{d}\lambda'+2\pi\delta(\lambda)+2\pi\delta(\lambda-\pi)+a(\lambda,2d+\eta)+a(\lambda,3\eta-2d)\nonumber\\
    &\quad\quad\quad+a(\lambda-\pi,\eta)+a(\lambda-\pi,3\eta)
\end{align}

Solving the above integral equation, we write the solution density in Fourier space

\begin{align}
    \tilde\rho_{\ket{0}_{bs,d}}(\omega)&=\frac{(2N_b+1)e^{-\eta|\omega|}+(-1)^\omega e^{-\eta|\omega|}-2(-1)^\omega e^{-\eta|\omega|}-(1+(-1)^\omega)}{4\pi(1+e^{-2\eta|\omega|})}-\frac{(-1)^\omega}{4\pi} \frac{e^{-\eta|\omega|}+e^{-3\eta|\omega|}}{(1+e^{-2\eta|\omega|})}\nonumber\\
    &+ \frac{e^{-(\eta-2b)|\omega|}+e^{-(\eta+2b)|\omega|}+e^{-(\eta+2d)|\omega|}-e^{-(2d-\eta)|\omega|}}{4\pi(1+e^{-2\eta|\omega|})}-\frac{e^{-(2d+\eta)|\omega|}+e^{-(3\eta-2d)|\omega|}}{4\pi(1+e^{-2\eta|\omega|})}
    \label{GSrootdens}
\end{align}

The total number of roots is
\begin{equation}
  M_{\ket{0}_{bs,d}}=  2+\int_{-\pi}^\pi\rho_{\ket{0}_{bs,d}}(\lambda)=1+2\pi\tilde\rho_{\ket{0}_{bs,d}}(0)=\frac{N_b+2}{2}
\end{equation}
Thus, the spin of this state is
\begin{equation}
     S_{\ket{0}_{bs,d}}=\frac{N_b+2}{2}- M_{\ket{0}_{bs,d}}=0
\end{equation}

We have already shown that the energy of the string solution $\lambda_{bs}$ is zero, and the energy of the string solution $\lambda_d$ is
\begin{equation}
    E_{d}=\frac{J \sinh ^2(\eta )}{\sinh (d) \sinh (d-\eta )}+J \sinh (\eta )\sum_{\omega=-\infty}^\infty e^{-2 \eta  | \omega | } \text{sech}(\eta   \omega  ) \cosh ( \omega   (2 d-\eta ))
\end{equation}
which is negative. Hence, the total energy of this state is
\begin{equation}
    E_{\ket{0}_{bs,d}}=E_{ar_{\ket{1}}}+\left[(N_b-1) J+\frac{J \sinh ^2 \eta}{\left(\sinh ^2 \eta-\sinh ^2 b\right)}+\frac{J \sinh ^2 \eta}{\left(\sinh ^2 \eta-\sinh ^2 d\right)}\right] \cosh \eta
\end{equation}
This is the ground state where the left impurity is screened by multiparticle screening (Kondo physics), and the right impurity is screened by a single particle bound mode. \\

We could remove the $\lambda_{bs}$ solution from the above state and add a hole propagating with rapidity $\theta$ such that the root density satisfies an integral equation of the form

\begin{align}
    &(2N_b+1) a(\lambda ,\eta )-2 a\left(\lambda -\pi,\eta \right)+ a(\lambda,\eta+2b )+ a(\lambda ,\eta-2b )+ a(\lambda ,\eta+2d )- a(\lambda,2d-\eta )+a(\lambda-\pi,\eta)\nonumber\\
    &=2\pi \rho(\lambda)+\int \rho(\lambda) \left[a(\lambda-\lambda',2\eta)+a(\lambda+\lambda',2\eta) \right]\mathrm{d}\lambda'+2\pi\delta(\lambda)+2\pi\delta(\lambda-\pi)+a(\lambda,2d+\eta)+a(\lambda,3\eta-2d)\nonumber\\
    &\quad\quad\quad+2\pi\delta(\lambda-\theta)+2\pi\delta(\lambda+\theta)
\end{align}

Solving for the root density in the Fourier space, we obtain
\begin{align}
    \tilde\rho_{\ket{1}_{\theta,d}}(\omega)&=\frac{(2N_b+1)e^{-\eta|\omega|}+(-1)^\omega e^{-\eta|\omega|}-2(-1)^\omega e^{-\eta|\omega|}-(1+(-1)^\omega)}{4\pi(1+e^{-2\eta|\omega|})}-\frac{2\cos(\omega\theta)}{4\pi(1+e^{-2\eta|\omega|})}\nonumber\\
    &+ \frac{e^{-(\eta-2b)|\omega|}+e^{-(\eta+2b)|\omega|}+e^{-(\eta+2d)|\omega|}-e^{-(2d-\eta)|\omega|}}{4\pi(1+e^{-2\eta|\omega|})}-\frac{e^{-(2d+\eta)|\omega|}+e^{-(3\eta-2d)|\omega|}}{4\pi(1+e^{-2\eta|\omega|})}
\end{align}

The total number of roots is
\begin{equation}
  M_{\ket{1}_{\theta,d}}=  1+\int_{-\pi}^\pi\rho_{\ket{1}_{\theta,d}}(\lambda)=1+2\pi\tilde\rho_{\ket{1}_{\theta,d}}(0)=\frac{N_b}{2}
\end{equation}
Thus, the spin of this state is
\begin{equation}
     S_{\ket{1}_{\theta,d}}=\frac{N_b+2}{2}- M_{\ket{1}_{\theta,d}}=1
\end{equation}
and the energy of the state is
\begin{equation}
    E_{\ket{1}_{\theta,d}}=E_d+E_{ar_{\ket{1}}}+\left[(N_b-1) J+\frac{J \sinh ^2 \eta}{\left(\sinh ^2 \eta-\sin ^2 b\right)}+\frac{J \sinh ^2 \eta}{\left(\sinh ^2 \eta-\sinh ^2 d\right)}\right] \cosh \eta+E_\theta
\end{equation}

Since the lowest energy of spinon is non-zero, this is an excited state where both the impurities are still screened.

Thus, in this phase, the ground state involves multiparticle screening of the left impurity and the right impurity is screened by an exponentially localized bound formed at the right end of the impurity. There are unique boundary excitations in this regime which involves the unscreening of the right impurity by removing the impurity boundary string solution. Thus, there are two distinct towers of excited states: one where both impurities are screened (one by multiparticle Kondo effect and another by exponentially localized bound mode), whereas another tower includes the excited states where one impurity is screened by multiparticle Kondo effect but the other impurity is unscreened. 

\subsection{Kondo-FBM and FBM-Kondo Phase}

Let us consider the case when $0<b<\frac{\eta}{2}$ , $\eta<d<\frac{3\eta}{2}$ and $N_b$ is even. The case where $0<d<\frac{\eta}{2}$ and $\eta<b<\frac{3\eta}{2}$ can then be obtained  by applying the transformation $L\leftrightarrow R$.

In this regime, Bethe equation takes the form
\begin{align}
   & \left(\frac{\sin \frac{1}{2}\left(\lambda _j-{i \eta }\right)}{\sin \frac12\left(\lambda _j+{i \eta }\right)}\right)^{2 N_b}
\frac{\cos^2\frac12\left(\lambda _j+{i \eta }\right)}{\cos^2\frac12\left(\lambda _j-{i \eta }\right)}\frac{\sin\frac12\left(\lambda_j -i(\eta-2b )\right)}{\sin\frac12\left(\lambda_j +i(\eta-2b )\right)}\frac{\sin\frac12\left(\lambda_j -i(2b+ \eta )\right)}{\sin\frac12\left(\lambda_j +i(2b+ \eta )\right)} \nonumber\\
&\frac{\sin\frac12\left(\lambda_j -i(2 d+ \eta )\right)}{\sin\frac12\left(\lambda_j +i(2 d+ \eta )\right)}\frac{\sin\frac12\left(\lambda_j +i(2d- \eta)\right)}{\sin\frac12\left(\lambda_j -i( 2d- \eta)\right)}=\prod_{k=1(\neq j)}^M \frac{\sin\frac12 \left(\lambda_j-\lambda_k-2i\eta\right) \sin \frac12\left(\lambda_j+\lambda_k-2i\eta\right)}{\sin\frac12 \left(\lambda_j-\lambda_k+2i\eta\right) \sin \frac12\left(\lambda_j+\lambda_k+2i\eta\right)}
\end{align}

Taking $\ln$ on both sides, we write
\begin{align}
   (2N_b+1) \phi\left(\lambda_j,\eta\right)&-2\psi\left(\lambda_j,\eta\right)+\phi\left(\lambda_j ,{\eta+2b}\right)+\phi\left(\lambda_j ,\eta-2b\right)+\phi\left(\lambda_j ,\eta+2d\right)-\phi\left(\lambda_j ,2d-\eta\right)\nonumber\\
   &+\psi(\lambda_j,\eta)=\pi i I_j +\sum_{k}\left[\phi(\lambda_j+\lambda_k,2\eta)+\phi(\lambda_j-\lambda_k,2\eta) \right]
\end{align}

Differentiating and removing the solutions $\lambda=0$ and $\lambda=\pi$, we obtain
\begin{align}
    (2N_b+1) &a(\lambda ,\eta )-2 a\left(\lambda -\pi,\eta \right)+ a(\lambda,\eta+2b )+ a(\lambda ,\eta-2b )+ a(\lambda ,\eta+2d )- a(\lambda,2d-\eta )+a(\lambda-\pi,\eta)\nonumber\\
    &=2\pi \rho(\lambda)+\int \rho(\lambda) \left[a(\lambda-\lambda',2\eta)+a(\lambda+\lambda',2\eta) \right]\mathrm{d}\lambda'+2\pi\delta(\lambda)+2\pi\delta(\lambda-\pi)
\end{align}
Solving the above integral equation, we write the solution density in Fourier space
\begin{align}
    \tilde\rho_{\ket{1}}(\omega)&=\frac{(2N_b+1)e^{-\eta|\omega|}+(-1)^\omega e^{-\eta|\omega|}-2(-1)^\omega e^{-\eta|\omega|}-(1+(-1)^\omega)}{4\pi(1+e^{-2\eta|\omega|})}\nonumber\\
    &+ \frac{e^{-(\eta-2b)|\omega|}+e^{-(\eta+2b)|\omega|}+e^{-(\eta+2d)|\omega|}-e^{-(2d-\eta)|\omega|}}{4\pi(1+e^{-2\eta|\omega|})}
\end{align}
The total number of Bethe roots is given by
\begin{equation}
    M_{\ket{1}}=\int_{-\pi}^\pi \rho_{\ket{1}}(\lambda)\mathrm{d}\lambda=2\pi \tilde \rho_{\ket{1}}(0)=\frac{N_b}{2}
\end{equation}

Thus, the total spin of the state is
\begin{equation}
    S^z_{\ket{1}}=\frac{N_b+2}{2}-M_{\ket 1}=1
\end{equation}
The energy of this state is
\begin{equation}
    E_{\ket1}=E_{ar_{\ket{1}}}+\left[(N_b-1) J+\frac{J \sinh ^2 \eta}{\left(\sinh ^2 \eta-\sinh ^2 b\right)}-\frac{J \sinh ^2 \eta}{\left(\sinh ^2 d-\sinh ^2 \eta\right)}\right] \cosh \eta
\end{equation}

where the energy of all real roots is given by
\begin{equation}
    E_{ar_{\ket{1}}}=-4\pi J \sinh \eta \sum_{\omega=-\infty}^\infty \tilde \rho_{\ket{1}}(\omega)\tilde a(\omega,\eta) 
\end{equation}

Because of the $\mathbb{Z}_2$ symmetry there is a state $\ket{-1}$ state which is degenerate to state $\ket 1$.\\

Adding the boundary string and a higher order boundary string (see detail construction in the isotropic case in \cite{kattel2023kondo}), we obtain another $S^z=0$ state with the same energy in the thermodynamic limit as the energy of the fundamental boundary string and the higher order boundary string exactly cancel in the thermodynamic limit. 

Since the bound mode has positive energy, this four-fold degenerate state is the ground state in this regime. Here, the left impurity is screened by multiple particles, and the right impurity is unscreened.

Adding the complex root $\lambda_{bs}$ and the purely imaginary root $\lambda_d$, we obtain Bethe equation of the form

\begin{align}
   &(2N_b+1) \phi\left(\lambda_j,\eta\right)-2\psi\left(\lambda_j,\eta\right)+\phi\left(\lambda_j ,\eta+2b\right)+\phi\left(\lambda_j ,\eta-2b\right)+\phi\left(\lambda_j ,\eta+2d\right)-\phi\left(\lambda_j ,2d-\eta\right)\nonumber\\
   &+\psi(\lambda_j,\eta)=\pi i I_j +\sum_{k}\left[\phi(\lambda_j+\lambda_k,2\eta)+\phi(\lambda_j-\lambda_k,2\eta) \right]+\phi(\lambda_j,2d+\eta)+\phi(\lambda_j,3\eta-2d)
\end{align}

Differentiating and removing the solutions $\lambda=0$ and $\lambda=\pi$, we obtain
\begin{align}
    &(2N_b+1) a(\lambda ,\eta )-2 a\left(\lambda -\pi,\eta \right)+ a(\lambda ,\eta+2b )+ a(\lambda ,\eta-2b )+ a(\lambda ,\eta+2d )- a(\lambda,2d-\eta )+a(\lambda-\pi,\eta)\nonumber\\
    &=2\pi \rho(\lambda)+\int \rho(\lambda) \left[a(\lambda-\lambda',2\eta)+a(\lambda+\lambda',2\eta) \right]\mathrm{d}\lambda'+2\pi\delta(\lambda)+2\pi\delta(\lambda-\pi)+a(\lambda,2d+\eta)+a(\lambda,3\eta-2d)\nonumber\\
    &\quad\quad\quad+a(\lambda-\pi,\eta)+a(\lambda-\pi,3\eta)
\end{align}

Solving the above integral equation, we write the solution density in Fourier space

\begin{align}
    \tilde\rho_{\ket{0}_{bs,d}}(\omega)&=\frac{(2N_b+1)e^{-\eta|\omega|}+(-1)^\omega e^{-\eta|\omega|}-2(-1)^\omega e^{-\eta|\omega|}-(1+(-1)^\omega)}{4\pi(1+e^{-2\eta|\omega|})}-\frac{(-1)^\omega}{4\pi} \frac{e^{-\eta|\omega|}+e^{-3\eta|\omega|}}{(1+e^{-2\eta|\omega|})}\nonumber\\
    &+ \frac{e^{-(\eta-2b)|\omega|}+e^{-(\eta+2b)|\omega|}+e^{-(\eta+2d)|\omega|}-e^{-(2d-\eta)|\omega|}}{4\pi(1+e^{-2\eta|\omega|})}-\frac{e^{-(2d+\eta)|\omega|}+e^{-(3\eta-2d)|\omega|}}{4\pi(1+e^{-2\eta|\omega|})}
\end{align}

The total number of roots is
\begin{equation}
  M_{\ket{0}_{bs,d}}=  2+\int_{-\pi}^\pi\rho_{\ket{0}_{bs,d}}(\lambda)=1+2\pi\tilde\rho_{\ket{0}_{bs,d}}(0)=\frac{N_b+2}{2}
\end{equation}
Thus, the spin of this state is
\begin{equation}
     S_{\ket{0}_{bs,d}}=\frac{N_b+2}{2}- M_{\ket{0}_{bs,d}}=0
\end{equation}

We have already shown that the energy of the string solution $\lambda_{bs}$ is zero, and the energy of the string solution $\lambda_d$ is
\begin{equation}
    E_{d}=\frac{J \sinh ^2(\eta )}{\sinh (d) \sinh (d-\eta )}+J \sinh (\eta )\sum_{\omega=-\infty}^\infty e^{-2 \eta  | \omega | } \text{sech}(\eta   \omega  ) \cosh ( \omega   (2 d-\eta ))
\end{equation}
which is positive. Hence, the total energy of this state is
\begin{equation}
    E_{\ket{0}_{bs,d}}=E_d+E_{ar_{\ket{1}}}+\left[(N_b-1) J+\frac{J \sinh ^2 \eta}{\left(\sinh ^2 \eta-\sinh ^2 b\right)}-\frac{J \sinh ^2 \eta}{\left(\sinh ^2 d-\sinh ^2 \eta\right)}\right] \cosh \eta
\end{equation}
This is an excited state where the left impurity is screened by multiparticle screening (Kondo physics), and the right impurity is screened by a single particle bound mode. \\
Because of the $\mathbb{Z}_2$ symmetry there is a state $\ket{-1}_{bs,d}$ which is degenerate to state $\ket{1}_{bs,d}$.

We could remove the $\lambda_{bs}$ solution from the above state and add a hole propagating with rapidity $\theta$ such that the root density satisfies an integral equation of the form

\begin{align}
    &(2N_b+1) a(\lambda ,\eta )-2 a\left(\lambda -\pi,\eta \right)+ a(\lambda,\eta+2b )+ a(\lambda ,\eta-2b )+ a(\lambda ,\eta+2d )- a(\lambda,2d-\eta )+a(\lambda-\pi,\eta)\nonumber\\
    &=2\pi \rho(\lambda)+\int \rho(\lambda) \left[a(\lambda-\lambda',2\eta)+a(\lambda+\lambda',2\eta) \right]\mathrm{d}\lambda'+2\pi\delta(\lambda)+2\pi\delta(\lambda-\pi)+a(\lambda,2d+\eta)+a(\lambda,3\eta-2d)\nonumber\\
    &\quad\quad\quad+2\pi\delta(\lambda-\theta)+2\pi\delta(\lambda+\theta)
\end{align}

Solving for the root density in the Fourier space, we obtain
\begin{align}
    \tilde\rho_{\ket{1}_{\theta,d}}(\omega)&=\frac{(2N_b+1)e^{-\eta|\omega|}+(-1)^\omega e^{-\eta|\omega|}-2(-1)^\omega e^{-\eta|\omega|}-(1+(-1)^\omega)}{4\pi(1+e^{-2\eta|\omega|})}-\frac{2\cos(\omega\theta)}{4\pi(1+e^{-2\eta|\omega|})}\nonumber\\
    &+ \frac{e^{-(\eta-2b)|\omega|}+e^{-(\eta+2b)|\omega|}+e^{-(\eta+2d)|\omega|}-e^{-(2d-\eta)|\omega|}}{4\pi(1+e^{-2\eta|\omega|})}-\frac{e^{-(2d+\eta)|\omega|}+e^{-(3\eta-2d)|\omega|}}{4\pi(1+e^{-2\eta|\omega|})}
\end{align}

The total number of roots is
\begin{equation}
  M_{\ket{1}_{\theta,d}}=  1+\int_{-\pi}^\pi\rho_{\ket{1}_{\theta,d}}(\lambda)=1+2\pi\tilde\rho_{\ket{1}_{\theta,d}}(0)=\frac{N_b}{2}
\end{equation}
Thus, the spin of this state is
\begin{equation}
     S_{\ket{1}_{\theta,d}}=\frac{N_b+2}{2}- M_{\ket{1}_{\theta,d}}=1
\end{equation}
and the energy of the state is
\begin{equation}
    E_{\ket{1}_{\theta,d}}=E_d+E_{ar_{\ket{1}}}+\left[(N_b-1) J+\frac{J \sinh ^2 \eta}{\left(\sinh ^2 \eta-\sinh ^2 b\right)}-\frac{J \sinh ^2 \eta}{\left(\sinh ^2 d-\sinh ^2 \eta\right)}\right] \cosh \eta+E_\theta
\end{equation}

There exists another state $\ket{-1}_{\theta,d}$ degenerate to the above state.\\

subsection{Kondo-Unscreened}
Consider the case when $0<b<\frac{\eta}{2}$ is , $d>\frac{3\eta}{2}$ and $N_b$ is even.
In this regime, the Bethe equation takes the form
\begin{align}
   & \left(\frac{\sin \frac{1}{2}\left(\lambda _j-{i \eta }\right)}{\sin \frac12\left(\lambda _j+{i \eta }\right)}\right)^{2 N_b}
\frac{\cos^2\frac12\left(\lambda _j+{i \eta }\right)}{\cos^2\frac12\left(\lambda _j-{i \eta }\right)}\frac{\sin\frac12\left(\lambda_j -i(\eta-2b )\right)}{\sin\frac12\left(\lambda_j +i(\eta-2b )\right)}\frac{\sin\frac12\left(\lambda_j -i(2b+ \eta )\right)}{\sin\frac12\left(\lambda_j +i(2b+ \eta )\right)} \nonumber\\
&\frac{\sin\frac12\left(\lambda_j -i(2 d+ \eta )\right)}{\sin\frac12\left(\lambda_j +i(2 d+ \eta )\right)}\frac{\sin\frac12\left(\lambda_j +i(2d- \eta)\right)}{\sin\frac12\left(\lambda_j -i( 2d- \eta)\right)}=\prod_{k=1(\neq j)}^M \frac{\sin\frac12 \left(\lambda_j-\lambda_k-2i\eta\right) \sin \frac12\left(\lambda_j+\lambda_k-2i\eta\right)}{\sin\frac12 \left(\lambda_j-\lambda_k+2i\eta\right) \sin \frac12\left(\lambda_j+\lambda_k+2i\eta\right)}
\end{align}

Taking $\ln$ on both sides, we write
\begin{align}
   (2N_b+1) \phi\left(\lambda_j,\eta\right)&-2\psi\left(\lambda_j,\eta\right)+\phi\left(\lambda_j,{\eta+2b}\right)+\phi\left(\lambda_j ,\eta-2b\right)+\phi\left(\lambda_j ,\eta+2d\right)-\phi\left(\lambda_j ,2d-\eta\right)\nonumber\\
   &+\psi(\lambda_j,\eta)=\pi i I_j +\sum_{k}\left[\phi(\lambda_j+\lambda_k,2\eta)+\phi(\lambda_j-\lambda_k,2\eta) \right]
\end{align}

Differentiating and removing the solutions $\lambda=0$ and $\lambda=\pi$, we obtain
\begin{align}
    (2N_b+1) &a(\lambda ,\eta )-2 a\left(\lambda -\pi,\eta \right)+ a(\lambda ,\eta+2b )+ a(\lambda ,\eta -2b)+ a(\lambda ,\eta+2d )- a(\lambda,2d-\eta )+a(\lambda-\pi,\eta)\nonumber\\
    &=2\pi \rho(\lambda)+\int \rho(\lambda) \left[a(\lambda-\lambda',2\eta)+a(\lambda+\lambda',2\eta) \right]\mathrm{d}\lambda'+2\pi\delta(\lambda)+2\pi\delta(\lambda-\pi)
\end{align}
Solving the above integral equation, we write the solution density in Fourier space
\begin{align}
    \tilde\rho_{\ket{1}}(\omega)&=\frac{(2N_b+1)e^{-\eta|\omega|}+(-1)^\omega e^{-\eta|\omega|}-2(-1)^\omega e^{-\eta|\omega|}-(1+(-1)^\omega)}{4\pi(1+e^{-2\eta|\omega|})}\nonumber\\
    &+ \frac{e^{-(\eta-2b)|\omega|}+e^{-(\eta+2b)\eta|\omega|}+e^{-(\eta+2d)|\omega|}-e^{-(2d-\eta)|\omega|}}{4\pi(1+e^{-2\eta|\omega|})}
\end{align}
The total number of Bethe roots is given by
\begin{equation}
    M_{\ket{1}}=\int_{-\pi}^\pi \rho_{\ket{1}}(\lambda)\mathrm{d}\lambda=2\pi \tilde \rho_{\ket{1}}(0)=\frac{N_b}{2}
\end{equation}

Thus, the total spin of the state is
\begin{equation}
    S^z_{\ket{1}}=\frac{N_b+2}{2}-M_{\ket 1}=1
\end{equation}
The energy of this state is
\begin{equation}
    E_{\ket1}=E_{ar_{\ket{1}}}+\left[(N_b-1) J+\frac{J \sinh ^2 \eta}{\left(\sinh ^2 \eta-\sinh ^2 b\right)}-\frac{J \sinh ^2 \eta}{\left(\sinh ^2 d-\sinh ^2 \eta\right)}\right] \cosh \eta
\end{equation}

where the energy of all real roots is given by
\begin{equation}
    E_{ar_{\ket{1}}}=-4\pi J \sinh \eta \sum_{\omega=-\infty}^\infty \tilde \rho_{\ket{1}}(\omega)\tilde a(\omega,\eta) 
\end{equation}

Because of the $\mathbb{Z}_2$ symmetry, there is a state $\ket{-1}$ which is degenerate to state $\ket 1$.

We can add the purely imaginary solution $\lambda_d=\pm i(\eta-2d)$ on top of all real root solution such that the Bethe equation becomes
\begin{align}
   &(2N_b+1) \phi\left(\lambda_j,\eta\right)-2\psi\left(\lambda_j,\eta\right)+\phi\left(\lambda_j,{\eta+2b}\right)+\phi\left(\lambda_j ,\eta-2b\right)+\phi\left(\lambda_j ,\eta+2d\right)-\phi\left(\lambda_j ,2d-\eta\right)\nonumber\\
   &+\psi(\lambda_j,\eta)=\pi i I_j +\sum_{k}\left[\phi(\lambda_j+\lambda_k,2\eta)+\phi(\lambda_j-\lambda_k,2\eta) \right]+\phi(\lambda_j,2d+\eta)-\phi(\lambda_j,2d-3\eta)
\end{align}

Differentiating and removing the solutions $\lambda=0$ and $\lambda=\pi$, we obtain
\begin{align}
    &(2N_b+1) a(\lambda ,\eta )-2 a\left(\lambda -\pi,\eta \right)+ a(\lambda ,\eta+2b )+ a(\lambda ,\eta -2b)+ a(\lambda ,\eta+2d )- a(\lambda,2d-\eta )+a(\lambda-\pi,\eta)\nonumber\\
    &=2\pi \rho(\lambda)+\int \rho(\lambda) \left[a(\lambda-\lambda',2\eta)+a(\lambda+\lambda',2\eta) \right]\mathrm{d}\lambda'+2\pi\delta(\lambda)+2\pi\delta(\lambda-\pi)+a(\lambda,2d+\eta)-a(\lambda,2d-3\eta)
\end{align}

Solving the above integral equation, we write the solution density in Fourier space

\begin{align}
    \tilde\rho_{\ket{0}_{d}}(\omega)&=\frac{(2N_b+1)e^{-\eta|\omega|}+(-1)^\omega e^{-\eta|\omega|}-2(-1)^\omega e^{-\eta|\omega|}-(1+(-1)^\omega)}{4\pi(1+e^{-2\eta|\omega|})}\nonumber\\
    &+ \frac{e^{-(\eta-2b)|\omega|}+e^{-(\eta+2b)|\omega|}+e^{-(\eta+2d)|\omega|}-e^{-(2d-\eta)|\omega|}}{4\pi(1+e^{-2\eta|\omega|})}-\frac{e^{-(2d+\eta)|\omega|}-e^{-(2d-3\eta)|\omega|}}{4\pi(1+e^{-2\eta|\omega|})}
\end{align}

The total number of Bethe roots is
\begin{equation}
    M_{\ket{0}_{d}}=1+\int_{-\pi}^\pi \rho_{\ket{0}_{d}}(\lambda)\mathrm{d}\lambda=1+2\pi \tilde\rho_{\ket{0}_{d}}(0)=\frac{N_b+2}{2}
\end{equation}

Thus, the spin of this state is
\begin{equation}
    S_{\ket{0}_{d}}=\frac{N_b+2}{2}-M_{\ket{0}_{d}}=0
\end{equation}

Now, we need to compute the energy of the wide boundary string. 

The bare energy of the string solution is
\begin{equation}
    E_{bare,d}=J \frac{\sinh^2\eta}{\sinh d \sinh(d-\eta)}
\end{equation}
and the energy associated with the change in density due to the string solution
\begin{equation}
    \Delta\tilde\rho_{d}=-\frac{e^{-(2d+\eta)|\omega|}-e^{-(2d-3\eta)|\omega|}}{4\pi(1+e^{-2\eta|\omega|})}
\end{equation}
is

\begin{equation}
    E_{\Delta\tilde\rho_{d}}=-4\pi J \sinh\eta \sum_{\omega=-\infty}^\infty \Delta\tilde\rho(\omega) \tilde a(\omega,\eta)= -\frac{J \sinh ^2(\eta )}{\sinh (d) \sinh (d-\eta )}
\end{equation}

Thus, the total energy of the wide boundary string is
\begin{equation}
    E_{d}=E_{bare,d}+ E_{\Delta\tilde\rho_{d}}=0
\end{equation}

Thus, the energy of this state is
The energy of this state is
\begin{equation}
    E_{\ket{0}_d}=E_{ar_{\ket{1}}}+\left[(N_b-1) J+\frac{J \sinh ^2 \eta}{\left(\sinh ^2 \eta+\sin ^2 \beta\right)}-\frac{J \sinh ^2 \eta}{\left(\sinh ^2 d-\sinh ^2 \eta\right)}\right] \cosh \eta
\end{equation}

Thus, in this phase, the ground state involves multiparticle screening of the left impurity, and the right impurity is unscreened. There are unique boundary excitations in this phase which involves the screening of the right impurity by an exponentially localized bound mode formed at the right end of the chain. Thus, there are two distinct towers of excited states: one where both impurities are screened (one by multiparticle Kondo effect and another by exponentially localized bound mode) whereas another tower includes the excited states where one impurity is screened by multiparticle Kondo effect but the other impurity is unscreened.

\subsection{Kondo-US and US-Kondo phases}
Let us consider the case when $0<b<\frac{\eta}{2}$, $d>\frac{3\eta}{2}$ and $N_b$ is even. The case where $0<d<\frac{\eta}{2}$, $b>\frac{3\eta}{2}$ can be obtained  by applying the transformation $L\leftrightarrow R$.

In this regime, the impurity boundary string solution $\lambda_d$ has zero energy, and hence the addition of such a solution can no longer screen the impurity. Thus, the ground state in these phases is similar to the Kondo-FBM and FBM-Kondo phases. The only difference is the structure in the excited state, as the impurity can no longer be screened in this phase at any energy scale.

\subsection{Other phases}
\subsubsection{ABM-ABM phase}
When both impurity parameters take values between $\eta/2$ and $\eta$, the model is in the ABM-ABM phase. 

The ground state is constructed by adding the boundary strings solution $\lambda_d$ and $\lambda_b$ on top of all the real roots of Bethe equations and the complex boundary solution $\lambda_{bs}$. Both impurities are screened by the exponentially localized bound mode formed at the edges of the chain. 

Since the impurity boundary string solutions can be removed, boundary excitations where impurities are unscreened at the edges can be constructed at the expense of the energy of the bound mode, which is always higher than the maximum possible energy of a single kink excitation $M_g$. 

The impurity string solutions could be removed just from the left or right, thereby unscreening only one of the impurities, or simultaneously from left and right such that both impurities are unscreened. All other excited states can be formed by adding an even number of spinons, bulk strings, quartets, etc, to each of these four kinds of states. Each of the four types of states is manyfold degenerate. For example, the state where both impurities are screened is two-fold degenerate (both $S^z=0$ states when $N_b$ is even and $S^z=\pm \frac{1}{2}$ when $N_b$ is odd), the state where only the left or right impurity is screened is each four-fold degenerate (two $S^z=\frac{1}{2}$ states and two $S^z=-\frac{1}{2}$ state when $N_b$ is odd and two $S^z=0$, one $S^z=1$, and one $S^z=-1$ state when $N_b$ is even), and finally, the state where both impurities are unscreened is eight-fold degenerate (three $S^z=\frac{1}{2}$ states, three $S^z=-\frac{1}{2}$ states, one $S^z=\frac{3}{2}$ state and one $S^z=-\frac{3}{2}$ states when $N_b$ is odd and four $S^z=0$ states, two $S^z=-1$ states, and two $S^z=1$ states when $N_b$ is even). Adding an even number of spinons, bulk strings, quarters, etc to each of these 18 distinct states, all the excitations in this phase can be sorted into four unique towers containing excited states with the following properties: one tower where both impurities are screened, one where the left impurity is screened and the right is unscreened, one where the left impurity is unscreened and the right impurity is screened, and finally, one where both left and right impurities are unscreened. 

\subsubsection{FBM-FBM phase}

When both impurity parameters take values between $\eta$ and $3\eta/2$, the model is in the FBM-FBM phase. 

The energy of the impurity boundary strings solutions $\lambda_b$ and $\lambda_d$ are positive in this region. Thus, these solutions are not included in the ground state root distribution.

The ground state is constructed by considering all real roots of Bethe equations and the complex boundary solution $\lambda_{bs}$. Both impurities are unscreened in the ground state, thereby making the degeneracy of the ground state eight-fold.

Since the impurity boundary string solutions can be added, boundary excitations where impurities are screened at the edges can be constructed at the expense of the energy of the bound mode, which is always higher than the maximum possible energy of a single kink excitation $M_g$. 

The impurity string solutions could be added just from the left or right, thereby screening only one of the impurities or simultaneously from left and right such that both impurities are screened. All other excited states can be formed by adding an even number of spinons, bulk strings, quartets, etc to each of these four kinds of (manyfold degenerate) states where either both impurities are screened or unscreened or only one of the impurities is screened thereby creating four distinct towers of excited states. Thus, the four towers have excited states with the following properties: one tower where both impurities are unscreened, one where the left impurity is screened and the right is unscreened, one where the left impurity is unscreened and the right impurity is screened, and finally one where both left and right impurities are screened.

\subsubsection{ABM-FBM and FBM-ABM phases}
When one of the impurity parameters take values between $\eta$ and $3\eta/2$, and another impurity parameters take values between $\eta/2$ and $\eta$ the model is in the ABM-FBM phase. Consider the case where $\eta/2<b<\eta$ and $\eta<d<3/2\eta$ such that the left impurity is in the ABM phase and the right one is in the FBM phase. The FBM-ABM case can be constructed by applying $L\leftrightarrow R$ transformation. 

In this case, the impurity string solution $\lambda_b$ has negative energy but the string solution $\lambda_d$ has positive energy. Thus, the ground state contains only the $\lambda_b$ root. In the ground state, the left impurity is screened by an exponentially localized bound mode formed at the left edge of the spin chain, whereas the right impurity is unscreened, thereby making it four-fold degenerate. 

In this regime, the spin chain can have boundary excitations on both ends. The boundary excitation in the left end involves removing the impurity string solution, thereby unscreening the left impurity, whereas the boundary excitation in the right end involves adding the impurity string solution such that the impurity is screened.

As before, there are a total of four towers that have excited states with the following properties: one tower where both impurities are unscreened, one where the left impurity is screened and the right is unscreened, one where the left impurity is unscreened and the right impurity is screened, and finally one where both left and right impurities are screened.

\subsubsection{ABM-US and US-ABM phases}
When one of the impurity parameters take values greater that $3\eta/2$, and another impurity parameters take the values between $\eta/2$ and $\eta$ the model is in the US-ABM phase. Consider the case where $\eta/2<b<\eta$ and $d>3/2\eta$ such that the left impurity is in the ABM phase and the right one is in the US phase. The US-ABM case can be constructed by applying $L\leftrightarrow R$ transformation. 

In this case, the impurity string solution $\lambda_b$ has negative energy but the string solution $\lambda_d$ has zero energy. In the ground state, the left impurity is screened by an exponentially localized bound mode formed at the left edge of the spin chain, whereas the right impurity is unscreened, thereby making it four-fold degenerate. 

In this regime, the spin chain can have boundary excitations only at the left end where impurity is screened in the ground state by an exponentially localized bound mode. The boundary excitation involves removing the impurity string solution, thereby unscreening the left impurity.

The two towers have excited states that have the following properties: one tower where both impurities are unscreened, one where the left impurity is screened and the right is unscreened.

\subsubsection{FBM-US and US-FBM phases}
When one of the impurity parameters take values greater that $3\eta/2$, and another impurity parameters take the values between $\eta$ and $3\eta/2$, the model is in the US-FBM phase. Consider the case where $\eta<b<3/2\eta$ and $d>3/2\eta$ such that the left impurity is in the FBM phase and the right one is in the US phase. The US-FBM case can be constructed by applying $L\leftrightarrow R$ transformation. 

In this case, the impurity string solution $\lambda_b$ has positive energy but the string solution $\lambda_d$ has zero energy. In the ground state, both impurities are unscreened thereby making it eight-fold degenerate. 

In this regime, the spin chain can have boundary excitations only at the left end where the impurity is. The boundary excitation involves adding the impurity string solution $\lambda_b$, thereby screening the left impurity.

The two towers of excitations in this phase have excited states with the following properties: one tower where both impurities are unscreened, one where the left impurity is screened and the right is unscreened.

\subsubsection{US-US phase}
When both impurity parameters take real values greater than $3\eta/2$, the model is in the US-US phase. 

The energy of the impurity boundary strings solutions $\lambda_b$ and $\lambda_d$ vanish in this regime. Thus, the impurities can not be screened in this phase.

Thus the ground state is an eight-fold state where both impurities are unscreened. There are no boundary excitations in this phase, and hence, the excitations form a single tower of excited states, which can be constructed by an even number of spinons, bulk string solutions, quartets, etc, to each of the eight-fold degenerate vacua.

\section{Impurity at the edge of Haldane chain and spin-1 XXZ chain.}\label{diff-haldane-xxz}
In the main text in section \ref{imp-effects-1}, we briefly discussed the impurity effect in the gapped phase of spin-$S$ XXZ chain and mentioned that it is different from the impurity physics in the topological phase of the bilinear-biquadratic chain. Here, we shall discuss those differences in more detail. 
Let us consider the bilinear-biquadratic spin-1 chain 
\begin{equation}
    H(\theta)=\sum_{j=1}^{N_b}\cos(\theta)\vec S_j\cdot\vec S_{j+1}+\sin\theta(\vec S_j\cdot\vec S_{j+1})^2,
\end{equation}
in the Haldane phase \textit{i.e.} between the two integrable points $\theta=-\frac{\pi}{4}$, the Takhtajan-Babujian model and $\theta=\frac{\pi}{4}$, the $SU(3)$ symmetric Sutherland-Lai model. In the Haldane phase $\left(-\frac{\pi}{4}<\theta<\frac{\pi}{4}\right)$, the model is gapped, has exponentially decaying correlation function, and a nonvanishing non-local string order parameter. Furthermore, the entire entanglement spectrum has exact double (or higher even) degeneracy in the Haldane phase. For example, at the AKLT point ($\theta=\arctan\frac{1}{3}$, the entanglement spectrum features only two non-zero degenerate values.

Moreover, for finite open boundary conditions, the Haldane phase has a fourfold degenerate ground state in the thermodynamic limit and features fractionalized spin-$\frac{1}{2}$ edge excitations. The two effectively free edge spins could form singlet or triplet, giving rise to the degenerate four-fold ground state as shown in Fig.\ref{haldane-fig-1} at the Haldane point $\theta=0$.

\begin{figure}[H]
    \centering
    \begin{subfigure}[b]{0.23\textwidth}
        \includegraphics[width=\textwidth]{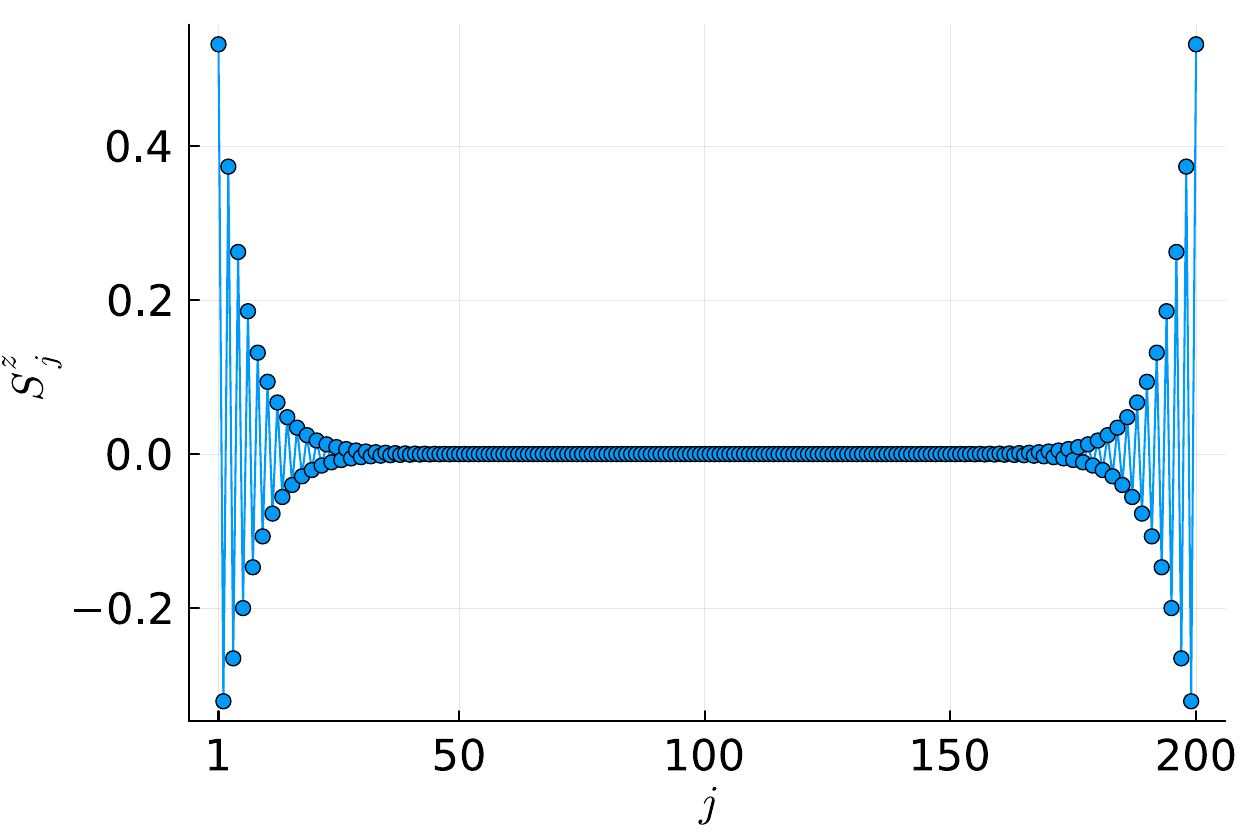}
    \end{subfigure}
    \begin{subfigure}[b]{0.23\textwidth}
        \includegraphics[width=\textwidth]{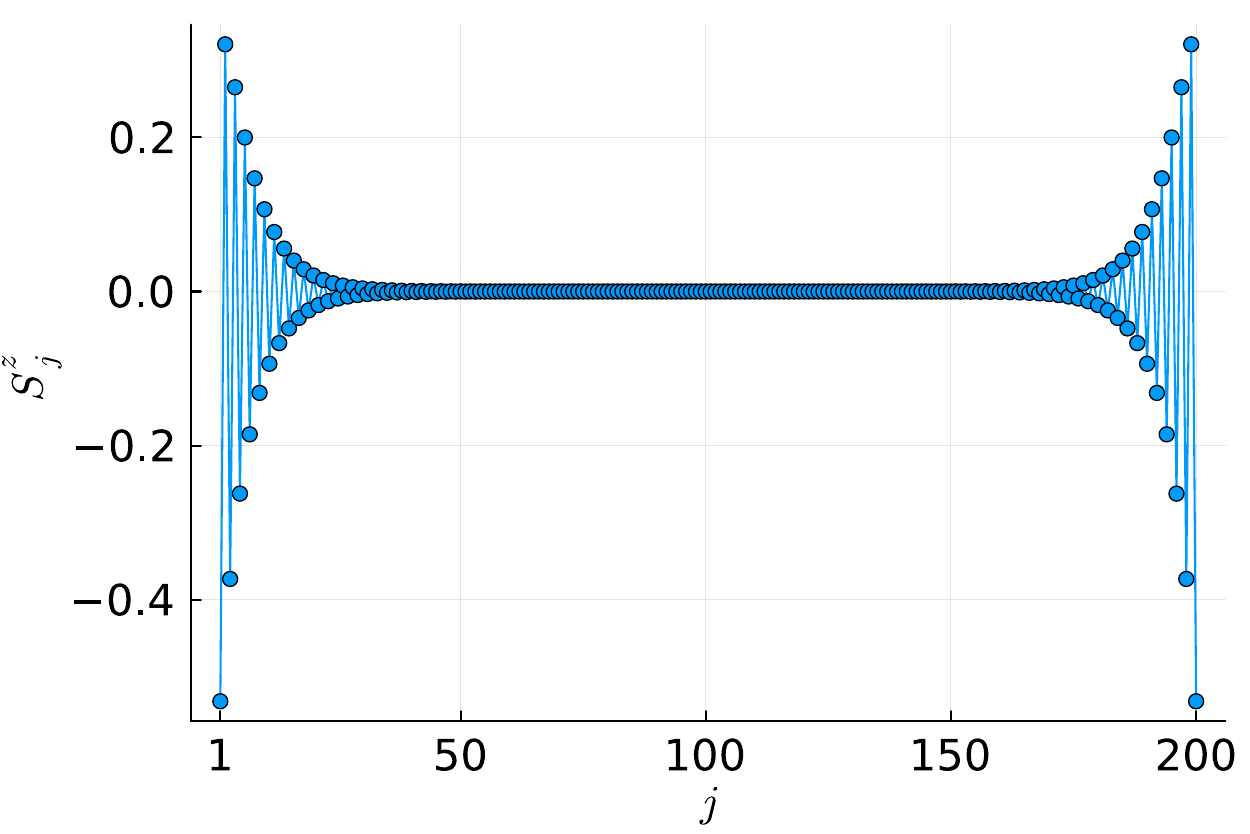}
    \end{subfigure}
    \begin{subfigure}[b]{0.23\textwidth}
        \includegraphics[width=\textwidth]{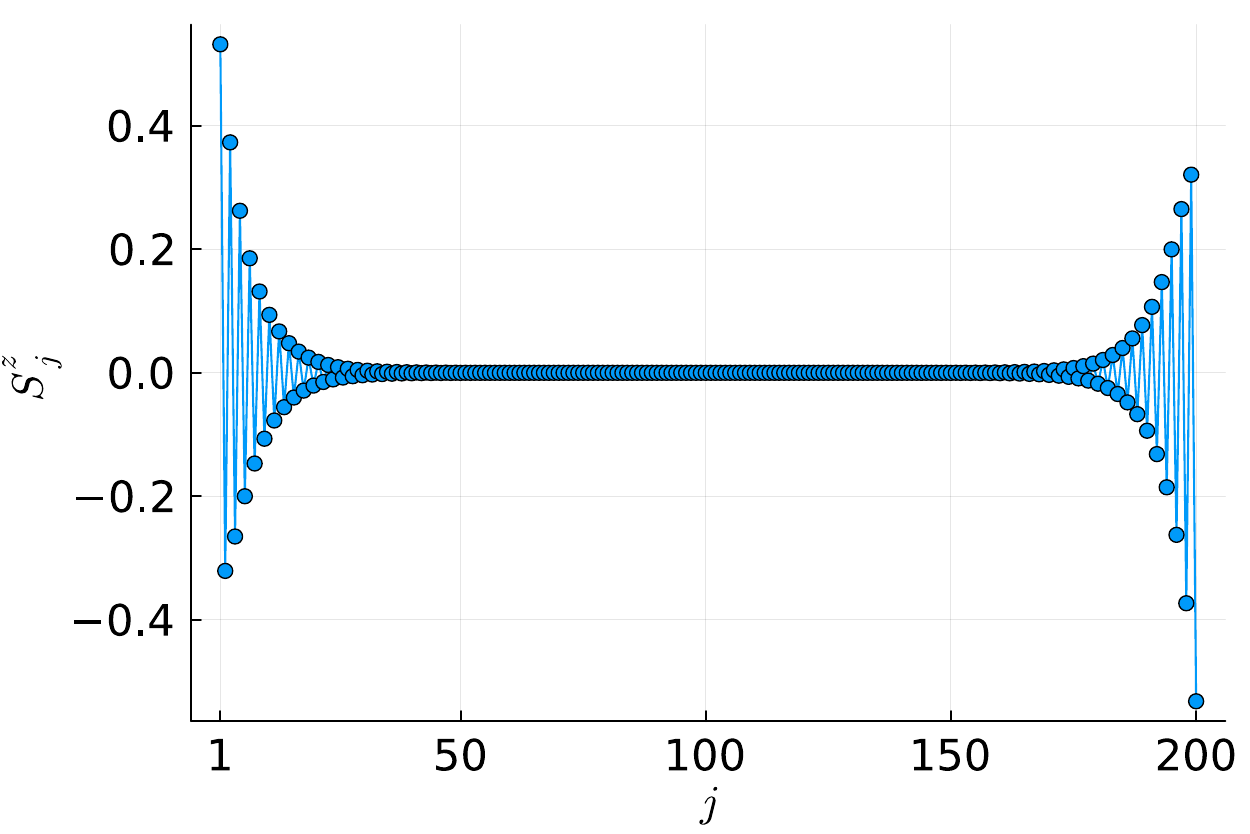}
    \end{subfigure}
    \begin{subfigure}[b]{0.23\textwidth}
        \includegraphics[width=\textwidth]{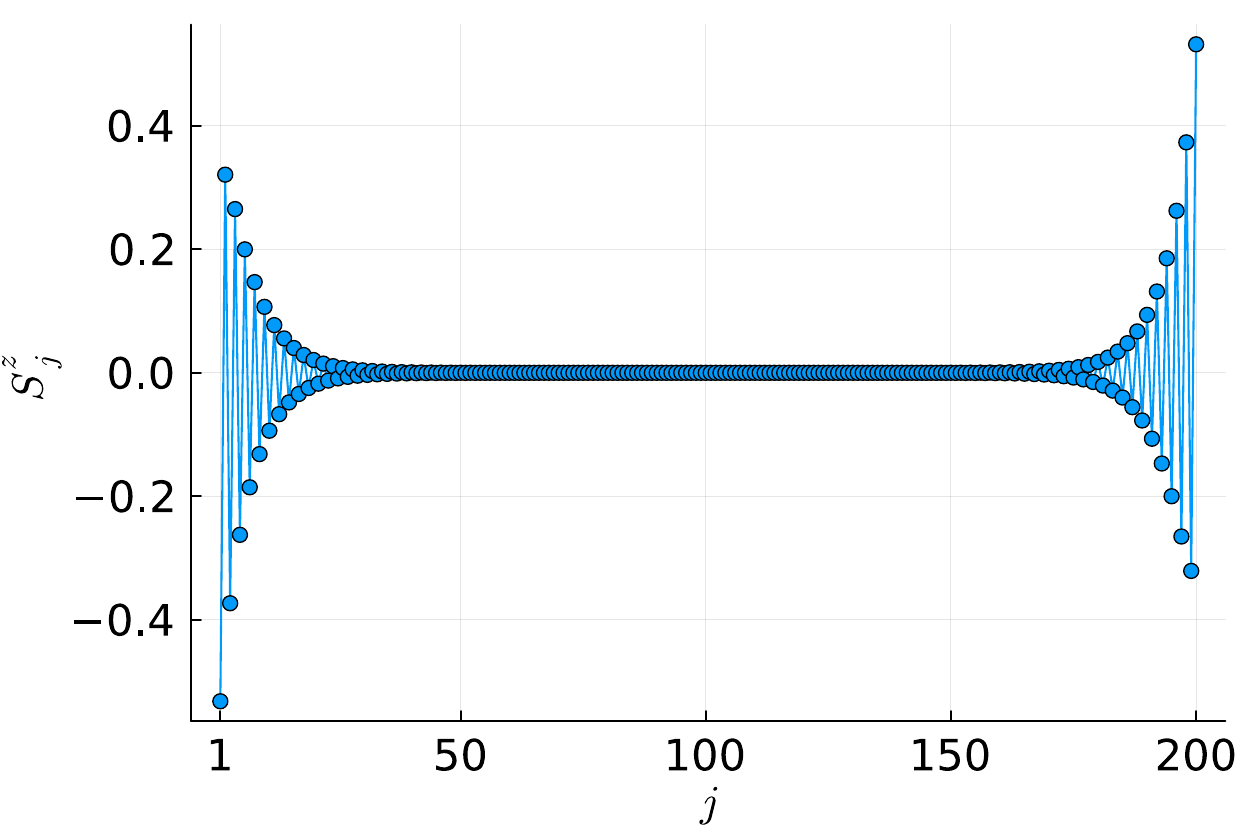}
    \end{subfigure}
    \begin{tikzpicture}[overlay, remember picture]
    \node at (-16.75,2.7) {\tiny{a)}};
    \node at (-12.5,2.7) {\tiny{b)}};
    \node at (-8.25,2.7) {\tiny{c)}};
    \node at (-4,2.7) {\tiny{d)}};
    \end{tikzpicture}
    \caption{The ground state of the Haldane chain with open boundary conditions is four-fold degenerate, with each of the two fractionalized $\frac{1}{2}$ edge modes acting as free spin-$\frac{1}{2}$ particles that can independently point up or down. In a), the two spin-$\frac{1}{2}$ edge modes both point up, resulting in a total spin of $S^z=1$ in the ground state. In b), both edge modes point down, giving a total spin of $S^z=-1$. The subfigures c) and d) depict the two ground states with $S^z=0$, where the edge modes at the two ends of the chain are anti-aligned: in c), the left edge mode points up while the right edge mode points down, and in d), the left edge mode points down while the right edge mode points up. The data shown is for the Haldane chain $H(0)$ with $200$ sites.}
    \label{haldane-fig-1}
\end{figure}

We shall now antiferromagnetically couple a single spin-$\frac{1}{2}$ spin to the left of the Haldane chain $H(0)$. This spin-$\frac{1}{2}$ impurity forms singlet with the effective spin-$\frac{1}{2}$ edge mode and hence reduces the degeneracy of the ground state to two-fold as shown in Fig.\ref{hald-imp-1}. 
\begin{figure}[H]
    \centering
    \begin{subfigure}[b]{0.46\textwidth}
        \includegraphics[width=\textwidth]{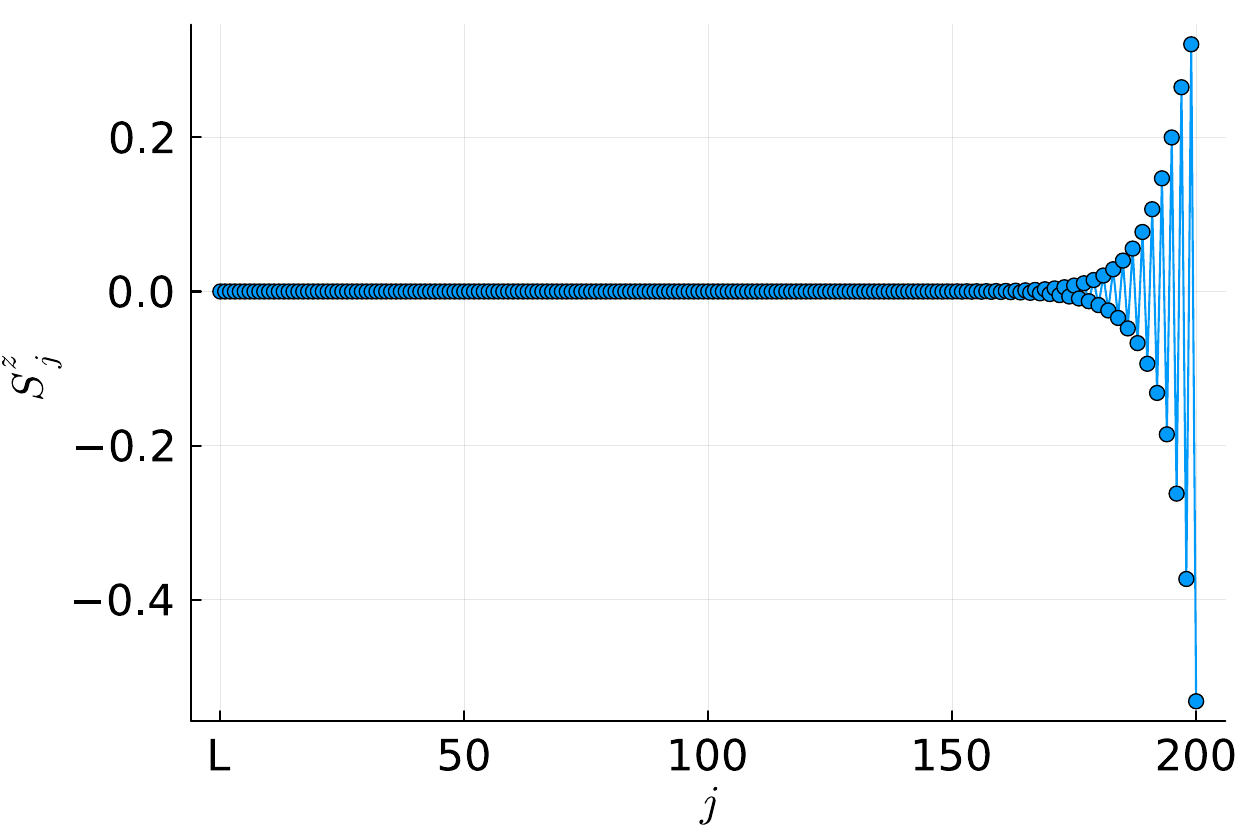}
    \end{subfigure}
    \begin{subfigure}[b]{0.46\textwidth}
        \includegraphics[width=\textwidth]{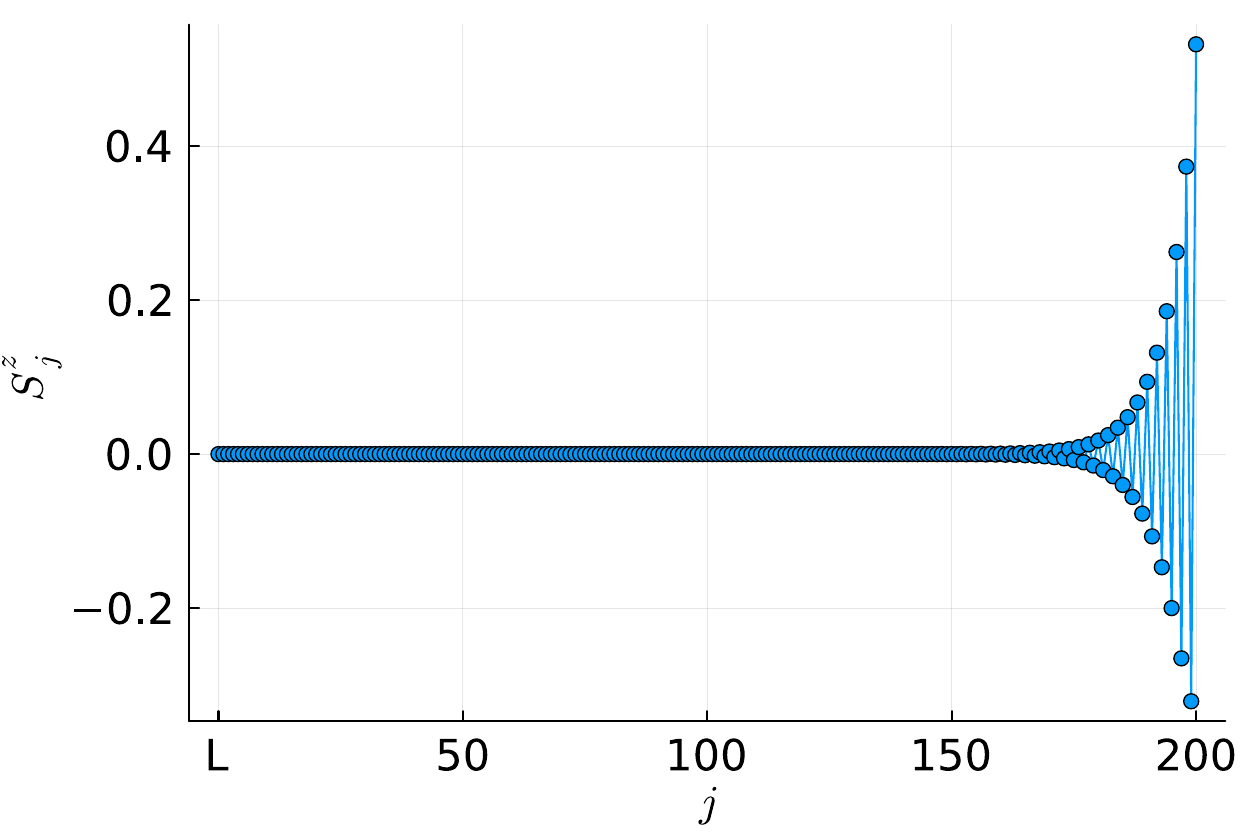}
    \end{subfigure}
    \caption{A single spin-$\frac{1}{2}$ impurity coupled to the Haldane with antiferromagnetic exchange coupling at its left edge forms a singlet with the edge mode at the left edge. The unbounded right edge mode can free point either up or down (as shown in the left and right panels, respectively), thereby giving rise to a two-fold degenerate ground state. Here, site $L$ is the impurity site left of site $1$ in the chain, and there are 200 sites in the bulk of the chain.}
    \label{hald-imp-1}
\end{figure}

Furthermore, adding two spin-$\frac{1}{2}$ impurities at the two edges of the Haldane chain $H(0)$ with antiferromagnetic exchange coupling gives rise to a unique ground state in the Haldane chain with open boundary conditions. The two impurities form singlets with the free spin-$\frac{1}{2}$ edge modes situated respectively at the edge they are attached to. Thus, the spin profile $S^z_j$ identically vanishes throughout the chain, as shown in Fig.\ref{hald-imp-2}.
\begin{figure}[H]
    \centering
    \includegraphics[width=0.5\linewidth]{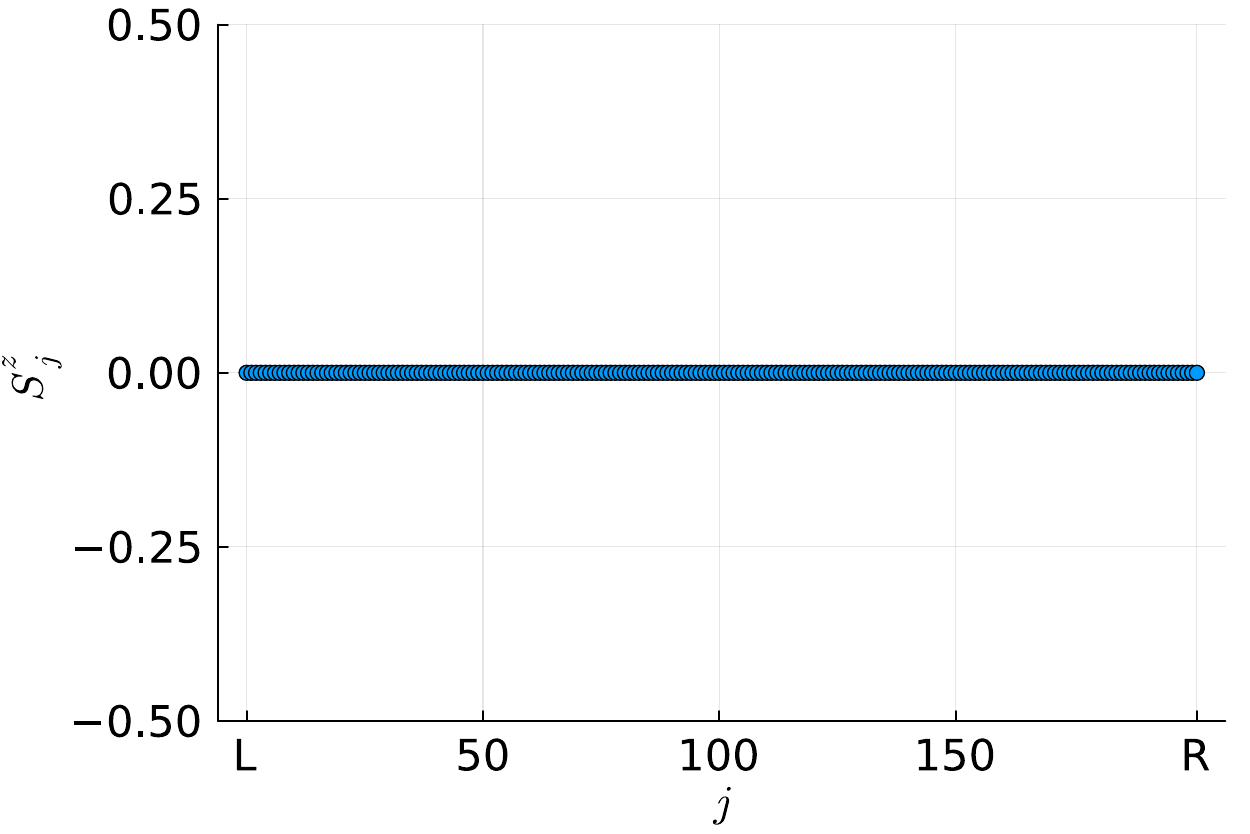}
    \caption{Unique ground state in Haldane chain with two impurities are the edges that are coupled antiferromagnetically to the chain. The impurities form singlets with the edge modes, and hence, the spin $S^z_j$ is vanishing in each site, including the sites near the boundary.}
    \label{hald-imp-2}
\end{figure}

Now we shall look at the edge modes in the spin-1 XXZ chain in the antiferromagnetic phase given by Hamiltonian
\begin{equation}
    H(\Delta)=\sum_{j=1}^{N_b-1}S^x_jS^x_{j+1}+S^y_jS^y_{j+1}+\Delta S^z_jS^z_{j+1},
\end{equation}
where we consider $\Delta \gtrapprox  1.185$ such that the model is in a gapped antiferromagnetic phase. As shown in \cite{kattel2024edge}, this model hosts fractionalized edge modes. It is important to note that this edge modes are not topological edge modes. We shall now discuss how both the bulk and boundary are different compared to the topological edge modes in the Haldane chain. Unlike in the Haldane chain, the half-cut entanglement spectra are not doubly degenerate in the spin-1 XXZ chain, as shown in Fig.\ref{ee-plots-xxz-hald}.
\begin{figure}[H]
    \centering
    \begin{subfigure}[b]{0.46\textwidth}
        \includegraphics[width=\textwidth]{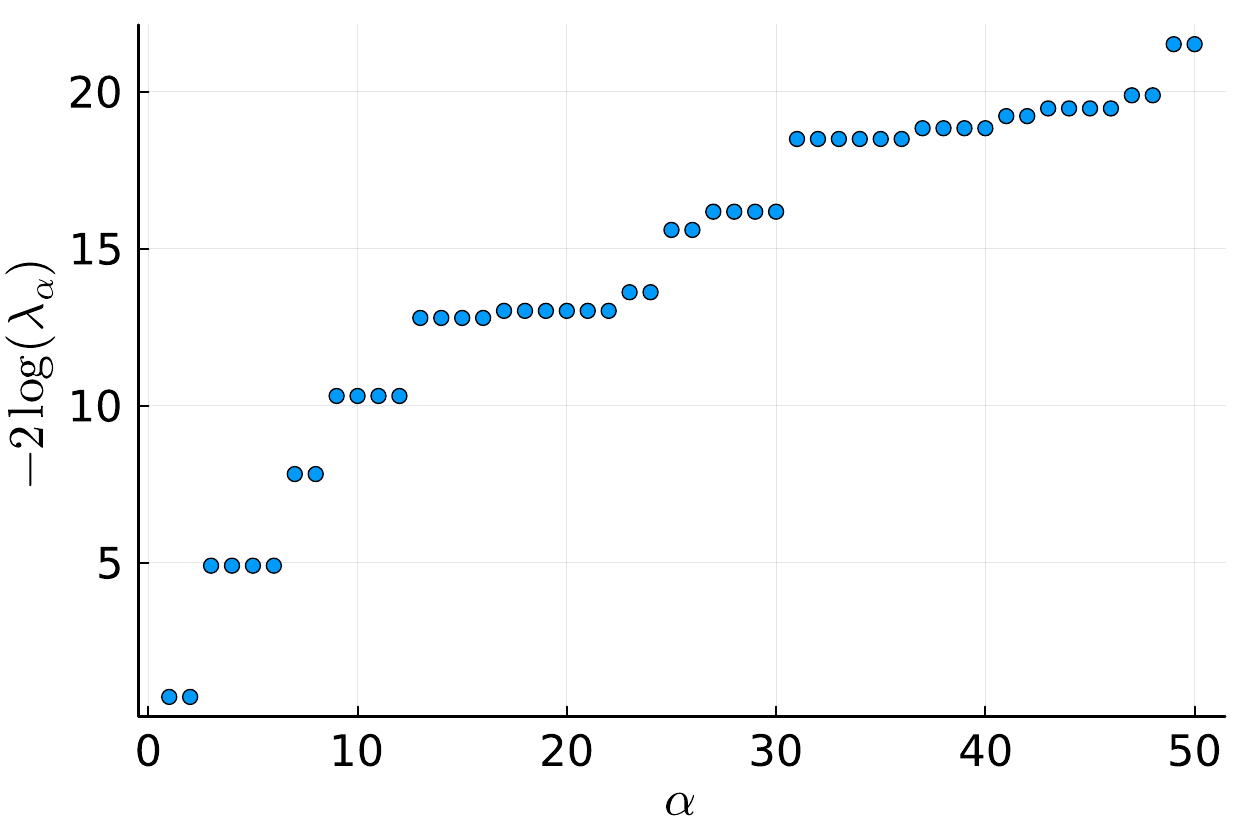}
    \end{subfigure}
    \begin{subfigure}[b]{0.46\textwidth}
        \includegraphics[width=\textwidth]{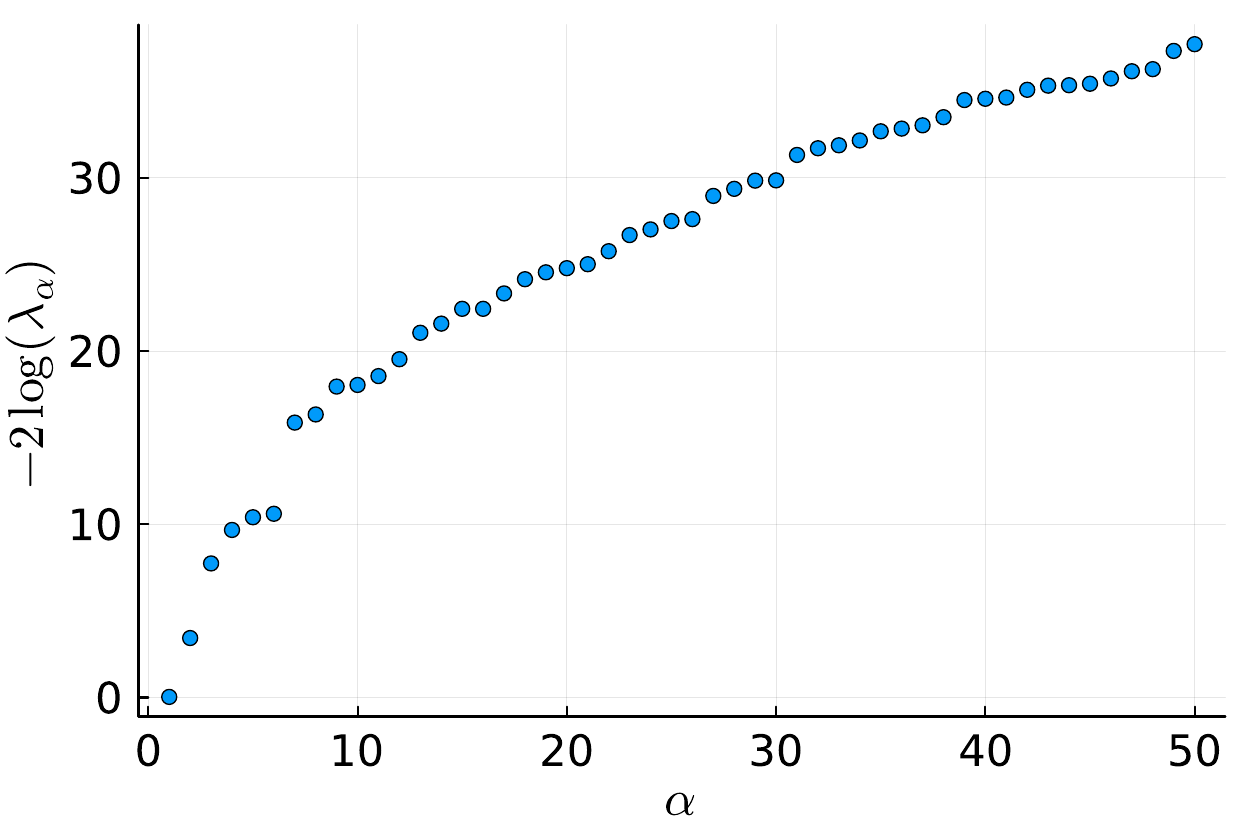}
    \end{subfigure}
    \begin{tikzpicture}[overlay, remember picture]
    \node at (-8,5.35) {b)};
    \node at (-16.5,5.35) {a)};
    \end{tikzpicture}
    \caption{The 50 low-lying entanglement spectra $EE=-2\ln(\lambda_\alpha)$ where $\lambda_\alpha^2$ are the eigenvalues of the reduced density matrix obtained by tracing out half of the system a) has even degeneracy throughout the spectra in the Haldane chain b) has no double degeneracy in the spectrum for the spin-1 XXZ chain with $\Delta=2$. These data are obtained using DMRG for 200 site chains.}
    \label{ee-plots-xxz-hald}
\end{figure}

Unlike in the topological Haldane phase, there is spontaneous symmetry breaking of the discrete $\mathbb{Z}_2$ spin flip symmetry, which leads to two-fold degenerate ground states. The two-fold degenerate ground state has a magnetic order. The spin profile is of the form
\begin{equation}
    S^z_j=(-1)^j\sigma+\Delta(S^z_L)(j)+\Delta(S^z_R)(j),
\end{equation}
where $\sigma$ is the staggered magnetization in bulk and $\Delta(S^z_L)(j)$ and $\Delta(S^z_R)(j), $ are the deviation at the left and right edges, respectively, as shown below.
\begin{figure}[H]
    \centering
    \begin{subfigure}[b]{0.46\textwidth}
        \includegraphics[width=\textwidth]{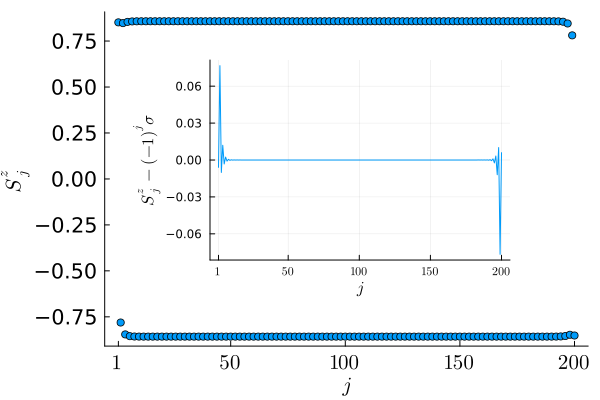}
    \end{subfigure}
    \begin{subfigure}[b]{0.46\textwidth}
        \includegraphics[width=\textwidth]{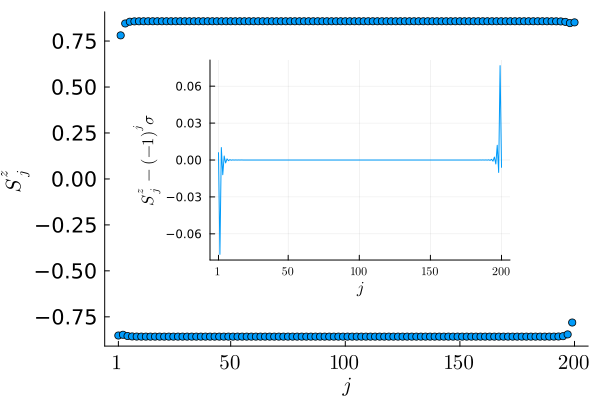}
    \end{subfigure}
    \begin{tikzpicture}[overlay, remember picture]
    \node at (-8,5.35) {b)};
    \node at (-16.5,5.35) {a)};
    \end{tikzpicture}
    \caption{The spin profile in the two-fold degenerate ground state of spin-1 XXZ chain, both of which have bulk staggered magnetization of $\sigma\approx 0.8576$. The insets show deviation $\Delta(S^z_L)$ and $\Delta(S^z_R)$ at the edges.}
    \label{XXZ-1-vacs}
\end{figure}

In the main text, we constructed the edge modes by using prescription $\hat S^z_L=\lim_{\alpha\to 0}\lim_{N\to \infty}\sum_{j=1}^N e^{-\alpha j}S^z_j$. The expectation value of this edge mode can also be computed by adding half of the bulk staggered magnetization to the sum of the edge deviation $\Delta(S^z_q)$ for $q=\{L,R\}$ in half of the chain. Defining
\begin{equation}
    \tilde \Delta S^z_L= \sum_{j=1}^{\frac{N}{2}}\Delta(S^z_L) \quad\quad \text{and}\quad\quad \tilde \Delta S^z_R= \sum_{j=\frac{N}{2}+q}^{N}\Delta(S^z_R),
\end{equation}
the eigenvalues of the edge mode operator can simply be obtained as
\begin{equation}
    S^z_q=  \tilde \Delta S^z_q+ \mathrm{sign}( \tilde \Delta S^z_q)\frac{\sigma}{2} \quad\quad\text{where} \quad \quad q=\{L,R\}.
\end{equation}

As shown in Fig.~\ref{XXZ-1-vacs}, the sum of the half chain edge deviations $|\tilde S^z_L|\approx 0.0712 \approx |\tilde S^z_R|$. Thus, we obtain that in one of the degenerate vacua shown in Fig.~\ref{XXZ-1-vacs} a), the left edge hosts spin-$\frac{1}{2}$ edge mode that point in up direction \textit{i.e.} $S^z_L=\frac{1}{2}$  and the right edge hosts spin-$\frac{1}{2}$ edge mode that point in down direction \textit{i.e.} $S^z_R=-\frac{1}{2}$. Likewise, in the other degenerate vacua shown in Fig.~\ref{XXZ-1-vacs} b), the left edge hosts spin-$\frac{1}{2}$ edge mode that point in down direction \textit{i.e.} $S^z_L=-\frac{1}{2}$  and the right edge hosts spin-$\frac{1}{2}$ edge mode that point in up direction \textit{i.e.} $S^z_R=\frac{1}{2}$.

We shall now attach a spin-$\frac{1}{2}$ impurity to the left edge of the spin-1 XXZ chain. The spin-$\frac{1}{2}$ impurity anti-aligns with the edge mode at the left edge of the spin-1 XXZ chain, and hence there is no longer edge mode at the edge of the chain. However, notice that unlike in the Haldane case, the degeneracy of the chain is not lifted by the presence of the impurity, and the degeneracy is due to the spontaneous breaking of the spin-flip $\mathbb{Z}_2$ symmetry, which remains spontaneously broken in the presence of the impurity. Thus, we now obtain a two-fold degenerate ground state with spin accumulations $S_L=0$ and $S_R=\frac{1}{2}$ as shown in Fig~\ref{XXZ-1-1imp-vacs}.
\begin{figure}[H]
    \centering
    \begin{subfigure}[b]{0.46\textwidth}
        \includegraphics[width=\textwidth]{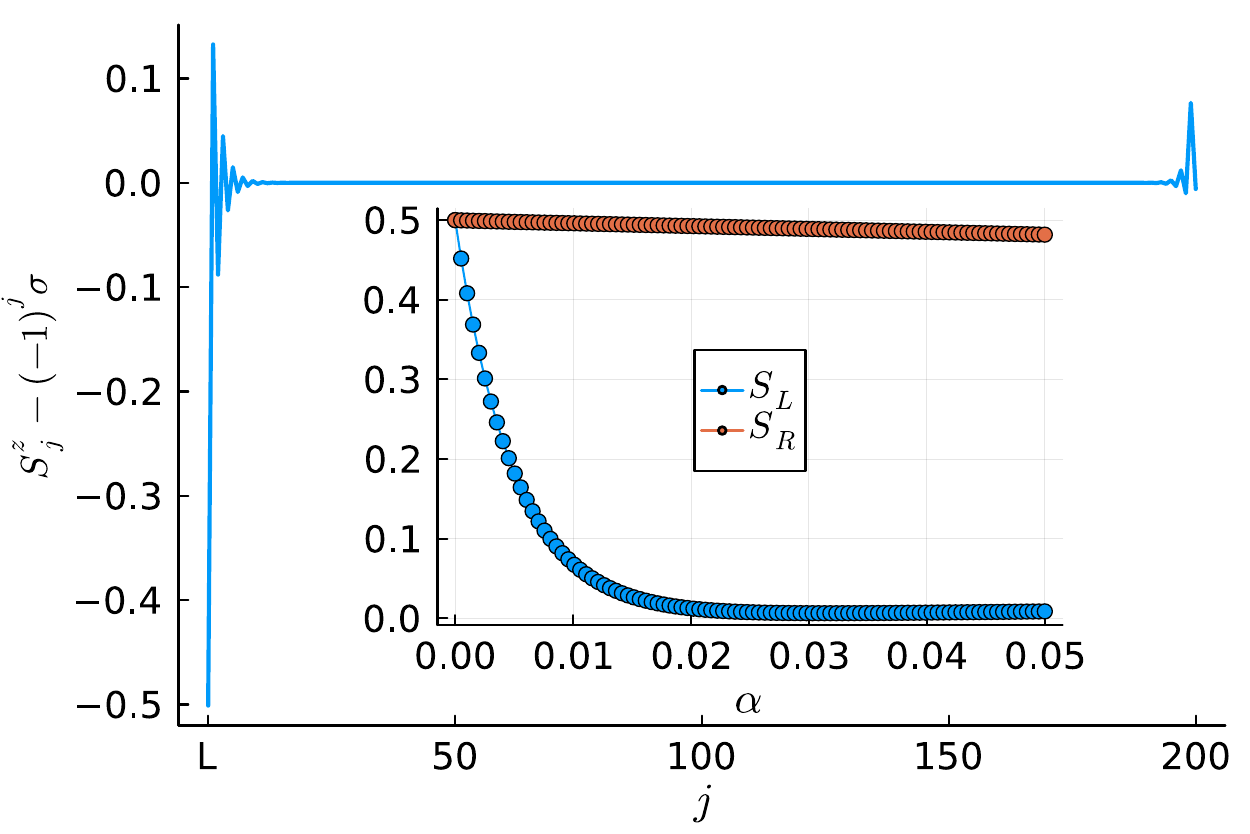}
    \end{subfigure}
    \begin{subfigure}[b]{0.46\textwidth}
        \includegraphics[width=\textwidth]{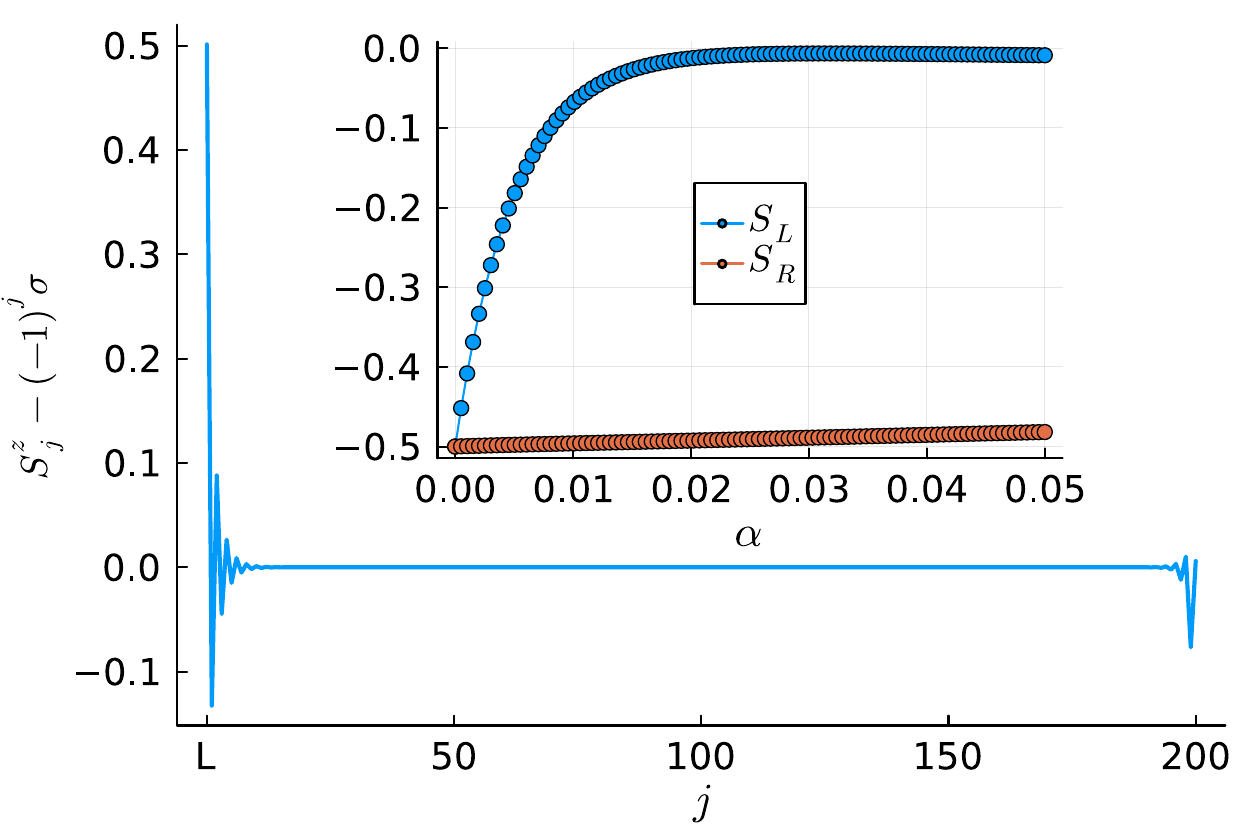}
    \end{subfigure}
    \begin{tikzpicture}[overlay, remember picture]
    \node at (-8,5.35) {b)};
    \node at (-16.5,5.35) {a)};
    \end{tikzpicture}
    \caption{The spin profile in two-fold degenerate spin-1 XXZ chain with spin-$\frac{1}{2}$ impurity coupled antiferromagnetically at its left edge. Each of the two-fold degenerate vacua has bulk staggered magnetization of $\sigma\approx 0.8576$ and edge deviations as shown in a) and b), respectively. The insets show that the edge mode at the left edge is $S_L=0$ as the edge spin deviation is exactly canceled by the half of the bulk staggered magnetization and the right edge hosts spin-$\frac{1}{2}$ spin accumulation that points respectively up and down in a) and b) as shown in the inset.}
    \label{XXZ-1-1imp-vacs}
\end{figure}

Finally, when two spin-$\frac{1}{2}$ impurities are coupled antiferromagnetically to both edges of the spin-1 XXZ chain, there are no edge modes on either edge as shown in Fig.\ref{XXZ-1-2imp-vacs}.

\begin{figure}[H]
    \centering
    \begin{subfigure}[b]{0.46\textwidth}
        \includegraphics[width=\textwidth]{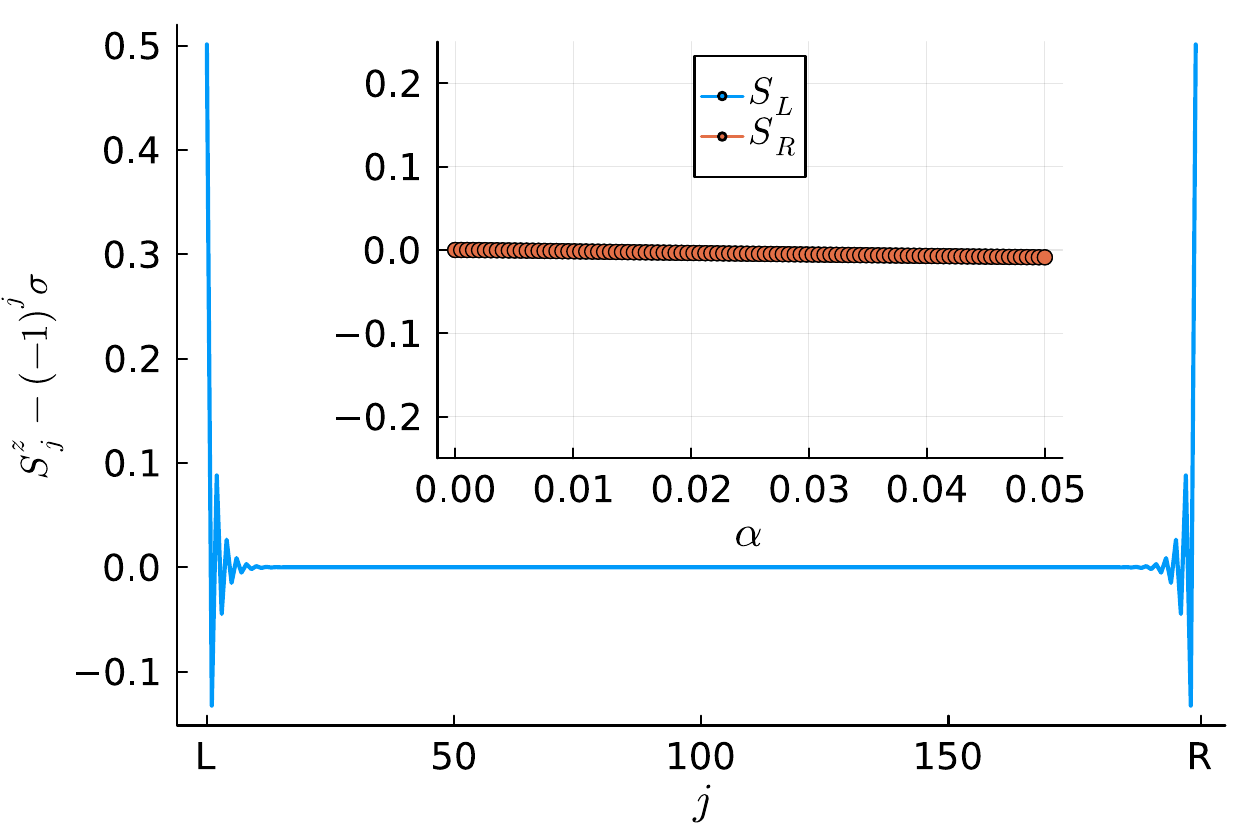}
    \end{subfigure}
    \begin{subfigure}[b]{0.46\textwidth}
        \includegraphics[width=\textwidth]{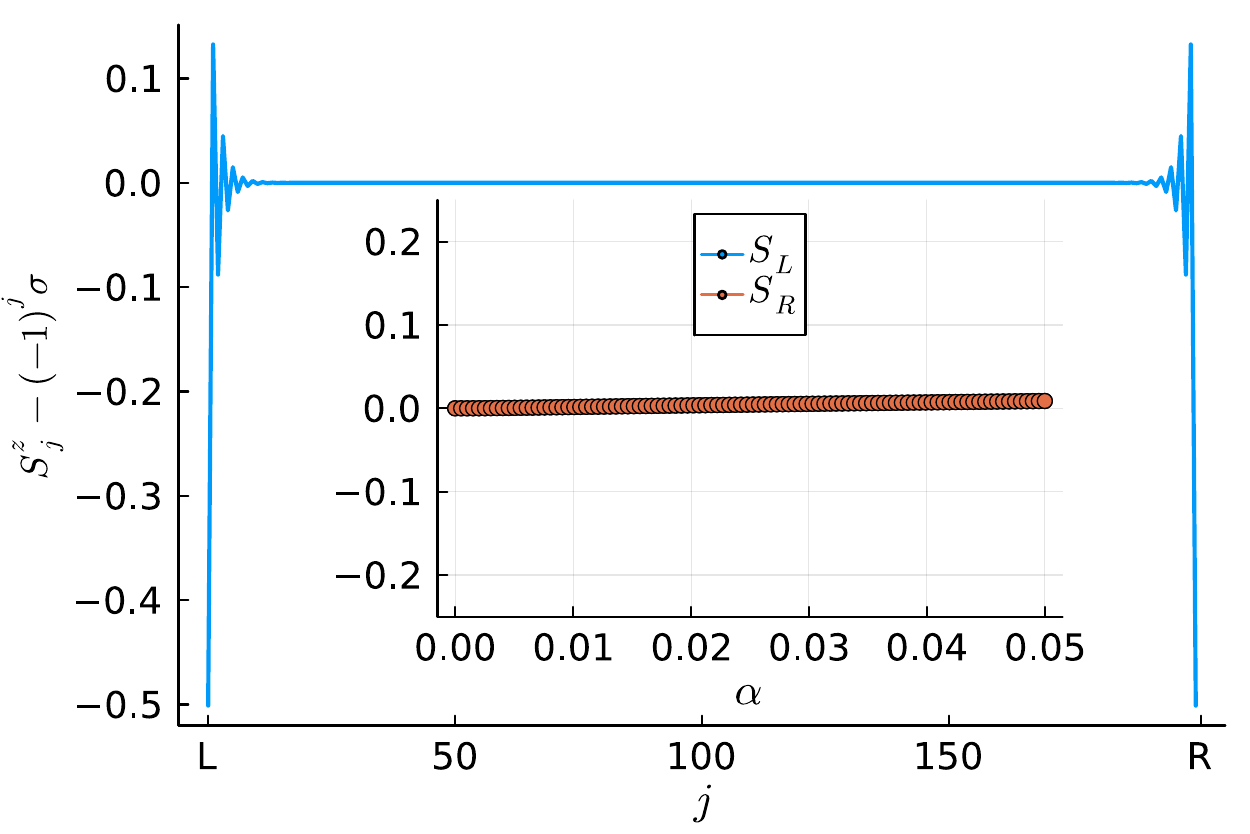}
    \end{subfigure}
    \begin{tikzpicture}[overlay, remember picture]
    \node at (-8,5.35) {b)};
    \node at (-16.5,5.35) {a)};
    \end{tikzpicture}
    \caption{The spin profile in two-fold degenerate spin-1 XXZ chain with spin-$\frac{1}{2}$ impurity coupled antiferromagnetically at its left edge. Each of the two-fold degenerate vacua has bulk staggered magnetization of $\sigma\approx 0.8576$ and edge deviations as shown in a) and b), respectively. The insets show that the edge mode at both the left and right edges are vanishing \textit{i.e.} $S_L=0=S_R$ as the edge spin deviations are exactly canceled by half of the bulk staggered magnetization.}
    \label{XXZ-1-2imp-vacs}
\end{figure}
\end{appendix}
\end{widetext}

\end{document}